\documentstyle[12pt,eqsection,indent,subeqnarray,epsf]{article}

\begin{document}

\date{October 14, 1996}

\font\sevenrm  = cmr7
%%%  
\def\fancyplus{\hbox{+\kern-6.65pt\lower3.2pt\hbox{\sevenrm --}%
\kern-4pt\raise4.6pt\hbox{\sevenrm --}%
\kern-9pt\raise0.65pt\hbox{$\tiny\vdash$}%
\kern-2pt\raise0.65pt\hbox{$\tiny\dashv$}}}
%%%  
\def\fancycross{\hbox{$\times$\kern-9.8pt\raise3.6pt\hbox{$\tiny \times$}%
\kern-1.2pt\raise3.6pt\hbox{$\tiny \times$}%
\kern-10.5pt\lower1.0pt\hbox{$\tiny \times$}%
\kern-1.2pt\lower1.0pt\hbox{$\tiny \times$}}}
%%%  
\def\fancysquare{%
\hbox{$\small\Box$\kern-9.6pt\raise7.1pt\hbox{$\vpt\backslash$}%
\kern+3.9pt\raise7.1pt\hbox{$\vpt /$}%
\kern-11.5pt\lower3.0pt\hbox{$\vpt /$}%
\kern+3.9pt\lower3.0pt\hbox{$\vpt\backslash$}}}
%%%  
\def\fancydiamond{\hbox{$\diamond$\kern-4.25pt\lower3.8pt\hbox{$\vpt\vert$}%
\kern-2.25pt\raise7pt\hbox{$\vpt\vert$}%
\kern-7.05pt\raise1.57pt\hbox{\vpt --}%
\kern+5.25pt\raise1.57pt\hbox{\vpt --}}}

%%%  % some definitions from the huge paper
\def\scrmtens{{\stackrel{\leftrightarrow}{\cal M}_T}}
\def\ttens{{\stackrel{\leftrightarrow}{T}}}
\def\scro{{\cal O}}
\def\smfrac#1#2{{\textstyle\frac{#1}{#2}}}
\def\smhalf{ {\smfrac{1}{2}} }

\def\btau{\mbox{\protect\boldmath $\tau$}}

\def\bn{{\bf n}}
\def\br{{\bf r}}
\def\bz{{\bf z}}
\def\bv{{\bf v}}
\def\bh{\mbox{\protect\boldmath $h$}}

\newcommand{\imag}{\mathop{\rm Im}\nolimits}
\newcommand{\real}{\mathop{\rm Re}\nolimits}

%%%  % "arrow" vectors
\def\vs{\vec{\mbox{$s$}}}
\def\vsigma{\vec{\mbox{$\sigma$}}}
\def\va{\vec{\mbox{$a$}}}
\def\vb{\vec{\mbox{$b$}}}
\def\hatp{\hat p}

\newcommand{\taum}{\tau_{int,\,\vec{\cal M}}}
\newcommand{\taux}{\tau_{int,\,{\cal M}_V^2}}
\newcommand{\tauA}{\tau_{int,\,A}}
\newcommand{\taue}{\tau_{int,\,{\cal E}}}
\newcommand{\taueF}{\tau_{int,\,{\cal E}_F}}
\newcommand{\taueA}{\tau_{int,\,{\cal E}_A}}
\newcommand{\taudele}{\tau_{int,\,({\cal E}-\overline{E})^2}}

\newcommand{\tauxexp}{\tau_{exp,{\cal M}_V^2}}

\newcommand{\plotdot}{\makebox(0,0){$\bullet$}}
\newcommand{\plotcross}{\makebox(0,0){{\Large $\times$}}}

\newcommand{\plota}{\makebox(0,0){$\circ$}}      % 512^2 lattice
\newcommand{\plotb}{\makebox(0,0){$\star$}}      % 256^2 lattice
\newcommand{\plotc}{\makebox(0,0){$\bullet$}}    % 128^2 lattice
\newcommand{\plotd}{\makebox(0,0){{\scriptsize $+$}}}       % 64^2  lattice
\newcommand{\plote}{\makebox(0,0){{\scriptsize $\times$}}}  % 32^2  lattice
\newcommand{\plotf}{\makebox(0,0){$\ast$}}       % 16^2  lattice

\newcommand{\plotA}{\makebox(0,0){$\triangleleft$}}   % 512^2 lattice
\newcommand{\plotB}{\makebox(0,0){$\triangleright$}}  % 256^2 lattice
\newcommand{\plotC}{\makebox(0,0){$\diamond$}}        % 128^2 lattice
\newcommand{\plotD}{\makebox(0,0){{\scriptsize $\oplus$}}} % 64^2  lattice
\newcommand{\plotE}{\makebox(0,0){{\scriptsize $\otimes$}}}% 32^2  lattice
\newcommand{\plotF}{\makebox(0,0){{\scriptsize $\ominus$}}}% 16^2  lattice

\def\reff#1{(\ref{#1})}
\newcommand{\csd}{critical slowing-down}
\newcommand{\be}{\begin{equation}}
\newcommand{\ee}{\end{equation}}
\newcommand{\<}{\langle}
\renewcommand{\>}{\rangle}
\newcommand{\half}{ {{1 \over 2 }}}
\newcommand{\quarter}{ {{1 \over 4 }}}
\newcommand{\fourth}{\quarter}
\newcommand{\eighth}{ {{1 \over 8 }}}
\newcommand{\sixteenth}{ {{1 \over 16 }}}
\def\var{ \hbox{var} }
\newcommand{\HB}{ {\hbox{{\scriptsize\em HB}\/}} }
\newcommand{\MGMC}{ {\hbox{{\scriptsize\em MGMC}\/}} }
\newcommand{\gtilde}{ {\widetilde{G}} }
\newcommand{\longto}{\longrightarrow}

%%%  \ltapprox and \gtapprox produce > and < signs with twiddle underneath
\def\spose#1{\hbox to 0pt{#1\hss}}
\def\ltapprox{\mathrel{\spose{\lower 3pt\hbox{$\mathchar"218$}}
 \raise 2.0pt\hbox{$\mathchar"13C$}}}
\def\gtapprox{\mathrel{\spose{\lower 3pt\hbox{$\mathchar"218$}}
 \raise 2.0pt\hbox{$\mathchar"13E$}}}

\newcommand{\scra}{{\cal A}}
\newcommand{\scrb}{{\cal B}}
\newcommand{\scrc}{{\cal C}}
\newcommand{\scrd}{{\cal D}}
\newcommand{\scre}{{\cal E}}
\newcommand{\scrf}{{\cal F}}
\newcommand{\scrg}{{\cal G}}
\newcommand{\scrh}{{\cal H}}
\newcommand{\scrk}{{\cal K}}
\newcommand{\scrl}{{\cal L}}
\newcommand{\scrm}{{\cal M}}
\newcommand{\scrmvec}{\vec{\cal M}}
\newcommand{\scrn}{{\cal N}}
\newcommand{\scrp}{{\cal P}}
\newcommand{\scrr}{{\cal R}}
\newcommand{\scrs}{{\cal S}}
\newcommand{\scru}{{\cal U}}

\def\bsigma{\mbox{\protect\boldmath $\sigma$}}
\def\bpi{\mbox{\protect\boldmath $\pi$}}
\def\bs{ {\bf s} }
\newcommand{\re}{\mathop{\rm Re}\nolimits}
\newcommand{\im}{\mathop{\rm Im}\nolimits}
\newcommand{\tr}{\mathop{\rm tr}\nolimits}
\newcommand{\CP}{ \hbox{\it CP\/} }
\def\T{{\rm T}}

\def\msbar{ {\overline{\hbox{\scriptsize MS}}} }
\def\normalmsbar{ {\overline{\hbox{\normalsize MS}}} }

\def\eff{ {\hbox{\scriptsize\em eff}} }
\def\B{ {\hbox{\scriptsize BNNW}} }

\def\hboxsans#1{ {\hbox{\scriptsize\sf #1}} } % small type for exponents
\def\hboxscript#1{ {\hbox{\scriptsize\it #1}} }

\font\specialroman=msym10 scaled\magstep1  % 12-point Special Roman (caps only)
\font\sevenspecialroman=msym7              % 7-point Special Roman (caps only)
\def\zed{\hbox{\specialroman Z}}
\def\szed{\hbox{\sevenspecialroman Z}}
\def\R{\hbox{\specialroman R}}
\def\sR{\hbox{\sevenspecialroman R}}
\def\Z{\hbox{\specialroman Z}}
\def\N{\hbox{\specialroman N}}
\def\C{\hbox{\specialroman C}}
\renewcommand{\emptyset}{\hbox{\specialroman ?}}
%%%  \newcommand{\zed}{{\bf \rm Z}}
%%%  \newcommand{\R}{\hbox{{\rm I}\kern-.2em\hbox{\rm R}}}
%%%  \font\srm=cmr7                 % to get seven roman
%%%  \def\szed{\hbox{\srm Z\kern-.45em\hbox{\srm Z}}}
%%%  \def\sR{\hbox{{\srm I}\kern-.2em\hbox{\srm R}}}
%%%  \def\C{{\bf C}}

\font\german=eufm10 scaled\magstep1     % 12-point Euler Fraktur (German)
\def\germang{\hbox{\german g}}
\def\germansu{\hbox{\german su}}
\def\germanso{\hbox{\german so}}

%%%  
%%%   Array for subscripts
%%%  
\newenvironment{sarray}{
          \textfont0=\scriptfont0
          \scriptfont0=\scriptscriptfont0
          \textfont1=\scriptfont1
          \scriptfont1=\scriptscriptfont1
          \textfont2=\scriptfont2
          \scriptfont2=\scriptscriptfont2
          \textfont3=\scriptfont3
          \scriptfont3=\scriptscriptfont3
        \renewcommand{\arraystretch}{0.7}
        \begin{array}{l}}{\end{array}}

\newenvironment{scarray}{
          \textfont0=\scriptfont0
          \scriptfont0=\scriptscriptfont0
          \textfont1=\scriptfont1
          \scriptfont1=\scriptscriptfont1
          \textfont2=\scriptfont2
          \scriptfont2=\scriptscriptfont2
          \textfont3=\scriptfont3
          \scriptfont3=\scriptscriptfont3
        \renewcommand{\arraystretch}{0.7}
        \begin{array}{c}}{\end{array}}

\title{Multi-Grid Monte Carlo via $XY$ Embedding  \\[4mm]
         II. Two-Dimensional $SU(3)$ Principal Chiral Model
      }
\author{
  \\[-0.5cm]
  {\small Gustavo Mana\thanks{Current e-mail: {\tt GUSTAVO@IMAGINE-SW.COM}} }
                                          \\[-0.2cm]
  {\small\it Department of Physics}       \\[-0.2cm]
  {\small\it New York University}         \\[-0.2cm]
  {\small\it 4 Washington Place}          \\[-0.2cm]
  {\small\it New York, NY 10003 USA}      \\[-0.2cm]
  {\small {\tt MANA@MAFALDA.PHYSICS.NYU.EDU}}        \\[-0.2cm]
  {\protect\makebox[5in]{\quad}}  % To force authors' names to be written
                                  %   vertically, one above another.
                                  % (\author seems to put them side-by-side
                                  %   if there is room.)
  \\[-0.35cm] \and
  {\small Andrea Pelissetto }        \\[-0.2cm]
  {\small\it Dipartimento di Fisica}        \\[-0.2cm]
  {\small\it Universit\`a degli Studi di Pisa}        \\[-0.2cm]
  {\small\it Pisa 56100, ITALIA}        \\[-0.2cm]
  {\small {\tt PELISSET@IBMTH1.DIFI.UNIPI.IT}}  \\[-0.2cm]
  {\protect\makebox[5in]{\quad}}  % To force authors' names to be written
                                  %   vertically, one above another.
                                  % (\author seems to put them side-by-side
                                  %   if there is room.)
   \\[-0.35cm] \and
  \small  Alan D. Sokal               \\[-0.2cm]
  \small\it Department of Physics     \\[-0.2cm]
  \small\it New York University       \\[-0.2cm]
  \small\it 4 Washington Place        \\[-0.2cm]
  \small\it New York, NY 10003 USA    \\[-0.2cm]
  \small {\tt SOKAL@NYU.EDU} \\[-0.2cm]
  {\protect\makebox[5in]{\quad}}  % To force authors' names to be written
                                  %   vertically, one above another.
                                  % (\author seems to put them side-by-side
%   if there is room.)
   \\
}

\vspace{0.2cm}
\maketitle
\thispagestyle{empty}   % Suppress page number on front page.

\vspace{-0.4cm}

\begin{abstract}

We carry out a high-precision simulation of the
two-dimensional $SU(3)$ principal chiral model
at correlation lengths $\xi$ up to $\sim\! 4 \times 10^5$,
using a multi-grid Monte Carlo (MGMC) algorithm
and approximately one year of Cray C-90 CPU time.
We extrapolate
the finite-volume Monte Carlo data to infinite volume
using finite-size-scaling theory,
and we discuss carefully the systematic and statistical errors
in this extrapolation.
We then compare the extrapolated data to the renormalization-group predictions.
The deviation from asymptotic scaling,
which is $\approx 12\%$ at $\xi \sim 25$,
decreases to $\approx 2\%$  at 
$\xi \sim 4 \times 10^5$.
We also analyze the dynamic critical behavior of the MGMC 
algorithm using lattices up to $256 \times 256$,
finding the dynamic critical exponent
$z_{int,{\cal M}^2} \approx 0.45 \pm 0.02$
(subjective 68\% confidence interval).
Thus, for this asymptotically free model,
critical slowing-down is greatly reduced compared to local algorithms,
but not completely eliminated.

\end{abstract}

\clearpage

%%%% \doublespace

\section{Introduction} \label{sec1}

This paper has two distinct objectives:
first, to study the dynamic critical behavior of the 
multi-grid Monte Carlo (MGMC)
algorithm for the two-dimensional $SU(3)$ principal chiral model;
and second, to apply this algorithm to obtain a high-precision 
test of asymptotic scaling for this model. We discuss these two
objectives in separate subsections.

\subsection{Multi-Grid Monte Carlo} \label{sec1.1}

By now it is widely recognized
\cite{Sokal_Lausanne,Adler_LAT88,Wolff_LAT89,Sokal_LAT90}
that better simulation algorithms,
with strongly reduced critical slowing-down,
are needed for high-precision Monte Carlo studies of
statistical-mechanical systems near critical points
and of quantum field theories (such as QCD) near the continuum limit.
One promising class of such algorithms is
{\em multi-grid Monte Carlo}\/ (MGMC)
\cite{MGMC_0,MGMC_1,MGMC_2,MGMC_O4,mgmco4d1,Mendes_LAT95,MGMC_ON,%
SU3_letter,Meyer-Ortmanns_85,Mack_Cargese,Mack-Meyer_90,%
Hasenbusch_LAT90,Hasenbusch-Meyer_PRL,Hasenbusch_CP3,Irving_92,%
Hulsebos_91,Laursen_91,Laursen_93,Grabenstein_92,Grabenstein_93a,%
Grabenstein_93b,Grabenstein_thesis,Grabenstein_94,Janke_multicanonical_MGMC}:
this is a collective-mode approach that introduces block updates
(of fixed shape but variable amplitude) on all length scales.
The basic ingredients of the method are\footnote{
   See \cite{MGMC_1} for details.
}:
 
1) {\em Interpolation operator:}\/
This is a rule specifying the shape of the block update.
The interpolations most commonly used are
{\em piecewise-constant}\/ (square-wave updates)
and {\em piecewise-linear}\/ (pyramidal-wave updates).
 
2) {\em Cycle control parameter $\gamma$:}\/
This is an integer number that determines the way in which the
different block sizes are visited.
In general, blocks of linear size $2^l$
are updated $\gamma^l$ times per iteration.
Thus, in the W-cycle ($\gamma=2$) more emphasis is placed
on large length scales than in the V-cycle ($\gamma=1$).

3) {\em Basic (smoothing) iteration:}\/
This is the local Monte Carlo update that is performed on each level.
Typically one chooses to use {\em heat-bath}\/ updating if the distribution
can be sampled in some simple way, and {\em Metropolis}\/ updating otherwise.
 
4) {\em Implementation:}\/
The computations can be implemented either in the
{\em recursive multi-grid}\/ style
using explicit coarse-grid fields
\cite{Hackbusch_85,Briggs_87,MGMC_0,MGMC_1,MGMC_2,MGMC_O4,mgmco4d1,%
Mendes_LAT95,MGMC_ON},
or in the {\em unigrid}\/ style using block updates acting directly
on the fine-grid fields
\cite{McCormick-Ruge,Mack_Cargese,Mack-Meyer_90,%
Hasenbusch_LAT90,Hasenbusch-Meyer_PRL,Hasenbusch_CP3}.
We use here the recursive multi-grid approach, in which
the computational labor per iteration
for a $d$-dimensional system of linear size $L$ is
\be
  \hbox{Work(MG)} \;\sim\;
  \cases{ L^d                & for $\gamma < 2^d$  \cr
          \noalign{\vskip 1mm}
          L^d \, \log L      & for $\gamma = 2^d$  \cr 
          \noalign{\vskip 1mm}
          L^{\log_2 \gamma}  & for $\gamma > 2^d$  \cr 
        }
\label{workMG_0}
\ee
%%%%  for the recursive multi-grid approach, and
%%%%  \be
%%%%    \hbox{Work(UG)} \;\sim\;
%%%%    \cases{ L^d \, \log L               & for $\gamma=1$ \cr
%%%%            \noalign{\vskip 1mm}
%%%%            L^{d \,+\, \log_2 \gamma}   & for $\gamma>1$ \cr
%%%%          }
%%%%  \label{workUG}
%%%%  \ee
%%%%  for the unigrid approach.
%%%%  Thus, the unigrid implementation is marginally more expensive
%%%%  for a V-cycle, but prohibitively more expensive for a W-cycle.

The efficiency of the MGMC method can be analyzed rigorously in the case
of the Gaussian (free-field) model, for which it can be proven
\cite{MGMC_0,MGMC_1,Goodman-Sokal_unpub}
that critical slowing-down is completely eliminated.\footnote{
   This holds for $\gamma \ge 2$ (i.e.\ W-cycle or higher) in the case of
   piecewise-constant interpolation, and for $\gamma \ge 1$ in the case of
   piecewise-linear interpolation.
}
That is, the {\em autocorrelation time}\/ $\tau$ is bounded as
the correlation length $\xi$ and the lattice size $L$ tend to infinity,
so that the {\em dynamic critical exponent}\/ $z$ is zero.\footnote{
   See \cite{Sokal_LAT90} for a pedagogical discussion of
   the various autocorrelation times and their associated
   dynamic critical exponents.
}
%%%%  More precisely, the algorithm with piecewise-linear interpolation
%%%%  exhibits $z = 0$ for both the V-cycle and the W-cycle, while the algorithm with
%%%%  piecewise-constant interpolation has $z = 0$ only for the W-cycle
%%%%  (the piecewise-constant V-cycle has $z = 1$).

One is therefore motivated to apply MGMC to ``nearly Gaussian''
systems, such as asymptotically free nonlinear $\sigma$-models;
one might hope that critical slowing-down would likewise
be completely eliminated (possibly modulo a logarithm)
or at least greatly reduced compared to the
$z \approx 2$ of local algorithms.
However, previous numerical study of MGMC in the
two-dimensional $N$-vector models
with $N = 3,4,8$ \cite{MGMC_O4,MGMC_ON} has shown, to our initial surprise,
that the dynamic critical exponent is {\em not}\/ zero.
Nevertheless, it is quite small ($z \approx 0.50$--0.70),
so these algorithms work reasonably well.
In view of these results for the $N$-vector models,
we want to investigate the performance of MGMC in 
other asymptotically free $\sigma$-models,
such as the two-dimensional $SU(N)$ principal chiral models.

Of course, for two-dimensional $N$-vector models,
Wolff's cluster algorithm \cite{Wolff_89a}
apparently succeeds in {\em eliminating}\/ the critical slowing-down 
\cite{Wolff_89a,Edwards_89,Wolff_90,CEPS_swwo4c2},
so there is no point in using MGMC in this case.
But there are strong reasons to believe \cite{CEPS_swwo4c2}
that Wolff-type embedding algorithms will {\em not}\/ achieve $z \ll 2$
for other $\sigma$-models, except perhaps the $RP^{N-1}$ models.
In particular, for $\sigma$-models taking values in the group $SU(N)$
with $N\ge 3$,
MGMC is the {\em only}\/ known collective-mode algorithm 
(except perhaps Fourier acceleration) that has a chance of achieving 
$z \ll 2$.

A major drawback of our group's standard MGMC
\cite{MGMC_0,MGMC_1,MGMC_2,MGMC_O4,mgmco4d1}
is that its implementation is
cumbersome and model-dependent, in the sense that the program
(and in particular the heat-bath subroutine)
has to be drastically rewritten for each distinct model.
With this problem in mind, we have recently developed
\cite{Mendes_LAT95,MGMC_ON}
a new implementation of MGMC that
can be used conveniently for a large class of $\sigma$-models
with very little modification of the program.\footnote{
   We devised this approach after extensive discussions with
   Martin Hasenbusch and Steffen Meyer at the Lattice '92 conference
   in Amsterdam.
   In particular, the idea of $XY$ embedding is made explicit in their work:
   see equations (5)/(6) in \cite{Hasenbusch-Meyer_PRL}
   and equations (5)--(9) in \cite{Hasenbusch_CP3}.
}
The idea is to {\em embed}\/ angular variables $\{ \theta_x \}$
into the given $\sigma$-model, and then update the resulting
induced $XY$ model by our standard (piecewise-constant, W-cycle,
heat-bath, recursive) MGMC method.

Consider, therefore, the $SU(N)$ principal chiral model:
the original variables $U_x$ of this model are $SU(N)$ matrices
living on the lattice sites $x$, and the original Hamiltonian is
\be
  {\cal H} \;=\;  -\beta  \sum_{\<xx'\>}
    \re \tr (U_x^{\dagger} \, U_{x'})
  \;.
\label{ham:SU(N)}
\ee
The global symmetry group is
$SU(N)_{\hboxscript{left}} \times SU(N)_{\hboxscript{right}}$.
The idea behind $XY$ embedding
is to choose randomly a $U(1)$ subgroup
$H \subset SU(N)_{\hboxscript{left}} \times SU(N)_{\hboxscript{right}}$,
and to apply a ``rotation'' $\theta_x$ in this subgroup
to the original spin variable $U_x \in SU(N)$.
Thus, the angular variables $\theta_x$ are {\em updates}\/
to the original variables $U_x$. Here we choose to exploit
only the left-multiplication subgroup.\footnote{ Actually,
   our program uses the left-multiplication at the odd-numbered
   iterations and the right-multiplication at the even-numbered
   iterations.
}
More precisely, we define the updated variable $U_x^{new}$ by
\be
  U_x^{new} \;=\; e^{i\theta_x RTR^{-1}} U_x^{old}
                  \;=\; R e^{i\theta_x T} R^{-1} U_x^{old}
  \;,
\label{update}
\ee
where $R$ is a random element of $SU(N)$,
and $T$ is a fixed nonzero element (to be specified later)
of the Lie algebra ${\germansu(N)}$ (i.e.\ a traceless Hermitian matrix).
The embedded $XY$ model consisting of the spins $\{ \theta_x \}$
is then simulated using the induced Hamiltonian
\be
   \scrh_{embed}( \{ \theta_x \} )   \;=\;
   \scrh( \{ U_x^{new} \} )
   \;,
 \label{ham_induced}
\ee
with initial condition $\theta_x = 0$
(i.e.\ $U_x^{new} = U_x^{old}$) for all $x$.
At each iteration of the algorithm, a new random matrix $R$ is chosen.

In general the induced $XY$ Hamiltonian \reff{ham_induced}
can be extremely complicated (and thus impractical to simulate by
true recursive MGMC).  However, if the original Hamiltonian $\scrh$
is sufficiently ``nice'' {\em and}\/ one makes a clever choice of the
generator $T$, then in some cases the induced $XY$ Hamiltonian can be
reasonably simple.  In particular, if we choose $T$ to have
all its eigenvalues in the set $\{-1,0,1\}$, it follows that
\be
  e^{i\theta T} \;=\; T^2\, \cos \theta \,+\, iT\,\sin \theta
    \,+\, (I \,-\, T^2)
  \;,
 \label{T_eigenvalues}
\ee
where $I$ is the identity matrix. Then the induced Hamiltonian is
of the simple form
\be
{\cal H}_{embed} \;=\;  - \sum_{\<xx'\>}
\left[ \alpha_{xx'}\,\cos(\theta_x \,-\, \theta_{x'}) \,+\,
   \beta_{xx'}\,\sin(\theta_x \,-\, \theta_{x'}) \right] \,+\, {\rm const}
   \;,
\label{ham_embed0}
\ee
where the induced couplings $\{\alpha_{xx'},\beta_{xx'}\}$
depend on the current configuration $\{ U_x^{old} \}$
of the original model:
\begin{subeqnarray}
\alpha_{xx'} &=& \beta\, \re\tr ( U_x^{old\dagger} R T^2 R^{-1} U^{old}_{x'})   \\
\beta_{xx'} &=&  \beta\, \re\tr ( U_x^{old\dagger} R (-iT) R^{-1} U^{old}_{x'})
           \;=\; \beta\, \im\tr ( U_x^{old\dagger} R T R^{-1} U^{old}_{x'})
\label{suncouplings}
\end{subeqnarray}
Such a ``generalized $XY$ Hamiltonian'' is easily simulated by MGMC;
indeed, the coarse-grid Hamiltonians in $XY$-model MGMC are inevitably
of the form \reff{ham_embed0},
even when the fine-grid Hamiltonian is the standard
$XY$ model $\alpha_{xx'} \equiv \alpha \ge 0$, $\beta_{xx'} \equiv 0$
\cite{MGMC_1,MGMC_2}.
So one may just as easily start from \reff{ham_embed0}
already on the finest grid.

Clearly $T$ must have $k$ eigenvalues $+1$,
$k$ eigenvalues $-1$, and $N-2k$ eigenvalues 0,
where $1 \le k \le \lfloor N/2 \rfloor$.
Here we shall choose $k=1$; without loss of generality we can take
\be
T  \;=\;    \left( \begin{array}{cccc}
           1 & 0 & 0 & \\
           0 & -1 & 0 & \cdots \\
           0 & 0 & 0 & \\
            & \vdots & & \ddots
    \end{array} \right)
\;.
\label{TSUN}
\ee
With the explicit choice \reff{TSUN} for $T$, 
the couplings are
\begin{subeqnarray}
\alpha_{xx'} &=& \beta\, 
    \left[ \re (R^{-1}\,U^{old}_{x'}\,U_x^{old\dagger}\,R)_{11} \,+\,
           \re (R^{-1}\,U^{old}_{x'}\,U_x^{old\dagger}\,R)_{22} \right]   \\
\beta_{xx'} &=& \beta\, 
    \left[ \im (R^{-1}\,U^{old}_{x'}\,U_x^{old\dagger}\,R)_{11} \,-\,
           \im (R^{-1}\,U^{old}_{x'}\,U_x^{old\dagger}\,R)_{22} \right] \;.   
  \label{SUN_couplings_explicit}
\end{subeqnarray}
Let us remark that the Hamiltonian
\reff{ham_embed0}/\reff{suncouplings} is not only non-ferromagnetic,
but is in fact typically {\em frustrated}\/ \cite{MGMC_ON}.\footnote{
    We call the Hamiltonian \reff{ham_embed0} {\em ferromagnetic}\/ if
    $\alpha_{xx'} \ge 0$ and $\beta_{xx'} = 0$ for all bonds $\< xx' \>$.
    We call it {\em unfrustrated}\/ if there exists a configuration
    $\{\theta_x\}$ that simultaneously minimizes the bond energy
    $- [\alpha_{xx'} \cos(\theta_x - \theta_{x'}) +
       \beta_{xx'} \sin(\theta_x - \theta_{x'})]$
    on all bonds $\< xx' \>$.
}
However, this frustration is weak when $\beta \gg 1$.
%%  {\bf Do we want to discuss this further in this paper???
%%     E.g.\ give perturbation calculation of frustration???}

\subsection{Asymptotic Scaling} \label{sec1.2}

A key tenet of modern elementary-particle physics is the asymptotic freedom
of four-dimensional nonabelian gauge theories \cite{Creutz_83,Rothe_92}.
However, the nonperturbative validity of asymptotic freedom has been
questioned \cite{Patrascioiu_85,Pat-Seil_91,Pat-Seil_LAT92,Pat-Seil_95ab};
and numerical studies of lattice gauge theory
have thus far failed to detect asymptotic scaling in the bare coupling
\cite{UKQCD_93,Ukawa_LAT92,Bali_93,Bitar_96}.
It is therefore useful to explore asymptotic scaling in a model
easier to simulate numerically than four-dimensional gauge theories, 
but still theoretically interesting.  A good candidate is the
two-dimensional $SU(N)$ principal chiral model \reff{ham:SU(N)},
which possesses the property of perturbative asymptotic freedom
\cite{Polyakov_75,McKane_80,Shigemitsu_81}
along with other interesting characteristics.\footnote{
   The $SU(N)$ chiral model has a $1/N$ expansion in terms
   of planar graphs, similar to that of the $SU(N)$ gauge theories
   \cite{tHooft_74,Rossi_96_review}.
   The $SU(N)$ chiral model also 
   has lattice Schwinger-Dyson equations and a
   high-temperature character expansion that are similar to those
   of the $SU(N)$ lattice gauge theories
   \cite{Rossi_96_review,Drouffe_83}.
   Finally, the Migdal-Kadanoff approximate renormalization group
   predicts the same recursion equations for
   the two-dimensional $SU(N)$ spin models as for the four-dimensional
   $SU(N)$ gauge theories \cite{Migdal,Kadanoff}.
   %%  The {\em four}\/-dimensional $SU(3)$ chiral model has been employed
   %%  as an effective low-energy theory of light mesons \cite{migdal_rg?????}.
   %%  {\bf Let's take a look at the references on that!!!}
}

Let us recall the logic underlying the conventional wisdom on
asymptotic freedom:
Renormalization-group (RG) calculations in
weak-coupling (large-$\beta$) perturbation theory
show that for two-dimensional $\sigma$-models
taking values in a {\em curved}\/ compact Riemannian manifold $M$,
the RG flow at large $\beta \equiv 1/g^2$ is toward {\em smaller}\/ $\beta$
\cite{Polyakov_75,McKane_80,Shigemitsu_81,%
Brezin_76,Bardeen_76,Kogut_79,Friedan_85}.
It is therefore natural to {\em conjecture}\/ that
this flow continues to the $\beta=0$ fixed point,
without encountering any other fixed point(s).
If this is indeed the case, then it follows that the theory
has exponential decay of correlations for all $\beta < \infty$;
and the RG then gives precise predictions for the scaling behavior of
the correlation length $\xi$ and the susceptibility $\chi$
as $\beta \to \infty$.
Moreover, for certain $\sigma$-models it is possible to calculate,
modulo some plausible hypotheses, the nonperturbative coefficient
in the asymptotic formula for the correlation length
\cite{Hasenfratz-Niedermayer_1,Hasenfratz-Niedermayer_2,%
Hasenfratz-Niedermayer_3,Balog_92,Hollowood_94,Evans_95}.
It should be emphasized, however, that all these results depend on a
conjecture which transcends perturbation theory and which has thus far
been neither proven nor disproven.
This is why we want to test the nonperturbative validity of
asymptotic freedom, using numerical simulations.

Let us clarify our use of the words ``scaling'' and ``asymptotic scaling''.
Consider a sequence $\{\scrh_n\} _{n=1}^\infty$ of lattice theories with
correlation lengths $\xi_n$ tending to infinity.
We say that this sequence exhibits {\em scaling}\/ if,
after rescaling lengths by $\xi_n$ and rescaling the spins
by appropriate values $\zeta_n$,
all the correlation functions $\< \; \cdots \; \> _{\scrh_n}$
converge to some continuum-limit values.
Equivalently, the sequence exhibits scaling
if all dimensionless ratios of long-distance observables tend to
constants.
More loosely, we say that a
{\em finite}\/ sequence of theories $\{\scrh_n\}_{n=1}^N$
exhibits scaling to within some given degree of accuracy if
all dimensionless ratios of long-distance observables are constant
within the given degree of accuracy.
(This latter notion is often used in Monte Carlo work, expressed by
 some phrase like
 ``we are in the scaling region'' or ``we are near the continuum limit''.)
Note that the parameters in $\scrh_n$ (such as $\beta$)
{\em play no role}\/ in the concept of scaling.

Now consider a sequence $\{\scrh_n\} _{n=1}^\infty$ of lattice theories with
correlation lengths $\xi_n$ tending to infinity, for which there exists
a {\em theoretical}\/ prediction for the asymptotic behavior of
long-distance observables {\em as a function of the parameters in $\scrh_n$}\/
(or as a function of short-distance observables like the energy).
[The example of interest is of course an asymptotically free theory
 in the limit $\beta \to \infty$, where the renormalization group predicts
 $\scro(\beta) = C e^{a\beta} \beta^b (1 + a_1/\beta + a_2/\beta^2 + \ldots)$
 for each long-distance observable $\scro$,
 with $a,b,a_1,a_2,\ldots$ computable in perturbation theory
 but $C$ usually unknown.]
We say that the given sequence exhibits {\em asymptotic scaling}\/ if
the theoretical predictions for the leading-order asymptotic behavior
are valid.
[In the asymptotically-free case this means that
 $\scro(\beta_n)/(e^{a\beta_n} \beta_n^b)$ tends to a constant
 as $n\to\infty$.]
More loosely, we say that a
{\em finite}\/ sequence of theories $\{\scrh_n\}_{n=1}^N$
exhibits asymptotic scaling to within some given degree of accuracy if
the theoretical predictions for the leading-order asymptotic behavior
are valid to within the given degree of accuracy.
[In the asymptotically-free case this means that
 $\scro(\beta_n)/(e^{a\beta_n} \beta_n^b)$ is constant
 to within the given degree of accuracy.]

Clearly, asymptotic scaling implies scaling
(if the observables behave correctly as a function of $\beta$,
then their dimensionless ratios necessarily converge), but not conversely.
Note also that even if asymptotic scaling does hold along the given path
in parameter space,
it may be necessary to go to much larger correlation lengths
to observe asymptotic scaling to some reasonable degree of accuracy
than to observe scaling to the same degree of accuracy.

In the renormalization-group language, deviations from scaling are caused
by irrelevant operators (so that the RG flow does not lie exactly on the
unstable manifold), while deviations from asymptotic scaling arise also from
higher-order corrections to the flow {\em on}\/ the unstable manifold.
In an asymptotically free theory, deviations from scaling are nonperturbative
effects (suppressed by powers of $\xi$ and hence exponentially small in
$\beta$),
while deviations from asymptotic scaling are perturbative effects
(a power series in $1/\beta \sim 1/\log\xi$, with coefficients that are
computable in lattice perturbation theory).
Therefore, scaling may be expected to set in at a rather modest correlation
length (e.g.\ $\xi \sim 10$ or even smaller),
because the corrections to scaling fall off like inverse powers of $\xi$.
On the other hand, asymptotic scaling is much more elusive,
because the corrections fall off like inverse powers of the {\em logarithm}\/
of $\xi$:
depending on the magnitude of the perturbative coefficients
(including unknown high-order ones), asymptotic scaling could set in
at correlation lengths as small as $\sim 10$
or could require correlation lengths as large as $\sim 10^{30}$.

Consequently it is not a surprise that
numerical studies of lattice gauge theory
have thus far failed to detect asymptotic scaling in the bare coupling.
Even in the simpler case of two-dimensional nonlinear $\sigma$-models,
numerical simulations at correlation lengths $\xi \sim 10$--100
have often shown discrepancies of order 10--50\%
from asymptotic scaling. In the $SU(3)$ chiral model, previous 
Monte Carlo studies
\cite{Dagotto_87:SU(3),Hasenbusch-Meyer_PRL,Drummond_94,Rossi_94a}
up to $\xi \approx 35$ have found that
the ratio $\xi(\beta)/[e^{a\beta} \beta^b (1+a_1/\beta)]$
is not approximately constant, nor does its value agree with the
predicted nonperturbative coefficient \cite{Balog_92};
on both points the discrepancy is of order 10--20\%.

These studies seem to show empirically that
to observe asymptotic scaling in the bare coupling
in the $SU(3)$ chiral model,
the numerical simulations will need to reach correlation lengths $\xi \gg 35$
(how large is not so clear).
Unfortunately, it is at present unfeasible to simulate lattices
of linear size $L$ bigger than $\sim 1000$;
so, if we want to do a direct ``infinite-volume'' simulation,
which requires $L/\xi \gtapprox 6$--8 to avoid significant finite-size effects,
we cannot hope to reach correlation lengths beyond about 150.
To circumvent this problem, we shall resign ourselves to
using lattices that are far from being ``infinite'',
and we shall attempt to understand the finite-size effects in such detail
that we can correct for them.
We do this by applying an extremely powerful method
\cite{fss_greedy,fss_greedy_fullpaper}
for the extrapolation of finite-size data to the infinite-volume limit,
due originally to L\"uscher, Weisz and Wolff \cite{Luscher_91}
(see also Kim \cite{Kim_93,Kim_LAT93,Kim_94a,Kim_94b,Kim_95}),
based on finite-size-scaling theory.
Using only lattices $L \le 256$,
we are able to obtain the infinite-volume correlation length
$\xi_\infty$ to an accuracy
of order 0.5\% (resp.\ 0.9\%, 1.1\%, 1.3\%, 1.5\%)
when $\xi_\infty \approx 10^2$
 (resp.\ $10^3$, $10^4$, $10^5$, $4 \times 10^5$).
We realize that this sounds crazy at first,
but we hope to convince the reader that we do in fact have
reliable control over all systematic and statistical errors
(see Section \ref{sec5} for details).\footnote{
   We have previously carried out a similar study of asymptotic scaling
   in the two-dimensional $O(3)$ $\sigma$-model
   \cite{o3_scaling_LAT94,o3_scaling_prl,lat95_4loop,o3_scaling_fullpaper}.
   See also the criticisms of this work by
   Patrascioiu and Seiler \cite{Pat-Seil_comment}
   and our reply \cite{CEPS_O3_reply_to_Pat-Seil}.
   We discuss these criticisms further in Section~\ref{sec5.4} below.
}

Finally, let us remark that other studies have used
different approaches to observe either
scaling or asymptotic scaling at smaller correlation lengths.
Thus, the various ``improved actions''
(Symanzik \cite{Symanzik_83,Luscher_86,Parisi_85,CP_Symanzik},
 Hasenfratz--Niedermayer \cite{Hasenfratz_perfect,Niedermayer_LAT96}, etc.)\ 
are aimed at reaching {\em scaling}\/ at the smallest possible
correlation length.
If they have any effect on asymptotic scaling, it is by coincidence
rather than by design.\footnote{
   A recent comparative study of the standard and Symanzik-improved
   actions for four-dimensional $SU(2)$ and $SU(3)$ lattice gauge theories
   found {\em no}\/ difference in the quality of asymptotic scaling
   between the two actions \cite{Cella_94a,Cella_94b}.
}
On the other hand, the various ``improved expansion parameters''
are aimed at reaching {\em asymptotic scaling}\/
at the smallest possible correlation length,
by redefining slightly the meaning of ``asymptotic scaling''
(using the energy as the parameter in place of $\beta$).

In the model treated here, scaling is reached (to within about one percent)
already at a correlation length of a few lattice spacings
\cite{Rossi_94a}.
Since we are able to go to much larger correlation lengths than this,
scaling is no problem at all for us;
we thus have no need of ``improved actions''.
On the other hand, asymptotic scaling is much more elusive,
and we are therefore very interested in trying out the proposed
``improved expansion parameters''.
But we have some reticence about the conceptual and theoretical basis
underlying this approach (see Section \ref{sec3.3}).

\subsection{Plan of this Paper} \label{sec1.3}

The plan of this paper is as follows:
In Section \ref{sec2} we set the notation.
In Section \ref{sec3} we summarize the perturbative predictions for the
two-dimensional $SU(N)$ principal chiral models.
In Section \ref{sec4} we present our raw data,
which are based on approximately one year of Cray C-90 CPU time.
In Section \ref{sec5} we carry out a detailed analysis of our static data,
making systematic use of the finite-size-scaling extrapolation method,
and we compare the extrapolated values with the perturbative predictions.
In Section \ref{sec6} we analyze our dynamic data using conventional 
finite-size-scaling plots to extract the dynamic critical 
exponents $z_{int, {\cal M}_F^2}$ and $z_{int, {\cal M}_A^2}$.
In Appendices \ref{appenA} and \ref{appenB}
we present some perturbative computations.

Parts of this work have appeared previously
in brief preliminary reports \cite{Mendes_LAT95,SU3_letter}.

\section{Notations and Preliminaries} \label{sec:results}  \label{sec2}

\subsection{Observables to be Measured}   \label{sec2.1}

We wish to study various correlation functions of
the fundamental-representation field $U_x$ and 
the adjoint-representation field $V_x$ defined by
\be
   (V_x)^{\alpha \,\cdot\, \,\cdot\, \delta}_{\,\cdot\, \beta \gamma \,\cdot\,}
   \;\equiv\;   
   (U_x)^{\alpha}_{\,\cdot\, \gamma}
   (\overline{U_x})^{\,\cdot\, \delta}_{\beta} \, - \,
   {1 \over N}\, \delta^{\alpha}_{\beta} \delta^{\delta}_{\gamma}  \;.
\label{eq2.1}
\ee
Note the relation between the traces in the 
fundamental and the adjoint representations,
\be
   \tr_A U   \,\equiv\, \tr V   \,\equiv\,
   (V_x)^{\alpha \,\cdot\, \,\cdot\, \beta}_{\,\cdot\, \beta \alpha \,\cdot\,}
   \;=\;  |\tr(U)|^2  - 1   \;,
\ee
which follows immediately from \reff{eq2.1}.
We thus define the fundamental and adjoint 2-point correlation functions
\begin{subeqnarray}
   G_F(x-y)   & = &  \< \tr ( U_{x}^\dagger \,  U_y ) \>            \\[2mm]
   G_A(x-y)   & = &  \< \tr_A ( U_{x}^\dagger \,  U_{y}) \>
              \;=\;   \< | \tr ( U_{x}^\dagger \,  U_y) |^{2} \> - 1 \;.
\label{green_func}
\end{subeqnarray}

All our numerical work will be done on an $L \times L$ lattice
with periodic boundary conditions.
We are interested in the following quantities:
%%%  
\begin{itemize}
%%%  
\item  The fundamental and adjoint energies\footnote{
   We have chosen this normalization in order to have $0 \le E_{F,A} \le 1$,
   with $E_{F,A} = 1$ for a totally ordered state.
   Several other normalizations are in use in the literature.
}
\begin{subeqnarray}
   E_F   & = & {1 \over N} \,
         \< \tr ( U_{{\bf e}}^\dagger \, U_{0} ) \>
     \;=\;   {1 \over N} \,  G_F({\bf e})                                      \\[2mm]
   E_A   & = & {1 \over N^2-1} \,
            (\< | \tr ( U_{{\bf e}}^\dagger \,  U_0) |^{2} \> \,-\, 1)
     \;=\;  {1 \over N^2-1} \, G_A({\bf e})
\label{def_energies}
\end{subeqnarray}
where ${\bf e}$ stands for any nearest neighbor of the origin.

 \item The fundamental, adjoint and mixed specific heats\footnote{
    Here we return to the standard normalization {\em per site}\/
    (albeit without the ``thermodynamic'' factor $\beta^2$).
 }
 \begin{subeqnarray}
    C_{FF}  & = &
      {d \over N} \sum\limits_{\<yz\>}
      \< \real\, \tr ( U^\dagger_{\bf e} \,  U_0) \,;\,\real\,\tr (  U^\dagger_y \,  U_z) \>
      \;=\;  d {\partial E_F  \over  \partial \beta}                  \\[4mm]
    C_{AA}  & = &
      {d \over N^2-1} \sum\limits_{\<yz\>}
      \< | \tr ( U^\dagger_{\bf e} \,  U_0) |^2 \,;\, |\tr (  U^\dagger_y \, U_z)|^2  \>       \\[4mm]
    C_{FA}  & = &
      {d \over \sqrt{N^2-N}} \sum\limits_{\<yz\>}
      \< | \tr ( U^\dagger_{\bf e} \,  U_0) |^2 \,; \,\real\,\tr (  U^\dagger_y \,  U_z) \>
 \end{subeqnarray}
 where ${\bf e}$ stands for any nearest neighbor of the origin,
 $d$ is the spatial dimension (in this paper $d=2$),
 and $\< A;B \>  \equiv \< AB \> - \<A\> \<B\>$.

\item  The fundamental and adjoint magnetic susceptibilities
\begin{eqnarray}
   \chi_\#   & = &   \sum\limits_x   G_\#(x)   \;,
  %               \\[2mm]
  % \chi_A   & = &   \sum\limits_x   G_A(x)
\end{eqnarray}
where $\#$ stands for $F$ or $A$.
%%%  
\item  The fundamental and adjoint correlation functions at the
smallest nonzero momentum:
\begin{eqnarray}
   F_\#   & = &   \sum\limits_x e^{ip_0 \cdot x} \,   G_\#(x)   \;,
  %               \\[2mm]
  % \chi_A   & = &   \sum\limits_x   G_A(x)
\end{eqnarray}
where $p_0 = (\pm 2\pi/L, 0) \hbox{ or } (0,\pm 2\pi/L)$.
%%%  
\item  The fundamental and adjoint second-moment correlation lengths
\begin{eqnarray}
  \xi^{(2nd)}_\#  & = &  {(\chi_\#/F_\# \,-\, 1)^{1/2} \over 2\sin(\pi/L)}
  \;.
 \label{corr_len_2mom}
\end{eqnarray}
In the infinite-volume limit this becomes
\be
  \xi^{(2nd)}_\#  \;=\;
       \left( {1 \over 2d} \,
              { \displaystyle  \sum\limits_x  |x|^2 G_\#(x)
                \over
                \displaystyle  \sum\limits_x  G_\#(x)
              }
       \right) ^{1/2}   \;.
\ee
\item  The fundamental and adjoint exponential correlation lengths
\begin{eqnarray}
  \xi^{(exp)}_\#  & = &   \lim\limits_{|x| \to\infty}
                           {-|x|  \over  \log G_\#(x)}
 %           \\[3mm]
 % \xi^{(exp)}_A  & = &   \lim\limits_{|x| \to\infty}
 %                          {-|x|  \over  \log G_A(x)}
 \label{corr_len_exp}
\end{eqnarray}
and the corresponding mass gaps $m_{\#} = 1/ {\xi^{(exp)}_{\#}}$.
[These quantities make sense only if the lattice is essentially infinite
 (i.e.\ $L \gg \xi^{(exp)}_\#$) in at least one direction.
 We will not {\em measure}\/ any exponential correlation lengths in this work;
 but we will use $\xi^{(exp)}_\#$ as a theoretical standard of comparison.]
\end{itemize}

All these quantities except $\xi^{(exp)}_\#$
can be expressed in terms of expectations involving the following observables:
\begin{subeqnarray}
   \scrm_F  & = &  \sum_x U_x        \\[2mm]
   \scrm_A  & = &  \sum_x V_x        \\[2mm]
   \scrm_F^2  & = & \tr ( \scrm_F^\dagger \scrm_F )        \\[2mm]
   \scrm_A^2  & = & \tr ( \scrm_A^\dagger \scrm_A )        \\[2mm]
   \scrf_F   &=&  {1\over2} \real \tr
      \left[ \hat{U}(0,2\pi/L) \hat{U}^{\dagger} (0,2\pi/L) \,+\,
             \hat{U}(2\pi/L,0) \hat{U}^{\dagger} (2\pi/L,0 ) \right]  \\[2mm]
   \scrf_A   &=&  {1\over2} \real \tr
      \left[ \hat{V}(0,2\pi/L) \hat{V}^{\dagger} (0,2\pi/L) \,+\,
             \hat{V}(2\pi/L,0) \hat{V}^{\dagger} (2\pi/L,0 ) \right]   \\[2mm]
   \scre_F   &=&  {1 \over N} \sum_{\< xy \>}
                              \real\tr ( U^{\dagger}_x  U_y ) \\[2mm]
   \scre_A   &=&  {1 \over N^2-1}
       \sum_{\< xy \>} \left[ | \tr ( U^{\dagger}_x  U_y ) |^2 \,-\, 1 \right]
\end{subeqnarray}
where $\hat{U}(p)$ and $\hat{V}(p)$ are
the Fourier transforms of $U_x$ and $V_x$.
Thus,
\begin{subeqnarray}
   E_\#     & = &   \smhalf V^{-1}  \< \scre_\# \>              \\[2mm]
   C_{FF}  & = &    N \,V^{-1} \left[ \< \scre_F^2 \> - \< \scre_F \>^2 \right]
                                                     \\[2mm]
   C_{AA}  & = &   (N^2-1)\, V^{-1} \left[ \< \scre_A^2 \> - \< \scre_A \>^2 \right]
                                                     \\[2mm]
   C_{FA}  & = &   \sqrt{N^2-N}\, V^{-1} \left[ \< \scre_F \scre_A \> -
                              \< \scre_F \> \< \scre_A \> \right]   \\[2mm]
   \chi_\#  & = &   V^{-1}  \< \scrm_\#^2 \>            \\[2mm]
   F_\#     & = &   V^{-1}  \< \scrf_\# \>    
\end{subeqnarray}
%where $\# = V \hbox{ or } T$;
where $V=L^2$ is the number of sites in the lattice.

\subsection{Autocorrelation Functions and Autocorrelation Times} \label{sec2.2}

Let us now define the quantities ---
autocorrelation functions and autocorrelation times ---
that characterize the Monte Carlo dynamics.
Let $A$ be an observable
(i.e.\ a function of the spin configuration $\{U_x\}$).
We are interested in the evolution of $A$ in Monte Carlo time,
and more particularly in the rate at which the system ``loses memory''
of the past.
We define, therefore, the {\em unnormalized autocorrelation function}\footnote{
   In the mathematics and statistics literature, this is called the
   {\em autocovariance function}\/.
}
\be
  C_{AA}(t)  \;=\;   \< A_s A_{s+t} \>   -  \< A \> ^2  \,,
\ee
where expectations are taken {\em in equilibrium\/}.
The corresponding {\em normalized autocorrelation function\/} is
\be
  \rho_{AA}(t)  \;=\;  C_{AA}(t) / C_{AA}(0) \,.
\ee
We then define the {\em integrated autocorrelation time}
\begin{subeqnarray}
\tau_{int,A}  & =&
 \half \sum_{{t} \,=\, - \infty}^{\infty} \rho_{AA} (t)  \\[1mm]
 &=&  \half \ +\  \sum_{{t} \,=\, 1}^{\infty} \  \rho_{AA} (t)
\end{subeqnarray}
[The factor of $\half$ is purely a matter of convention;  it is
inserted so that $\tau_{int,A} \approx \tau$ if
$\rho_{AA}(t) \approx e^{-|t|/ \tau}$ with $\tau \gg 1$.]
Finally,
the {\em exponential autocorrelation time}\/ for the observable $A$ is
defined as
\be
\tau_{{\rm exp},A}  \;=\;
   \limsup_{{t}  \to \infty}  {|t| \over -  \log  | \rho_{AA}(t)|}
   \;,
 \label{def_tau_exp}
\ee
and the exponential autocorrelation time (``slowest mode'')
for the system as a whole is defined as
\be
\tau_{{\rm exp}} \;=\; \sup_A \,  \tau_{{\rm exp},A}  \;.
\ee
Note that $\tau_{{\rm exp}} = \tau_{{\rm exp},A}$
whenever the observable $A$ is not orthogonal to the
slowest mode of the system.

The integrated autocorrelation time controls the statistical error
in Monte Carlo measurements of $\< A \>$.  More precisely,
the sample mean
\begin{equation}
\bar A \ \ \equiv\ \ {1 \over n }\  \sum_{t=1}^n \ A_t
\end{equation}
has variance
\begin{subeqnarray}
\var( \bar A )  &= &
  {1 \over n^2} \ \sum_{r,s=1}^n \ C_{AA} (r-s)   \\[1mm]
 &=& {1 \over n }\ \sum_{{t} \,=\, -(n-1)}^{n-1}
  (1 -  {{|t| \over n }} ) C_{AA} (t) \slabel{var_observa}  \\[1mm]
 &\approx&  {1 \over n }\ (2 \tau_{int,A} ) \ C_{AA} (0)
   \qquad {\rm for}\ n\gg \tau \slabel{var_observb}
\end{subeqnarray}
Thus, the variance of $\bar{A}$ is a factor $2 \tau_{int,A}$
larger than it would be if the $\{ A_t \}$ were
statistically independent.
Stated differently, the number of ``effectively independent samples''
in a run of length $n$ is roughly $n/2 \tau_{int,A}$.
The autocorrelation time $\tau_{int,A}$ (for interesting observables $A$)
is therefore a ``figure of (de)merit'' of a Monte Carlo algorithm.

The integrated autocorrelation time $\tau_{int,A}$ can be estimated
by standard procedures of statistical time-series analysis
\cite{Priestley_81,Anderson_71}.
These procedures also give statistically valid {\em error bars}\/
on $\< A \>$ and $\tau_{int,A}$.
For more details, see \cite[Appendix C]{Madras_88}.
In this paper we have used a self-consistent truncation window of width
$c \tau_{int,A}$, where $c=8$ for $\scrm^2_F$ and $\scrm^2_A$ and $c=10$ for
the other observables. We made these choices because the autocorrelation
functions for $\scrm_F^2$ and $\scrm_A^2$ appear to decay roughly
like a pure exponential, while those for the other observables exhibit 
somewhat heavier long-time tails.
We have checked the dependence of $\tau_{int,A}$ on the window
width, and found that in all cases the estimated $\tau_{int,A}$ 
changes by less than $0.1\%$ for  $5 \le c \le 15$.

\section{Perturbative Predictions for $SU(N)$ Chiral Models}
\label{sec:perturb}  \label{sec3}

In this section we review the perturbative (large-$\beta$) predictions
for the two-dimensional $SU(N)$ principal chiral models. 
Most of these results are old \cite{Rossi_94a,Rossi_94b};
the results concerning the adjoint sector,
as well as those concerning the finite-size-scaling functions, are new. 
The calculations leading to the new results are summarized in 
Appendices \ref{appenA} and \ref{appenB}.

\subsection{Short-Distance Quantities} \label{sec3.1}

Modulo some conceptual problems arising from infrared divergencies
in dimension $d\le 2$, the calculation of the
perturbation expansion for {\em local}\/ quantities such as the
energies $E_F$ and $E_A$ is straightforward but tedious.
For the $SU(N)$ chiral model \reff{ham:SU(N)} in dimension $d=2$,
$E_F$ has been calculated through three-loop order \cite{Rossi_94a}:
\be
   E_F(\beta) \;=\;  1 \,-\, {N^{2}-1 \over 4N\beta}  \, \left[ 1
                   \,+\, {N^2-2 \over 16N\beta}
          \,+\, {0.0756 - 0.0634\,{N^2} + 0.01743\,{N^4}
                   \over  {N^2}\beta^2}
                   \,+\, O(1/\beta^3) \right]   \;.
\label{energy_F}
\ee
We have calculated $E_A$ through a trivial two-loop order
(see Appendix \ref{appenA}), obtaining
\be
   E_A(\beta) \;=\;  1 \,-\, {N \over 2\beta} \,+\,
                     {N^2 + 4 \over 32\beta^2} \,+\, O(1/\beta^3)
   \;.
\label{energy_A}
\ee
The large-$\beta$ expansions for the specific heats $C_{FF}$ and 
$C_{FA}$ can be obtained by differentiating \reff{energy_F} and \reff{energy_A}.

\subsection{Asymptotic Scaling of Correlation Lengths and
Susceptibilities} \label{sec3.2}

Renormalization-group calculations in the low-temperature expansion
($\equiv$ weak-coupling perturbation theory)
\cite{Polyakov_75,McKane_80,Shigemitsu_81}
suggest that the models \reff{ham:SU(N)} are asymptotically free,
i.e.\ that their only critical point is at $\beta = \infty$.
%Let us remark that the Renormalizion-group Gamma functions
%for the fundamental sector and adjoint sector differ, while the
%Beta functions are equal. In Appendix B we show the 
%calculations leading to the Gamma function for the adjoint sector
%up to order $1/beta^2$.
The renormalization group further predicts that 
the second-moment correlation lengths $\xi_F^{(2nd)}, \xi_A^{(2nd)}$,
the exponential correlation lengths $\xi_F^{(exp)}, \xi_A^{(exp)}$
and the susceptibilities $\chi_F, \chi_A$ behave as
\begin{eqnarray}
\xi_\#(\beta)    & = &   \widetilde{C}_{\xi_\#} \, \Lambda^{-1}
         \left[ 1 + {a_1 \over \beta} + \cdots \right]
                                           \label{xi_predicted2}  \\[3mm]
\chi_F(\beta)   & = &   \widetilde{C}_{\chi_F} \, \Lambda^{-2} \,
       \left( {4\pi\beta \over N} \right)^{\! -2(N^2-1)/N^2}  \,
         \left[ 1 + {b_1 \over \beta} + \cdots \right]
                                           \label{chiF_predicted2} \\[3mm]
\chi_A(\beta)   & = &   \widetilde{C}_{\chi_A} \, \Lambda^{-2} \,
          \left( {4\pi\beta \over N} \right)^{\! -4}  \,
         \left[ 1 + {d_1 \over \beta} + \cdots \right]
                                           \label{chiA_predicted2}
\end{eqnarray}
as $\beta \to \infty$, where
\be
   \Lambda  \;\equiv\;    e^{-4\pi\beta/N} \,
       \left( {4\pi\beta \over N} \right) ^{\! 1/2}   \,
       2^{5/2}\, \exp\!\left(\pi {N^2 - 2  \over  2N^2} \right)
\label{lambda_parameter}
\ee
is the fundamental mass scale\footnote{
   In (\protect\ref{lambda_parameter}),
   the exponential and power of $\beta$ are universal.
   The remaining factor is chosen so as make the $\beta\to\infty$ limit
   of the lattice theory agree with the standard continuum $\sigma$-model
   in the $\overline{\protect\hbox{MS}}$ normalization;
   this factor is special to the
   standard nearest-neighbor action (\protect\ref{ham:SU(N)}),
   and comes from a one-loop lattice calculation
   \protect\cite{Shigemitsu_81}.
},
and $\xi_\#$ denotes any one of
$\xi_F^{(2nd)}, \xi_A^{(2nd)}, \xi_F^{(exp)}, \xi_A^{(exp)}$.
Here
$\widetilde{C}_{\xi_\#}$, $\widetilde{C}_{\chi_F}$ and $\widetilde{C}_{\chi_A}$
are {\em universal}\/ (albeit nonperturbative) quantities characteristic
of the continuum theory (and thus depending only on $N$),
while the $a_k$, $b_k$ and $d_k$ are nonuniversal constants
(depending on $N$ and on the lattice Hamiltonian)
that can be computed in weak-coupling perturbation theory on the lattice
at $k+2$ loops.
It is worth emphasizing that the {\em same}\/ coefficients $a_k$
occur in all four correlation lengths:
this is because the ratios of these correlation lengths
take their continuum-limit values plus corrections that are powers
of the mass $m = 1/\xi^{(exp)}$, hence exponentially small in $\beta$.

When analyzing the susceptibilities,
it is convenient to study instead the ratios
\begin{eqnarray}
{ \chi_F(\beta)   \over   \xi_\#(\beta)^2 }
   & = &
   { \widetilde{C}_{\chi_F}   \over   \widetilde{C}_{\xi_\#}^2 }
   \,
   \left( {4\pi\beta \over N} \right)^{\! -2(N^2-1)/N^2}  \,
   \left[ 1 + {c_1 \over \beta} + {c_2 \over \beta^2} + \cdots \right]
  \label{chiFoverxiFsquared_predicted}   
\\[3mm]
  { \chi_A(\beta)   \over   \xi_\#(\beta)^2 }
   & = &
   { \widetilde{C}_{\chi_A}   \over   \widetilde{C}_{\xi_\#}^2 }
   \,
   \left( {4\pi\beta \over N} \right)^{\! -4}  \,
   \left[ 1 + {e_1 \over \beta} + {e_2 \over \beta^2} + \cdots \right]
  \label{chiAoverxiAsquared_predicted}
\end{eqnarray}
The advantage of this formulation in the case of $\chi_F$
is that one additional term of perturbation theory is available
(i.e.\ $c_2$ but not $a_2$ or $b_2$). 

For the standard nearest-neighbor action \reff{ham:SU(N)},
the perturbative coefficients
$a_1$, $b_1$, $c_1$, and $c_2$ can be easily recovered
from the lattice renormalization-group functions calculated
through three loops \cite{Rossi_94a,Rossi_94b};
and we computed $d_1$ and $e_1$ (see Appendix \ref{appenA}).
The results are:
\begin{eqnarray}
   a_1   & = &   - {3 \pi \over 8} N^{-3} +
                   \left({13 \pi \over 48} - {1 \over 8}\right) N^{-1} +
		   \left({1 \over 16 \pi} + {1 \over 16} - {\pi \over 24}
		   - {\pi \over 2} G_1\right) N
\label{a1}
   \\[2mm]
   b_1   & = &  \left({1\over 2} - {3\over 4 \pi}\right) {1\over {N^3}} + 
	        \left({1 \over 4 \pi} - 1 + {13 \pi\over 24}\right) {1\over N} +
		\left(-{1 \over 8\pi} + {3 \over 8} - {\pi\over 12} - \pi G_1\right) \, N
\label{b1}
   \\[2mm]
   c_1  & = &   (N^2 - 1) \left[ - {1 \over 2N^3} + {1 \over 4N} 
                                  - {1 \over 4\pi N} \right] 
\label{c1}
   \\[2mm]
   c_2 & = &    (N^2 - 1) \Biggl[ - {1 \over 8N^6}
                    + {1 \over 2N^4} \left( 1 - {1 \over 4\pi} \right)
                    - {1 \over 4N^2} \left( {17 \over 12} - {1 \over \pi} 
                                            + {1 \over 8\pi^2} \right) 
         \nonumber \\
   & & \qquad  + {13 \over 192} - {3 \over 32\pi} + {1 \over 32\pi^2} 
               + {G_1 \over 4}  \Biggr] 
%%   c_2  & = &   {1\over {8N^6}} - {{0.585211}\over {{N^4}}} + 
%%                {0.737966\over {{N^2}}} -0.330329 + 0.052573\,{N^2}  
\label{c2}
   \\[2mm]
   d_1  & = &   -{3\over 4 \pi} {1\over {N^3}} +
                \left({13 \pi \over 24} -  {5 \over 4}\right) {1\over N} +
                \left(-{3 \over 8\pi} + {5 \over 8} - {\pi\over 12} - \pi G_1 \right) \, N
\label{d1}
   \\[2mm]
   e_1  & = &  - {{1}\over N} + \left({1\over 2} - {1\over 2 \pi}\right)\,N
\label{e1}
\end{eqnarray}
where $G_1 \approx 0.04616363$.
Perturbation theory predicts trivially --- or rather, {\em assumes}\/ ---
that the lowest mass in the $SU(N)$ adjoint channel is the scattering state of
two fundamental particles, i.e.\ there are no adjoint bound states\footnote{
   For $N \ge 4$ there are bound states in {\em other}\/ channels,
   namely those corresponding to the completely antisymmetrized product
   $(f \otimes \ldots \otimes f)_{antisymm}$
   of $k$ fundamental representations, where $2 \le k \le N-2$
   \cite{Abdalla_84,Wiegmann_84a,Wiegmann_84b}.
}:
\be
   \widetilde{C}_{\xi^{(exp)}_A}/\widetilde{C}_{\xi^{(exp)}_F}
   \;=\;
   \half \;.
 \label{mT=2mV}
\ee
The nonperturbative universal quantity
$\widetilde{C}_{\xi_F^{(exp)}} \equiv \Lambda_\msbar/m_F$
for the standard continuum $SU(N)$ $\sigma$-model
has been computed exactly by Balog, Naik, Niedermayer and Weisz 
(BNNW) \cite{Balog_92}
using the thermodynamic Bethe Ansatz:  it is
\be
   \widetilde{C}_{\xi_F^{(exp)}}
   \;=\;
   \widetilde{C}_{\xi_F^{(exp)}}^{\hbox{\scriptsize (BNNW)}}
   \;\equiv\;
   \left( {e \over 8\pi} \right) ^{\! 1/2}   \,
   {\pi/N    \over  \sin(\pi/N)}   
   \;\,.
 \label{exact_Cxi}
\ee
The other nonperturbative constants are unknown,
but Monte Carlo studies suggest that
$\widetilde{C}_{\xi_F^{(2nd)}}/\widetilde{C}_{\xi_F^{(exp)}}$
lies between $\approx\! 0.985$ and 1 for all $N \ge 2$;
for $N=3$ it is $0.987 \pm 0.002$ \cite{Rossi_94a}.\footnote{
   The $SU(2)$ principal chiral model is equivalent
   to the $4$-vector model; and the $1/N$ expansion of the latter model,
   evaluated at $N=4$, indicates that
   $\widetilde{C}_{\xi_F^{(2nd)}}/\widetilde{C}_{\xi_F^{(exp)}} \approx 0.9992$
   \protect\cite{Flyvbjerg_91c}.
}

For future reference we define the ``theoretical predictions \`a la BNNW'':
\begin{subeqnarray}
\xi_{F,\hbox{\scriptsize BNNW},2-loop}^{(exp)}(\beta)    & = &
   \widetilde{C}_{\xi_F^{(exp)}}^{\hbox{\scriptsize (BNNW)}}
   \, \Lambda^{-1}
 \slabel{xiF_B_2loop}   \\[1mm]
\xi_{A,\hbox{\scriptsize BNNW},2-loop}^{(exp)}(\beta)    & = &
   \smhalf \widetilde{C}_{\xi_F^{(exp)}}^{\hbox{\scriptsize (BNNW)}}
   \, \Lambda^{-1}
 \slabel{xiA_B_2loop}   \\[1mm]
\xi_{F,\hbox{\scriptsize BNNW},3-loop}^{(exp)}(\beta)    & = &
   \widetilde{C}_{\xi_F^{(exp)}}^{\hbox{\scriptsize (BNNW)}}
   \, \Lambda^{-1}
   \left[ 1 + {a_1 \over \beta} \right]
 \slabel{xiF_B_3loop}   \\[1mm]
\xi_{A,\hbox{\scriptsize BNNW},3-loop}^{(exp)}(\beta)    & = &
   \smhalf \widetilde{C}_{\xi_F^{(exp)}}^{\hbox{\scriptsize (BNNW)}}
   \, \Lambda^{-1}
   \left[ 1 + {a_1 \over \beta} \right]
 \slabel{xiA_B_3loop}
 \label{xi_B_formulae}
\end{subeqnarray}
where $\Lambda$ is defined in \reff{lambda_parameter}.

\subsection{``Improved Expansion Parameters''} \label{sec3.3}

There have recently been a variety of proposals in the literature for
``improved expansion parameters'' to be employed in place of the
bare coupling constant $1/\beta$:
the goal of all these schemes is to observe perturbative asymptotic scaling
at the smallest possible correlation length,
by redefining slightly the meaning of ``asymptotic scaling''.
In this subsection we would like to analyze critically the
logic behind these proposals,
and analyze in particular the application to the $SU(N)$ chiral models.

When one fails to observe $k$-loop asymptotic scaling
in some given expansion parameter and some given range of $\beta$,
there are two possible causes:
\begin{itemize}
   \item[(a)]  The perturbative contribution at $l$-loop order
 is {\em large}\/ (in the range of $\beta$ in question) for one or more
 of the terms $l = k+1, k+2, \ldots\;$.
 In this case one {\em expects}\/ large deviations from
 $k$-loop asymptotic scaling.
 We call this the ``perturbative'' obstruction to asymptotic scaling.
   \item[(b)]  The perturbative contributions at $l$-loop order
 ($l \ge k+1$) are all individually small, but in spite of this,
 $k$-loop asymptotic scaling has not been reached.
 This could be due to the higher-order terms having a large ``sum''
 in spite of their individual smallness, or it could be due to
 ``nonperturbative'' contributions.  Whatever the ultimate explanation,
 we call this the ``nonperturbative'' obstruction to asymptotic scaling.
\end{itemize}
Of course, in the strict sense these concepts are ill-defined,
because we are dealing here with {\em non-convergent}\/
(and indeed usually non-Borel-summable \cite{tHooft,Duncan_95})
asymptotic series.  As a result, the {\em very}\/-high-order terms
in perturbation theory will {\em always}\/ be large.
But in practice this will not pose a significant problem,
since we are dealing with $k=2$ or 3 or (in rare cases) 4,
while the ultimate growth of the perturbative contributions
usually occurs at much larger values of $l$.

Each of these two possible obstructions to asymptotic scaling
gives rise to a distinct intuition regarding
``improved expansion parameters'',
and a distinct logic by which their use can be justified:

\bigskip

{\em Perturbative justification.}\/
Since the weak-coupling
perturbation expansion is a power series in $1/\beta \sim 1/\log\xi$,
it follows that the perturbative corrections decay extremely slowly
as $\xi\to\infty$.
In particular, these corrections could be large at all accessible
correlation lengths (say, $\xi \ltapprox 10^2$--$10^6$)
if the perturbative coefficients are sufficiently large (say, 5--10).
The ``perturbative'' logic governing the choice of
expansion parameters has been summarized very clearly by
Lepage and Mackenzie \cite{Lepage_93}:
\begin{quote}
   If an expansion parameter $\alpha_{good}$ produces well-behaved
   perturbation expansions for a variety of quantities,
   using an alternate expansion parameter
   $\alpha_{bad} \equiv \alpha_{good} (1 - 10000 \alpha_{good})$
   will lead to second-order corrections that are uniformly large,
   each roughly equal to $10000 \alpha_{bad}$ times the first-order
   contribution.  Series expressed in terms of $\alpha_{bad}$,
   although formally correct, are misleading if truncated and 
   compared with data.
\end{quote}
Conversely, they argue,
\begin{quote}
   The signal for a poor choice of expansion
   parameter is the presence in a variety of calculations of large
   second-order coefficients that are all roughly equal relative
   to first order.
\end{quote}
Indeed, this latter is precisely the condition under which one can
define a new expansion parameter
$\alpha_{new} \equiv \alpha_{old} (1 + C \alpha_{old})$
with respect to which the second-order coefficients,
for a variety of observables,
are all significantly smaller than they were relative to $\alpha_{old}$.

However, while this is a {\em necessary}\/ condition for the
perturbation series in $\alpha_{new}$
to be better-behaved than that in $\alpha_{old}$,
it is not a {\em sufficient}\/ condition.
The trouble, of course, is that the coefficients at third and higher orders
may become large after the change of variables,
{\em even if they were small before the change of variables.}\/
Different changes of variable that are equivalent at second order,
for example
$\alpha_{new} \equiv \alpha_{old} (1 + C \alpha_{old})$
and
$\alpha'_{new} \equiv \alpha_{old} / (1 - C \alpha_{old})$,
can produce vastly different effects at third and higher orders.
The decision to use one variable $\alpha_{new}$ rather than another
is inherently a guess about approximate magnitudes and signs
of the uncomputed high-order corrections ---
that is, it is an attempt to {\em resum}\/ perturbation theory.
Clearly this is a hazardous enterprise, especially when one has in hand
only the first one or two terms of the perturbation series as guidance.
In our opinion a proposed resummation method ---
if it is to be more than mere numerology --- must be based on
some {\em theoretical}\/ input which suggests
the approximate magnitudes and signs of the dominant contributions
to the high-order corrections.
Moreover, a valid claim of ``success'' cannot be based simply on having found
{\em one}\/ expansion parameter that yields good agreement between
``theory'' and ``experiment'' (while other expansion parameters,
equally sensible {\em a priori}\/, yield poor agreement).
Rather, one can claim to {\em understand}\/ the situation only when
one can exhibit a {\em systematic correspondence}\/ between the
degree of agreement between ``theory'' and ``experiment'' 
and some plausible {\em theoretical}\/ measure of the reliability
of the expansion.

A minimal demand for a $k$-loop ``improved expansion parameter''
is that the $(k+1)$-loop correction term be smaller in the new variable
than in the old.  Unfortunately, this criterion can be checked only after
the $(k+1)$-loop terms have been computed --- at which point one is more
likely to be interested in $(k+1)$-loop ``improved expansion parameters''
and thus in the relative size of the $(k+2)$-loop corrections!

For models that are exactly solvable in the limit $N \to \infty$,
some guidance concerning the choice of ``improved expansion parameters''
can be obtained from the $N=\infty$ solution.
For example, for the mixed isovector/isotensor $\sigma$-models
in two dimensions,
several ``improved expansion parameters'' related to the
isovector and isotensor energies lead to the
{\em vanishing}\/ of the perturbative corrections,
at {\em all}\/ orders of perturbation theory,
in the limit $N \to\infty$ \cite{CEPS_RPN}.
Of course, this fact
does not establish the relevance of these ``improved expansion parameters''
for small $N$.
Moreover, for our $SU(N)$ models we unfortunately lack an
exact solution at $N=\infty$ \cite{Rossi_96_review}.

\bigskip

{\em Nonperturbative justification.}\/
In some models the specific heat has a sharp bump at some finite $\beta$,
due presumably to a nearby singularity in the complex $\beta$-plane.
For example, this behavior is observed empirically \cite{Rossi_94a,Rossi_94b}
in the two-dimensional $SU(N)$ $\sigma$-models for $N \gtapprox 6$;
indeed, in this case the singularity appears to pinch the real axis
(and thus become a true second-order phase transition)
in the limit $N \to\infty$ \cite{Campostrini_95}.
In such a situation it is natural to expect that other
observables, such as the correlation length and the susceptibilities,
may show similar bumps and singularities.
Indeed, for the $SU(N)$ $\sigma$-models it is observed empirically
\cite{Rossi_94a,Rossi_94b} that the correlation length shows
large deviations from asymptotic scaling precisely in the
weak-to-strong-coupling crossover regime where the specific heat
has its peak;  this behavior is particularly pronounced for large $N$.

If, by a change of variables $\beta \to f(\beta)$
one could move the complex singularity farther away from the real axis,
one would expect to observe a flatter specific-heat curve and
--- to the extent that this same singularity appears in long-distance
observables such as the correlation length ---
also a smoother approach to asymptotic scaling.
One possible choice is to take $f(\beta)$ equal to the energy $E(\beta)$:
assuming that the energy {\em diverges}\/ at the complex singularity,
this would move the singularity to {\em infinity}\/ in the new variable.
[Of course, one could alternatively take $f(\beta)$ equal to the correlation
length $\xi(\beta)$, but this is cheating:  ``asymptotic scaling''
would not have the same {\em physical meaning}\/ in the new variable
as it did in the old.  The energy, by contrast, is a {\em short-distance}\/
observable, and is thus a plausible substitute for the bare parameter $\beta$.]
%% One idea is to use a {\em short}\/-distance observable, such as the 
%% energy $E_F$, as the ``improved expansion parameter'' to be used
%% instead of the bare parameter $\beta$.
This choice can alternatively be justified
on the plausible heuristic grounds that the ``nonperturbative effects''
and/or high-order perturbative effects
responsible for the sharp crossover from strong to weak coupling
are likely to have the same qualitative effect on correlations at both
short and long distances.
%% Thus, we may expect that the ``nonperturbative''
%% and/or perturbative correction terms cancel each other, yielding
%% a resummed perturbation theory with much smaller higher order 
%% corrections.

These arguments are admittedly somewhat vague, but they give some grounds
for trying an ``improved expansion parameter'' based on the energy $E(\beta)$,
as was long ago suggested (for somewhat different reasons)
by Parisi \cite{Parisi_Madison,Martinelli-Parisi-Petronzio}
and others \cite{Chen_82,Samuel_82,Makeenko_82,Makeenko_83,Makeenko_84,%
Karsch_84,Samuel_85,Wolff_O4_O8,Lepage_LAT90,Lepage_LAT91,Lepage_93}.

The implementation of this ``improved expansion parameter'' is as follows:
We first revert the perturbation expansion \reff{energy_F} for $E_F$,
yielding $\beta$ as a power series in $x_F \equiv 1-E_F$:
\begin{subeqnarray}
 	\beta(x_F) &=& 
        \alpha_{-1} x_F^{-1} + \alpha_0 + \alpha_{1} x_F + O(x_F^2)  \\[2.5mm]
 &=&  {N^2-1  \over 4N} \,  x_F^{-1}  \;+\;
      \left( {N \over 16} - {1 \over 8N} \right)        \nonumber \\
 & &  \qquad \;+\;  { 0.05409696 N^3 - 0.1910544 N + 0.2398288 N^{-1}
                \over N^2-1}  \, x_F    \nonumber \\
 & &  \qquad \;+\; O(x_F^2) \;.   
\label{beta_energy}
\end{subeqnarray}
% {\bf Double-check this!!!!}
Then, to obtain the ``energy-improved expansion'' of 
any long-distance observable $\scro$, we just insert \reff{beta_energy}
into the standard perturbation prediction 
\reff{xi_predicted2}--\reff{chiAoverxiAsquared_predicted} 
and expand in $x_F$ to the relevant order.
For example, for $\xi_\#$ we have
\be
	\xi_\#(\beta)    \;=\;   \widetilde{C}'_{\xi_\#}  \,
        e^{(4\pi\alpha_{-1}/N) x_F^{-1}} \,
       	\left( {N x_F \over 4\pi\alpha_{-1} } \right) ^{\!1/2}
        \left[ 1 \,+\, a'_1 x_F \,+\, \ldots \right]  \;,
\label{improv_2}
\ee
where
\begin{subeqnarray}
   \widetilde{C}'_{\xi_\#}   & = &   \widetilde{C}_{\xi_\#} \,
        e^{4\pi\alpha_0/N} \,
        2^{-5/2} \, \exp\!\left(-\pi {N^2 - 2  \over  2N^2} \right)   \\[2mm]
   a'_1  & = & {{-\alpha_{0}}\over {2\,\alpha_{-1}}} + 
               {a_1 \over \alpha_{-1} } + 
	       {{4\,\alpha_{1}\,\pi }\over N}
\label{improv_3}
\end{subeqnarray}
For the other observables we shall proceed similarly.

Let us now apply our {\em perturbative}\/ test of
the goodness of the 2-loop expansion variables
--- standard versus ``energy-improved'' ---
by comparing the relative magnitudes of the 3-loop
perturbative coefficients
$a_1/\beta$ and $a'_1 x_F \approx a'_1 \alpha_{-1}/\beta$, respectively.
We have
\begin{eqnarray}
   a_1   & = &   -1.178097 N^{-3} + 0.725848 N^{-1} - 0.121019 N   \\[2mm]
   a'_1 \alpha_{-1}   & = &   -0.424651 N^{-3} + 0.188133 N^{-1}
                                                    + 0.0176814 N
\end{eqnarray}
In Table~\ref{table_improved} we show these coefficients
(divided by $N$ so as to have a good $N\to\infty$ limit)
for $N=2,3,\ldots,20,\infty$.
We see that the 2-loop ``energy-improved'' scheme is
a factor of $\approx 7$ better than standard perturbation theory
for large $N$;  the advantage drops to a factor of $\approx 3$ for $N=4$,
and a factor of $\approx 1.6$ for $N=3$.
Only for $N=2$ (which is isomorphic to the 4-vector model)
is the ``energy-improved'' scheme actually worse than
standard perturbation theory (by a factor of $\approx 3$).\footnote{
   The opposite conclusions in \cite[p.~1623]{Rossi_94a}
   are due to an algebraic error:  the final term in their equation (20)
   should have a minus sign.  The same error infects equations (148) and (150)
   of \cite{Rossi_94b}.
   We thank Ettore Vicari for double-checking this computation.
}

\subsection{Finite-Size Scaling of Correlation Lengths and Susceptibilities}
   \label{sec3.4}

Since Monte Carlo simulations are carried out in systems of
finite size, it is important to understand how to connect these
measurements with infinite-volume physics.
Let us work on a periodic lattice of linear size $L$.
Then finite-size-scaling theory
\cite{Barber_FSS_review,Cardy_FSS_book,Privman_FSS_book}
predicts quite generally that
\be
   {\scro(\beta,sL) \over \scro(\beta,L)}   \;=\;
   F_{\scro} \Bigl( \xi(\beta,L)/L \,;\, s \Bigr)
   \,+\,  O \Bigl( \xi^{-\omega}, L^{-\omega} \Bigr)
   \;,
\ee
where $\scro$ is any long-distance observable,
$s$ is any fixed scale factor,
$\xi(\beta,L)$ is a suitably defined finite-volume correlation length,
$L$ is the linear lattice size,
$F_{\scro}$ is a scaling function characteristic of the universality class,
and $\omega$ is a correction-to-scaling exponent.
Here we will use $\xi_F^{(2nd)}$ in the role of $\xi(\beta,L)$;
for the observables $\scro$ we will use the four
``basic observables'' $\xi^{(2nd)}_F$, $\xi^{(2nd)}_A$, $\chi_F$, $\chi_A$
as well as certain combinations of them such as $\chi_F/(\xi^{(2nd)}_F)^2$,
$\chi_A/(\xi^{(2nd)}_A)^2$ and $\xi^{(2nd)}_F/\xi^{(2nd)}_A$.
%{\bf Do we actually use the two $\chi/\xi^2$ combinations?????}

In an asymptotically free model,
the functions $F_{\scro}(x;s)$ at $x \gg 1$
can be computed in perturbation theory
in powers of $1/x^2$, where $x \equiv \xi_F^{(2nd)}(\beta,L)/L$.
We obtain the following expansions (see Appendix \ref{appenB} for details):
\begin{subeqnarray}
   {\xi_F^{(2nd)}(\beta,sL) \over \xi_F^{(2nd)}(\beta,L)}   & = &
   s \left[ 1 \,-\,  {\log s \over 8\pi} \, {N^2 \over N^2-1} \, x^{-2}
       \,-\,  {N^4 \over (N^2 -1)^2}
              \Biggl( {\log s \over 64\pi^2} + {\log^2 s \over 128\pi^2}
              \Biggr) x^{-4}
       \,+\,  O(x^{-6}) \right]
   \nonumber \\ \slabel{xiF_FSS_PT} \\[3mm]
   {\xi_A^{(2nd)}(\beta,sL) \over \xi_A^{(2nd)}(\beta,L)}
   & = &
   s \Biggl\{ 1 \,-\, {\log s \over 8\pi} \, {N^2 \over N^2-1} \, x^{-2}
   \nonumber \\
   & & \qquad
   - \; {N^2 \over (N^2-1)^2}
        \left[
        \left( (N^2+1) \left( {\pi \over 4} I_{3,\infty} 
                              + {1 \over 32\pi^3} \right) \right. \right.
   \nonumber \\
   & & \qquad \qquad \qquad \qquad
      \left.  \left. + {N^2 \over 64\pi^2}
        \right) {\log s}
        \,+\,  {N^2 \log^2 s \over 128\pi^2}
        \right]  x^{-4}  \;  + \; O(x^{-6}) \Biggr\}
   \slabel{xiA_FSS_PT}    \\[3mm]
   {\chi_F(\beta,sL) \over \chi_F(\beta,L)}   & = &
   s^2 \Biggl[ 1 \,-\, {\log s \over 2\pi} \, x^{-2}            \nonumber \\
   & & \qquad +\, {N^2-2 \over N^2-1} \left(
               {\log^2 s \over 16\pi^2} +
               \Bigl( {\pi \over 2} I_{3,\infty} + 
		      {1 \over 16\pi^3} \Bigr) \log s
               \right)  x^{-4}  \,+\, O(x^{-6}) \Biggr]
   \slabel{chiF_FSS_PT}  \\[3mm]
   {\chi_A(\beta,sL) \over \chi_A(\beta,L)}   & = &
   s^2 \Biggl[ 1 \,-\, {\log s \over \pi} \, {N^2 \over N^2-1} \, x^{-2}
                                                               \nonumber \\
   & & \qquad +\, {N^2 \over (N^2-1)^2} \left(
               {3 N^2 \over 8\pi^2} \log^2 s  + (N^2-2)
               \Bigl( \pi I_{3,\infty} + {1 \over 8\pi^3} \Bigr) \log s
               \right)  x^{-4}  
   \nonumber \\
   & & \qquad +\, O(x^{-6}) \Biggr]
   \slabel{chiA_FSS_PT}
   \label{all_FSS_PT}
\end{subeqnarray}
and also
\be
   {\xi_F^{(2nd)}(\beta,L) \over \xi_A^{(2nd)}(\beta,L)}   \;=\;
   \left( {2N^2 \over N^2-1} \right) ^{\! 1/2} \,
   \left[ 1 \,+\,
     {N^2+1 \over N^2-1} \Bigl( \pi^2 I_{3,\infty} + {1 \over 8\pi^2} \Bigr)
         x^{-2}
              \,+\, O(x^{-4}) \right]
   \label{xiFoverxiA_FSS_PT}
\ee
where
%%% $A = [2N/(N^2 - 1)]^{1/2}$, $w_0 = N/(8 \pi)$, $w_1 = N^2/(128 \pi^2)$ and
\be
   I_{3,\infty}  \approx  (2 \pi)^{\! -4} \times 3.709741314407459
   \;.
\ee

\section{Numerical Results} \label{sec4}

We have carried out extensive Monte Carlo runs on the two-dimensional
$SU(3)$ chiral model, on periodic $L \times L$ lattices of size
$L = 8, 16, 32, 64, 128, 256$,
at 264 different pairs $(\beta,L)$
in the range $1.65 \le \beta \le 4.35$.
The results of these computations are shown in
Tables~\ref{su3_static} (static data)
and \ref{su3_dyndata} (dynamic data).
Five of our $(\beta,L)$ pairs coincide with those studied
previously by Hasenbusch and Meyer \cite{Hasenbusch-Meyer_PRL},
and three with those by Horgan and Drummond \cite{Drummond_94};
in all these cases the static data are in good agreement.

For most of our $\beta$ values
we have made runs at four, five or even six different lattice sizes.
In this way we have obtained detailed information on the finite-size effects,
covering densely the interval
$0.1 \ltapprox x \equiv \xi_F^{(2nd)}(\beta,L)/L \ltapprox 1.1$.
Using a finite-size-scaling extrapolation method
(see Section \ref{sec5.1}), we are able to extrapolate 
$\xi_F^{(2nd)}$, $\chi_F$, $\xi_A^{(2nd)}$ and $\chi_A$
to the $L=\infty$ limit with good control over the statistical 
and systematical errors (see Section \ref{sec5.2}).

These runs employed the $XY$-embedding MGMC algorithm described in 
Section~\ref{sec1.1} (see \cite{MGMC_ON} for details).
The induced $XY$ model \reff{ham_embed0}
was updated using our standard $XY$-model MGMC program \cite{MGMC_2}
with $\gamma = 2$ (W-cycle) and $m_1 = 1$, $m_2 = 0$ 
(one heat-bath pre-sweep and no heat-bath post-sweeps).
In all cases the coarsest grid is taken to be $2 \times 2$.
All runs used a disordered initial configuration (``hot start'').
Because the measurement of the observables (particularly the adjoint
observables) was very time-consuming compared to the MGMC updating,
the observables were measured once every two MGMC iterations.
All times (run lengths and autocorrelation times)
are therefore specified in units of {\em measurements}\/,
i.e.\ in units of {\em two}\/ MGMC iterations.

These runs were performed partly on a Cray C-90 and partly on an IBM SP2
(in both cases using only a single processor).
In Table~\ref{cpu_timings} we show the CPU time per measurement,
as a function of $L$, for each of these two machines:
each timing thus includes {\em two}\/ MGMC iterations followed by one
measurement of all observables.\footnote{
	The CPU time spent in the measurement of the observables
	is roughly 28\%,22\%,15\%,12\%,7\%,5\% of the total CPU time
	for $L=8$,16,32,64,128,256, respectively, when the runs are performed 
	on the CRAY C-90; it is roughly 22\%,20\%,18\%,17\%,5\%,3\% 
        for $L=8$,16,32,64,128,256 when the runs are performed
        on the IBM SP2.
}
Observe that the timings on the Cray C-90
grow {\em sublinearly}\/ in the volume,
in contrast to the theoretical prediction \reff{workMG_0},
because the vectorization is more effective on the larger lattices.\footnote{
   The heat-bath subroutine uses von Neumann rejection to
   generate the desired random variables
   \protect\cite[Appendix A]{MGMC_2}.
   The algorithm is vectorized by gathering all the sites of one sublattice
   (red or black) into a single Cray vector, making one trial of the
   rejection algorithm, scattering the ``successful'' outputs,
   gathering and recompressing the ``failures'', and repeating until
   all sites are successful.
   Therefore, although the original vector length in this subroutine is
   $L^2/2$,
   the vector lengths after several rejection steps are much smaller.
   It is thus advantageous to make the original vector length
   as large as possible.
}
But the ratio $\hbox{time}(2L)/\hbox{time}(L)$
is increasing with $L$, and appears very roughly to be approaching
the theoretical value of 4 as $L \to\infty$.
On the other hand, the timings on the IBM SP2
grow {\em superlinearly}\/ in the volume,
presumably as a result of the increased frequency of cache misses
for larger $L$.  
Because of these opposite variations in the CPU time,
the runs with $L=128,256$ were performed on the Cray
while those with $L=8,16,32$ were done on the IBM;
the runs with $L=64$ were divided between the two machines.

The running speed on the Cray C-90
for our $XY$-embedding MGMC program at $L=256$ was approximately 259 MFlops.
The total CPU time for the runs reported here was about
0.85 Cray C-90 years plus 0.7 IBM SP2 years.

\section{Finite-Size-Scaling Analysis: Static Quantities}
\label{sec:fss_static}  \label{sec5}

In this section we analyze the static data reported in
Table~\ref{su3_static}.
First, we review the finite-size-scaling extrapolation method
(Section \ref{sec5.1}).
Next, we apply this method to extrapolate
$\xi_F^{(2nd)}$, $\chi_F$, $\xi_A^{(2nd)}$ and $\chi_A$
to the $L=\infty$ limit, taking great care to analyze the
systematic errors arising from correction to scaling (Section \ref{sec5.2}).
Then we compare both the raw and the extrapolated values with the
perturbative predictions (Section \ref{sec5.3}).
We conclude by discussing further the conceptual foundations
of our method, and replying to some criticisms that have been
leveled against it (Section \ref{sec5.4}).

\subsection{Finite-Size-Scaling Extrapolation Method} \label{sec5.1}

\subsubsection{Basic Ideas}  \label{sec5.1.1}

We will extrapolate our finite-$L$ data to $L=\infty$
using an extremely powerful and general method
\cite{fss_greedy,fss_greedy_fullpaper}
due originally to L\"uscher, Weisz and Wolff \cite{Luscher_91}
(see also Kim \cite{Kim_93,Kim_LAT93,Kim_94a,Kim_94b,Kim_95}),
based on the theory of {\em finite-size scaling}\/ (FSS)
\cite{Barber_FSS_review,Cardy_FSS_book,Privman_FSS_book}.
We have successfully employed this method in previous works
on different models \cite{swaf2d,o3_scaling_prl,o3_scaling_fullpaper}.

Consider, for simplicity, a model controlled by a renormalization-group
fixed point having {\em one}\/ relevant operator.
Let us work on a periodic lattice of linear size $L$.
Let $\xi(\beta,L)$ be a suitably defined finite-volume correlation length,
such as the second-moment correlation length $\xi_F^{(2nd)}(\beta,L)$
defined by \reff{corr_len_2mom},
and let $\scro$ be any long-distance observable
(e.g.\ the correlation length or the susceptibility).
Then finite-size-scaling theory
\cite{Barber_FSS_review,Cardy_FSS_book,Privman_FSS_book}
predicts that
\be
   {\scro(\beta,L) \over \scro(\beta,\infty)}   \;=\;
   f_{\scro} \Bigl( \xi(\beta,\infty)/L \Bigr)
   \,+\,  O \Bigl( \xi^{-\omega}, L^{-\omega} \Bigr)
   \;,
 \label{eq1}
\ee
where $f_{\scro}$ is a universal function
and $\omega$ is a correction-to-scaling exponent.\footnote{
   This form of finite-size scaling
   assumes hyperscaling, and thus is expected to hold only below the upper
   critical dimension of the model.
   See e.g.\ \cite[Chapter I, section 2.7]{Privman_FSS_book}.
   %% The above is a citation to Privman's review article.
   %% {\bf Cite Binder et al re hyperscaling in $d=5$ Ising?????}
}
It follows that if
$s$ is any fixed scale factor (usually we take $s=2$), then
\be
   {\scro(\beta,sL) \over \scro(\beta,L)}   \;=\;
   F_{\scro} \Bigl( \xi(\beta,L)/L \,;\, s \Bigr)
   \,+\,  O \Bigl( \xi^{-\omega}, L^{-\omega} \Bigr)
   \;,
 \label{eq2}
\ee
where $F_\scro$ can easily be expressed in terms of $f_\scro$ and $f_\xi$.
%%% {\bf (see footnote for details???)}
(Henceforth we shall suppress the argument $s$
 if it is clear from the context.)
In other words, if we make a plot of $\scro(\beta,sL) / \scro(\beta,L)$
versus $\xi(\beta,L)/L$, then all the points should lie on a single curve,
modulo corrections of order $\xi^{-\omega}$ and $L^{-\omega}$.

Our extrapolation method works as follows\footnote{
   See \cite[note 8]{o3_scaling_prl} for further history of this method.
}:
We make Monte Carlo runs at numerous pairs $(\beta,L)$ and $(\beta,sL)$.
We then plot $\scro(\beta,sL) / \scro(\beta,L)$ versus $\xi(\beta,L)/L$,
using those points satisfying both $\xi(\beta,L) \ge$ some value $\xi_{min}$
and $L \ge$ some value $L_{min}$.
If all these points fall with good accuracy on a single curve ---
thus verifying the Ansatz \reff{eq2} for
 $\xi \ge \xi_{min}$, $L \ge L_{min}$ ---
we choose a smooth fitting function $F_{\scro}$.
Then, using the functions $F_\xi$ and $F_\scro$,
we extrapolate the pair $(\xi,\scro)$ successively from
$L \to sL \to s^2 L \to \ldots \to \infty$.

We have chosen to use functions $F_{\scro}$ of the form\footnote{ 
   In performing this fit, one may use any basis one pleases in
   the space spanned by the functions $\{e^{-k/x}\}_{1 \le k \le n}$;
   the final result (in exact arithmetic) is of course the same.
   However, in finite-precision arithmetic the calculation may become
   numerically unstable if the condition number of the least-squares matrix
   gets too large.  In particular, this disaster occurs if we use
   as a basis the monomials $t^k$ (where $t = e^{-1/x}$).
   The trouble is that these monomials are ``almost collinear''
   in the relevant Hilbert space $L^2(\mu)$
   defined by $\mu(t) = \sum_i w_i \, \delta(t-t_i)$,
   where $t_i$ are the values of $t \equiv e^{-L/\xi(\beta,L)}$
   arising in the data pairs and
   $w_i = 1/[\hbox{error on } \scro(2L)/\scro(L)]^2$
   are the corresponding weights.
   To avoid this disaster, we should seek to use a basis
   that is closer to orthogonal
   in $L^2(\mu)$.  Of course, exactly orthogonalizing in $L^2(\mu)$
   is equivalent to diagonalizing the least-squares matrix,
   which is unfeasible;  but we can do well enough by
   using polynomials with zero constant term that are orthogonal
   with respect to the simple measure
   $w(t) = t^a (t_{max} - t)^b$ on $[0,t_{max}]$,
   where $a$ and $b$ are some chosen numbers $> -1$.
   These polynomials are Jacobi polynomials
   $f_k(t) = t P_{k-1}^{(b,a+2)}(2t/t_{max} \,-\, 1)$
   for $1 \le k \le n$ \cite[pp.~321--328]{Hassani_91}.
   The idea here is that the measure $w(t) = t^a (t_{max} - t)^b$
   should roughly approximate the measure $\mu(t)$.
   Empirically (for our data) the measure $\mu(t)$ seems to have a little peak
   near $t=0$ followed by a dip,
   and a big peak near $t=t_{max}$;
   for this reason we have chosen $a=0$, $b=-3/4$.
   But the performance is very insensitive to the choices of $a$ and $b$.
   This cleverness in the choice of basis
   {\em vastly} improves the numerical stability of the result,
   by reducing the condition number of the matrix arising in the fit.
   Typical condition numbers using Jacobi polynomials are  
   $\approx 75$ for $n=11$ and  $\approx 123$ for $n=15$. 
   Typical condition numbers using monomials (and 100-digit arithmetic!) 
   are  $7.5 \times 10^{11}$ for $n=11$ and  $6.5 \times 10^{12}$ for $n=15$.
%   {\bf GUSTAVO: Can you quote a typical condition number with vs. without???
%      Perhaps do it for two different values of $n$, e.g.\ 10 and 15.
%      Make sure to use very high-precision arithmetic, because if you don't,
%      you'll UNDERESTIMATE the condition number of the bad matrix!!!}
}
\be
   F_\scro(x)   \;=\;
   1 + a_1 e^{-1/x} + a_2 e^{-2/x} + \ldots + a_n e^{-n/x}   \;.
 \label{fss:gen}
\ee
(Other forms of fitting functions can be used instead.)
This form is partially motivated by theory, which tells us
that in some cases
$F_\scro(x) \to 1$ exponentially fast as $x\to 0$ \cite{CEPS_RPN}.\footnote{
   The finite-size corrections to {\em Euclidean}\/ correlation functions
   in an $L^d$ box are expected to behave as $e^{-mL}$, where
   $m \equiv 1/\xi_F^{(exp)}$ is the lightest mass in the theory.
   (This can be proven to all orders in perturbation theory
    \cite{Neuberger_89} and presumably also holds nonperturbatively.)
   This is slightly different from our $e^{-1/x}$
   because we have defined $x$ as $\xi_F^{(2nd)}/L$
   rather than $\xi_F^{(exp)}/L$,
   but the difference is expected to be very small,
   since $\xi_F^{(2nd)}/\xi_F^{(exp)} \approx 0.987$ \cite{Rossi_94a}.

   It follows from this that the finite-size-scaling functions
   for the susceptibilities $\chi_F$ and $\chi_A$
   tend to 1 exponentially fast as $x\to 0$.
   However, this is {\em not}\/ the case for finite-size-scaling functions
   for the correlation lengths $\xi_F^{(2nd)}$ and $\xi_A^{(2nd)}$,
   because
   the definition of these correlation lengths contains an explicit
   $L$-dependence, so that one expects corrections of order
   $(\xi/L)^2 \sim x^2$.
   Nevertheless, for $\xi_F^{(2nd)}$ 
   one expects the
   correction $\sim\! x^2$ to be {\em extremely}\/ small, because $G_F$
   is almost exactly a free field.
   For $\xi_A^{(2nd)}$ this reasoning is no longer valid,
   but in any case we find empirically that the form \reff{fss:gen}
   gives an adequate fit over the range of interest
   ($0.1 \ltapprox x \ltapprox 1.1$).
   \protect\label{footnote_small_x}
}

Typically a fit of order $5 \le n \le 15$ is sufficient;
the required order depends on the range of $x$ values covered by the data
and on the shape of the curve.
Empirically, we increase $n$ until the $\chi^2$ of the fit
becomes essentially constant.
The resulting $\chi^2$ value provides
a check on the systematic errors arising from
corrections to scaling and/or from the inadequacies of the form \reff{fss:gen}.

The {\em statistical}\/ error on the extrapolated value of
$\scro_\infty(\beta) \equiv \scro(\beta,\infty)$ comes from three sources:
\begin{itemize}
   \item[(i)]  Error on $\scro(\beta,L)$, which gets multiplicatively
       propagated to $\scro_\infty$.
   \item[(ii)]  Error on $\xi(\beta,L)$, which affects the argument
       $x \equiv \xi(\beta,L)/L$ of the scaling functions
       $F_\xi$ and $F_\scro$.
   \item[(iii)]  Statistical error in our estimate of the coefficients
       $a_1, \ldots, a_n$ in $F_\xi$ and $F_\scro$.
\end{itemize}
The errors of type (i) and (ii) depend on the statistics available
at the single point $(\beta,L)$, while the error of type (iii) depends
on the statistics in the whole set of runs.
Errors (i)+(ii) [resp.\ (i)+(ii)+(iii)]
can be quantified by performing a Monte Carlo experiment in which
the input data at $(\beta,L)$ [resp.\ the whole set of input data]
are varied randomly within their error bars
and then extrapolated.\footnote{
   In principle, $\xi$ and $\scro$ should be generated
   from a {\em joint}\/ Gaussian with the correct covariance.   We ignored this
   subtlety and simply generated {\em independent}\/ fluctuations on
   $\xi$ and $\scro$.
}

The discrepancies between the extrapolated values from different
lattice sizes at the same $\beta$ --- to the extent that these exceed
the estimated statistical errors --- can serve as a rough estimate
of the remaining systematic errors.
More precisely,
let $\scro_i$ ($i=1,\dots,m$) be the extrapolated values
at some given $\beta$, and let $C = (C_{ij})_{i,j=1}^m$
be the estimated covariance matrix for their statistical errors.\footnote{
   This covariance matrix is computed from the auxiliary Monte Carlo
   experiment mentioned in the preceding paragraph.
   Since this $C$ is only a statistical estimate,
   the values of $\bar{\scro}$, $\bar{\sigma}$ and $\scrr$
   will vary {\em slightly}\/ from one analysis run to the next.
}
[Errors of type (iii) induce off-diagonal terms in $C$.]
Then we form the weighted average
\be
   \bar{\scro}   \;=\;
   \left( \sum\limits_{i,j=1}^m  (C^{-1})_{ij} \scro_j \right)
   \Biggl/
   \left( \sum\limits_{i,j=1}^m  (C^{-1})_{ij} \right)
   \;,
 \label{scrobar}
\ee
the error bar on the weighted average
\be
   \bar{\sigma}   \;=\;
   \left( \sum\limits_{i,j=1}^m  (C^{-1})_{ij} \right) ^{\! -1/2}
   \;,
 \label{sigmabar}
\ee
and the residual sum-of-squares
\be
   \scrr   \;=\;
   \sum\limits_{i,j=1}^m  (\scro_i - \bar{\scro}) (C^{-1})_{ij}
                          (\scro_j - \bar{\scro})
   \;.
  \label{residual}
\ee
Under the assumptions that
\begin{itemize}
   \item[(a)] the fluctuations among the $\scro_1,\ldots,\scro_m$ are purely
      statistical [i.e.\ there are {\em no}\/ systematic errors in the
      extrapolation], and
   \item[(b)]  the statistical error bars are correct,
\end{itemize}
$\scrr$ should be distributed as a $\chi^2$ random variable
with $m-1$ degrees of freedom.
Moreover, the sum of $\scrr$ over all the values of $\beta$
should be distributed as a $\chi^2$ random variable
with $\sum (m-1)$ degrees of freedom.\footnote{ 
   This latter statement is not quite correct, as it ignores the
   correlations between the various $\scro_i$ at {\em different}\/
   $\beta$, which are induced by errors of type (iii).
   [Correlations between different $\scro_i$ at the {\em same}\/
    $\beta$, which are also induced by errors of type (iii),
    {\em are}\/ included in \reff{scrobar}--\reff{residual}.]
   \protect\label{footnote24}
}
In this way, we can search for values of $\beta$ for which the extrapolations
from different lattice sizes are mutually inconsistent (``dati schifosi'');
and we can test the overall self-consistency of the extrapolations.

A figure of (de)merit of the method is the relative variance on the
extrapolated value $\scro_\infty(\beta)$,
multiplied by the computer time needed to obtain it.\footnote{
   At fixed $(\beta,L)$,
   this variance-time product tends to a constant as the CPU time
   tends to infinity.  However, if the CPU time used is too small,
   then the variance-time product can be significantly larger than
   its asymptotic value, due to nonlinear cross terms between error sources
   (i) and (ii).
}
We expect this {\em relative variance-time product}\/
[for errors (i)+(ii) only] to scale as
\be
   \hbox{RVTP}(\beta,L)   \;\approx\;
   \xi_\infty(\beta)^{d+z_{int,\scro}}
      \, G_{\scro} \Bigl( \xi_\infty(\beta)/L \Bigr)
   \;,
\label{eq:rvtp}
\ee
where $d$ is the spatial dimension
and $z_{int,\scro}$ is the dynamic critical exponent
of the Monte Carlo algorithm being used;
here $G_\scro$ is a combination of several static and dynamic
finite-size-scaling functions,
and depends both on the observable $\scro$ and on the algorithm
but not on the scale factor $s$.
As $\xi_\infty/L$ tends to zero,
we expect $G_\scro$ to diverge as $(\xi_\infty/L)^{-d}$
(since it is wasteful to use a lattice $L \gg \xi_\infty$).
As $\xi_\infty/L$ tends to infinity,
we expect $G_\scro \sim (\xi_\infty/L)^p$ for some power $p$
(see \cite{fss_greedy_fullpaper} for details).
Note that {\em the power $p$ can be either positive or negative}\/.
If $p>0$, there is an optimum value of $\xi_\infty/L$;
this determines the best lattice size at which to perform runs
for a given $\beta$.
If $p<0$, it is most efficient to use the {\em smallest}\/ lattice size
for which the corrections to scaling are negligible compared to the
statistical errors.
[Of course, this analysis neglects errors of type (iii).
 The optimization becomes much more complicated if errors of type (iii)
 are included, as it is then necessary to optimize the set of runs
 as a whole.]

%% {\bf Comment re ``elimination of volume factor'' (modulo logarithms)
%%      in AF theories (cite Brandt)!!!  But only in $d=2$???
%%      E.g. in $d=4$ LGT, do we still have $p=2$ only???}

Finally, let us note that this method can also be applied to extrapolate
the exponential correlation length (inverse mass gap)
$\xi^{(exp)}(L) = m(L)^{-1}$
defined in a cylinder $L^{d-1} \times \infty$.
For this purpose one must work in a system of size $L^{d-1} \times T$
with $T \gtapprox 6\xi^{(exp)}(\beta,L)$ (compare \cite{Luscher_91}).

\subsubsection{Theory of Error Propagation}  \label{sec5.1.2}

When the statistical error of type (iii) is neglected,
it is possible to work out analytically the theory of error propagation,
and in particular to compute the statistical error on the extrapolated values.

Let us consider first the correlation length.
The standard error-propagation formula gives 
\be
{{\rm Var} (\xi(\beta,sL))\over \xi(\beta,sL)^2}   \;=\;
\left[ 1 + {x\over F_\xi(x;s)} {\partial F_\xi(x;s)\over \partial x}
\right]^2  \,  {{\rm Var} (\xi(\beta,L))\over \xi(\beta,L)^2}
\label{xiproperrors}
\ee
where $x\equiv \xi(\beta,L)/L$.
[Here, by abuse of notation, we write ${\rm Var} (\xi(\beta,L))$
 for the variance of {\em our Monte Carlo estimate}\/ of $\xi(\beta,L)$.
 We shall use the same convention also for other observables.]
If we now introduce $z\equiv \xi_\infty(\beta)/L$
and 
\be
{\cal F}_\xi (z) \;\equiv\; z f_\xi (z)
 \label{def_calF_xi}
\ee
[so that $x = {\cal F}_\xi (z)$ and
 $F_\xi(x;s) = s {\cal F}_\xi (z/s) / {\cal F}_\xi (z)$],
we can rewrite \reff{xiproperrors} as
\be
{{\rm Var} (\xi(\beta,sL))\over \xi(\beta,sL)^2}   \;=\;
\left[ {1\over s} {{\cal F}'_\xi (z/s)\over {\cal F}_\xi (z/s)}
    {{\cal F}_\xi (z)\over {\cal F}'_\xi (z)} \right]^2
\, {{\rm Var} (\xi(\beta,L))\over \xi(\beta,L)^2}   \;.
\ee
Iterating this formula and using the relation
\be
\lim_{n\to\infty} {1\over s^n} {{\cal F}'_\xi (z/s^n)\over {\cal F}_\xi (z/s^n)}
    \;=\;  {1\over z}
\ee
(which follows from the fact that ${\cal F}_\xi (z) \approx z$ for 
$z\to 0$), we get
\be
  {{\rm Var} (\xi_\infty(\beta))\over \xi_\infty(\beta)^2}  \;=\;
  K_\xi(z)^2 \, {{\rm Var} (\xi(\beta,L))\over \xi(\beta,L)^2}   \;,
 \label{eq5.12}
\ee
where we have defined
\be
   K_\xi(z) \;\equiv\;  {{\cal F}_\xi (z) \over z {\cal F}'_\xi (z)}   \;.
 \label{def_Kxi}
\ee
It is worth noticing that the error on the extrapolated $\xi_\infty(\beta)$
is independent of the chosen scale factor $s$. 

Let us now compute the large-$z$ expansion of $K_\xi(z)$
for the case of an asymptotically free theory.
Perturbation theory (Appendix \ref{appenB.1})
%% {\bf But maybe we have to restore a little of the deleted stuff
%%   in Appendix \ref{appenB.1}, concerning the extrapolation!!!}
predicts that, for $x\to\infty$, we have
\be
{\xi_\infty(\beta)\over L}  \;=\; D 
    \left( {x\over A} \right)^{\! -2 w_1 / w_0^2}  \,
    \exp\!\left[ {1\over w_0} \left({x\over A}\right)^2\right] 
    \, \left[ 1 + O(x^{-2})\right]
    \;,
\label{pertfss}
\ee
where $w_0$ and $w_1$ are the first two coefficients of the
renormalization-group beta-function,
$A$ is a constant that depends on the explicit definition of $\xi(\beta,L)$,
and $D$ is a nonperturbative coefficient related to $\widetilde{C}_\xi$.
For $\xi^{(2nd)}_F(\beta,L)$ in the $SU(N)$ $\sigma$-model, we have
\be
   A \;=\;  \left( {N\over N^2-1} \right) ^{\! 1/2}   \;.
\ee
{}From \reff{pertfss} we can derive the large-$z$ expansion of 
${\cal F}_\xi (z)$:  we get
\be
{\cal F}_\xi (z)  \;=\; A \left[ w_0 \log {z\over D} \right]^{1/2}
   \left[1 + {w_1\over 2 w_0^2} 
    {\log \log (z/D)\over \log (z/D)} +
                   O\!\left( { 1 \over \log (z/D)} \right) \right]
 \label{largez_Fxi}
\ee
and thus 
\be
K_\xi (z) \;=\;  2 \log {z\over D} 
   \left[ 1 + O\!\left( {\log \log (z/D)\over \log (z/D)} \right) \right]
  \;.
 \label{largez_Kxi}
\ee
We conclude that the statistical errors [of types (i) + (ii)]
increase under extrapolation only logarithmically with $z$.

We still have to take account of the finite-size-scaling behavior of
the variance of the raw data point $\xi(\beta,L)$.
If for $\xi(\beta,L)$ we take
the second-moment correlation length defined in \reff{corr_len_2mom},
we have
\be
{\rm Var} (\xi(\beta,L))  \;=\;
    {1\over 64 \sin^4 (\pi/L)} \, {1\over \xi(\beta,L)^2} \,
    {\rm Var} \!\left( {\chi(\beta,L)\over F(\beta,L)}\right)
\ee
and thus, in the limit $L \gg 1$,
\be
{{\rm Var} (\xi(\beta,L)) \over \xi(\beta,L)^2}  \;=\;
  {1\over 64 \pi^4} \left( {L\over \xi(\beta,L)} \right)^4  
  {\rm Var} \!\left( {\chi(\beta,L)\over F(\beta,L)}\right)
 \;.
\ee
Let us now define the observable
\be
\Delta  \;=\; {\chi\over \< \chi\>} - {F\over \< F\>}   \;,
 \label{def_Delta}
\ee
which controls the statistical error in measurements of $\chi/F$.
%% {\bf Strictly speaking we should write everywhere $\scrm^2$ and $\scrf$
%%    in place of $\chi$ and $F$ --- but then we get confusion regarding
%%    the notation $\scrf$!!!}
We then have, for a Monte Carlo run of $N_{iter}$ iterations,
\be
{\rm Var} \!\left( {\chi(\beta,L)\over F(\beta,L)}\right)  \;=\;
 {2 \tau_{int,\Delta}(\beta,L)\over N_{iter}}  \,
{\rm var} \!\left( {\chi(\beta,L)\over F(\beta,L)}\right)
\ee
where ${\rm var} (X)$ is the {\em static}\/ variance of $X$.
In the finite-size-scaling limit we have
\begin{eqnarray}
\tau_{int,\Delta} (\beta,L) &=& \xi_\infty(\beta)^{z_{int,\Delta}} 
  \, \bar{g}_\Delta (z)  \label{def_bar_g}  \\[2mm]
{\rm var} \!\left( \Delta(\beta,L) \right) &=& 
    \bar{v}(z)           \\[2mm]
{\rm var} \!\left( {\chi(\beta,L)\over F(\beta,L)}\right) &=& 
    v(z)  \;=\; \left( {\chi(\beta,L) \over F(\beta,L)} \right) ^{\! 2}
                \bar{v}(z)
   \;=\; \left[ 1 + 4\pi^2 {\cal F}_\xi(z)^2 \right]^2  \bar{v}(z)  \quad
 \label{var_chioverF}
\end{eqnarray}
where $z_{int,\Delta}$ is a dynamic critical exponent,
and $\bar{g}_\Delta (z)$, $\bar{v}(z)$ and $v(z)$
are scaling functions.
It follows that
\be
{{\rm Var} (\xi(\beta,L)) \over \xi(\beta,L)^2}  \;=\;
  {\xi_\infty(\beta)^{z_{int,\Delta}}\over 32 \pi^4 N_{iter}}  \,
   {\bar{g}_\Delta (z) v(z) \over {\cal F}_\xi(z)^4}
 \;.
\ee
Now the total CPU time is proportional to $N_{iter} L^d$,
so the relative variance-time product for $\xi$ is
\be
\hbox{RVTP}_\xi(\beta,L) \;=\; \xi_\infty(\beta)^{d+z_{int,\Delta}} \,
   G_\xi(z)
\ee
with
\be
  G_\xi (z) \;=\; z^{-d} K_\xi(z)^2  \,\times\,
          {\bar{g}_\Delta (z) v(z) \over 32\pi^4 {\cal F}_\xi(z)^4}   \;.
\ee
Here the second factor on the right-hand side comes from the
variance of the raw data point $\xi(\beta,L)$,
while the first factor comes from the extrapolation process.

Let us now discuss the large-$z$ behavior of $G_\xi (z)$
in an asymptotically free theory.
We have already seen that $K_\xi(z)$ and ${\cal F}_\xi(z)$ increase as 
powers of $\log z$, and that $K_\xi(z)^2/{\cal F}_\xi(z)^4$
tends to a nonzero constant.
The functions $\bar{v}(z)$ and $v(z)$ are static variances,
hence in principle computable at large $z$ in perturbation theory;
we have not bothered to carry out this computation,
but we find empirically
(see Section~\ref{sec5.2.3})
%% (see Figure~\ref{fig_bar_v_AND_v})
that $\bar{v}(z)$ tends to a nonzero constant as $z \to\infty$,
and hence that $v(z) \sim \log^2 z$
[cf.\ \reff{var_chioverF}/\reff{largez_Fxi}].
Finally, for $\bar{g}_\Delta (z)$ our numerical data 
indicate that $\bar{g}_\Delta (z) \sim z^{-z_{int,\Delta}}$
for the MGMC algorithm
(see again Section~\ref{sec5.2.3});
indeed, for fixed $L$ and large $z$,
each $\tau_{int,A}(\beta,L)$ approaches a constant 
$\overline{\tau}_{int,A}(L)$ which (as expected) scales approximately as 
$\overline{\tau}_{int,A}(L) \sim L^{z_{int,A}}$. 
Putting this all together, we predict that
\be
   G_\xi(z)  \;\sim\;  z^{-(d+z_{int,\Delta})} \, (\log z)^2   \;.
 \label{RVTP_prediction_large_z}
\ee
This means that large values of $z$ are {\em vastly more efficient}\/
than small values of $z$;
at any given $\beta$,
it is most efficient to use the {\em smallest}\/ lattice size
for which the corrections to scaling are negligible compared to the
statistical errors, and the gain from doing so is {\em enormous}\/.

Let us now extend the foregoing results to generic observables.
Consider a set of observables ${\cal O}_i$
($i=1,\ldots,n$) and the relative covariance matrix 
$C_{AB}$ ($A,B=0,\ldots,n$) defined by
\begin{subeqnarray}
C_{00}(\beta,L) &=& {{\rm Var} (\xi(\beta,L))\over \xi(\beta,L)^2} \\
C_{0i}(\beta,L) = C_{i0}(\beta,L)
       &=& {{\rm Cov} (\xi(\beta,L),{\cal O}_i(\beta,L))\over 
           \xi(\beta,L){\cal O}_i(\beta,L)} \\
C_{ij}(\beta,L) &=& {{\rm Cov} ({\cal O}_i(\beta,L), {\cal O}_j(\beta,L))
     \over {\cal O}_i(\beta,L) {\cal O}_j(\beta,L) }
\end{subeqnarray}
where ${\rm Var}$ and ${\rm Cov}$ denote, as before,
the variances and covariances of our Monte Carlo estimates.
A little algebra then yields the following generalization of \reff{eq5.12}:
\be
   C(\beta,\infty) \;=\;  K(z) \, C(\beta,L) \, K(z)^{\rm T}   \;,
\label{Cmatrix}
\ee
where $K(z)$ is an $(n+1)\times(n+1)$ matrix given by
\be
\left( \begin{array}{cc}
      K_\xi ( z ) &   0    \\[1mm]
      \overline{K}_{\cal O} ( z )  &   I
   \end{array} \right)
   \;\;;
\label{Kmatrix}
\ee
here $I$ is an $n\times n$ identity matrix, and 
\be
\overline{K}_{{\cal O}_i} ( z )   \;=\;
      - \, {f'_{ {\cal O}_i} (z) \over 
            f_{ {\cal O}_i} (z) }
        \, { {\cal F}_\xi (z) \over 
             {\cal F}'_\xi (z) }
        \;\; .
\ee
For an asymptotically-free theory, if ${\cal O}_i$ is an 
observable of canonical dimension $\delta_i$ (for instance 
$\delta=2$ for the susceptibilities) and leading anomalous dimension
$\gamma_{i,0}$, we have the following asymptotic behavior as $z\to\infty$:
\be
f_{ {\cal O}_i} (z)    \;=\;
    E z^{-\delta_i} \left(\log {z\over D}\right)^{\! -\gamma_{i,0} / w_0 }
   \left[1 + O\!\left({\log\log (z/D)\over\log (z/D)}\right) \right]
\ee
where $E$ is a nonperturbative coefficient and $D$ is defined by 
\reff{pertfss}.
It follows that
\be
\overline{K}_{ {\cal O}_i}(z) \;\approx\; \delta_i K_\xi(z) \;\approx\;
   2 \delta_i \log {z\over D}
\label{KbarOas}
\ee
whenever $\delta_i \neq 0$.
In case $\delta_i=0$ (this happens, for instance, for ${\cal O} = \chi/\xi^2$),
we have instead 
\be
\overline{K}_{ {\cal O}_i}(z) \;\approx\; 
     {\gamma_{i,0}\over w_0} {1\over \log (z/D)} K_\xi(z) 
     \;\approx\; {2 \gamma_{i,0}\over w_0}
   \;.
\ee

Let us now write explicitly our result \reff{Cmatrix}/\reff{Kmatrix}
for $\hbox{\rm Var}({\cal O}_{i,\infty})$. We have
\begin{eqnarray}
{\hbox{\rm Var}({\cal O}_{i,\infty}(\beta)) \over 
                {\cal O}_{i,\infty}(\beta)^2 } &=& 
     \overline{K}_{ {\cal O}_i}(z)^2 
    {\hbox{\rm Var}({\xi}(\beta,L)) \over {\xi}(\beta,L)^2}  \,+\,
    2 \overline{K}_{ {\cal O}_i}(z) 
   {\hbox{\rm Cov}({\cal O}_i(\beta,L),\xi(\beta,L)) \over 
                   {\cal O}_i(\beta,L)\xi(\beta,L)} \nonumber \\
  && \qquad +\, {\hbox{\rm Var}({\cal O}_{i}(\beta,L)) \over
                {\cal O}_{i}^2(\beta,L)}
  \;.
\label{varcalO}
\end{eqnarray}
The last term on the right-hand side represents the error of type (i),
while the first two terms constitute the error of type (ii).
Asymptotically for large $z$, the first term dominates (unless $\delta_i=0$): 
the final statistical error on ${\cal O}_{i,\infty}(\beta)$
is controlled by the error on $\xi(\beta,L)$
and {\em not}\/ by the error on ${\cal O}_i (\beta,L)$.
In other words, the error of type (ii) dominates that of type (i).
Notice, moreover,
that \reff{varcalO} reduces to \reff{eq5.12} when 
${\cal O}_i = \xi$, since $\overline{K}_{ \xi}(z) = K_\xi(z) - 1$.

It is also immediate to verify that different observables become 
perfectly correlated for $z\to\infty$ (if their canonical dimension is 
not zero). Indeed, using \reff{Cmatrix}/\reff{Kmatrix} and 
\reff{KbarOas} we get
\be
{\hbox{\rm Cov}({\cal O}_{i,\infty}(\beta),{\cal O}_{j,\infty}(\beta))
  \over 
  [\hbox{\rm Var}({\cal O}_{i,\infty}(\beta))
   \hbox{\rm Var}({\cal O}_{j,\infty}(\beta))]^{1/2}  }
  \;=\;   1 - O\!\left({1\over \log (z/D)}\right) \;\; .
\ee
This again occurs expresses the dominance of errors of type (ii),
all of which arise from the statistical fluctuations on the {\em same}\/
random variable $\xi(\beta,L)$.

\subsection{Data Analysis: Extrapolation to Infinite Volume} \label{sec5.2}

In this subsection we apply the finite-size-scaling extrapolation
procedure to our data for the $SU(3)$ chiral model.
We begin by showing in some detail how the method works for $\xi_F^{(2nd)}$;
this allows us to illustrate the treatment of statistical and
systematic errors and to show the quality of results that can be obtained.
Then we show more briefly the results for
$\chi_F$, $\xi_A^{(2nd)}$ and $\chi_A$. Finally, we discuss the ratio
$\xi^{(2nd)}_F/\xi^{(2nd)}_A$ and the relative variance-time product.

\subsubsection{Basic Observables}   \label{sec5.2.1}

We shall always use a scale factor $s=2$.
Out of our 264 data points $(\beta,L)$, we are able to form 203 pairs
$(\beta,L)$/$(\beta,2L)$;
these pairs cover the range
$0.08 \ltapprox x \equiv \xi_F^{(2nd)}(L)/L \ltapprox 1.12$.
In what follows, we shall sometimes omit for simplicity the superscript
${}^{(2nd)}$ on the correlation lengths;
and when we write $\xi(L)$ {\em tout court}\/ we shall always mean
$\xi_F^{(2nd)}(L)$.

We found tentatively that for $\scro = \xi_F^{(2nd)}$
a thirteenth-order fit \reff{fss:gen} is indicated:
see the last few rows of Table~\ref{xiF_chisq_tab}.
We next sought to investigate the strength of the corrections to scaling:
we performed the fit with the conservative choices $L_{min} = 64$,
$\xi_{min} = 10$ and $n=13$, and plotted on an expanded vertical scale
the {\em deviations}\/ from this fit.
The results are shown in Figure~\ref{fig_su3_dev_xif}.
Clearly, there are significant corrections to scaling
in the regions $x \ltapprox 0.84$ (resp.\ 0.64, 0.52, 0.14)
when $L=8$ (resp.\ 16, 32, 64);
but the corrections to scaling become negligible
(within statistical error) at $x$ larger than this.
To take account of this $x$-dependence of the corrections to scaling,
we adopted a modified scheme for imposing lower cutoffs on $\xi(L)$ and $L$,
as follows:
For each lattice size $L$, we choose a value $x_{min}(L)$,
and we allow into the fit only those data pairs $(\beta,L)$/$(\beta,2L)$
satisfying $x \equiv \xi(L)/L \ge x_{min}(L)$.
Our method is thus specified by the five cut points
$x_{min}(8)$, $x_{min}(16)$, $x_{min}(32)$, $x_{min}(64)$, $x_{min}(128)$
along with the interpolation order $n$.
We shall always choose $x_{min}(64) = 0.14$
and $x_{min}(128) = 0$, and shall thus omit them from the tables.

We next sought to investigate systematically the $\chi^2$ of the fits,
as a function of the cut points $x_{min}(L)$ and the interpolation order $n$;
some typical results are collected in Table~\ref{xiF_chisq_tab}.
A reasonable $\chi^2$ is obtained when $n \ge 13$ and
$x_{min} \ge (0.80,0.70,0.60,0.14,0)$ for $L=(8,16,32,64,128)$.
Our preferred fit is $n=13$ and $x_{min} = (\infty,0.90,0.65,0.14,0)$:
see Figure \ref{fig_su3_fss_xif}, where we compare also with the
order-$1/x^2$ and order-$1/x^4$ perturbative predictions \reff{xiF_FSS_PT}.
This fit has $\chi^2 = 60.85$ (90 DF, level = 99.2\%).

We then used this preferred fit to extrapolate the data to infinite volume.
The extrapolated values $\xi_{F,\infty}^{(2nd)}$ from different lattice sizes
at the same $\beta$ are consistent within statistical errors:
only one of the 58 $\beta$ values has an ${\cal R}$
that is too large at the 5\% level;
and summing all $\beta$ values we have ${\cal R} = 64.28$
(103 DF, level = 99.9\%).

Both the $\chi^2$ and ${\cal R}$ values are unusually small;
we don't know why.
Perhaps we have somewhere overestimated our statistical errors by
about 25\%.

In Table~\ref{estrap_xi_fund} we show the extrapolated values
$\xi_{F,\infty}^{(2nd)}$
from our preferred fit and from some alternative fits, together with the 
propagated statistical error bars [including errors of type (i)+(ii)+(iii)].
The deviations between the different acceptable fits
(those in italics or sans-serif),
if larger than the statistical errors,
can serve as a rough estimate of the remaining systematic errors due
to corrections to scaling.
The statistical errors on $\xi_{F,\infty}^{(2nd)}$ 
in our preferred fit are of order
0.6\% (resp.\ 0.8\%, 1.2\%, 1.4\%, 1.5\%)
at $\xi_{F,\infty} \approx 10^2$
(resp.\ $10^3$, $10^4$, $10^5$, $4 \times 10^5$).
The systematic errors are smaller 
than the statistical errors (anywhere from 0.1 to 0.9 times as large)
for $\beta \ltapprox 3.60$, and slightly larger than the statistical errors
(by a factor 1--2 times as large) for $\beta \gtapprox 3.60$.
% I checked
%that few $\beta$'s have slightly larger sistematic than statistical errors
The statistical errors at different $\beta$ are strongly positively correlated.

Now we report the results for the observables
$\chi_F$, $\xi_A^{(2nd)}$ and $\chi_A$.
%% The resulting extrapolated values are reported here in
%% Tables~\ref{estrap_chi_fund}, \ref{estrap_xi_adj} and \ref{estrap_chi_adj}.

For $\scro = \chi_F$ we observed tentatively that 
a fifteenth-order fit \reff{fss:gen} is indicated:
see Table~\ref{chiF_chisq_tab}.
There are significant corrections to scaling
for all $x$ when $L=8$,
and in the regions $x \ltapprox 0.85$ (resp.\ 0.50)
when $L=16$ (resp.\ 32):
see the deviations plotted in Figure~\ref{fig_su3_dev_chif}.
Our preferred fit is $n=15$ and $x_{min} = (\infty,\infty,0.80,0.14,0)$
for $L=(8,16,32,64,128)$:
see Figure \ref{fig_su3_fss_chif}, where we compare also with
the order-$1/x^2$ and order-$1/x^4$ perturbative predictions
\reff{chiF_FSS_PT}.
This fit has $\chi^2 = 62.74$ (66 DF, level = 59.1\%).
In order to extrapolate $\chi_F(L)$ to infinite volume,
we have to know both $F_{\xi_F}(x;s)$ and $F_{\chi_F}(x;s)$;
but our preferred fit for
$\chi_F$ requires a more stringent cut in $x_{min}$ than does our
preferred fit for $\xi_F$. Therefore,
to ensure the trustworthiness of the extrapolated values $\chi_{F,\infty}$,
we enforce the more stringent cut on both observables: 
$x_{min} = (\infty,\infty,0.80,0.14,0)$. For $\xi_F$ we use the
interpolation order $n=13$, while for $\chi_F$ we use $n=15$.
%to extrapolate $\chi_F(L)$ to infinite volume.
%{\bf Is this what you did???
%   Or did you use the fit $x_{min} = (\infty,\infty,0.80,0.14,0)$
%   for BOTH observables???   And $n=13$ for $\xi$ but $n=15$ for $\chi$???
%   EXPLAIN!!!}
The extrapolated values from different lattice sizes
at the same $\beta$ are consistent within statistical errors:
only one of the 58 $\beta$ values has an ${\cal R}$
that is too large at the 5\% level;  
and summing all $\beta$ values we have
$\scrr = 58.32$ (81 DF, level = 97\%).
In Table~\ref{estrap_chi_fund} we show the extrapolated values
$\chi_{F,\infty}$
from our preferred fit and from some alternative fits.
The statistical errors on $\chi_{F,\infty}$ in our preferred fit
are of order 0.6\% (resp.\ 2.2\%, 2.9\%, 3.6\%, 4.1\%)
at $\xi_{F,\infty} \approx 10^2$
(resp.\ $10^3$, $10^4$, $10^5$, $4\times10^5$).
The systematic errors are smaller than the statistical errors
(anywhere from 0.07 to 0.8 times as large)
for $\beta \ltapprox 3.45$, and slightly larger than the statistical errors
(by a factor 1--2.4 times as large) for $\beta \gtapprox 3.60$.
%\ltapprox 3.7$, and slightly larger than
%the statistical errors for $\beta \gtapprox 3.7$.
%{\bf Check this!!!!}
% Checked it!!

For $\scro = \xi_A^{(2nd)}$ we observed tentatively that 
a thirteenth-order fit \reff{fss:gen} is indicated:
see Table~\ref{xiA_chisq_tab}.
There are significant corrections to scaling
for all $x$ when $L=8,16$ and probably also when $L=32$
(the correction is strongly negative for $x \ltapprox 0.55$
and weakly positive when $x \gtapprox 0.55$):
see the deviations plotted in Figure~\ref{fig_su3_dev_xia}.
Our preferred fit is therefore $n=13$ and
$x_{min} = (\infty,\infty,\infty,0.14,0)$ for $L=(8,16,32,64,128)$:
see Figure \ref{fig_su3_fss_xia}, where we compare also with
the order-$1/x^2$ and order-$1/x^4$ perturbative predictions \reff{xiA_FSS_PT}.
This fit has $\chi^2 = 32.82$ (50 DF, level = 97.1\%).
To extrapolate $\xi_A^{(2nd)}(L)$ to infinite volume we use the more stringent
cut $x_{min} = (\infty,\infty,\infty,0.14,0)$ for both 
$\xi_F^{(2nd)}$ and $\xi_A^{(2nd)}$. The proper order
of interpolation for this cut for both $\xi_F^{(2nd)}$ and $\xi_A^{(2nd)}$
is $n=13$ (see Tables \ref{xiF_chisq_tab} and \ref{xiA_chisq_tab}). 
%to extrapolate $\xi_A^{(2nd)}(L)$ to infinite volume
%We now use the preferred fits for $\xi_F^{(2nd)}$ and $\xi_A^{(2nd)}$
%to extrapolate $\xi_A^{(2nd)}(L)$ to infinite volume.
%{\bf Is this what you did???
%   Or did you use the fit $x_{min} = (\infty,\infty,0.80,0.14,0)$
%   for BOTH observables???
%   EXPLAIN!!!}
The extrapolated values from different lattice sizes
at the same $\beta$ are consistent within statistical errors:
only one of the 58 $\beta$ values has a $\scrr$
that is too large at the 5\% level;
and summing all $\beta$ values we have
$\scrr = 38.12$ (63 DF, level = 99.4\%).
In Table~\ref{estrap_xi_adj} we show the extrapolated values
$\xi_{A,\infty}^{(2nd)}$
from our preferred fit and from some alternative fits.
The statistical errors on $\xi_{A,\infty}^{(2nd)}$ in our preferred fit
are of order 0.5\% (resp.\ 1.3\%, 2.3\%, 2.9\%, 3.5\%)
at $\xi_{F,\infty} \approx 10^2$
(resp.\ $10^3$, $10^4$, $10^5$, $4\times10^5$).
Since our preferred fit is the most conservative one possible
(and all less conservative fits are bad),
we are unable to say anything about the systematic errors.

For $\scro = \chi_A$ we observed tentatively that
a fourteenth-order fit \reff{fss:gen} is indicated:
see Table~\ref{chiA_chisq_tab}.
There are significant corrections to scaling
for all $x$ when $L=8$,
and in the regions $x \ltapprox 0.84$ (resp.\ 0.64)
when $L=16$ (resp.\ 32):
see the deviations plotted in Figure~\ref{fig_su3_dev_chia}.
Our preferred fit is $n=14$ and $x_{min} = (\infty,\infty,0.90,0.14,0)$:
see Figure \ref{fig_su3_fss_chia}, where we compare also with
the order-$1/x^2$ and order-$1/x^4$ perturbative predictions
\reff{chiA_FSS_PT}.
This fit has $\chi^2 = 40.31$ (62 DF, level = 98.5\%).
To extrapolate $\chi_A(L)$ to infinite volume
we use more stringent fit $x_{min} = (\infty,\infty,0.90,0.14,0)$
for both  $\xi_F^{(2nd)}$ and $\chi_A$, using $n=13$ for $\xi_F^{(2nd)}$
and $n=14$ for $\chi_A$.
%We now use the preferred fits for $\xi_F^{(2nd)}$ and $\chi_A$
%to extrapolate $\chi_A(L)$ to infinite volume.
%{\bf Is this what you did???
%   Or did you use the fit $x_{min} = (\infty,\infty,0.80,0.14,0)$
%   for BOTH observables???   And $n=13$ for $\xi_F$ but $n=14$ for $\chi_A$???
%   EXPLAIN!!!}
The extrapolated values from different lattice sizes
at the same $\beta$ are consistent within statistical errors:
none of the 58 $\beta$ values has an $\scrr$
that is too large at the 5\% level;
and summing all $\beta$ values we have
$\scrr = 46.86$ (75 DF, level = 99.6\%).
In Table~\ref{estrap_chi_adj} we show the extrapolated values $\chi_{A,\infty}$
from our preferred fit and from some alternative fits.
The statistical errors on $\chi_{A,\infty}$ in our preferred fit
are of order 0.3\% (resp.\ 1.6\%, 3.1\%, 3.9\%, 4.3\%)
at $\xi_{F,\infty} \approx 10^2$
(resp.\ $10^3$, $10^4$, $10^5$, $4\times10^5$).
The systematic errors are smaller than the statistical errors
(anywhere from 0.05 to 0.5 times as large)
for $\beta \ltapprox 3.825$, and comparable to the statistical errors
(anywhere from 0.75 to 1.9 times as large) for $\beta \gtapprox 3.90$.

We also extrapolated the quantities 
$\chi_F/(\xi^{(2nd)}_F)^2$ and $\chi_A/(\xi^{(2nd)}_A)^2$.
The reason for doing these extrapolations is that the
errors in the infinite-volume estimates of the ratios
are much smaller (at least 15 times for the fundamental sector and
7 times for the adjoint sector) than those obtained by direct
extrapolation of numerator and denominator assuming independent
errors. Besides, knowing the covariance of the statistical fluctuations on 
our estimates of $\chi(\beta,L)$ and $\xi^{(2nd)}(\beta,L)$,
we can compute correctly the error bars of the extrapolated ratios;
by contrast, if we extrapolate $\chi$ and $\xi^{(2nd)}$ separately,
we are obliged either to make the false assumption of independent errors
or else involve the triangle inequality --- both of which lead to
error bars that are gross overestimates.  In any case, we observe
that the central values are consistent within error bars with those 
obtained by separate extrapolation of the numerator and denominator.

%{\bf Discuss briefly
%     $\chi_F/(\xi^{(2nd)}_F)^2$ and $\chi_A/(\xi^{(2nd)}_A)^2$????
%     If we use these later, then we should at least SAY what we did,
%     even if we don't give detailed tables.  We could mention e.g.
%     that the results are consistent with those obtained by direct
%     extrapolation of numerator and denominator (IS THIS TRUE???),
%     but the error bars are slightly smaller (IS THIS TRUE???).
%}

For $\scro = \chi_F/(\xi^{(2nd)}_F)^2$ we observed tentatively that
a thirteenth-order fit \reff{fss:gen} is indicated.
%see Table~\ref{chiA_chisq_tab}.
There are significant corrections to scaling
for all $x$ when $L=8,16$,
and in the regions $x \ltapprox 0.6$ and $x \gtapprox 0.8$ 
when $L=32$.
%see the deviations plotted in Figure~\ref{fig_su3_dev_chia}.
Then, our preferred fit is $n=13$ and 
$x_{min} = (\infty,\infty,\infty,0.14,0)$.
%see Figure \ref{fig_su3_fss_chia}, where we compare also with
%the order-$1/x^2$ and order-$1/x^4$ perturbative predictions
%\reff{chiA_FSS_PT}.
This fit has $\chi^2 = 22.37$ (50 DF, level = 99.7\%).
To extrapolate $\chi_F/(\xi^{(2nd)}_F)^2$ to infinite volume
we use the more stringent fit $x_{min} = (\infty,\infty,\infty,0.14,0)$ 
for both $\xi_F^{(2nd)}$ and $\chi_F/(\xi^{(2nd)}_F)^2$,
using $n=13$ for both.
The extrapolated values from different lattice sizes
at the same $\beta$ are consistent within statistical errors:
no one of the 58 $\beta$ values has a $\scrr$
that is too large at the 5\% level;
and summing all $\beta$ values we have
$\scrr = 36.38$ (63 DF, level $>99.9\%$).
%In Table~\ref{estrap_xi_adj} we show the extrapolated values
%$\xi_{A,\infty}^{(2nd)}$
%from our preferred fit and from some alternative fits.
The statistical errors on $\chi_F/(\xi^{(2nd)}_F)^2$ in our preferred fit
are of order 0.4\% (resp.\ 0.5\%, 0.6\%, 0.7\%, 0.7\%)
at $\xi_{F,\infty} \approx 10^2$
(resp.\ $10^3$, $10^4$, $10^5$, $4\times10^5$).
Since our preferred fit is the most conservative one possible
(and all less conservative fits are bad),
we are unable to say anything about the systematic errors.

For $\scro = \chi_A/(\xi^{(2nd)}_A)^2$ we observed tentatively that
a sixteenth-order fit \reff{fss:gen} is indicated.
%see Table~\ref{chiA_chisq_tab}.
There are strong corrections to scaling
for all $x$ when $L=8,16,32$,
%see the deviations plotted in Figure~\ref{fig_su3_dev_chia}.
so our preferred fit is $n=16$ and
$x_{min} = (\infty,\infty,\infty,0.14,0)$.
%see Figure \ref{fig_su3_fss_chia}, where we compare also with
%the order-$1/x^2$ and order-$1/x^4$ perturbative predictions
%\reff{chiA_FSS_PT}.
This fit has $\chi^2 = 18.92$ (50 DF, level $> 99.9\%$).
To extrapolate $\chi_A/(\xi^{(2nd)}_A)^2$ to infinite volume
we use the more stringent fit $x_{min} = (\infty,\infty,\infty,0.14,0)$
for both $\xi_F^{(2nd)}$ and $\chi_A/(\xi^{(2nd)}_A)^2$, using 
$n=13$ for $\xi_F^{(2nd)}$ and $n=16$ for $\chi_A/(\xi^{(2nd)}_A)^2$.
The extrapolated values from different lattice sizes
at the same $\beta$ are consistent within statistical errors:
no one of the 58 $\beta$ values has a $\scrr$
that is too large at the 5\% level;
and summing all $\beta$ values we have
$\scrr = 29.38$ (63 DF, level $>99.9\%$).
%In Table~\ref{estrap_xi_adj} we show the extrapolated values
%$\xi_{A,\infty}^{(2nd)}$
%from our preferred fit and from some alternative fits.
The statistical errors on $\chi_A/(\xi^{(2nd)}_A)^2$ in our preferred fit
are of order 0.7\% (resp.\ 0.9\%, 1.2\%, 1.4\%, 1.4\%)
at $\xi_{F,\infty} \approx 10^2$
(resp.\ $10^3$, $10^4$, $10^5$, $4\times10^5$).
Since our preferred fit is the most conservative one possible
(and all less conservative fits are bad),
we are unable to say anything about the systematic errors.

\subsubsection{Ratio $\xi_F^{(2nd)}(L)/\xi_A^{(2nd)}(L)$} \label{sec5.2.2}

\begin{figure}
\vspace*{0cm} \hspace*{-0cm}
\begin{center}
\epsfxsize = 0.9\textwidth
\leavevmode\epsffile{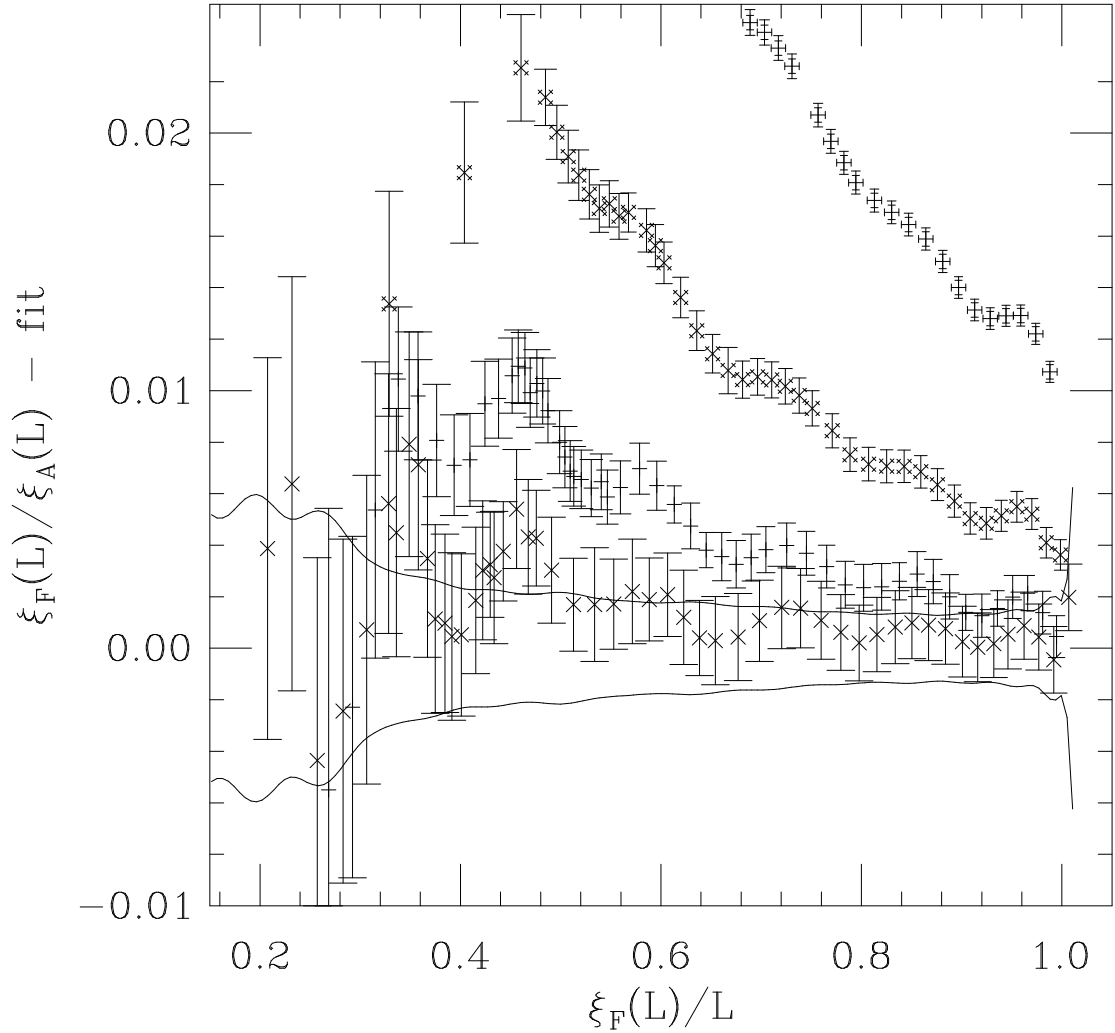}
%%\quad \vspace{5cm}  %% TEMPORARY UNTIL FILE IS THERE
\end{center}
\vspace*{-2cm}
\caption{
   Deviation of points from fit \protect\reff{fit_xi_ratio_exp}
   to $\xi_F(L)/\xi_A(L)$ with $n=15$,
   $\xi_{min}=10$ and $L_{min} = 128$. Note the difference between this fit
   and previous ones: here we plot the finite-size-scaling curve
   for the ratio of $\xi_F$ and $\xi_A$ at the {\it same}\/ $L$.
   Symbols indicate $L=8$ ($\protect\fancyplus$), 16 ($\protect\fancycross$),
   32 ($+$), 64($\times$).
   Error bars are one standard deviation.
   Curves near zero indicate statistical error bars
   ($\pm$ one standard deviation) on the fitting function.
}
\label{fig_su3_dev_xif_ov_xia}
\end{figure}

\begin{figure}
\vspace*{0cm} \hspace*{-0cm}
\begin{center}
\epsfxsize = 0.9\textwidth
\leavevmode\epsffile{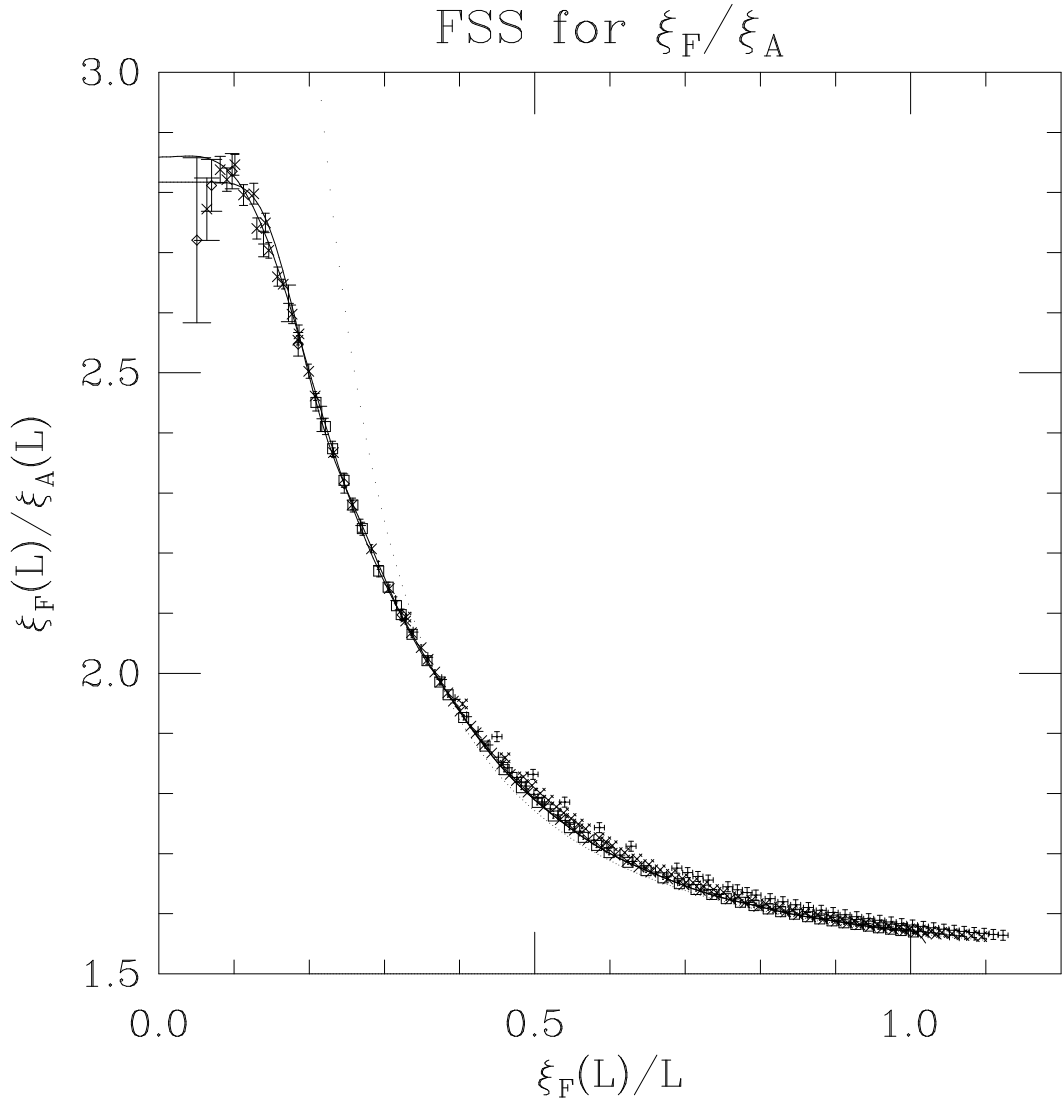}
%%\quad \vspace{5cm}  %% TEMPORARY UNTIL FILE IS THERE
\end{center}
\vspace*{-2cm}
\caption{
   $\xi_F(\beta,L)/\xi_A(\beta,L)$ versus $\xi_F(\beta,L)/L$.
   Symbols indicate $L=8$ ($\protect\fancyplus$), 16 ($\protect\fancycross$),
   32 ($+$), 64 ($\times$), 128 ($\Box$), 256 ($\Diamond$).
   Error bars are one standard deviation.
   Solid curve reaching 2.817 at $x=0$
   is a fourteenth-order fit in (\protect\ref{fit_xi_ratio_exp}),
   with $L_{min} = 128$, $\xi_{min} = 10$.
   %%%  $x_{min} = (\infty,\infty,\infty,\infty,0,0)$
   %%% for $L = (8,16,32,64,128,256)$.
   Solid curve reaching 2.859 at $x=0$
   is a sixteenth-order fit in (\protect\ref{fit_xi_ratio_pol}),
   also with $L_{min} = 128$, $\xi_{min} = 10$.
   Dotted curve is the perturbative prediction
   (\protect\ref{xiFoverxiA_FSS_PT}) through order $1/x^2$.
}
\label{fig_su3_fss_xif_ov_xia}
\end{figure}

In this subsection we discuss the finite-size-scaling curve for the ratio
$\xi_F^{(2nd)}(L)/\xi_A^{(2nd)}(L)$.
We fit to the Ansatz
\be
   {\xi_F^{(2nd)}(L) \over \xi_A^{(2nd)}(L)}   \;=\;
   a_0 + a_1 e^{-1/x} + a_2 e^{-2/x} + \ldots + a_n e^{-n/x}   \;,
 \label{fit_xi_ratio_exp}
\ee
using $n = 15$, $L_{min} = 128$ and $\xi_{min} = 10$.
There are strong corrections to scaling
for all $x$ when $L=8,16,32$ (Figure~\ref{fig_su3_dev_xif_ov_xia}): 
these corrections are of positive sign and behave roughly as $L^{-\Delta}$ 
with $1 \ltapprox \Delta \ltapprox 2$. For $L=64$ these corrections to
scaling are on the borderline of statistical significance, but the fact
that they are nearly all positive (and are of the magnitude expected
from extrapolation of the $L=32$ corrections) suggests that they are
real.\footnote{
   Moreover, we have here treated the Monte Carlo data for
   $\xi_F^{(2nd)}(L)$ and $\xi_A^{(2nd)}(L)$ as if they were
   {\em independent}\/ random variables.
   In fact they are presumably {\em positively}\/ correlated,
   which means that we have overestimated the error bar on the ratio.
   So the corrections to scaling are in fact more statistically
   significant than they appear to be.
}
For all these reasons, we have chosen $L_{min}=128$.
%{\bf Is this in fact what you did???
%    Was it NECESSARY to use $L_{min} = 128$???
%    If so, explain why, at least briefly in words.
%    Let's look at a $\chi^2$ table, at least for ourselves.}
%%%  and $x_{min} = (\infty,\infty,\infty)$
The resulting fit is shown in Figure~\ref{fig_su3_fss_xif_ov_xia}
(lower solid curve);
it has $\chi^2 = 21.45$ (47 DF, level $> 99.9\%$).\footnote{
   If in fact we have overestimated the error bar on the ratio,
   then we have underestimated the $\chi^2$ of the fit.
   This explains the unusually low value of $\chi^2$.
}
We thus estimate the limiting value as the value of $a_0$
\be
   {\xi_{F,\infty}^{(2nd)}  \over  \xi_{A,\infty}^{(2nd)}}
   \;=\;
   a_0
   \;=\;   2.817  \pm 0.001
  \label{eqxiratio}
\ee
(68\% confidence interval, statistical errors only).
This estimate needs to be accompanied by one caveat:
the paucity of our data with $L \le 128$ in the region $x \ltapprox 0.08$
makes the fit extremely sensitive to the {\em assumed}\/ behavior
at small $x$. 
%{\bf I think this does not apply. We have a lot of
%equally dense data for $x>0.08$!!!}
Now, for $\xi_A^{(2nd)}$
(and hence also for the ratio $\xi_A^{(2nd)}/\xi_F^{(2nd)}$)
there {\em may}\/ be significant finite-size corrections of order $x^2$
at small $x$
(see footnote \ref{footnote_small_x} in Section \ref{sec5.1} above),
which could be much larger than the $e^{-1/x}$ corrections assumed here.
So we tried the alternative Ansatz
\be
   {\xi_F^{(2nd)}(L) \over \xi_A^{(2nd)}(L)}   \;=\;
   a'_0 + a'_1 x^2 + a'_2 x^4 + \ldots + a'_n x^{2n}   \;.
 \label{fit_xi_ratio_pol}
\ee
Using $n=16$, $L_{min} = 128$ and $\xi_{min} = 10$
we obtain an equally good fit 
($\chi^2 = 13.86$, 46 DF, $\hbox{level} > 99.9 $\%),
which is shown as the upper solid curve in
Figure~\ref{fig_su3_fss_xif_ov_xia}.
Note the different value of the limiting constant:
\be
   {\xi_{F,\infty}^{(2nd)}  \over  \xi_{A,\infty}^{(2nd)}}
   \;=\;
   a'_0  \;=\;  2.859 \pm 0.002
    \;.
  \label{eqxiratio_power}
\ee
%%%%   This value seems much too high (see below).
%
%A, but it does not
%seem highly significant considering the strong dependence
%on the order of interpolation.
%We also must nevertheless remain open to the possibility that there are
%significant corrections of order $x^2$,
%which could alter the estimate \reff{eq3a1}. 
%[Note, however, that the $N=\infty$ solution \reff{xiVoverxiT_FSS_Ninf_smallx}
% has a small {\em upward}\/ curvature at small $x$,
% namely $a'_1 = \sqrt{6}\pi^2/15 \approx 1.61$,
% in contrast to the strong {\em downward}\/ curvature
% predicted by the fit \reff{fit_xi_ratio_pol},
% which has $a'_1 = -29.59$.
% This is another reason to be skeptical of the latter fit.]

An alternative way of estimating the universal ratio
${\xi_{F,\infty}^{(2nd)} / \xi_{A,\infty}^{(2nd)}}$
is to use the separately extrapolated values for
$\xi_{A,\infty}^{(2nd)}$ and $\xi_{F,\infty}^{(2nd)}$
(Tables~\ref{estrap_xi_fund} and \ref{estrap_xi_adj})
and simply form the ratio. Note that the deviations
from constancy in ${\xi_{F,\infty}^{(2nd)} / \xi_{A,\infty}^{(2nd)}}$
are corrections to {\it scaling} (not to asymptotic scaling), and
thus fall off as an inverse power of $\xi_{F,\infty}$
(most likely $\xi_{F,\infty}^{-2}$). 
Experience with other similar quantities suggests that good
scaling will be observed for $\xi_{F,\infty} \gtapprox 10$
(i.e.\ $\beta \gtapprox 1.80$) or even smaller.
Surprisingly, this does {\it not} occur here: if we use
all data with $\xi_{F,\infty} \ge 10$ (i.e.\ $\beta \ge 1.80$),
we obtain the estimate
\be
   {\xi_{F,\infty}^{(2nd)}  \over  \xi_{A,\infty}^{(2nd)}}
   \; = \;
   2.8111 \pm 0.0023
   \;,
 \label{eq3a3bad}
\ee
but with a very poor goodness of fit ($\chi^2 = 127.68$, 55 DF, 
$\hbox{level}=10^{-7}$). In order to obtain a reasonable $\chi^2$, 
we have to restrict the fit to $\xi_{F,\infty} \ge 70$
(i.e.\ $\beta \ge 2.25$): we then get
\be
   {\xi_{F,\infty}^{(2nd)}  \over  \xi_{A,\infty}^{(2nd)}}
   \; = \;
   2.798 \pm 0.006
 \label{eq3a3}
\ee
with $\chi^2 = 16.28$, 31 DF, $\hbox{level}=98.7$\%.
The discrepancy between \reff{eq3a3bad} and \reff{eq3a3} appears to be
a real correction to scaling: its magnitude is very small ($\approx\! 0.013$)
and is consistent with a correction term $A \xi_{F,\infty}^{-2}$
with $A\sim 1$. We have two possible explanations for the 
horrible $\chi^2$ in \reff{eq3a3bad}:
\begin{itemize}
\item We have a large number of data points, each of which has a very 
  small error bar; so very small corrections to scaling {\it can} became 
  statistically significant.
\item The extrapolated values $\xi_{F,\infty}^{(2nd)} / \xi_{A,\infty}^{(2nd)}$
  at different $\beta$ are presumably positively correlated as a result 
  of errors of type (iii) in the extrapolation, but we {\it not} taken 
  account of this correlation here (see footnote \ref{footnote24});
  this could be causing the $\chi^2$ to appear larger than it really is.
  [On the other hand, we have overestimated the error bar on the
   ratio $\xi_{F,\infty}^{(2nd)} / \xi_{A,\infty}^{(2nd)}$ 
   by assuming {\it independent} errors on 
   $\xi_{F,\infty}^{(2nd)}$ and $\xi_{A,\infty}^{(2nd)}$, when in
   fact they are probably positively correlated; this would cause
   the $\chi^2$ to appear {\it smaller} than really is.]
\end{itemize}
In any case, the magnitude of this correction-to-scaling effect is very small,
and we can simply fold the uncertainties into an enlarged error bar.
One possible advantage of this method over the preceding one
is that in the fit \reff{fss:gen} to
$\xi_A^{(2nd)}(2L)/\xi_A^{(2nd)}(L)$ we {\em know}\/ the correct value at $x=0$
--- namely, 1 ---
in contrast to the fit \reff{fit_xi_ratio_exp} where $a_0$ is unknown.
As a result, the former fit is somewhat less sensitive to the assumed form
of the small-$x$ corrections:
if in Figures~\ref{fig_su3_fss_xif} and \ref{fig_su3_fss_xia}
we had fit to powers of $x^2$
instead of powers of $e^{-1/x}$,
the resulting curve would have changed only slightly.
%{\bf Is this true??? Check it!!!}
% I checked !!!

\begin{figure}
\vspace*{0cm} \hspace*{-0cm}
\begin{center}
\epsfxsize = 0.9\textwidth
\leavevmode\epsffile{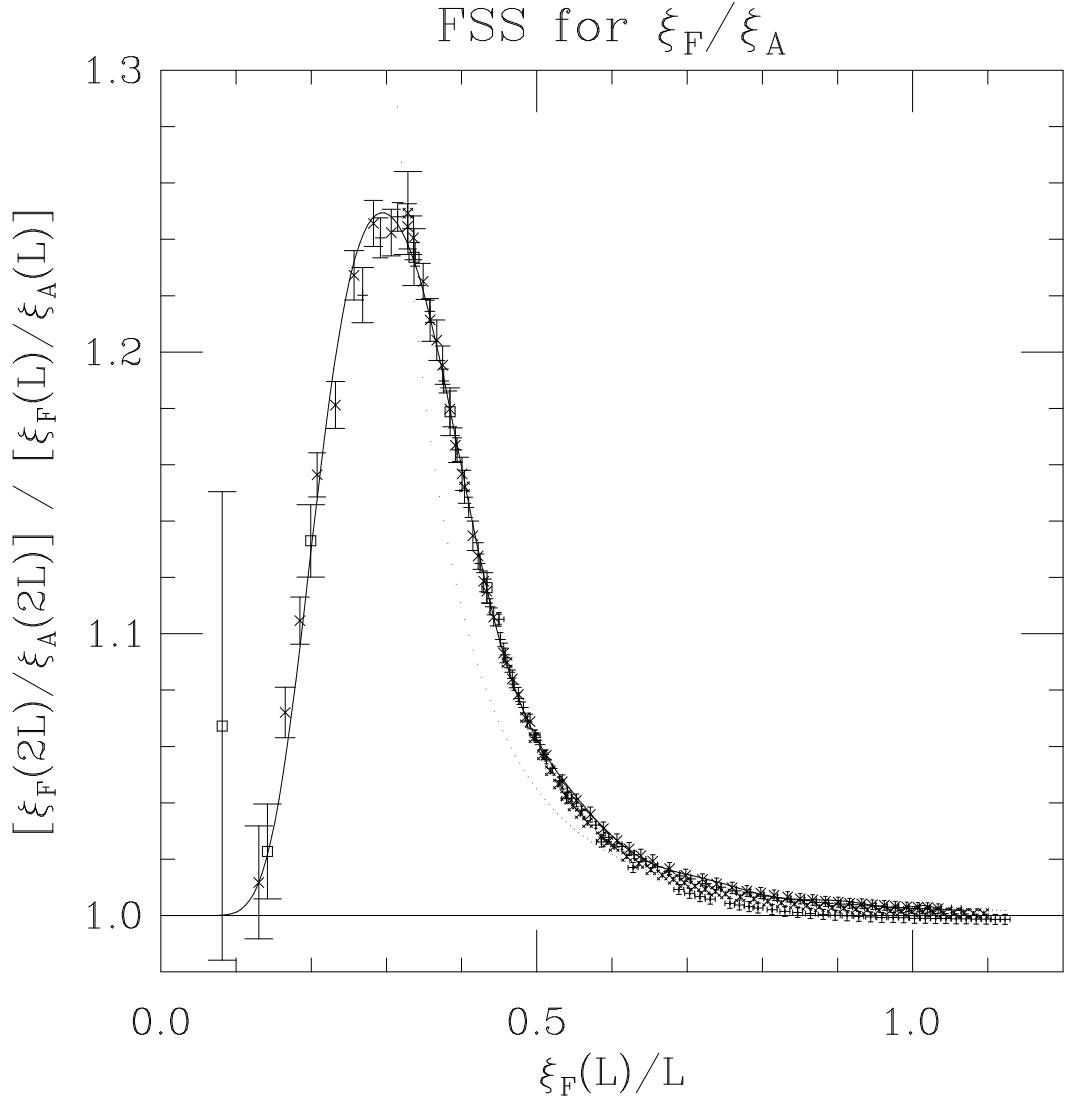}
%%\quad \vspace{5cm}  %% TEMPORARY UNTIL FILE IS THERE
\end{center}
\vspace*{-2cm}

\caption{
   $R_{\xi}(\beta,2L)/R_{\xi}(\beta,L)$ versus
   $\xi_F^{(2nd)}(\beta,L)/L$, where
   $R_{\xi} \equiv \xi_F^{(2nd)}/\xi_A^{(2nd)}$.
   Symbols indicate $L=8$ ($\protect\fancyplus$), 16 ($\protect\fancycross$),
   32 ($+$), 64 ($\times$), 128 ($\Box$).
   Error bars are one standard deviation.
   Solid curve is a twelve-order fit in \protect\reff{fss:gen},
   with $x_{min} = (\infty,\infty,\infty,0.14,0)$ for
   $L = (8,16,32,64,128)$.
   Dotted curve is the perturbative prediction
   (\protect\ref{all_FSS_PT}a,b) through
   order $1/x^4$ [at order $1/x^2$ it is identically 1].
}
\label{fig_su3_fss_xif_ov_xia_extrap}
\end{figure}

Yet a third way of estimating
${\xi_{F,\infty}^{(2nd)} / \xi_{A,\infty}^{(2nd)}}$
is to treat the ratio ${\xi_{F}^{(2nd)} / \xi_{A}^{(2nd)}}$
as an observable $\scro$ in its own right,
and perform the fit \reff{fss:gen} to $\scro(2L)/\scro(L)$
directly on it.
This procedure is very similar to the preceding one,
but has the advantage that the errors of type (iii) in the extrapolation
--- which are particularly important for the points at larger $\beta$ ---
are likely to partially cancel between $\xi_{F}^{(2nd)}$ and $\xi_{A}^{(2nd)}$.
There are significant corrections to scaling for all $x$ when $L=8,16$;
and for $L=32$ the corrections to scaling are positive and at least
0.5 standard deviations. Having
presumably overestimated the error bars (see footnote \ref{footnote24}),
we assume that also the corrections to scaling are significant
also for $L=32$. We therefore choose $x_{min} = (\infty,\infty,\infty)$,
and use $n=12$: the resulting fit has $\chi^2 = 25.06$ (51 DF, level = 99.9\%)
and is shown in Figure~\ref{fig_su3_fss_xif_ov_xia_extrap}.
The extrapolated values from different lattice sizes
at the same $\beta$ are consistent within statistical errors:
two of the 58 $\beta$ values have a $\chi^2$
too large at the 5\% level;
and summing all $\beta$ values we have
$\scrr = 35.66$ (63 DF, level = 99.8\%).
Comparing the estimates of ${\xi_{F,\infty}^{(2nd)} / \xi_{A,\infty}^{(2nd)}}$
from different $\beta$ for consistency with a constant,
we find a very large $\chi^2$ (confidence level $<2$\%) 
no matter what cutoff $\beta_{min}$ we impose. 
Presumably these discrepancies arise from the correction to scaling in the
$L=64$ points, which we discarded in the fits \reff{fit_xi_ratio_exp} 
and \reff{fit_xi_ratio_pol} but
cannot afford to discard here. For this reason we believe the result
obtained by this approach
\be
   {\xi_{F,\infty}^{(2nd)}  \over  \xi_{A,\infty}^{(2nd)}}
   \;\approx\;
   2.79 \pm 0.01
   \;,
 \label{eq3a4}
\ee
to be less reliable than the estimate \reff{eq3a3}.
%{\bf Make sure to check the RELATIVE $\chi^2/DF$
%   for different orders of interpolation:
%   since $\xi_{A}^{(2nd)}$ and $\xi_{F}^{(2nd)}$ have opposite signs of
%   corrections, an interpolation of order $n>10$ may be needed;
%   and the ABSOLUTE $\chi^2$ are meaningless as a signal, because
%   we've overestimated the error bars on the raw-data ratio.}

It is not clear to us whether
\reff{eqxiratio}, \reff{eqxiratio_power} or \reff{eq3a3}
%%% or \reff{eq3a4}
is the more reliable estimate.
A reasonable compromise would be to take
\be
   {\xi_{F,\infty}^{(2nd)}  \over  \xi_{A,\infty}^{(2nd)}}
   \; \approx \;
    2.82 \pm 0.05
 \label{eq3a5}
\ee
as our ``best estimate'';
here we have increased the error bar to take account of the
systematic uncertainties in the extrapolation.

\subsubsection{Relative Variance-Time Product}   \label{sec5.2.3}

Finally, let us discuss the efficiency of our extrapolation method for
this model, as reflected in the scaling behavior \reff{eq:rvtp} of
the relative variance-time product (RVTP).
We would like to test the theoretical predictions presented
in Section~\ref{sec5.1.2},
and in particular to determine the scaling functions
${\cal F}_\xi(z)$, $K_\xi(z)$, $\bar{v}(z)$, $v(z)$,
$\bar{g}_\Delta(z)$ and $G_\xi(z)$ arising in that theory.

The functions ${\cal F}_\xi(z)$ and $K_\xi(z)$
[defined in \reff{def_calF_xi}/\reff{def_Kxi}]
can be easily obtained from the
fitted finite-size-scaling function $F_\xi(x;2)$.
Indeed, using the obvious recursion relation
\be
F_\xi(x;s^2)  \;=\;  F_\xi(F_\xi(x;s)/s;s) \, F_\xi(x;s)
\ee
one can compute numerically $F_\xi(x;\infty)$.
Then, from $z = x F_\xi(x;\infty)$, one determines $x={\cal F}_\xi(z)$
and thence $K_\xi(z)$.
%%%  {\bf Show graphs of ${\cal F}_\xi(z)$ and $K_\xi(z)$?????}
Of course, the functions ${\cal F}_\xi(z)$ and $K_\xi(z)$
have the predicted logarithmic growths
\reff{largez_Fxi}/\reff{largez_Kxi} at large $z$,
because $F_\xi(x;2)$ has the predicted perturbative behavior at large $x$.
 
Next we determined the functions $\bar{v}(z)$ and $v(z)$
controlling the static variance of the observable $\Delta$
[defined in \reff{def_Delta}]:
see Figure~\ref{fig_bar_v_AND_v}.
Again we observe an excellent scaling,
modified only by small corrections to scaling for the smallest lattices.
We see that $\bar{v}(z)$ tends to a nonzero constant as
$z\to\infty$, while $v(z) \sim \log^2 z$.
(A plot of $v(z)^{1/2}$ versus $\log z$ shows an excellent straight line
 at large $z$.)

\begin{figure}
\vspace*{0cm} \hspace*{-0cm}
\begin{center}
\epsfxsize = 0.9\textwidth
\leavevmode\epsffile{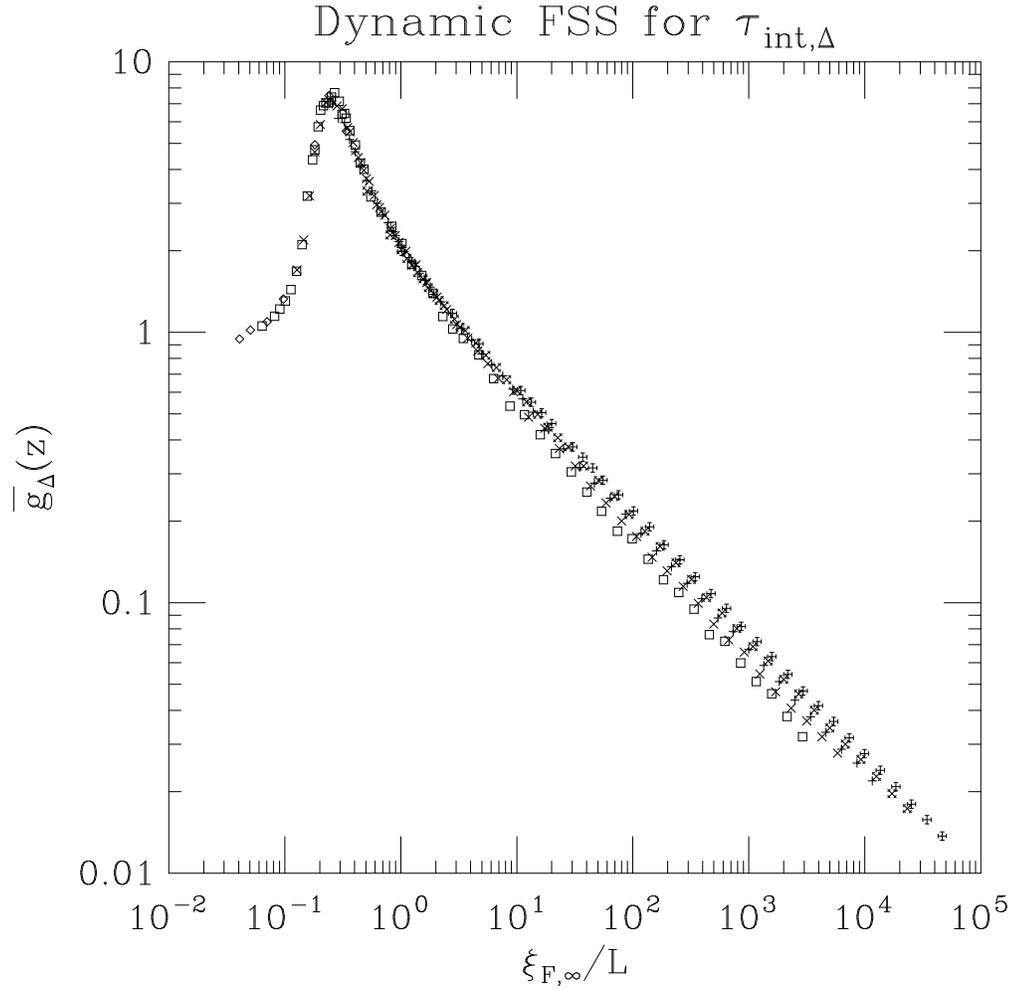}
%%\quad \vspace{5cm}  %% TEMPORARY UNTIL FILE IS THERE
\end{center}
\vspace*{-2cm}
\caption{
   The dynamic scaling function $\bar{g}_\Delta(z)$
   plotted versus $z \equiv \xi_{F,\infty}/L$,
   for the $SU(3)$ chiral model using the MGMC algorithm.
   Here $z_{int,\Delta} = 0.45$.
   Symbols indicate $L=8$ ($\protect\fancyplus$), 16 ($\protect\fancycross$),
   32 ($+$), 64 ($\times$), 128 ($\Box$), 256 ($\Diamond$).
   Note that $\bar{g}_\Delta(z) \sim z^{-0.45}$ as $z \to\infty$.
}
\label{fig_bar_g}
\end{figure}

Next we studied $\bar{g}_\Delta(z)$
[defined in \reff{def_bar_g}],
which is the dynamic finite-size-scaling function
for the autocorrelation time $\tau_{int,\Delta}$
in the MGMC algorithm.
We varied the dynamic critical exponent $z_{int,\Delta}$
until we got a good fit:
see Figure~\ref{fig_bar_g},
where we have taken $z_{int,\Delta} = 0.45$.
We observe an excellent scaling, albeit with moderately strong
corrections to scaling for the smallest lattices at large $z$.
The large-$z$ behavior is approximately
$\bar{g}_\Delta(z) \sim z^{-0.45}$, as predicted.

\begin{figure}
\vspace*{0cm} \hspace*{-0cm}
\begin{center}
\epsfxsize = 0.9\textwidth
\leavevmode\epsffile{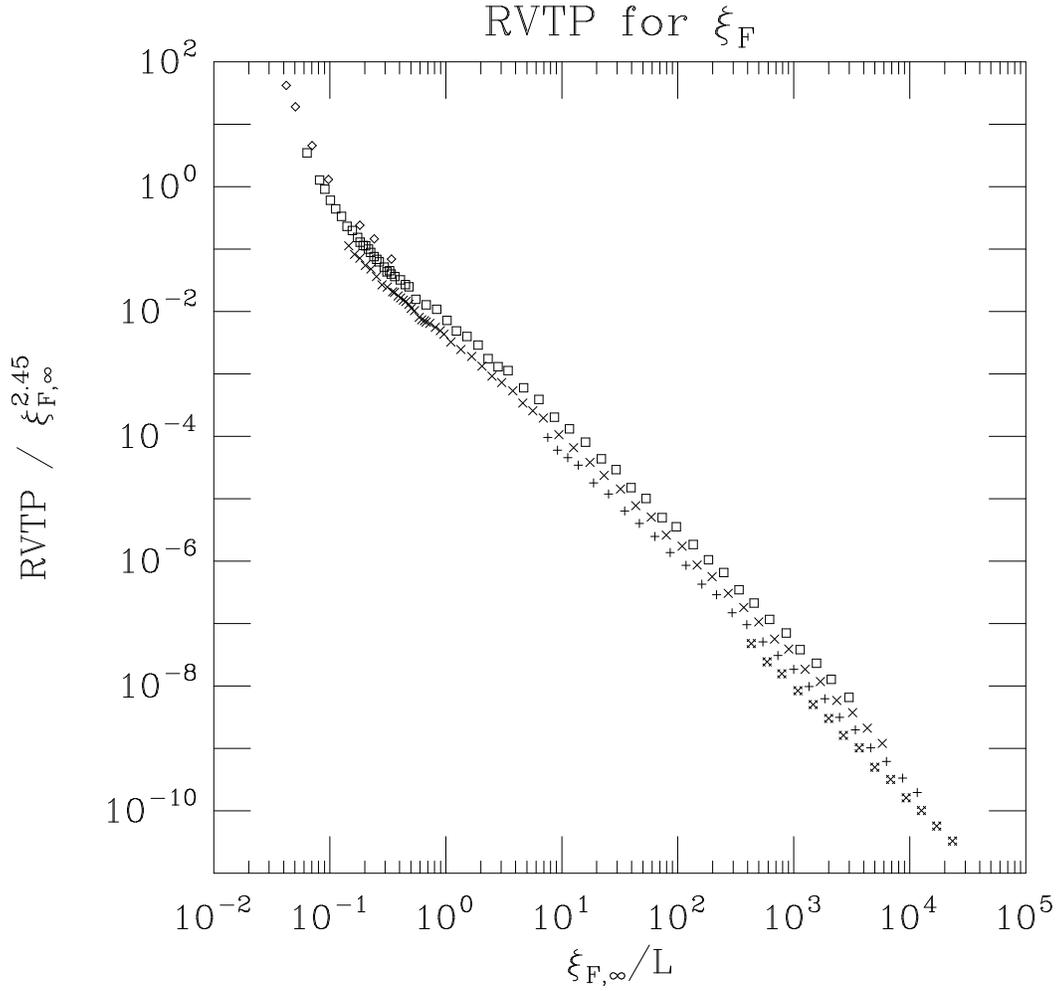}
%%\quad \vspace{5cm}  %% TEMPORARY UNTIL FILE IS THERE
\end{center}
\vspace*{-2cm}
\caption{
   Scaling plot \protect\reff{eq:rvtp} for the relative variance-time
   product (RVTP), obtained using the fit $F_{{\xi}_F}$ with $s=2$, $n=13$,
   $x_{min} = (\infty,0.90,0.65,0.14,0)$.
   Symbols indicate $L=16$ ($\protect\fancycross$),
   32 ($+$), 64 ($\times$), 128 ($\Box$), 256 ($\Diamond$).
}
\label{fig_rvtp}
\end{figure}

Finally, we determined the RVTP scaling function $G_\xi(z)$ using the relation
\be
G_\xi(z)  \;=\;
 {N_{iter} K_\xi(z)^2\over z^2 \xi_\infty(\beta)^{z_{int,\Delta}} } \,
   {\hbox{\rm Var}(\xi(\beta,L))\over \xi(\beta,L)^2}
 \;,
\ee
where $\hbox{\rm Var}(\xi(\beta,L))$ is the variance of
our Monte Carlo estimate of $\xi(\beta,L)$
as obtained from a run of $N_{iter}$ iterations.
The resulting function $G_\xi(z)$ for $z_{int,\Delta}=0.45$ is shown in 
Figure~\ref{fig_rvtp}.
The scaling is reasonably good, though far from perfect.
The large-$z$ behavior is in fairly good agreement with the prediction
\reff{RVTP_prediction_large_z} that $G_\xi(z) \sim z^{-2.45} (\log z)^2$,
but there are some discrepancies:
indeed, a somewhat better fit at large $z$ is obtained
with $z^{-2.45} (\log z)^4$.
It is therefore possible that our analysis in Section \ref{sec5.1.2}
has somewhere overlooked an additional source of logarithms.

As a practical matter, the rapid decrease of $G_\xi(z)$
means that runs at $\xi_\infty/L \sim 10^4$ using the extrapolation method
are roughly a factor $10^9$ more efficient
[as regards statistical errors of types (i) + (ii)]
than the traditional approach using runs at $\xi_\infty/L \approx 1/6$.

\subsection{Data Analysis: Comparison with Perturbation Theory}   \label{sec5.3}

Let us now compare our data with the predictions of weak-coupling
perturbation theory, and in particular with the asymptotic-freedom scenario.
In Section \ref{sec5.3.1} we look at the local quantities (viz.\ the energies).
In Section \ref{sec5.3.2} we compare the raw (finite-$L$) data for the
long-distance quantities (correlation lengths and susceptibilities)
with the predictions of finite-volume perturbation theory
[cf.\ \reff{all_finiteV_PT}].
Finally, in Sections \ref{sec5.3.3}--\ref{sec5.3.6} we compare the
extrapolated ($L=\infty$) data for the
long-distance quantities with the asymptotic-freedom predictions
\reff{xi_predicted2}--\reff{lambda_parameter}.

\subsubsection{Local Quantities}   \label{sec5.3.1}

%{\bf Discuss finite-size effects first!!!!!  Were these graphs simply using
%   the largest lattice for each $\beta$???  Or are the finite-size
%   corrections too small to see on the graphs???
%   We should say quantitatively how big they are!!!}

We can compare the fundamental energy $E_F$ with the one-loop, two-loop
and three-loop perturbative predictions \reff{energy_F},
and the adjoint energy $E_A$ with the one-loop and two-loop
predictions \reff{energy_A}.
In each case we use the value measured on the largest lattice available
(which is usually $L=128$);
we define the error bar to be the statistical error (one standard deviation)
on the largest lattice
{\em plus}\/ the discrepancy between the values on the largest and
second-largest lattices (this is a conservative estimate of the
systematic error due to finite-size effects).
For $E_F$ the finite-size corrections are between
0.000035 and 0.000132 (10--20 times larger than the statistical errors)
for $\beta \gtapprox 2.60$, and between 0.0001 and 0.0005
(20--50 times larger than the statistical errors) for
$1.95 \ltapprox \beta \ltapprox 2.60$.
For $E_A$ the finite-size corrections are between
0.000065 and 0.000153 (10--20 times larger than the statistical errors)
for $\beta \gtapprox 2.60$, and between 0.0002 and 0.0006
(20--50 times larger than the statistical errors) for
$1.95 \ltapprox \beta \ltapprox 2.60$.

Both the fundamental and adjoint energies
are in reasonably good agreement with the perturbative predictions:
see Figures \ref{fig_su3_enerF}(a) and \ref{fig_su3_enerA}(a). Furthermore,
we can use the observed deviations from these perturbative
predictions to obtain crude estimates of the next perturbative coefficients 
(which we hope someone will calculate in the near future). In 
Figure~\ref{fig_su3_enerF}(b) we plot $E_F - E_F^{(3-loop)}$ 
versus $1/\beta^4$.\footnote{
          The symbols in Figure~\ref{fig_su3_enerF}(b) indicate 
	  $L=128$ ($\Box$) and $L=256$ ($\Diamond$). 
	  The finite-size corrections in $E_F$ appear to be negligible
	  compared to the deviation from the three-loop perturbative 
	  prediction.}
The limiting slope suggests a four-loop coefficient of order $-0.05$ to $-0.1$.
If we fit $E_F - E_F^{(3-loop)} = k_4 \beta^{-4} +
 k_5 \beta^{-5}$, a reasonable fit is obtained if we restrict
attention to the points with $\beta \ge 2.35$, and we get
$k_4 = -0.02430 \pm  0.00942$, $k_5 = -0.222 \pm 0.026$.
Unfortunately, this fit would imply that $|k_5/\beta^5|$ is more
than twice as large as $|k_4/\beta^4|$ even at our largest
value of $\beta$ ($=4.35$), so that the extrapolation to $\beta = \infty$
cannot be taken seriously. All we can conclude is that:
(a) $k_4$ is somewhere in the range from $-0.02$ to $-0.10$;
and (b) {\it if} $k_4$ turns to be closer to the former value, {\it then} 
$k_5$ must be negative and of rather large magnitude 
(of order $-0.10$ or $-0.20$). These estimates can be compared to the 
known values of 
$k_1 = -2/3$, $k_2 = -0.0972222$, $k_3 = -0.0679225$. 

We proceed similarly for the adjoint energy. In
Figure~\ref{fig_su3_enerA}(b) we plot $E_A - E_A^{3-loop}$ versus 
$1/\beta^3$.\footnote{
   The symbols in Figure~\ref{fig_su3_enerA}(b) indicate 
   $L=128$ ($\Box$) and $L=256$ ($\Diamond$). In this case the
   finite-size corrections are clearly significant:
   the $L=256$ points lie noticeably above the $L=128$
   points.
}
The limiting slope suggests a three-loop coefficient of order $-0.02$.
If we fit $E_A - E_A^{3-loop} = l_3 \beta^{-3} +
 l_4 \beta^{-4}$, a reasonable fit is obtained if we restrict
attention to the points with $\beta \ge 2.35$, and we get
$l_3 = -0.0361 \pm 0.0046 $, $l_4 = 0.054 \pm 0.013$.
However, these error bars should not be taken seriously, as we are
neglecting terms of order $\beta^{-5}$, $\beta^{-6}$, etc.
In any case, it is worth comparing these estimates to the known values
$l_1 = -3/2$, $l_2 = 0.40625$.

\subsubsection{Comparison of Long-Distance Quantities with
    Finite-Volume Perturbation Theory}   \label{sec5.3.2}

Let us next compare the finite-volume Monte Carlo data $\scro(\beta,L)$
for the long-distance observables $\scro = \xi_\#^{(2nd)}$ and $\chi_\#$
with the predictions \reff{all_finiteV_PT} of
finite-volume perturbation theory ($\beta\to\infty$ at fixed $L<\infty$).
The expansions \reff{all_finiteV_PT} give $\chi_\#$ through order $1/\beta^2$,
and $\xi_\#$ through order $1/\beta$;
they are derived from the expansions \reff{hasenfratz_GFA_finite_L},
which give $G_\#$ through order $1/\beta^2$.
We restrict attention to $8 \le L \le 128$,
as our very few $L=256$ data points are all far from the perturbative regime
(they all have $\beta \le 2.30$ and $x \equiv \xi_F^{(2nd)}(L)/L < 0.33$).

We begin with the correlation lengths $\xi_F^{(2nd)}$ and $\xi_A^{(2nd)}$.
For these observables, the expansion is of the form
\be
   \xi_\#^{(2nd)}(\beta,L)   \;=\;
     A \, \beta^{1/2} \, L
     \left[ 1 \,-\, {A_1(L) \over \beta} \,-\, O(\beta^{-2}) \right]
\ee
with $A_1(L) \sim \log L$ at large $L$ [cf.\ \reff{fss:xi1}].
Luckily, $A_1(L)$ is not too large for these two expansions:
for $\xi_F^{(2nd)}$ it ranges from $\approx\! 0.48$ at $L=8$
to $\approx\! 0.80$ at $L=128$,
while for $\xi_A^{(2nd)}$ it ranges from $\approx\! 0.61$ at $L=8$
to $\approx\! 0.92$ at $L=128$.
As a result, the first-order perturbation correction
in the range of interest ($2 \ltapprox \beta \ltapprox 4$)
is of modest size, namely 10--40\%.
Furthermore, the discrepancy between the Monte Carlo data and the
perturbative predictions,
\be
   R_2(\beta,L)   \;\equiv\;
      {\xi_\#^{(2nd)}(\beta,L) \over A \, \beta^{1/2} \, L}
      \,-\, \left[ 1 \,-\, {A_1(L) \over \beta} \right]
   \;,
\ee
is smaller than this by a factor of 2--10:
see Tables~\ref{table_R2_xiF} and \ref{table_R2_xiA}.

Let us now examine more closely the behavior of the remainder term
$R_2(\beta,L)$.
Heuristically we would expect the remainder in
first-order perturbation theory to be of the
same order of magnitude as the second-order perturbative correction,
i.e.\ of order $-A_2(L)/\beta^2 \sim (\log L)^2/\beta^2$.
Let us therefore define
\be
   \widetilde{R}_2(\beta,L)   \;\equiv\;
      {\beta^2 \over (\log L)^2} \, R_2(\beta,L)
   \;.
\ee
We would like to know whether or not $|\widetilde{R}_2(\beta,L)|$ is
{\em uniformly bounded}\/ in $\beta \ge \hbox{some } \beta_0$
and $L \ge \hbox{some } L_0$,
as it should be if the perturbation series is to be
``well-behaved''.\footnote{
   This question has been raised forcefully by
   Patrascioiu and Seiler \cite{Patrascioiu_85,Pat-Seil_91,Pat-Seil_95ab}.
}

We can say something {\em rigorously}\/ in three different regimes:
\begin{itemize}
   \item[i)]  As $\beta\to\infty$ at fixed $L<\infty$, we have
      $\lim\limits_{\beta\to\infty} \beta^2 R_2(\beta,L) = -A_2(L)$
      [with corrections of order $1/\beta$],
      and hence
      $\lim\limits_{L\to\infty} \lim\limits_{\beta\to\infty}
       \widetilde{R}_2(\beta,L) = -A_{22} = -A_{11}^2/2$
      [cf. \reff{fss:AB_relations_A}].
   \item[ii)]  As $L\to\infty$ at fixed $\beta<\infty$, we have
      $\lim\limits_{L\to\infty} (\log L)^{-1} R_2(\beta,L) = A_{11}/\beta$,
      and hence $\lim\limits_{L\to\infty} \widetilde{R}_2(\beta,L) = 0$
      [with corrections of order $1/\log L$].
   \item[iii)]  Since $\xi(\beta,L) \ge 0$, we have
      $R_2(\beta,L) \ge [A_1(L)/\beta] - 1$ for all $\beta,L$.
      In particular, along any curve $\beta = c A_1(L)$ with $c>0$
      [which makes sense at least for large $\beta$ since $A_{11} > 0$],
      we can conclude that
      $\widetilde{R}_2(\beta,L) \ge c(1-c) A_1(L)^2/\log^2 L$,
      which for large $L$ tends to $c(1-c) A_{11}^2$.
      For $0<c<1$ this proves that $\widetilde{R}_2(\beta,L) > 0$
      and provides a {\em lower}\/ bound on its magnitude;
      for $c>1$ it constrains only how negative $\widetilde{R}_2(\beta,L)$
      can get.
\end{itemize}
Furthermore, under the {\em assumptions}\/ of the conventional wisdom,
we can say something analytically in one sub-case of regime (iii):
\begin{itemize}
   \item[iii${}'$)]  As $\beta,L \to\infty$
      at fixed $x \equiv \xi_F^{(2nd)}(\beta,L)/L \neq 0,\infty$,
      we have
\be
    \lim\limits_{\begin{scarray}
                    \beta,L \to\infty \\
                    x \, \hbox{\scriptsize fixed}\, \neq 0,\infty
                 \end{scarray}}
    \! {\beta \over \log L}   \;=\;
    w_0  \;=\;  2A_{11}
 \label{lim_Rtilde_fixed_x_1}
\ee
[corresponding to $c=2$ in regime (iii)]
and
\be 
    \lim\limits_{\begin{scarray}
                    \beta,L \to\infty \\
                    x \, \hbox{\scriptsize fixed}\, \neq 0,\infty
                 \end{scarray}}
      \! \widetilde{R}_2(\beta,L)   \;=\; -2A_{11}^2
      \;.
 \label{lim_Rtilde_fixed_x_2}
\ee
\end{itemize}
[This shows that the lower bound for regime (iii) is sharp when $c=2$.]
{\sc Proof of \reff{lim_Rtilde_fixed_x_1}/\reff{lim_Rtilde_fixed_x_2}:}
{}From asymptotic scaling \reff{xi_predicted2} we have
$\log\xi_\infty(\beta) = \beta/w_0 + O(\log\beta)$ where $w_0 = 2A_{11}$
is the first coefficient of the RG beta-function
[cf.\ \reff{fss:RGpertxi}/(\ref{fss:RGpertvalues}a)].
According to finite-size scaling \reff{eq1},
taking $\beta,L \to\infty$ at fixed $x \equiv \xi(\beta,L)/L \neq 0,\infty$
implies that $\xi_\infty(\beta)/L$ also converges to a limit $\neq 0,\infty$,
so that in particular $\log\xi_\infty(\beta) / \log L \to 1$.
This implies \reff{lim_Rtilde_fixed_x_1}.
On the other hand, $\xi(\beta,L)/[A \beta^{1/2} L] = x/[A \beta^{1/2}]$,
which tends to zero as $\beta\to\infty$ at fixed $x$.
It follows that $R_2(\beta,L) \to -{1 \over 2}$,
which together with \reff{lim_Rtilde_fixed_x_1}
implies \reff{lim_Rtilde_fixed_x_2}. Q.E.D.
[This is somewhat strange:  the limiting value of $\widetilde{R}_2(\beta,L)$
 must behave in a highly non-monotonic way as we pass from regime (ii)
 through regime (iii)/(iii${}'$) to regime (i).]

We can use our Monte Carlo data to study the behavior of 
$\widetilde{R}_2(\beta,L)$.
First of all, let us look at limit (i):
We know that
\be
   A_2(L)   \;=\;  A_{22} \log^2 L \,+\, A_{21} \log L \,+\, A_{20} \,+\,
                      O\!\left( {\log^2 L \over L^2} \right)
 \label{A2_MC1}
\ee
with
\begin{subeqnarray}
   A_{22}   & = &   {N^2 \over 128\pi^2}   \\
   A_{21}   & = &   {N^2 \over 64\pi^2} \,+\, {N \over 8\pi} A_{10}
 \label{A2_MC2}
\end{subeqnarray}
[cf.\ (\ref{fss:RGpertvalues}a,b) and \reff{ap:w0}/\reff{ap:w1}],
where we have
\begin{subeqnarray}
  A_{10}^{(F)}  & = &
    {1 \over 4} \left[ N I_{1,fin} \,+\, {N^2 -2 \over 4N} \,+\,
       {N^2 -2 \over 2N} \left( 4\pi^2 I_{3,\infty} + {1 \over 2\pi^2} \right)
    \right]
  \\[1mm]
  A_{10}^{(A)}  & = &
    {1 \over 4} \left[ N I_{1,fin} \,+\, {N^2 -2 \over 4N} \,+\,
       {3N \over 2} \left( 4\pi^2 I_{3,\infty} + {1 \over 2\pi^2} \right)
    \right]
 \label{A2_MC3}
\end{subeqnarray}
for the fundamental and adjoint sectors
[see \reff{I123_defs}--\reff{I123_numer} for definitions].
%%  In particular, for $N=3$ we have numerically
%%  $A_{10}^{(F)} \approx 0.223965$ and $A_{10}^{(A)} \approx 0.342700$.
Unfortunately $A_{20}^{(F)}$ and $A_{20}^{(A)}$ are unknown.
We can use our Monte Carlo data to estimate $A_2^{(\#)}(L)$
for $L=8,16,32,64,128$;
in this way we can test \reff{A2_MC1}--\reff{A2_MC3}
and obtain a rough estimate of $A_{20}^{(\#)}$.
The data are approximately converged for $L=8$,
and suggest very roughly
$A_{20}^{(F)} \approx 0.09$ and $A_{20}^{(A)} \approx 0.14$
(see the next-to-last column of
 Tables~\ref{table_R2_xiF} and \ref{table_R2_xiA}).
The data for larger $L$ are at least consistent with convergence
to these values.

In limit (ii), we see $|\widetilde{R}_2(\beta,L)|$
slowly decreasing as a function of $L$ when $L \ll \xi_{F,\infty}(\beta)$,
then beginning to grow slightly when $L \sim \xi_{F,\infty}(\beta)$.
We know that $|\widetilde{R}_2(\beta,L)|$ must ultimately decrease
again to zero as $L \to\infty$, but our data do not allow us
to observe this decrease (this is not surprising since the rate of
convergence is only $1/\log L$).

Along the curves $x = {\rm constant}$ [limit (iii${}'$)],
$\widetilde{R}_2(\beta,L)$ stays bounded
and is roughly consistent with convergence to the predicted value
$-2A_{11}^2 = N^2/(32\pi^2) \approx -0.028497$.

In summary, our data in Tables~\ref{table_R2_xiF} and \ref{table_R2_xiA}
give no evidence of $\widetilde{R}_2(\beta,L)$
becoming unbounded in any region of the $(\beta,L)$-plane.
It is of course still possible that $\widetilde{R}_2(\beta,L)$
does become unbounded in some region far from the one we have studied;
this question needs ultimately to be resolved by a rigorous
mathematical proof.

%  {\bf Delete this paragraph now?????}
%  Things seem empirically, however, to be even better-behaved than this:
%  the remainder appears to be bounded by something of order $\log L/\beta^2$,
%  uniformly in $\beta \ge ????$ and $L \ge 8$,
%  as shown in the last column of
%  Tables~\ref{table_pert_xiF} and \ref{table_pert_xiA}.
%  In fact this {\em cannot}\/ be the true behavior:
%  as $\beta\to\infty$ at fixed $L$, the remainder multiplied by $\beta^2$
%  {\em must}\/ equal the next-order perturbative coefficient $-A_2(L)$.
%  Perhaps this indicates that the coefficient of $(\log L)^2$
%  in the second-order perturbation coefficient $A_2(L)$ is very small.
%  In any case, our Monte Carlo data are fully consistent with the
%  belief that the remainder is bounded by a constant times
%  $(\log L)^2/\beta^2$,
%  uniformly in $\beta \ge ????$ and $L \ge 8$.

Let us now look at the susceptibilities $\chi_F$ and $\chi_A$,
for which the expansion is of the form
\be
   \chi_\#(\beta,L)   \;=\;
     B \, L^2
     \left[ 1 \,-\, {B_1(L) \over \beta} \,-\, {B_2(L) \over \beta^2}
              \,-\, O(\beta^{-3}) \right]
\ee
with $B_1(L) \sim \log L$ and $B_2(L) \sim (\log L)^2$ at large $L$.
For $\chi_F$ the coefficient $B_1(L)$ is rather large,
ranging from $\approx\! 1.01$ at $L=8$ to $\approx\! 2.19$ at $L=128$,
while $B_2(L)$ is fairly small,
ranging from $\approx\! 0.04$ at $L=8$ to $\approx -0.43$ at $L=128$.
As a result, the first-order perturbation corrections are quite large
in the range of interest, but the second-order corrections are small;
and the discrepancy between the Monte Carlo data and the
second-order perturbative prediction is a factor 1--10 smaller than
the second-order correction (see Table~\ref{table_S3_chiF}).
For $\chi_A$, by contrast, both coefficients are very large:
$B_1(L)$ ranges from $\approx\! 2.28$ at $L=8$ to $\approx\! 4.93$ at $L=128$,
while $B_2(L)$ ranges from $\approx -1.39$ at $L=8$ to $\approx -7.75$
at $L=128$.
As a result, both the first-order and second-order perturbation corrections
are huge;  nevertheless, the discrepancy between the Monte Carlo data
and the second-order perturbative prediction is surprisingly small
(see Table~\ref{table_S3_chiA}).

Let us define the discrepancy between the exact values and the
second-order perturbative predictions,
\be
   S_3(\beta,L)   \;\equiv\;
      {\chi_\#(\beta,L) \over B L^2}
      \,-\, \left[ 1 \,-\, {B_1(L) \over \beta} \,-\, {B_2(L) \over \beta^2}
            \right]
   \;,
\ee
and the corresponding rescaled quantity
\be
   \widetilde{S}_3(\beta,L)   \;\equiv\;
      {\beta^3 \over (\log L)^3} \, S_3(\beta,L)
   \;.
\ee
Just as for the correlation lengths, we can prove rigorously
the behavior in regimes (i)--(iii):
\begin{itemize}
   \item[i)]  As $\beta\to\infty$ at fixed $L<\infty$, we have
      $\lim\limits_{\beta\to\infty} \beta^3 S_3(\beta,L) = -B_3(L)$,
      and hence
      $\lim\limits_{L\to\infty} \lim\limits_{\beta\to\infty}$  % BREAK LINE!!!
      $\widetilde{S}_3(\beta,L) = -B_{33}$.
   \item[ii)]  As $L\to\infty$ at fixed $\beta<\infty$, we have
      $\lim\limits_{L\to\infty} (\log L)^{-2} S_3(\beta,L) = B_{22}/\beta$,
      and hence $\lim\limits_{L\to\infty} \widetilde{S}_3(\beta,L) = 0$.
   \item[iii)]  Since $\chi(\beta,L) \ge 0$, we have
      $S_3(\beta,L) \ge [B_1(L)/\beta] + [B_2(L)/\beta^2] - 1$
      for all $\beta,L$.
      In particular, along any curve $\beta = c A_1(L)$ with $c>0$,
      we can conclude that
      (writing for simplicity only the $L\to\infty$ limit)
      we have
      $\widetilde{S}_3(\beta,L) \gtapprox
       c^2 A_{11}^2 B_{11} + c A_{11} B_{22} - c^3 A_{11}^3  =
       (c^2 + c) A_{11}^2 B_{11} - \half c A_{11} B_{11}^2 - c^3 A_{11}^3$
      [cf.\ (\ref{fss:AB_relations}b)].
      Unfortunately, the sign of this lower bound is far from obvious.
\end{itemize}
Furthermore, the assumptions of the conventional wisdom imply that
\begin{itemize}
   \item[iii${}'$)]  As $\beta,L \to\infty$
      at fixed $x \equiv \xi_F^{(2nd)}(\beta,L)/L \neq 0,\infty$,
      we have
\be 
    \lim\limits_{\begin{scarray}
                    \beta,L \to\infty \\
                    x \, \hbox{\scriptsize fixed}\, \neq 0,\infty
                 \end{scarray}}
      \! \widetilde{S}_3(\beta,L)   \;=\;
      6 A_{11}^2 B_{11} - A_{11} B_{11}^2 - 8 A_{11}^3
      \;.
 \label{lim_Stilde_fixed_x}
\ee
\end{itemize}
[This shows that the lower bound for regime (iii) is sharp when $c=2$.]
{\sc Proof of \reff{lim_Stilde_fixed_x}:}
We have \reff{lim_Rtilde_fixed_x_1} as before.
On the other hand, let us write
\be
    {\chi(\beta,L) \over L^2}   \;=\;
    {\chi(\beta,L) \over \chi(\beta,\infty)}  \,
    {\chi(\beta,\infty) \over \xi(\beta,\infty)^2}  \,
    \left({\xi(\beta,\infty) \over L} \right)^2
    \;.
\ee
By finite-size-scaling theory, the first and third factors
on the right-hand side tend to finite constants as
$\beta,L \to\infty$ at fixed $x$;
while asymptotic scaling implies that
the second factor scales as $\beta^{-\gamma_0/w_0}$,
hence vanishes as $\beta\to\infty$ because $\gamma_0/w_0 > 0$.
Using \reff{lim_Rtilde_fixed_x_1},
we easily deduce \reff{lim_Stilde_fixed_x}. Q.E.D.

We can use our Monte Carlo data to study the behavior of 
$\widetilde{S}_3(\beta,L)$.
First of all, let us look at limit (i):
We know that
\be
   B_3(L)   \;=\;  B_{33} \log^3 L \,+\, B_{32} \log^2 L \,+\,
                   B_{31} \log L \,+\, B_{30} \,+\,
                      O\!\left( {\log^3 L \over L^2} \right)
 \label{B3_MC1}
\ee
with
\begin{subeqnarray}
   B_{33}   & = &   {1 \over 6}
                    (\gamma_0^3 - 3 w_0 \gamma_0^2 + 2 w_0^2 \gamma_0)  \\
   B_{32}   & = &   {1 \over 2}
             (w_1 \gamma_0 + 2w_0 \gamma_1^{lat} - 2 \gamma_0 \gamma_1^{lat})
           \,+\,  {B_{10} \over 2} (2w_0^2 + \gamma_0^2 - 3 w_0 \gamma_0)  \\
   B_{31}   & = &   \gamma^{lat}_2 \,+\, B_{10} (w_1 - \gamma^{lat}_1)
                                   \,+\, B_{20} (2w_0 - \gamma_0)
 \label{B3_MC2}
\end{subeqnarray}
where the RG coefficients $w_0, w_1, \gamma_0, \gamma_1^{lat}, \gamma^{lat}_2$
can be found in Appendix \ref{appenA},
and the constants $B_{10}$ and $B_{20}$ can be extracted from
Appendix \ref{appenB} (the latter formulae are somewhat lengthy).
We consider first $\chi_F$, because it is only for the
fundamental sector that we know the value of $\gamma^{lat}_2$.
Numerically, for $N=3$ we have $B_{33}^{(F)} \approx  -0.0006967924$,
$B_{32}^{(F)} \approx -0.0175782344$ and $B_{31}^{(F)} \approx 0.0679511385$.
Unfortunately $B_{30}^{(F)}$ is unknown.
We can use our Monte Carlo data to test \reff{B3_MC1}--\reff{B3_MC2}
and obtain a rough estimate of $B_{30}^{(F)}$.
The first thing to note is that $S_3(\beta,L)$ undergoes a curious change
of sign as $L$ varies (see Table~\ref{table_S3_chiF});
but this increase is almost entirely accounted for
by the known terms $B_{33} \log^3 L + B_{32} \log^2 L + B_{31} \log L$,
which exhibit a similar sign change.
The difference is much smaller in magnitude,
and as a result the estimated $B_{30}^{(F)}$ is much smaller
in magnitude than $\beta^3 S_3(\beta,L)$
[see the next-to-last column of Table~\ref{table_S3_chiF}].
Unfortunately the estimates of $B_{30}^{(F)}$ are not well converged
(the fluctuations at larger $L$ are statistical error);
all we can say is that $B_{30}^{(F)}$ is very small,
probably somewhere between $-0.1$ and $0.1$.

Likewise, in limits (ii) and (iii${}'$), we are unable to see the
predicted convergence of $\widetilde{S}_3(\beta,L)$
to a limiting value,
or to say whether $|\widetilde{S}_3(\beta,L)|$
appears to remain bounded.
In any case,
$|\widetilde{S}_3(\beta,L)|$ stays extremely small.

Finally, let us consider the adjoint susceptibility $\chi_A$,
for which the known numerical values (for $N=3$)
are $B_{33}^{(A)} \approx 0.0544244644$
and $B_{32}^{(A)} \approx -0.1358424235$.
For $\chi_A$, the first-order perturbative corrections are enormous
(50--110\% even at our largest $\beta$),
and the second-order corrections are quite large
(7--40\% even at the largest $\beta$):
see Table~\ref{table_S3_chiA}.
In view of this, the deviations from second-order perturbation theory
are amazingly small:
e.g.\ a fraction of a percent when the second-order term is as large as 20\%,
or 10--40\% when the second-order term is 100\% or more.
Furthermore, these deviations are almost perfectly explained
by the $B_3(L)/\beta^3$ term,
as can be seen from the almost-constancy of $\widetilde{S}_3(\beta,L)$
as a function of $\beta$ at each fixed $L$.
The values of $\widetilde{S}_3(\beta,L)$ vary significantly with $L$,
but it is plausible that they are approaching their predicted limiting value
$-B_{33}^{(A)} \approx -0.054$ as $L\to\infty$
(the corrections are, after all, of order $1/\log L$
 with a relatively large coefficient $B_{32}^{(A)}$).
We do not have any explanation for these incredibly accurate predictions
from an {\em a priori}\/ badly behaved perturbation series.

\subsubsection{Fundamental Sector: Correlation Length} \label{sec5.3.3}

In the next four subsections we shall compare the
extrapolated {\em infinite-volume}\/ values ${\cal O}_\infty(\beta)$
for the long-distance observables ${\cal O} = \xi_\#^{(2nd)}$ and $\chi_\#$,
as generated in Section \ref{sec5.2.1},
with the asymptotic-freedom predictions.

We begin by comparing the fundamental correlation length $\xi_F^{(2nd)}$
with the two-loop and three-loop perturbative predictions
\reff{xi_predicted2}/\reff{a1}.
In all cases, we use the extrapolated data from our preferred fit:
see Table \ref{estrap_xi_fund}, estimate $(\infty,0.90,0.65)$.

Let us recall that perturbation theory \reff{xi_predicted2}/\reff{a1}
combined with the nonperturbative (BNNW) prefactor \reff{exact_Cxi}
give a quantitative prediction
for the {\em exponential}\/ correlation length $\xi_F^{(exp)}$
[cf.\ (\ref{xi_B_formulae}a,c)].
The factor $\xi_F^{(2nd)}/\xi_F^{(exp)}$ is unknown,
but a high-precision Monte Carlo study of the $SU(3)$ chiral model
yields the value $0.987 \pm 0.002$ \cite{Rossi_94a}.
We shall therefore plot $\xi_F^{(2nd)}(\beta)$
divided by $\xi_{F,\hbox{\scriptsize BNNW},k-loop}^{(exp)}(\beta)$
for $k=2,3$, and look for convergence as $\beta\to\infty$
to a value $\approx 0.987$.
The results are shown in Figure \ref{fig_su3_scal_xif}(a)
(points $+$ and $\times$).
The discrepancy from two-loop asymptotic scaling,
which is $\approx 20\%$ at $\beta=2.0$ ($\xi_{F,\infty} \approx 25$),
decreases to 5--6\% at $\beta=4.35$
($\xi_{F,\infty} \approx 3.7 \times 10^5$).
The discrepancy from three-loop asymptotic scaling,
which is $\approx 12\%$ at $\beta=2.0$,
decreases to 2--3\% at $\beta=4.35$.
Furthermore, if we fit 
$\xi_F^{(2nd)}/\xi_{F,\B,3-loop}^{(exp)} = \kappa_0 + \kappa_2 \beta^{-2}$,
a good fit is obtained if we restrict attention to the points with 
$\beta \ge 2.60$ ($\xi_{F,\infty} \gtapprox 300$), and we obtain the estimates 
\begin{subeqnarray}
	\widetilde{C}_{\xi_F^{(2nd)}}/\widetilde{C}_{\xi_F^{(exp)}} 
	\equiv \kappa_0 & = & 0.989 \pm 0.007 \\
	a_2 \equiv \kappa_2/\kappa_0 & = & -0.38 \pm 0.06 
	\slabel{a2:fit}
\end{subeqnarray}
[see Figure \ref{fig_su3_scal_xif}(b), points $\times$].
Of course, this estimate of $a_2$ should not be taken too
seriously, as we have neglected corrections of order
$\beta^{-3}$ and higher; it is in any case of the same order of magnitude
as the known value $a_1 = -0.164$.
Moreover, the error bar on $\kappa_0$ is probably
significantly underestimated, because in this fit we have ignored
the {\em correlations}\/ between the estimated values $\xi_{F,\infty}^{(2nd)}$
at different $\beta$ [which arise from errors of type (iii)].
Still, the agreement with the predicted value 0.987 is remarkable.

We can also try ``improved expansion parameters''
(see Section \ref{sec3.3}).
For example, we can use $x_F \equiv 1 - E_F$ as a substitute for $\beta$,
and compare to the prediction \reff{improv_2}/\reff{improv_3} for
$\xi_F^{(exp)}$ as a function of $1-E_F$.
For $E_F$ we use the value measured on the largest lattice (which is usually
$L = 128$);
the statistical errors and finite-size corrections on $E_F$
are less than $5\times 10^{-4}$,
 and they induce an error less than $0.85\%$
on the predicted $\xi_{F,\infty}$ (less than 0.55\% for $\beta \geq 2.2$).
In Figure \ref{fig_su3_scal_xif}(a) (points $\Box$ and $\Diamond$)
we show $\xi_F^{(2nd)}$ divided by the
two-loop and three-loop perturbative predictions 
\reff{improv_2}/\reff{improv_3} for $\xi_F^{(exp)}$.
The data agree with the two-loop prediction to within better than
5\% for $\beta \ge 2.10$ ($\xi_{F,\infty} \gtapprox 40$). 
The agreement with the three-loop prediction is excellent:
the discrepancy is $\ltapprox 1$--2\% for $\beta \ge 1.75$ 
($\xi_{F,\infty} \gtapprox 8$).
Furthermore, the ``improved'' 3-loop prediction is extremely flat, and 
to a constant $\kappa'_0$ if we restrict attention to the points with 
$\beta \ge 2.60$ ($\xi_{F,\infty} \gtapprox 300$), yielding the estimate
\be
\widetilde{C}_{\xi_F^{(2nd)}}/\widetilde{C}_{\xi_F^{(exp)}}
\equiv \kappa'_0  =  0.983 \pm 0.002
\ee
[see Figure \ref{fig_su3_scal_xif}(b), points $\Diamond$].\footnote{
	If, instead, we fit
	$\xi_F^{(2nd)}/\xi_{F,\B,improved \, 3-loop}^{(exp)} = 
	\kappa'_0 + \kappa'_2 \beta^{-2}$, 
	a good fit is obtained if we restrict attention to the points 
	with $\beta \ge 2.45$ ($\xi_{F,\infty} \gtapprox 160$), 
	and we obtain the estimates
	$\widetilde{C}_{\xi_F^{(2nd)}}/\widetilde{C}_{\xi_F^{(exp)}}
	\equiv \kappa'_0 = 0.979 \pm 0.006 $ and
	$a^{(imp)}_2 \equiv \kappa'_2/\kappa'_0  =  -0.04 \pm 0.05$.
	But since the $a_2^{(imp)}$ is consistent with zero,
	we may as well use a constant fit. \protect\label{foot:a2_imp_fund}}

In conclusion,
the three-loop perturbative prediction in the bare parameter
agrees with the Monte Carlo data to within about 2--3\% for
$\beta \ge 4.05$ ($\xi_{F,\infty} \gtapprox 10^5$), and 
three-loop ``improved'' perturbative prediction is even better ($<1$\%).
Furthermore, both the bare parameter and the ``improved''
perturbative predictions are extremely flat for $\beta \ge 3.15$
and $\beta \ge 2.35$, respectively. A good compromise for the limiting value
would be
\be
\widetilde{C}_{\xi_F^{(2nd)}}/\widetilde{C}_{\xi_F^{(exp)}}
= 0.985 \pm 0.007 \; ,
\ee
which is in excellent agreement with the Rossi-Vicari 
prediction $0.987 \pm 0.002$ \cite{Rossi_94a}.

\subsubsection{Fundamental Sector: Susceptibility} \label{sec5.3.4}

For the susceptibility $\chi_F$ we proceed in two different ways: using either
$\chi_F$ directly or else using the ratio $\chi_F/{\xi_F^{(2nd)}}^2$. 
The advantage of the latter approach is that one additional term
of perturbation theory is available. 

In Figure~\ref{fig_su3_scal_chif}(a) 
we plot $\chi_{F,\infty,estimate\,(\infty,\infty,0.80)}$
divided by the theoretical prediction \reff{chiF_predicted2}/\reff{b1}
{\it with the prefactor} $\widetilde{C}_{\chi_F}$ {\it omitted};
the $\beta \rightarrow \infty$ limit of this curve thus gives an
estimate of $\widetilde{C}_{\chi_F}$. Here we have two-loop
and three-loop predictions (points $+$, $\times$) as well
as ``improved'' two-loop and three-loop predictions ($\Box$, $\Diamond$).
The estimates from two-loop and three-loop standard perturbation theory
(which are virtually identical since $b_1 \approx -0.023$ is so small)
are strongly rising for $\beta \ltapprox 2.4$ and weakly rising 
thereafter. If we fit 
$\chi_F/(\chi_{F,3-loop}\hbox{ without the prefactor }$
$  \widetilde{C}_{\chi_F} )
= \kappa_0 + \kappa_2 \beta^{-2}$,
a good fit is obtained if we restrict attention to the points with $\beta
\ge 2.65$ ($\xi_{F,\infty} \gtapprox 360$), and we obtain the estimates
\begin{subeqnarray}
\widetilde{C}_{\chi_F}
\equiv \kappa_0 & = & 16.75 \pm 0.31 \slabel{C_chi_F:1} \\
b_2 \equiv \kappa_2/\kappa_0 & = & -0.72 \pm 0.12
\end{subeqnarray}
[see Figure \ref{fig_su3_scal_chif}(b), points $\times$].
The estimates from ``improved'' perturbation theory are
rather flatter, particularly the three-loop one which is
virtually constant for $\beta \gtapprox 2.45$ 
($\xi_{F,\infty} \gtapprox 160$). The ``improved'' 3-loop values 
can be fit well to a constant $\kappa'_0$ 
if we restrict attention to $\beta \ge 2.55$ 
($\xi_{F,\infty} \gtapprox 240$), and we obtain the estimate
\be
\widetilde{C}_{\chi_F}
\equiv \kappa'_0  =  16.30 \pm 0.07
\ee
[see Figure \ref{fig_su3_scal_chif}(b), points $\Diamond$].
This estimate is slightly lower than \reff{C_chi_F:1}, but consistent with it.

Similarly, we could plot 
$(\chi_F/{\xi_F^{(2nd)}}^2)_{\infty,estimate\,(\infty,\infty,\infty)}$
divided by the theoretical prediction 
\reff{chiFoverxiFsquared_predicted}/\reff{c1}/\reff{c2}
{\it with the prefactors $\widetilde{C}_{\chi_F}$ and
$\widetilde{C}_{\xi_F^{(2nd)}}$  omitted}; 
the $\beta \rightarrow \infty$ limit of this curve would thus give an
estimate of $\widetilde{C}_{\chi_F}/(\widetilde{C}_{\xi_F^{(2nd)}})^2$.
However, in order to make the vertical scale of this graph more
directly comparable to that of Figure~\ref{fig_su3_scal_chif}, 
we have multiplied
the quantity being plotted by $(\widetilde{C}^{(\B)}_{\xi_F^{(exp)}})^2$.
Note that this does not in any way alter the {\it logic} of the
analysis, as $\widetilde{C}^{(\B)}_{\xi_F^{(exp)}}$ is an explicit number
defined in \reff{exact_Cxi}. The resulting curve is plotted in 
Figure~\ref{fig_su3_scal_chif_ov_xif2}(a); 
its $\beta \rightarrow \infty$ limit gives
an estimate of $\widetilde{C}_{\chi_F} \times
(\widetilde{C}_{\xi_F^{(2nd)}}/\widetilde{C}^{(\B)}_{\xi_F^{(exp)}})^{-2}$.
In this case we have two-loop,
three-loop and four-loop predictions ($+$, $\times$, $\protect\fancyplus$)
as well as ``improved'' two-loop, three-loop and four-loop predictions 
($\Box$, $\Diamond$, $\bigcirc$). To convert this number to an estimate for
$\widetilde{C}_{\chi_F}$ itself, we need to multiply by
\be
   \left( {\widetilde{C}_{\xi_F^{(2nd)}} \over
\widetilde{C}^{(\B)}_{\xi_F^{(exp)}} } \right) ^{\! 2} \, = \, \left(
{\widetilde{C}_{\xi_F^{(2nd)}} \over \widetilde{C}_{\xi_F^{(exp)}} }
\right) ^{\! 2} \left( {\widetilde{C}_{\xi_F^{(exp)}} \over
\widetilde{C}_{\xi_F^{(exp)}}^{(\B)} } \right) ^{\! 2} \;.
\label{CchiF_stuff_a}
\ee
The first factor on the right side has been estimated by Monte Carlo
simulations \cite{Rossi_94a}, yielding $0.987^2 = 0.974$;
moreover, our data for $\xi_F^{(2nd)}$ itself 
(Figure~\ref{fig_su3_scal_xif}) are consistent with this prediction. 
The second factor on the right side is presumably equal to 1 (exactly).
So we can take the factor \reff{CchiF_stuff_a} to be $\approx 0.974$.
The estimates from the three-loop standard perturbation theory are
rapidly decreasing for $\beta \ltapprox 3.2$ and slowly decreasing 
thereafter; if we fit them to $\kappa_0 + \kappa_2 \beta^{-2}$,
a good fit is obtained if we restrict attention to $\beta
\ge 2.30$ ($\xi_{F,\infty} \gtapprox 85$), and we obtain the estimates
\begin{subeqnarray}
        \widetilde{C}_{\chi_F} \left( {\widetilde{C}_{\xi_F^{(2nd)}} \over
        \widetilde{C}^{(\B)}_{\xi_F^{(exp)}} } \right) ^{\! -2}
	\equiv \kappa_0 & = & 16.61 \pm 0.04 \\
	c_2 \equiv \kappa_2/\kappa_0 & = & 0.34 \pm 0.02 
	\slabel{c2:chi_F_stuff}
\end{subeqnarray}
and hence
\be
        \widetilde{C}_{\chi_F} = 16.17 \pm 0.04 \;.
\ee
The estimate \reff{c2:chi_F_stuff} is in excellent agreement with 
the known value $c_2 = 0.306$. If we now fit the standard 
four-loop values to $\kappa_0 + \kappa_3 \beta^{-3}$, we have a 
good fit for $\beta \ge 2.60$ ($\xi_{F,\infty} \gtapprox 300$), 
and we obtain the estimates
\begin{subeqnarray}
        \widetilde{C}_{\chi_F} \left( {\widetilde{C}_{\xi_F^{(2nd)}} \over
        \widetilde{C}^{(\B)}_{\xi_F^{(exp)}} } \right) ^{\! -2}
        \equiv \kappa_0 & = & 16.74 \pm 0.04 \\
        c_3 \equiv \kappa_3/\kappa_0 & = & 0.19 \pm 0.07
\end{subeqnarray}
[see Figure \ref{fig_su3_scal_chif_ov_xif2}(b), points $\protect\fancyplus$], 
and hence
\be
        \widetilde{C}_{\chi_F} = 16.30 \pm 0.04 \;.
\ee
The estimates from ``improved'' perturbation theory are
much flatter, particularly the four-loop one which is
virtually constant for $\beta \gtapprox 2.3$.
If we fit the ``improved'' three-loop values 
to $\kappa'_0 + \kappa'_2 \beta^{-2}$
for $\beta \ge 2.25$ ($\xi_{F,\infty} \gtapprox 70$), we get
\begin{subeqnarray}
        \widetilde{C}_{\chi_F} \left( {\widetilde{C}_{\xi_F^{(2nd)}} \over
        \widetilde{C}^{(\B)}_{\xi_F^{(exp)}} } \right) ^{\! -2}
        \equiv \kappa'_0 & = & 16.75 \pm 0.04 \\
        c^{(imp)}_2 \equiv \kappa'_2/\kappa'_0 & = & -0.3 \pm 0.2 
	\slabel{c2_imp:chi_F_stuff}
\end{subeqnarray}
and hence
\be
        \widetilde{C}_{\chi_F} = 16.31 \pm 0.04 \;.
\ee
The estimate \reff{c2_imp:chi_F_stuff} suggests that $c^{(imp)}_2$ is 
very close to zero, consistent with
the known value $c^{(imp)}_2 = 0.0010$. If we now fit
the ``improved'' four-loop values to 
$\kappa'_0 + \kappa'_3 \beta^{-3}$, we have a
good fit for $\beta \ge 2.075$ ($\xi_{F,\infty} \gtapprox 34$),
and we obtain the estimates
\begin{subeqnarray}
        \widetilde{C}_{\chi_F} \left( {\widetilde{C}_{\xi_F^{(2nd)}} \over
        \widetilde{C}^{(\B)}_{\xi_F^{(exp)}} } \right) ^{\! -2}
        \equiv \kappa'_0 & = &  16.85 \pm  0.03 \\
        c_3^{(imp)} \equiv \kappa'_3/\kappa'_0 & = & 0.10 \pm 0.02
\end{subeqnarray}
[see Figure \ref{fig_su3_scal_chif_ov_xif2}(b), points $\bigcirc$], and hence
\be
        \widetilde{C}_{\chi_F} = 16.41 \pm 0.03 \;.
\ee

In summary, all methods yield consistent results, but the ones based on
$\chi_F/{\xi_F^{(2nd)}}^2$ show an earlier convergence to the
limiting constant. Therefore, a reasonable compromise would be
\be 
\widetilde{C}_{\chi_F} \approx 16.35 \pm 0.20 \; .
\ee
%, most likely due to the knowledge of the four-loop
%perturbation term.  

\subsubsection{Adjoint Sector: Correlation Length} \label{sec5.3.5}

Now let us look at the adjoint sector.  We start with the correlation
length $\xi_A^{(2nd)}$, which we can compare with the two-loop and
three-loop perturbative predictions \reff{xi_predicted2}/\reff{a1}.
Combined with the
nonperturbative (BNNW) prefactor \reff{mT=2mV}/\reff{exact_Cxi},
these formulae
give a quantitative prediction for the {\em exponential}\/ correlation
length $\xi_A^{(exp)}$ [cf.\ (\ref{xi_B_formulae}b,d)].
By plotting $\xi_A^{(2nd)}/\xi^{(exp)}_{A,\B,k-loop}$ for $k=2,3$,
we can test asymptotic scaling and estimate the universal nonperturbative
ratio $\widetilde{C}_{\xi_A^{(2nd)}} / \widetilde{C}_{\xi^{(exp)}_{A}}$.

In Figure \ref{fig_su3_scal_xia}(a) we plot
$\xi_{A,\infty,estimate \, (\infty,\infty,\infty)}^{(2nd)} / 
 \xi^{(exp)}_{A,\B,k-loop}$ versus $\beta$ (points $+$ and $\times$).
We see that the behavior is similar to that observed in the fundamental
channel (Figure \ref{fig_su3_scal_xif}); this is inevitable since the
ratio $\xi_{F,\infty}^{(2nd)}/\xi_{A,\infty}^{(2nd)}$ 
is empirically close to constant ($\approx
2.80$) in this region.  
However, in the adjoint channel the estimates show strange
(and presumably spurious) pseudo-periodic oscillations;
we do not understand their cause, but they presumably arise
from some quirks in the extrapolation to infinite volume.\footnote{
	Note that the corrections to scaling in the
	adjoint channel are somewhat stronger than those in the
	fundamental channel [compare Figure \protect\ref{fig_su3_dev_xif}
	to Figure \protect\ref{fig_su3_dev_xia}]. 
	Since the
	extrapolation of $\xi_A^{(2nd)}$ uses the most
	stringent fit ($\infty$,$\infty$,$\infty$), we are
	unable to say anything about the remaining systematic errors
	due to corrections to scaling.}
Furthermore, these values present an apparent change of slope at 
$\beta \approx 3.15$, suggesting a positive coefficient $a_2$,
which is in total disagreement with the
result \reff{a2:fit} predicted by fitting the fundamental-sector 
quantities. Perhaps we have grossly underestimated the systematic errors
in the extrapolation to infinite volume, especially for
for $\beta \gtapprox 3.15$. The estimate of 
$\xi_A^{(2nd)}/\xi_{A,\B,3-loop}^{(exp)}$ will depend
whether or not we trust our extrapolation for 
$\beta \gtapprox 3.15$. If we do not trust it, we
may discard all those points with $\beta \gtapprox 3.15$
and fit
$\xi_A^{(2nd)}/\xi_{A,\B,3-loop}^{(exp)} = \kappa_0 + \kappa_2 \beta^{-2}$.
A good fit is obtained for $3.15 \ge \beta \ge 2.40$
($1.1 \times 10^3 \gtapprox \xi_F \gtapprox 130$), 
and we obtain the estimates
\begin{subeqnarray}
	\widetilde{C}_{\xi_A^{(2nd)}}/\widetilde{C}_{\xi_A^{(exp)}}
	\equiv \kappa_0 & = & 0.71 \pm 0.02 \slabel{C_xiA:1} \\
	a_2 \equiv \kappa_2/\kappa_0 & = & -0.43 \pm 0.10
\label{C_xiA_extr_wrong}
\end{subeqnarray}
[see Figure \ref{fig_su3_scal_chif}(b), points $\times$].
It is interesting to note the rough agreement between 
this prediction of $a_2$ and \reff{a2:fit}. On the other hand,
if we trust our extrapolation of $\xi_{A,\infty}$, 
we try a similar fit including all values of $\beta$.
We obtain a good fit if we restrict attention
to the points with $\beta \ge 3.15$ ($\xi_F \gtapprox 2.7 \times 10^3$), 
yielding the estimates
\begin{subeqnarray}
	\widetilde{C}_{\xi_A^{(2nd)}}/\widetilde{C}_{\xi_A^{(exp)}} 
	\equiv \kappa_0 & = & 0.63 \pm 0.02 \\
	a_2 \equiv \kappa_2/\kappa_0 &= & 0.8 \pm 0.4 
\end{subeqnarray}
The total disagreement between this prediction
of $a_2$ and \reff{a2:fit} seems to indicate that \reff{C_xiA_extr_wrong}
is more trustworthy.

The ``improved'' three-loop estimates of course show the same
pseudo-periodic oscillations. If we fit
$\xi_A^{(2nd)}/\xi_{A,\B,3-loop}^{(exp)} = \kappa'_0 + \kappa'_2 \beta^{-2}$,
a good fit is obtained for $\beta \ge 2.925$ 
($\xi_F \gtapprox 1.1 \times 10^3$), and we obtain the estimates
\begin{subeqnarray}
	\widetilde{C}_{\xi_A^{(2nd)}}/\widetilde{C}_{\xi_A^{(exp)}}
	\equiv \kappa'_0 & = & 0.65 \pm 0.02 \slabel{C_xiA:2} \\
	a^{(imp)}_2 \equiv \kappa'_2/\kappa'_0 & = & 0.8 \pm 0.3
\end{subeqnarray}
[see Figure \ref{fig_su3_scal_chif}(b), points $\Diamond$].
However, this is not consistent with the estimate $a_2^{(imp)}=-0.04 \pm 0.05$
obtained from the fundamental sector (see footnote \ref{foot:a2_imp_fund}).

These estimates can be compared with the combination
\be
{\widetilde{C}_{\xi_A^{(2nd)}} \over \widetilde{C}_{\xi^{(exp)}_{A}}} \,=\,
\left( {\widetilde{C}_{\xi_A^{(2nd)}} \over \widetilde{C}_{\xi_F^{(2nd)}}} \right) \,
\left( {\widetilde{C}_{\xi_{F}^{(2nd)}} \over \widetilde{C}_{\xi_{F}^{(exp)}}  } \right) \,
\left( {\widetilde{C}_{\xi_{F}^{(exp)}} \over \widetilde{C}_{\xi_{A}^{(exp)}} } \right) \,
\ee
of our previous estimates.
Our estimate of the first term of the right side is
$1/(2.80 \pm 0.05)=0.357 \pm 0.006$ [see \reff{eq3a5}]; Monte Carlo simulations 
\cite{Rossi_94a}
predict that the second term of the right side is $0.987 \pm 0.002$;
and the theoretical prediction for the third term of the
right side is exactly 2 [see \reff{mT=2mV}].
This approach yields
\be
	{ { \widetilde{C}_{\xi_A^{(2nd)}}} \over 
	  \widetilde{C}_{{\xi^{(exp)}_{A,3-loop}}} } = 0.70 \pm 0.01 \;.
	\label{C_xiA:3}
\ee
Let us recall that this approach suffers from various difficulties
in the estimation of $\xi_A^{(2nd)} / \xi_F^{(2nd)}$
(see Section \ref{sec5.2.2}); but, in spite of that, the result
\reff{C_xiA:3} is consistent with \reff{C_xiA:1}, and 
marginally consistent with \reff{C_xiA:2}.
A reasonable compromise would be to take
\be
        { \widetilde{C}_{\xi_A^{(2nd)}} \over 
  	  \widetilde{C}_{\xi^{(exp)}_{A}} }
\approx 0.69 \pm 0.04 \; .
\label{C_xiA_factor}
\ee

\subsubsection{Adjoint Sector: Susceptibility} \label{sec5.3.6}

The story for $\chi_A$ is similar to that of $\chi_F$:
we can either use
$\chi_A$ directly or else use the ratio $\chi_A/{\xi_A^{(2nd)}}^2$.

In Figure~\ref{fig_su3_scal_chia}(a)
we plot $\chi_{A,\infty,estimate\,(\infty,\infty,0.90)}$
divided by the theoretical prediction \reff{chiA_predicted2}/\reff{d1}
{\it with the prefactor} $\widetilde{C}_{\chi_A}$ {\it omitted};
the $\beta \rightarrow \infty$ limit of this curve thus an
estimate of $\widetilde{C}_{\chi_A}$. Here we have two-loop
and three-loop predictions from standard perturbation theory
(points $+$, $\times$) as well
as ``improved'' two-loop and three-loop predictions ($\Box$, $\Diamond$).
At each order (two-loop or three-loop), the standard and the ``improved''
estimates are virtually identical for $\beta \gtapprox 3$; but for 
$\beta \ltapprox 3$ the {\it standard} estimates are much flatter (in
marked contrast to what is observed for $\xi_F$, $\chi_F$ and $\xi_A$:
see Figures~\ref{fig_su3_scal_xif}--\ref{fig_su3_scal_xia}).
This casts some doubts on whether the ``improved'' 
perturbation theory is always an improvement! If we fit
the ``standard'' 3-loop perturbation values 
$\chi_A/(\chi_{A,3-loop}\hbox{ without the prefactor }$
$ \widetilde{C}_{\chi_A} )
= \kappa_0 + \kappa_2 \beta^{-2}$,
a good fit is obtained if we restrict attention to the points with $\beta
\ge 2.25$ ($\xi_{F,\infty} \gtapprox 70$), and we obtain the estimates
\begin{subeqnarray}
	\widetilde{C}_{\chi_A}
	\equiv \kappa_0 & = & 196 \pm 2  \slabel{pred_chi_a_bare} \\
	d_2 \equiv \kappa_2/\kappa_0 & = & 0.14 \pm 0.50 
\end{subeqnarray}
[see Figure \ref{fig_su3_scal_chia}(b), points $\times$].
Similarly, if we fit
the ``improved'' 3-loop perturbation values
$\chi_A/(\chi_{A,3-loop}\hbox{ without the prefactor } \widetilde{C}_{\chi_A})
= \kappa'_0 + \kappa'_2 \beta^{-2}$, a good fit is obtained for $\beta
\ge 2.55$ ($\xi_{F,\infty} \gtapprox 110$), and we get
\begin{subeqnarray}
	\widetilde{C}_{\chi_A}
	\equiv \kappa'_0 & = & 191 \pm 3 \slabel{pred_chi_a_imp} \\
	d^{(imp)}_2 \equiv \kappa'_2/\kappa'_0 & = & 0.53 \pm 0.12
\end{subeqnarray}
[see Figure \ref{fig_su3_scal_chia}(b), points $\Diamond$].

Similarly, we could plot 
$(\chi_A/{\xi_A^{(2nd)}}^2)_{\infty,estimate\,(\infty,\infty,\infty)}$
divided by the theoretical prediction 
\reff{chiAoverxiAsquared_predicted}/\reff{e1}
{\it with the prefactors $\widetilde{C}_{\chi_A}$ and
$\widetilde{C}_{\xi_A^{(2nd)}}$  omitted}; 
the $\beta \rightarrow \infty$ limit of this curve would thus give an
estimate of $\widetilde{C}_{\chi_A}/(\widetilde{C}_{\xi_A^{(2nd)}})^2$.
However, in order to make the vertical scale of this graph more
directly comparable to that of Figure~\ref{fig_su3_scal_chia}(a), 
we have multiplied
the quantity being plotted by $(\widetilde{C}^{(\B)}_{\xi_A^{(exp)}})^2$.
Note that this does not in any way alter the {\it logic} of the
analysis, as $\widetilde{C}^{(\B)}_{\xi_A^{(exp)}}$ is an explicit number
obtained from \reff{mT=2mV}/\reff{exact_Cxi}.
The resulting curve is plotted in Figure~\ref{fig_su3_scal_chia_ov_xia2}(a); 
its $\beta \rightarrow \infty$ limit gives
an estimate of $\widetilde{C}_{\chi_A}\,(\widetilde{C}^{(\B)}_{\xi_A^{(exp)}}
/\widetilde{C}_{\xi_A^{(2nd)}})^2$. In this case we just have two-loop and
three-loop predictions ($+$, $\times$)
as well as ``improved'' two-loop and three-loop predictions 
($\Box$, $\Diamond$). To convert this number to an estimate for
$\widetilde{C}_{\chi_A}$ itself, we need to multiply by
\be
	\left( {\widetilde{C}_{\xi_A^{(2nd)}} \over
	\widetilde{C}^{(\B)}_{\xi_A^{(exp)}} } \right) ^{\! 2} 
	\approx (0.69 \pm 0.04)^2 = 0.476 \pm 0.055 
\label{const_chia_xia}
\ee
[see \reff{C_xiA_factor}].
The estimates from two-loop and three-loop standard perturbation theory
(which are virtually identical since $e_1 \approx 0.0075$ is so small)
are strongly decreasing for $\beta \ltapprox 3.0$ and weakly decreasing
thereafter. If we fit
the 3-loop values to $\kappa_0 + \kappa_2 \beta^{-2}$,
a good fit is obtained if we restrict attention to the points with $\beta
\ge 3.075$ ($\xi_{F,\infty} \gtapprox 2 \times 10^3$), 
and we obtain the estimates
\begin{subeqnarray}
        \widetilde{C}_{\chi_A} \left( {\widetilde{C}_{\xi_A^{(2nd)}} \over
        \widetilde{C}^{(\B)}_{\xi_A^{(exp)}} } \right) ^{\! -2}
        \equiv \kappa_0 & = & 456 \pm 8 \\
        e_2 \equiv \kappa_2/\kappa_0 & = & 1.8 \pm 0.3 
\end{subeqnarray}
and hence
\be
        \widetilde{C}_{\chi_A} = 217 \pm 16 \;
\ee
[see Figure~\ref{fig_su3_scal_chia_ov_xia2}(b), points $\times$].
This estimate is in reasonable agreement with the previous predictions 
\reff{pred_chi_a_bare} and \reff{pred_chi_a_imp}, but exhibits much {\it larger}
uncertainties (in contrast to the situation observed for the fundamental
susceptibility, where $\chi_F/\xi_F^{(2nd)}$ is better behaved
than $\chi_F$).
The estimates from ``improved'' three-loop perturbation theory are
are virtually identical to those from three-loop standard perturbation 
theory. If we fit the ``improved'' three-loop estimates 
to $\kappa'_0 + \kappa'_2 \beta^{-2}$,
a good fit is obtained for $\beta \ge 3.075$
($\xi_{F,\infty} \gtapprox 2 \times 10^3$),
and we obtain the estimates
\begin{subeqnarray}
        \widetilde{C}_{\chi_A} \left( {\widetilde{C}_{\xi_A^{(2nd)}} \over
        \widetilde{C}^{(\B)}_{\xi_A^{(exp)}} } \right) ^{\! -2}
        \equiv \kappa'_0 & = & 459 \pm 7 \\
        e^{(imp)}_2 \equiv \kappa'_2/\kappa'_0 & = & 1.5 \pm 0.3
\end{subeqnarray}
and hence
\be
        \widetilde{C}_{\chi_A} = 218 \pm 15 \;
\ee
[see Figure~\ref{fig_su3_scal_chia_ov_xia2}(b), points $\Diamond$].

All methods yield consistent results, but the ones based on
$\chi_A$ alone show an earlier convergence to the limiting constant;
furthermore, as they do not require information about the constant
\reff{const_chia_xia}.
Thus, a reasonable compromise would be 
\be
   \widetilde{C}_{\chi_A} \approx 195 \pm 20 \;.
\ee

\subsection{Discussion}   \label{sec5.4}

Let us summarize our findings.
First, we hope to have convinced the reader 
that the extrapolation method has allowed us to obtain the infinite-volume 
behavior of long-distance observables with
reliable control over all systematic and statistical errors,
using only lattices $L \le 256$.
For example, we obtain the infinite-volume correlation length
$\xi_{F,\infty}^{(2nd)}$ to a statistical accuracy
of order 0.5\% (resp.\ 0.9\%, 1.1\%, 1.3\%, 1.5\%)
when $\xi_{F,\infty}^{(2nd)} \approx 10^2$
(resp.\ $10^3$, $10^4$, $10^5$, $4 \times 10^5$);
and the systematic errors arising from
corrections to scaling are of the same order or smaller.
The situation is similar for $\chi_F$ and $\chi_A$.
Only for $\xi_{A}^{(2nd)}$
and the two ratios $\chi_\#/(\xi_{\#}^{(2nd)})^2$
do we have severe worries about the possible remaining
corrections to scaling;
for these observables it would be useful to carry out simulations
at larger $L$, so that our fits can be checked against alternative fits
with larger $L_{min}$.

It is important to remark that the validity of these extrapolated
data rests on the {\em assumption}\/ that if finite-size-scaling \reff{eq2}
is found empirically to be satisfied (with a given function $F_\scro$)
for some range of $L$, say $L_{min} \le L \le L_{max}$,
then it will {\it also}\/ hold (with the same $F_\scro$) for $L > L_{max}$.
Now, we have found that
the data for 16--$32 \le L \le 128$ {\em are}\/ in good agreement with
rapid convergence as $L$ grows at fixed $x \equiv \xi_F^{(2nd)}(L)/L$ to a 
finite-size-scaling function $F_\scro(x;s)$.
It is then reasonable to {\em assume}\/ that the
limiting function empirically obtained for $L \le 128$ is close to
the true limiting function as $L \rightarrow \infty$,
i.e.\ that the systematic errors are in fact as small
as they seem empirically to be.
Of course, it is possible that apparent convergence of $F_\scro(x;s)$
for 16--$32 \le L \le 128$ is misleading ---
i.e.\ a ``false plateau'' ---  and that for very large 
values of $L$ the convergence is to a very different function
(or there is convergence at all).
This caveat is not special to our work,
but is inherent in {\em any}\/ numerical work that
attempts to evaluate a limit (here $L\to\infty$) by taking the relevant
parameter {\em almost}\/ to the limit (here $L$ large but finite).
In particular, this caveat is inherent in {\em all}\/ numerical (as well
as experimental) work in the fields of critical phenomena
and quantum field theory.

In any case, there is no evidence that this unfortunate situation
occurs in our model.  Indeed, what is remarkable in our model
is the extreme {\em weakness}\/ of the corrections to scaling:
for example, for $\scro = \xi_F^{(2nd)}$, even at $L=8$ the corrections to
scaling are almost unobservable for $x \gtapprox 0.8$. 
Even at smaller values of $x$, where the corrections to scaling
are clearly visible, they are perfectly consistent with a behavior
roughly of the form\footnote{
  We do not claim that the leading correction-to-scaling term
  is exactly of order $1/L^2$.
  Indeed, in $n^{th}$-order perturbation theory
  we know that terms of the form $\log^p L/L^2$ with $0 \le p \le n$
  are present, but we do not know how to resum these logarithms
  except in one exactly soluble case:
  In the $N$-component mixed isovector/isotensor model at $N=\infty$,
  these terms resum to give a correction of the form
  $L^{-2} (c_1 \log L + c_0 + c_{-1}/\log L + c_{-2}/\log^2 L + \cdots)$
  \cite{CP_LAT96,CP_1overN},
  as discussed further around \reff{large_N_FSS_exact} below.
  In the present case, the correction to scaling might be of the
  form $L^{-2} \times \hbox{logarithms}$, or it might be of the form
  $L^{-\omega}$ with $\omega \neq 2$ --- we don't know.
}
\be
   {{\cal O}(\beta,2L) \over {\cal O}(\beta,L)}
   \;=\;  F_{\cal O}(x) \,+\, {1 \over L^2} G_{\cal O}(x) \,+\, \ldots
   \;,
\ee
at least in the range $8 \le L \le 128$ where we have data.
If all hell breaks loose for larger $L$,
we certainly see no hint of it at $L \le 128$.

Our second principal observation is that the
finite-size-scaling functions $F_\scro$,
as determined from our Monte Carlo data by the above analysis,
agree well with the perturbative predictions for large $x$
[cf.\ \reff{all_FSS_PT}].
More precisely, we found good agreement for
$x \gtapprox 0.6$--0.9 (depending on the observable) up to the largest $x$
observed in our data ($x_{max}\approx 1.2$).

We would like to clarify the logic concerning this point,
which has caused some controversy
\cite{Pat-Seil_comment,CEPS_O3_reply_to_Pat-Seil}.
There are two {\em very different}\/
limits that can be taken in a two-dimensional $\sigma$-model:
\begin{itemize}
   \item[(a)] the finite-volume perturbative limit
       $\beta\to\infty$ at {\em fixed}\/ $L < \infty$;
   \item[(b)] the finite-size-scaling limit
       $\beta\to\infty$ and $L\to\infty$ such that the ratio
       $x \equiv \xi(\beta,L)/L$ is held fixed.
\end{itemize}
There is no doubt that conventional perturbation theory
[cf.\ \reff{fss:xichi1}]  is valid in limit (a):
it concerns, after all, a finite-dimensional integral.
The deep question is how the {\em remainder terms}\/
in this asymptotic expansion behave as a function of $L$;
one wants in particular to know whether the perturbation theory derived
from the study of limit (a) is {\em also}\/ correct in the double limit
obtained by first taking limit (b) and then taking $x\to\infty$
[cf.\ \reff{fss:Fxichi}].
The conventional wisdom says {\em yes}\/:
indeed, this or a similar interchange of limits
underlies the conventional derivations of asymptotic freedom
(a point that unfortunately has not always
 been clearly acknowledged by advocates of the conventional wisdom).
By contrast,
Patrascioiu and Seiler \cite{Patrascioiu_85,Pat-Seil_91,Pat-Seil_95ab}
say {\em no}\/:
they suspect that asymptotic freedom is false \cite{Pat-Seil_LAT92}.
At present, no rigorous proof is available to settle this question
one way or the other.

Patrascioiu and Seiler \cite{Pat-Seil_comment} have objected
that our data in the large-$x$ region are essentially perturbative,
in the sense that they are well reproduced by the finite-volume
perturbation theory [limit (a)], whose validity is not in question.\footnote{
   Their objection concerned our earlier work on the $O(3)$ $\sigma$-model
   \cite{o3_scaling_prl}, but it can be considered also
   in the present context.
}
For this reason, they contend, we are implicitly {\em assuming}\/
asymptotic scaling.  Our reply is twofold:

On the one hand, it is not quite true that our data 
in the large-$x$ region are ``essentially perturbative'':
as noted in Section \ref{sec5.3.2},
our raw data $\scro(\beta,L)$ deviate significantly from the
finite-volume perturbation expansions \reff{all_finiteV_PT}.
In the fundamental sector, the deviation $|S_3|$ between
$\chi_F$ and second-order perturbation theory is less than
1.0\% (resp.\ 0.3\%) for $x \ge 0.6$ (resp.\ $x \ge 0.9$);
while the deviation $|R_2|$ between
$\xi_F^{(2nd)}$ and first-order perturbation theory is less than
9.8\% (resp.\ 4.1\%) for these same intervals of $x$.\footnote{
   The deviations would have been considerably larger if we had defined them
   as $(\scro_{exact} - \scro_{pert})/\scro_{exact}$ instead of
   $(\scro_{exact} - \scro_{pert})/\scro_{zeroth-order}$;
   this is because the first-order perturbation corrections
   are large and {\em negative}\/.
}
All these deviations are observed at $L=128$, which is the largest value of $L$
for which we have data in the specified intervals of $x$;
rather smaller deviations are obtained at the same $x$ and smaller $L$.
For example, for $L=8$ we have discrepancies of
1.1\% (resp.\ 0.2\%) for $\chi_F$, and
6.5\% (resp.\ 2.2\%) for $\xi_F^{(2nd)}$.\footnote{
   This nonuniformity with $L$ at fixed $x$ can be at least roughly explained:
   One expects $k$-loop finite-volume perturbation theory
   to have an error term of order $(\log L)^{k+1}/\beta^{k+1}$.
   On the other hand, from \reff{xiV_finiteV_PT} we see that
   $$
     x^2  \;\sim\;  \beta \,-\, \log L \,-\, {\log^2 L \over \beta}
                                       \,-\, \cdots
   $$                                    
   (omitting all constants and subleading terms),
   so that $\beta \sim \log L$ if we keep $x$ fixed.
   It follows that if we try to use finite-volume perturbation theory
   in the limit $L\to\infty$ at $x$ {\em fixed}\/
   --- where it is not intended to be used!\ ---
   the error term will be of order 1 as $L\to\infty$.
 
   It is not clear to us why the error term appears to be {\em growing}\/
   as $L\to\infty$ at fixed $x$.  This {\em could}\/ be a sign that
   the coefficient of the $O(1/\beta^{k+1})$ remainder term
   grows {\em more}\/ rapidly than $(\log L)^{k+1}$,
   as contended by Patrascioiu and Seiler
   \cite{Patrascioiu_85,Pat-Seil_91,Pat-Seil_95ab}.
   However, our data do {\em not}\/ support this interpretation
   (see Section~\ref{sec5.3.2}).
   More likely, this is a preasymptotic effect
   arising from the fact that (reinserting the constants) we have
   from first-order perturbation theory
   $$
     \beta \;\approx\;   {8 \over 3} x^2  \,+\, {3 \over 4\pi} \log L
   $$
   [see also \reff{lim_Rtilde_fixed_x_1}];
   and even at $L=128$, the term $(3/4\pi) \log L$ is by no means
   dominant compared to $(8/3)x^2$.
}
For the adjoint sector, the agreement with perturbation theory is much worse:
at $L=128$, for $\chi_A$ we have discrepancies as large as
17.3\% (resp.\ 6.2\%) for $x \ge 0.6$ (resp.\ $x \ge 0.9$),
while for $\xi_A^{(2nd)}$ the discrepancies reach 12.2\% (resp.\ 5.0\%).
At $L=8$ things are much better for $\chi_A$,
with discrepancies reaching only 1.3\% (resp.\ 0.3\%);
but for $\xi_A^{(2nd)}$ they still reach 8.7\% (resp.\ 3.0\%).

Thus, it is only for $\chi_F$ (and $\chi_A$ on small lattices)
that our data are in any sense ``essentially perturbative'';
all the other observables show significant deviations
from finite-volume perturbation theory, even on the smallest lattices.
[To be sure, the agreement for $\xi_F^{(2nd)}$ and $\xi_A^{(2nd)}$
 would probably be improved dramatically if the two-loop
 (order-$1/\beta^2$) correction were available to us.]

Secondly, and more importantly,
we have {\em always}\/ analyzed our data in the sense of limit (b):
that is, at each {\em fixed}\/ $x \equiv \xi_F^{(2nd)}(\beta,L)/L$,
we have asked whether
the ratios ${\cal O}(\beta,2L)/{\cal O}(\beta,L)$ have a good limit
as $L\to\infty$, and if so we have attempted to evaluate this limit
numerically as explained above.
Thus, modulo the caveats discussed in the preceding paragraphs,
we believe we have determined (within statistical errors)
the true finite-size-scaling functions $F_\scro$.
There is no contradiction between Patrascioiu--Seiler's observation
and ours:
the same point $(\beta,L)$ may well lie within the range of validity
(to some given accuracy) of two distinct expansions.
The fact that some of our data points at large $x$ are consistent with
finite-volume perturbation theory [limit (a)]
does not constitute evidence against their {\em also}\/ being consistent with
nonperturbative finite-size scaling [limit (b)].
For this reason, we disagree with Patrascioiu--Seiler's claim
\cite{Pat-Seil_comment}
that our method has implicitly {\em assumed}\/ asymptotic scaling.
Quite the contrary:  our data at $x \gtapprox 0.6$--0.9,
interpreted in the sense of limit (b),
constitute a {\em test}\/ of the key assumptions
underlying the derivation of asymptotic scaling.

Finally, we have made in Sections~\ref{sec5.3.3}--\ref{sec5.3.6}
a direct test of asymptotic scaling for the various infinite-volume
long-distance observables $\scro_\infty(\beta)$.
As noted in the Introduction, this test involves two distinct questions:
\begin{itemize}
  \item[(i)]  Does the ratio between the extrapolated values
     and the $l$-loop asymptotic-freedom prediction,
\be
        C_\scro(\beta)  \;\equiv\;  {\scro_\infty(\beta) \over
                e^{a\beta} \beta^b (1 + k_1/\beta + \cdots + k_l/\beta^l)}\;,
\ee
     converge to a constant in the limit $\beta \to \infty$,
     modulo corrections of order $1/\beta^{l+1}$?
  \item[(ii)] In the case of $\scro = \xi_F^{(2nd)}$,
     does this constant equal the nonperturbative value predicted by
     the thermodynamic Bethe Ansatz \cite{Balog_92}
     combined with the best Monte Carlo estimate \cite{Rossi_94a}
     of $\xi_F^{(2nd)} / \xi_F^{(exp)}$?
\end{itemize}
Patrascioiu and Seiler \cite{Pat-Seil_comment} are right to point out
that our affirmative answer to question (i) is in some sense
a foregone conclusion:
since our Monte Carlo data for
$F_{\cal O}(x)$ at $x \gtapprox 0.6$--0.9 do in fact agree reasonably well
with the two-loop perturbative formula \reff{all_FSS_PT},
and our data for ${\cal O}(\beta,L)$ also agree at least roughly
with the fixed-$L$ perturbation expansion \reff{all_finiteV_PT},
it is then inevitable that our extrapolated values $\scro_\infty(\beta)$
at the largest values of $\beta$ will be consistent with
two-loop asymptotic scaling in the sense that
$\scro_\infty(\beta)/[e^{a\beta} \beta^b]$
will be roughly constant.\footnote{
   This statement is not strictly correct, as the fixed-$L$ perturbation
   expansion \reff{all_finiteV_PT} is only a ``one-loop'' expansion,
   in the sense that it is sufficient
   to obtain the one-loop renormalization-group coefficient $w_0$
   (as well as $\gamma_0$) but not subsequent coefficients
   [see \reff{fss:RGpertxichi} for definitions].
   To obtain the two-loop coefficient $w_1$ from an expansion of this type,
   it would be necessary to go to one higher order.
}
Therefore, our observation in Sections~\ref{sec5.3.3}--\ref{sec5.3.6}
of asymptotic scaling in sense (i)
contains no significant information
{\em beyond}\/ our observation in Section~\ref{sec5.2.1}
that the finite-size-scaling functions $F_{\cal O}(x)$ agree with
the perturbative predictions at $x \gtapprox 0.6$--0.9;
this latter observation already contains the essence of
asymptotic scaling in sense (i).
On the other hand, asymptotic scaling in sense (ii)
is truly an additional observation:
it is by no means inevitable that the observed constant value
of $C_{\xi_F^{(2nd)}}(\beta)$ at large $\beta$ will agree
with the thermodynamic Bethe Ansatz prediction to within a fraction
of a percent, as we have found here (Section~\ref{sec5.3.3}).
It seems to us that this apparent coincidence is significant evidence
in favor of the asymptotic-freedom picture.

We have found empirically that the ``energy-improved'' perturbative
expansion usually exhibits asymptotic scaling (to a given accuracy)
at lower values of $\beta$ than its bare-parameter counterpart.
The exceptions to this behavior are the observables $\chi_A$,
for which standard perturbation theory has an incredibly flat behavior
and the ``energy-improved'' perturbation theory is {\em less}\/ well behaved,
and $\chi_A/\xi_A^2$, for which the two expansions exhibit nearly
identical behavior.
This generally good behavior of the ``energy-improved'' expansion
confirms similar observations in other models,
such as the $N$-vector models for $N=3,4,8$
\cite{Wolff_O4_O8,o3_scaling_LAT94,o3_scaling_prl,lat95_4loop,%
o3_scaling_fullpaper,MGMC_ON},
the $CP^{N-1}$ $\sigma$-models for $N=2,4,10$ \cite{Campostrini_CPN},
the $SU(N)$ chiral models for $N=4,6,9,15,21,30$
\cite{Drummond_94,Rossi_94a,Rossi_94b},
and the $SU(2)$ and $SU(3)$ lattice gauge theories
\cite{Lepage_93,El-Khadra_LAT93,Schilling_LAT93,Michael_LAT94,Weisz_LAT95}.
What we lack is a good {\em theoretical}\/ explanation of this empirically
observed behavior (see Section \ref{sec3.3}).
It is not clear to us whether ``improved'' perturbation theory
can {\em systematically}\/ be expected to reach asymptotic scaling faster
than standard perturbation theory (except for some unusual cases in which 
standard perturbation theory has small high-order terms), or whether 
its apparent success is illusory.

\section{Finite-Size-Scaling Analysis: Dynamic Quantities}
\label{section:finite-size-scaling_dynamic}  \label{sec6}

%\subsubsection{Integrated Autocorrelation Times}  \label{sec6.1}

Of all the observables we studied, the slowest mode (by far)
is the squared fundamental magnetization $\scrm_F^2$:
this quantity measures the relative rotations of the spins
in different parts of the lattice,
and is the prototypical $SU(N)$-invariant ``long-wavelength observable''.
The autocorrelation time $\tau_{int,\scrm_F^2}$
has the following qualitative behavior:
as a function of $\beta$ it first rises to a peak and then falls;
the location of this peak shifts towards $\beta=\infty$ as $L$ increases;
and the height of this peak grows as $L$ increases.
A similar but less pronounced peak is observed in $\tau_{int,\scrm_A^2}$.
By contrast,
the integrated autocorrelation times of the energies, $\taueF$ and $\taueA$,
are much smaller and vary only weakly with $\beta$ and $L$.
This is because the energies are primarily ``short-wavelength observables'',
and have only weak overlap with the modes responsible for critical slowing-down.

Let us now make these considerations quantitative, by applying
finite-size scaling to the dynamic quantities
$\tau_{int,\scrm_F^2}$ and $\tau_{int,\scrm_A^2}$.
We use the Ansatz
\be
  \tau_{int,A}(\beta,L)  \;\sim\;
     \xi(\beta,L)^{z_{int,A}} \, g_A \Bigl( \xi(\beta,L)/L  \Bigr)
 \label{dyn_FSS_Ansatz}
\ee
for $A = \scrm_F^2$ and $\scrm_A^2$.
Here $g_A$ is an unknown scaling function, and
$g_A(0) = \lim_{x \downarrow 0} g_A(x)$
is supposed to be finite and nonzero.\footnote{
   It is of course equivalent to use the Ansatz
   $$\tau_{int,A}(\beta,L)  \;\sim\;
     L^{z_{int,A}} \, h_A \Bigl( \xi(\beta,L)/L  \Bigr)  \;,$$
   and indeed the two Ans\"atze are related by $h_A(x) = x^{z_{int,A}} g_A(x)$.
   However, to determine whether
   $\lim_{x \downarrow 0} g_A(x) = \lim_{x \downarrow 0} x^{-z_{int,A}} h_A(x)$
   is nonzero, it is more convenient to inspect a graph of $g_A$
   than one of $h_A$.
}
We determine $z_{int,A}$ by plotting $\tau_{int,A}/ \xi_F(L)^{z_{int,A}}$
versus $\xi_F(L)/L$ and adjusting $z_{int,A}$ until the points fall as closely
as possible onto a single curve (with priority to the larger $L$ values).
We emphasize that the dynamic critical exponent $z_{int,A}$
is in general {\em different}\/ from the exponent $z_{exp}$
associated with the exponential autocorrelation time $\tau_{exp}$
\cite{Sokal_Lausanne,CPS_90,Sokal_LAT90}.

Using the procedure just described, we find
\be
   z_{int,\scrm_F^2}   \;=\;  0.45 \pm 0.02 
 \label{z_estimates}
\ee
(subjective 68\% confidence limits).
In Figure \ref{fig_su3_dynamic_fss}
we show the ``best'' finite-size-scaling plot.
Note that the corrections to scaling are very weak:
only the $L=16$ points clearly deviate from the asymptotic 
scaling curve; the  $L=32$ and $L=64$ points show barely
significant deviations.

It is worth noting that the finite-size effects on dynamic quantities
are {\em very strong}\/ at $\xi/L$ as small as 0.1 or even 0.05,
whereas the finite-size effects on static quantities are negligible 
already when $\xi(L)/L \ltapprox 0.15$:
compare Figure~\ref{fig_su3_dynamic_fss} with Figure~\ref{fig_su3_dev_xif}.
Indeed, in Figure~\ref{fig_su3_dynamic_fss} it is far from clear
what is the limiting value of the scaling function,
$g_{\scrm_F^2}(0) = \lim_{x \downarrow 0} g_{\scrm_F^2}(x)$, and 
whether it is nonzero.
This extremely strong dynamic finite-size effect (here a factor
of order 5--10 for $\xi(L)/L$ between 0 and 0.2) seems to occur rather
frequently in collective-mode Monte Carlo algorithms: see e.g.\
\cite{MGMC_O4} for multi-grid in the two-dimensional 4-vector model,
and \cite{CEPS_RPN_dynamics} for the Swendsen-Wang-Wolff algorithm
in the two-dimensional $RP^{N-1}$ models.
 We conclude that finite-size corrections to dynamic critical behavior
 can be surprisingly strong;
 therefore, serious studies of dynamic critical phenomena
 {\em must}\/ include a finite-size-scaling analysis.
 It can be very misleading to assume that the finite-size corrections
 to dynamic quantities are small simply because $\xi/L$ is small,
 or because the finite-size corrections to {\em static}\/ quantities
 are small.

We can also analyze the dynamic critical behavior for the adjoint sector.
Proceeding as before, we obtain
\be
   z_{int,\scrm_A^2}   \;=\;  0.45 \pm 0.03
\ee
(subjective 68\% confidence limits). In Figure \ref{fig_su3_dynamic_fss_adj}
we show the ``best'' finite-size-scaling plot. Note that both
the magnitude and shape of the finite-size-scaling plot are similar for
$\scrm_F^2$ and $\scrm_A^2$, although the details are slightly different.
Note also that $z_{int,\scrm_F^2}$ and $z_{int,\scrm_A^2}$ are equal within
error bars; this contrasts with the behavior observed in MGMC for the 3-vector
model \cite{MGMC_ON}, 
in which the isotensor dynamic critical exponent
$z_{int,\scrm_T^2}$ appears to be {\em strictly smaller}\/
than the isovector exponent $z_{int,\scrm_V^2}$.
More work will clearly be required to sort out what is going on here.

\section*{Acknowledgments}

We wish to thank Sergio Caracciolo,
Martin Hasenbusch, Tereza Mendes, Steffen Meyer and Ettore Vicari
for helpful discussions.
The computations reported here were carried out
on the Cray C-90 at the Pittsburgh Supercomputing Center (PSC)
and on the IBM SP2 cluster at the Cornell Theory Center (CTC).
This work was supported in part 
by the U.S.\ National Science Foundation grants DMS-9200719
and PHY-9520978 (G.M.\ and A.D.S.),
and by NSF Metacenter grant MCA94P032P.

\appendix

\section{Perturbation Theory for the Non-Derivative Irreducible Operators}
   \label{appenA}

In this section we will compute the perturbative 
(large-$\beta$) predictions for a general two-point correlation 
function\footnote{ 
	Here we have inserted a factor $1/d_r$, as in \reff{def_energies}
	but contrary to \reff{green_func}. We hope this will not cause any
	confusion.}
\be
G_r (x;\beta) \;=\;  {1\over d_r} \< \chi_r (U_0 U_x^\dagger)\>
\label{funzionediGreen}
\ee
where the index $r$ labels an irreducible representation of $SU(N)$,
$\chi_r$ is the associated character and $d_r$ its dimension.
The perturbative expansion of $G_r(x;\beta)$ is obtained by 
setting\footnote{ In this appendix we use the summation convention 
	for repeated indices.}
\be
   U_x \;=\; \exp(iA_x) \qquad\hbox{with}\qquad
   A_x \;=\; A_x^a T^a
\ee
and then expanding in powers of $A$;
here $T^a$ are the generators of the Lie algebra ${\germansu(N)}$,
normalized so that $\hbox{Tr}(T^a T^b) = {1 \over 2} \delta^{ab}$,
and $A_x^a$ are $N^2 -1$ real fields.
We must also take into account the contributions from the integration 
measure. A straightforward calculation \cite{Kawai_81,Rothe_92}
shows that the Haar measure on $SU(N)$ is
\begin{subeqnarray}
dU_x &=& dA_x \exp\left\{ {1\over2} \hbox{Tr} \log
\left[ 2 (1 - \cos A^a_x T^a_A)\over (A^a_x T^a_A)^2 \right] \right\}  \\[2mm]
   &=& dA_x \exp \left[ - {N\over24} A^a_x A^a_x + O(A^4)\right]   \;,
  \slabel{app:eqA.3b}
\end{subeqnarray}
where $(T^a_A)_{bc} = - i f^{abc}$ are the $SU(N)$ generators 
in the adjoint representation,
for which $\hbox{Tr}(T^a_A T^b_A) = N \delta^{ab}$.

To compute the Green function \reff{funzionediGreen} we need the 
perturbative expansion of $\chi_r(U_0 U^\dagger_x)$. Let us first 
introduce $\Omega_x = \Omega^a_x T^a$ defined by
\be
e^{i \Omega_x} \equiv U_0 U_x^\dagger = e^{iA_0} e^{- i A_x} \; .
\ee
This $\Omega_x$ can be easily computed in terms of $A_0$ 
and $A_x$, using the Baker-Campbell-Hausdorff
formula. In terms of $\Omega_x$ we will now parametrize
\begin{eqnarray} 
{1\over d_r}  \chi_r(U_0 U^\dagger_x)   &=& 
   1 + \alpha_0 \hbox{Tr} \Omega_x^2 + 
       \alpha_{11} \hbox{Tr} \Omega_x^4 +
       \alpha_{12} (\hbox{Tr} \Omega_x^2)^2 +  
       \alpha_{21} \hbox{Tr} \Omega_x^6 
\nonumber \\
    && \qquad
      + \alpha_{22} (\hbox{Tr} \Omega_x^4)\, 
                     (\hbox{Tr} \Omega_x^2) +
       \alpha_{23} (\hbox{Tr} \Omega_x^2)^3 +
       \alpha_{24} (\hbox{Tr} \Omega_x^3)^2 
\nonumber \\
    && \qquad
	\, + \,    O(\Omega_x^8)
\end{eqnarray}
where the various constants depend on the representation 
$r$. Here $\alpha_0$ will be necessary in a calculation at
order $1/\beta$, 
$\alpha_{11}$ and $\alpha_{12}$ will appear at order $1/\beta^2$, while
$\alpha_{21},\ldots,\alpha_{24}$ will appear at order $1/\beta^3$. 
Let us notice 
that for low values of $N$ not all these invariants are independent.
Indeed it is easy to check that for $N=2$, $\hbox{Tr} \Omega_x^3$
vanishes; for $N=2,3$ we have
\be
     (\hbox{Tr} \Omega_x^2)^2 - 
     2 \hbox{Tr} \Omega_x^4 \;=\;  0 \; ;
\label{tracce1}
\ee
while for $N\le 5$ we have
\be
- (\hbox{Tr} \Omega_x^2)^3 + 
{8\over3}  (\hbox{Tr} \Omega_x^3)^2 + 
6 (\hbox{Tr} \Omega_x^4) \, 
  (\hbox{Tr} \Omega_x^2) -
8 \hbox{Tr} \Omega_x^6 \;=\;  0 \; .
\label{tracce2}
\ee
Before proceeding further let us give the explicit values of the 
various constants for the simplest representations:
\begin{enumerate} 
\item {\em Fundamental representation:}\/ In this case
$\chi_F(U_0 U^\dagger_x) = \hbox{Tr} (U_0 U^\dagger_x)$, $d_F=N$, and thus
\be
\alpha_0 =\, - {1\over2 N} \; ,\qquad
\alpha_{11} =\, {1\over 24 N} \; , \qquad
\alpha_{21} =\, - {1\over 720 N} \;\; ;
\label{alphafundrep}
\ee
all other coefficients are zero.
\item {\em Adjoint representation:}\/ We can consider the product
$f\otimes\overline{f} = (f\otimes\overline{f})_{traceless} \oplus {\bf 1}$,
where $f$ denotes the fundamental representation,
$\overline{f}$ denotes its complex conjugate,
and ${\bf 1}$ denotes the trivial representation. 
The representation $(f\otimes\overline{f})_{traceless}$,
whose dimension is $d_A = N^2 -1$, is the adjoint representation.
In this case 
$\chi_A(U_0 U^\dagger_x) = |\hbox{Tr} (U_0 U^\dagger_x)|^2 - 1$, so that
\begin{eqnarray}
\alpha_0 = - {N\over d_A} \; , \qquad
\alpha_{11} =  {N\over 12 d_A} \; , \qquad
\alpha_{12} =  {1\over 4 d_A} \; , \qquad \nonumber \\
\alpha_{21} = - {N\over 360 d_A} \; , \qquad
\alpha_{22} = - {1\over 24 d_A} \; , \qquad
\alpha_{23} = 0 \; , \qquad
\alpha_{24} = {1\over 36 d_A} \;\; .
\end{eqnarray}
\item We can also consider the product
$f\otimes f = (f\otimes f)_{symm} \oplus (f\otimes f)_{antisymm}$.
The latter two representations have dimensions
$d_{\pm} = N(N\pm 1)/2$, and
\be
\chi_{\pm}(U_0 U_x^\dagger) \;=\;
    {1\over2} \left( \hbox{Tr} (U_0 U_x^\dagger)\right)^2 \pm
    {1\over2} \hbox{Tr} ( U_0 U_x^\dagger U_0 U_x^\dagger)   \;.
\ee
We then have in the two cases
\begin{eqnarray}
\alpha_0 = - {1\over 2 d_{\pm}} (N\pm 2) \; ,\qquad 
\alpha_{11} =  {1\over 24 d_{\pm}} (N\pm 8) \; ,\qquad 
\alpha_{12} =  {1\over 8 d_{\pm}}  \; ,\qquad \nonumber \\
\alpha_{21} = -{1\over720 d_{\pm}} (N\pm 32) \; ,\qquad 
\alpha_{22} = -{1\over 48 d_{\pm}}  \; ,\qquad 
\alpha_{23} = 0  \; ,\qquad 
\alpha_{24} = -{1\over 72 d_{\pm}}  \; . \nonumber \\  \label{eqA10.a}
\end{eqnarray}
Notice that for $N=2$ the antisymmetric product is the identity representation,
while the symmetric product is the adjoint representation;
using \reff{tracce1} and \reff{tracce2} it is easy to show that
\reff{eqA10.a} are equivalent to the corresponding values
[$\alpha\equiv 0$ and \reff{alphafundrep}, respectively].
Similarly, for $N=3$ we have $(f\otimes f)_{antisymm} = \overline{f}$,
and it can again be checked that \reff{eqA10.a} is equivalent to
the complex conjugate of \reff{alphafundrep}.
\end{enumerate}

We want now to compute \reff{funzionediGreen} up to and
including terms of order $1/\beta^2$. In order to obtain this expression
we need to compute three different mean values, i.e.
$\< \hbox{Tr} \Omega_x^2\>$, 
$\< \hbox{Tr} \Omega_x^4\>$ and
$\< (\hbox{Tr} \Omega_x^2)^2\>$.
A simple Feynman-diagram calculation gives 
\begin{subeqnarray}
\< \hbox{Tr} \Omega_x^2 \> &=& 
    (N^2 - 1) \left[ {2\over\beta} J(x) + 
        {N^2 - 2\over 4 N \beta^2} J(x) + 
        {N\over 6\beta^2} J(x)^2 \right] +\, O(\beta^{-3})  \\
\< \hbox{Tr} \Omega_x^4 \> &=& 
  {4 (N^2-1)(2 N^2-3)\over N\beta^2} J(x)^2 +\, O(\beta^{-3}) \\
\< (\hbox{Tr} \Omega_x^2)^2 \> &=& 
  {4 (N^4 - 1)\over \beta^2} J(x)^2 + O(\beta^{-3})
\end{subeqnarray}
where
\be
   J(x) \;=\; \int\limits_{[-\pi,\pi]^2} \! {d^2 p\over (2 \pi)^2} \;
               {1 \,-\, \cos(p\cdot x) \over \hatp^2}
      \;.
\ee
A useful check is provided by the identity \reff{tracce1} for $N=2,3$.

We can now compute $G_r(x;\beta)$:
\begin{eqnarray}
&& G_r(x;\beta) \;=\; 1 + {2 (N^2-1) \alpha_0\over\beta} J(x)  \,+\,
          {N^2-1\over N\beta^2} \Biggl\{ 
          {(N^2-2)\alpha_0\over4} J(x)    \nonumber \\
&& \qquad + \left[ {1\over6}N^2\alpha_0 + 
    4 \alpha_{11} (2 N^2-3) + 4 \alpha_{12}N(N^2+1)\right] J(x)^2 \Biggr\}
   \;+\; O(\beta^{-3}) \;.
  \label{eqA.14}
\end{eqnarray}
In particular, for the fundamental and adjoint representations, we get
\begin{eqnarray}
G_F(x;\beta) &=& 1 - {N^2-1\over N\beta} J(x) \,
     + {(N^2 - 1) (N^2 - 2)\over 8 N^2 \beta^2} 
     \left[2 J(x)^2 - J(x)\right] + O(\beta^{-3}) \nonumber \\ \\[3mm]
G_A(x;\beta) &=& 1 - {2 N\over \beta} J(x) + 
       {3 N^2\over 2\beta^2} J(x)^2 - {N^2-2\over 4\beta^2} J(x) 
       + \, O(\beta^{-3})
\end{eqnarray}
The expression for $G_F(x)$ coincides with that given in 
\cite{Rossi_94b} apart from a different normalization of $\beta$.

{}From these expressions  it is immediate to derive expressions 
for the energies. Since $J({\bf e_1}) = \smfrac{1}{4}$, we have
\begin{eqnarray}
E_F(\beta) &=& 1 - {N^2 - 1\over 4 N \beta} - 
         {(N^2-1)(N^2-2)\over 64 N^2 \beta^2} + O(\beta^{-3}) \\
E_A(\beta) &=& 1 - {N\over 2\beta} + {N^2 +4\over 32 \beta^2} + 
         O(\beta^{-3})
\end{eqnarray}

We want now to derive the renormalization-group equations for 
the correlation function 
\reff{funzionediGreen}. As we are considering an irreducible 
representation, the Green function renormalizes multiplicatively 
and thus satisfies (for $a\to0$ or equivalently for $|x|\to\infty$)
a renormalization-group equation of the form
\be
\left[-a{\partial\over\partial a}+\, W^{lat}(\beta)
{\partial\over\partial(\beta^{-1})}+\, \gamma_r^{lat}(\beta)
\right]\, G_r(x^{cont}/a; \beta) \;=\; 0  \;,
\label{RGGreen}
\ee
where $W^{lat}(\beta)$ stands for the RG beta-function
of the lattice theory,
and $\gamma_r^{lat}(\beta)$ is the anomalous dimension for the 
representation $r$;
here $x^{cont}$ is a distance in centimeters,
$a$ is the lattice spacing in centimeters,
and $x \equiv x^{cont}/a$ is a lattice distance.
The function $W^{lat}$
is well known through order $1/\beta^4$ \cite{Rossi_94a}:
\be
W^{lat}(\beta)\;=\; 
- {w_0\over\beta^2} - {w_1\over\beta^3} -
{w_2^{lat} \over\beta^4} +\, O(\beta^{-5})
\ee
where
\begin{eqnarray}
w_0 & = & {N \over 4 \pi}  \\
\label{ap:w0}
w_1 & = & {N^2 \over 32 \pi^2}  \\
\label{ap:w1}
w_2^{lat} & = & {N^3 \over 128\pi^3} 
 \left[1 + {N^2-2 \over 2N^2} \pi - \pi^2 
   \left( { 2N^4-13N^2+18 \over 6N^4} + 4G_1 \right) \right] 
\label{ap:w2}
\end{eqnarray}
and
\be
   G_1 \;\approx\; 0.04616363   \;.
\ee
We have not bothered to add the superscript {\em lat}\/ to $w_0$ and $w_1$,
because these coefficients are universal
in the sense that they do not depend on the details of the lattice action.

We want now to obtain the function $\gamma_r^{lat}(\beta)$ 
through the term of order $1/\beta^2$. Expanding
\be
\gamma_r^{lat}(\beta)\;=\;  {\gamma_{r 0}\over\beta}+\,
{\gamma_{r 1}^{lat}\over\beta^2}+\, O(1/\beta^3) \; ,
\ee
we shall compute $\gamma_{r0}$ and $\gamma_{r1}^{lat}$. As $\gamma_{r0}$
does not depend on the specific lattice action we have not added
the superscript {\em lat}\/. To perform the computation we need 
the large-$|x|$ expansion of $J(x)$, which is given by
\cite[Sect. 4.2]{Itzykson}
\be
J(x) \;=\; {1\over 2\pi} \log |x|  \,+\,
           {1\over 2\pi} \left( \gamma_E + {3\over2} \log 2\right)  \,+\,
           o(1)
\ee
where $\gamma_E$ is the Euler constant.\footnote{
   Actually, the additive constant plays no role in the computation
   of the RG beta- and gamma-functions, at least up to the order
   we are considering here;
   all we need to know is that the coefficient of $\log |x|$ is $1/(2\pi)$.
}
Inserting \reff{eqA.14} into \reff{RGGreen} and comparing coefficients,
we obtain
\begin{eqnarray}
\gamma_{r 0} &=&  - {N^2-1\over\pi}\alpha_0  \\
\gamma_{r 1}^{lat} &=& - {(N^2-1)(N^2-2)\over 8\pi N}\alpha_0
\end{eqnarray}
Moreover, \reff{RGGreen} is satisfied only if the following 
non-linear relation among the $\alpha$
holds:
\be
- {N\over3} \alpha_0 - 2 (N^2-1)\alpha_0^2 + {4\over N} (2 N^2-3) \alpha_{11}
+ 4 (N^2+1)\alpha_{12} = 0
\ee
This identity should be satisfied by all {\em irreducible} representations
of $SU(N)$. We have explicitly verified it for the four representations
we have introduced at the beginning of this section.

For the fundamental representation we get
\begin{eqnarray}
\gamma_{F 0} &=&  {N^2-1\over 2\pi N}  \\
\gamma_{F 1}^{lat} &=& {(N^2-1)(N^2-2)\over 16\pi N^2}
\end{eqnarray}
while for the adjoint we have
\begin{eqnarray}
\gamma_{A 0} &=&   {N\over \pi} \\
\gamma_{A 1}^{lat} &=& {N^2-2\over 8\pi}
\end{eqnarray}
Of course, for $\gamma_F$ we reproduce the results of \cite{Rossi_94b}
after taking into account the different normalization of $\beta$.
Finally, we note that Rossi and Vicari \cite{Rossi_94b} have also calculated
$\gamma^{lat}_{F2}$; in our normalization of $\beta$ it is
\be
%% \gamma^{lat}_{F2} =   {{0.024868}\over {{N^3}}} - {{0.0455913}\over N} 
%%                       + 0.0252824\,N -   0.00455913\,{N^3} \; .
\gamma^{lat}_{F2}  \;=\;
   {N^2 -1 \over 384\pi^3}  \left[ (3+5\pi^2 + 24\pi^2 G_1) N 
                                   -25\pi^2 N^{-1} + 30\pi^2 N^{-3} \right]
\ee
A check on these results is provided by the 
fact that the $SU(2)$ chiral model is equivalent to the
4-vector model. Taking into account the different normalizations,
we have checked that $\gamma_F$ and $\gamma_A$, evaluated at $N=2$,
agree with the anomalous
dimensions of the spin-1 and spin-2 operators, respectively, 
in the 4-vector model \cite{CP_3loops}.

We can now use $\gamma_r^{lat}(\beta)$ and $W^{lat}(\beta)$
to determine the $\beta$-dependence of the representation-$r$
susceptibility $\chi_r = \sum_x G_r(x)$.\footnote{
	We apologize for using the same notation $\chi_r$ for
	both the character and the susceptibility; we trust 
	that it will not cause any confusion.}
{}From \reff{RGGreen} we have
\begin{subeqnarray}
\chi_r &=& C_{\chi_r} e^{2\beta/w_0}
  \left( {w_0 \over \beta}\right)^{2 w_1/w_0^2+\gamma_{r0}/w_0} 
  \; \times \nonumber \\
& & \quad \exp\left[\int_0^{1/\beta} dt\, \left({2\over W^{lat}(1/t)} +
  \, {2\over w_0 t^2} -\, {2w_1\over w_0^2 t}-\,
  {\gamma^{lat}_r(1/t)\over W^{lat}(1/t)}
  - \, {\gamma_{r0}\over w_0 t}\right)\right] \; \\[3mm]
&=&  C_{\chi_r} e^{2\beta/w_0}
  \left( {w_0\over \beta}\right)^{2 w_1/w_0^2+\gamma_{r0}/w_0}
  \left[ 1  + {b_1^{(r)} \over \beta} + {b_2^{(r)} \over \beta^2} 
  + \cdots \right]
\end{subeqnarray}
where $C_{\chi_r}$ is a non perturbative constant,
\be
b_1^{(r)} \,=\, 2 \left( {w_2^{lat}\over w_0^2} - {w_1^2\over w_0^3} \right)
  +  {\gamma_{r0} \over w_0} 
     \left( {\gamma_{r1}^{lat}\over \gamma_{r0}} 
     - {w_1\over w_0} \right) \; ,
\label{ap:b1}
\ee
and $b_2^{(r)}, b_3^{(r)}, \cdots$ can be determined analogously.
Likewise, for the correlation lengths we have
\begin{subeqnarray}
\xi_{\#}  &=&    C_{\xi_{\#}} e^{ {\beta / w_0} } \,
\left({w_0\over \beta}\right)^{w_1/w_0^2}
\exp \left[ \int_0^{1/\beta} dt\left({1\over W^{lat}(1/t)} +
{1\over w_0t^2} -
{w_1\over w_0^2 t}\right) \right] \\
&=& C_{\xi_{\#}} e^{\beta / w_0} \,
\left({w_0\over \beta}\right)^{w_1/w_0^2}
  \left[ 1  + {a_1 \over \beta} + {a_2 \over \beta^2}
  + \cdots \right]
\label{RG-eq-latt}
\end{subeqnarray}
where $C_{\xi_{\#}}$ is a non-perturbative constant,
\be
a_1 \,=\, {w_2^{lat}\over w_0^2} - {w_1^2\over w_0^3} \; ,
\label{ap:a1}
\ee
and $a_2, a_3, \cdots$ can be determined analogously. Finally,
for the ratio $\chi_r/\xi_{\#}^2$  we have
\be
{{\chi_r } \over {\xi_{\#}}^2}  
= { C_{\chi_r} \over C_{\xi_{\#}}^2 } 
  \, \left({w_0\over \beta}\right)^{\gamma_{r0}/w_0}
  \left[ 1  + {c_1^{(r)} \over \beta} + {c_2^{(r)} \over \beta^2}
  + \cdots \right]
\ee
with 
\be 
c_1^{(r)} = {\gamma_{r0} \over w_0}
     \left( {\gamma_{r1}^{lat}\over \gamma_{r0}}
     - {w_1\over w_0} \right) \; 
\ee
and so forth.

{}From \reff{ap:w0}--\reff{ap:w2} and \reff{ap:a1} we get:
\be
   a_1   \, = \,   - {3 \pi \over 8} N^{-3} +
                   \left({13 \pi \over 48} - {1 \over 8}\right) N^{-1} +
                   \left({1 \over 16 \pi} + {1 \over 16} - {\pi \over 24}
                   - {\pi \over 2} G_1\right) N \;.
  \label{appendix:a1}
\ee
Similarly, for the fundamental representation we get
\begin{eqnarray}
   b_1   &\equiv & b_1^{(F)}
       \; = \;  \left({1\over 2} - {3\over 4 \pi}\right) {1\over {N^3}} +
                \left({1 \over 4 \pi} - 1 
		+ {13 \pi\over 24}\right) {1\over N} +
                \left(-{1 \over 8\pi} + {3 \over 8} - {\pi\over 12} -
                         \pi G_1\right) \, N
   \nonumber \\[1mm] \\[2mm]
   c_1 &\equiv & c_1^{(F)} 
       \; = \;   (N^2 - 1) \left[ - {1 \over 2N^3} + {1 \over 4N}
                                  - {1 \over 4\pi N} \right]
   \\[2mm]
   c_2 &\equiv & c_2^{(F)}
       \; = \;   (N^2 - 1) \Biggl[ - {1 \over 8N^6}
                    + {1 \over 2N^4} \left( 1 - {1 \over 4\pi} \right)
                    - {1 \over 4N^2} \left( {17 \over 12} - {1 \over \pi}
                                            + {1 \over 8\pi^2} \right)
         \nonumber \\
   & & \qquad\qquad\qquad\qquad
       + {13 \over 192} - {3 \over 32\pi} + {1 \over 32\pi^2}
               + {G_1 \over 4}  \Biggr]
%%   {1\over {8N^6}} - {{0.585211}\over {{N^4}}} +
%%                {0.737966\over {{N^2}}} -0.330329 + 0.052573\,{N^2} 
\end{eqnarray}
while for the adjoint representation we get
\begin{eqnarray}
   d_1  &\equiv & b_1^{(A)}
       \; = \;   -{3\over 4 \pi} {1\over {N^3}} +
                \left({13 \pi \over 24} -  {5 \over 4}\right) {1\over N} +
                \left(-{3 \over 8\pi} + {5 \over 8} - {\pi\over 12} - \pi G_1\right) \, N
   \\[2mm]
   e_1  &\equiv & c_1^{(A)}
        \; = \;  - {{1}\over N} + \left({1\over 2} - {1\over 2 \pi}\right)\,N
\end{eqnarray}

\section{Perturbation Theory for Finite-Size-Scaling Functions}  \label{appenB}

\subsection{Theoretical Basis}   \label{appenB.1}

We work on a periodic lattice $\Lambda_L$ of linear size $L$.
The second-moment correlation length is defined by
\be
  \xi_\#^{(2nd)} (\beta,L)  \;=\;
                      { \left( { \displaystyle \chi_\#(\beta,L)
                                 \over
                                 \displaystyle F_\#(\beta,L)
                               } \,-\, 1
                        \right) ^{1/2}
                        \over
                        \displaystyle  2 \sin (\pi/L)
                      }
 \label{def:xibetaL}
\ee
where
\begin{subeqnarray}
   \chi_\#(\beta,L)   & = &   \sum\limits_{x\in\Lambda_L}
        G_\#(x;\beta,L)    \\[2mm]
   F_\#(\beta,L)      & = &   \sum\limits_{x\in\Lambda_L}
        G_\#(x;\beta,L) \, e^{ip_0\cdot x}
\end{subeqnarray}
with $\# = F$ or $A$;
here $p_0 \equiv (2\pi/L,0)$ is the smallest nonzero momentum.
Let $\scro$ be any long-distance observable
(e.g.\ the correlation length or the susceptibility).
Finite-size-scaling theory
\cite{Barber_FSS_review,Cardy_FSS_book,Privman_FSS_book}
then predicts quite generally that
\be
   {\scro(\beta,sL) \over \scro(\beta,L)}   \;=\;
   F_{\scro} \Bigl( \xi(\beta,L)/L \,;\, s \Bigr)
   \,+\,  O \Bigl( \xi^{-\omega}, L^{-\omega} \Bigr)
   \;,
 \label{FSS:eq2}
\ee
where $s$ is any fixed scale factor,
$F_\scro$ is a function characteristic of the
universality class,
and $\omega$ is a correction-to-scaling exponent.

In an asymptotically free model,
the functions $F_{\scro}$ can be computed in perturbation theory.
The starting point is a perturbation expansion
in powers of $1/\beta$ at fixed $L < \infty$:
\begin{subeqnarray}
   \xi(\beta,L)   & = &
     A \, \beta^{1/2} \, L
     \left[ 1 \,-\, {A_1(L) \over \beta} \,-\, {A_2(L) \over \beta^2}
              \,-\, O(\beta^{-3}) \right]
   \slabel{fss:xi1} \\[3mm]
   \chi(\beta,L)   & = &
     B \, L^2
     \left[ 1 \,-\, {B_1(L) \over \beta} \,-\, {B_2(L) \over \beta^2}
              \,-\, O(\beta^{-3}) \right]
   \slabel{fss:chi1}
   \label{fss:xichi1}
\end{subeqnarray}
where the functions $A_n(L)$ and $B_n(L)$ have the following
asymptotic behavior at large $L$:
\begin{subeqnarray}
   A_1(L)   & = &  A_{11} \log L \,+\, A_{10} \,+\,
                      O\!\left( {\log L \over L^2} \right)   \\[1mm]
   A_2(L)   & = &  A_{22} \log^2 L \,+\, A_{21} \log L \,+\, A_{20} \,+\,
                      O\!\left( {\log^2 L \over L^2} \right)   \\[1mm]
   B_1(L)   & = &  B_{11} \log L \,+\, B_{10} \,+\,
                      O\!\left( {\log L \over L^2} \right)   \\[1mm]
   B_2(L)   & = &  B_{22} \log^2 L \,+\, B_{21} \log L \,+\, B_{20} \,+\,
                      O\!\left( {\log^2 L \over L^2} \right)
 \label{fss:expansionAB}
\end{subeqnarray}
If we now {\em assume}\/ that the expansions \reff{fss:xichi1}
are valid also in the finite-size-scaling limit
$\beta,L \to\infty$ with $x \equiv \xi(\beta,L)/L$ fixed
followed by expansion in powers of $1/x^2$, we can obtain
\begin{subeqnarray}
   F_\xi(x;s)   & = &
    s\, \Biggl\{ 1 -  (A_{11} \log s) \left( {A \over x} \right)^{\! 2} 
                                                                \nonumber \\
    & & \qquad  \,-\,  \left[ \smhalf A_{11}^2 \log^2 s
                               + (A_{21} - A_{10} A_{11}) \log s \right]
                       \left( {A \over x} \right)^{\! 4}
                \,+\,  O(x^{-6})
       \Biggr\}
   \slabel{fss:Fxi} \\[3mm]
   F_\chi(x;s)   & = &
   s^2\,\Biggl\{ 1 -  (B_{11} \log s) \left( {A \over x} \right)^{\! 2} 
                                                                \nonumber \\
    & & \qquad  + \left[ (\smhalf B_{11}^2 - B_{11} A_{11}) \log^2 s
                 + (2A_{10} B_{11} - B_{10} B_{11} - B_{21}) \log s \right]
                       \left( {A \over x} \right)^{\! 4}
                                                                \nonumber \\
    & & \qquad   +\,  O(x^{-6})
       \Biggr\}
   \slabel{fss:Fchi}
   \label{fss:Fxichi}
\end{subeqnarray}
provided that
\begin{subeqnarray}
   A_{22} & = &  \smhalf A_{11}^2     \slabel{fss:AB_relations_A}  \\[1mm]
   B_{22} & = &  A_{11} B_{11} \,-\, \smhalf B_{11}^2
 \label{fss:AB_relations}
\end{subeqnarray}
Of course, the relations \reff{fss:AB_relations},
which guarantee the cancellation of
all divergent $L$-dependence in \reff{fss:Fxichi},
will be verified in the explicit calculation!

The foregoing expressions can be related to the renormalization-group
functions $W^{lat}$ and $\gamma^{lat}$, defined by
\reff{RGGreen} or equivalently by
\begin{subeqnarray}
   \left[ W^{lat}(t) {d \over dt} \,-\, 1 \right]   \xi_\infty(t^{-1})
      & = & 0
   \slabel{fss:RGdefxi}   \\[3mm]
   \left[ W^{lat}(t) {d \over dt} \,+\, \gamma^{lat}(t) \,-\, 2 \right]
      \chi_\infty(t^{-1}) & = & 0
   \slabel{fss:RGdefchi}
   \label{fss:RGdefxichi}
\end{subeqnarray}
where $t \equiv 1/\beta$,
$\xi_\infty(\beta) \equiv \xi(\beta,\infty)$
and $\chi_\infty(\beta) \equiv \chi(\beta,\infty)$.
Then, we can apply the RG equations \reff{fss:RGdefxichi}
to the finite-size-scaling Ans\"atze \reff{FSS:eq2},
yielding
\begin{subeqnarray}
   \left[ W^{lat}(t) {\partial \over \partial t} \,+\,
          L {\partial \over \partial L}  \right]
   {\xi(t^{-1},L) \over L}
      & = & 0 \,+\, O(L^{-\omega})
   \slabel{fss:RGeqLxi}   \\[3mm]
   \left[ W^{lat}(t) {\partial \over \partial t} \,+\,
          \gamma^{lat}(t) \,+\,
          L {\partial \over \partial L}  \right]
   {\chi(t^{-1},L) \over L^2}
      & = & 0 \,+\, O(L^{-\omega})
   \slabel{fss:RGeqLchi}
   \label{fss:RGeqLxichi}
\end{subeqnarray}
Imposing these equations on \reff{fss:xichi1}, and defining as
usual
\begin{subeqnarray}
   W^{lat}(t)   & = &   -w_0 t^2 \,-\, w_1 t^3  \,-\,
                         w_2^{lat} t^4 \,-\, w_3^{lat} t^5 \,-\, \ldots
   \slabel{fss:RGpertxi}  \\[1mm]
   \gamma^{lat}(t)   & = &   \gamma_0 t \,+\, \gamma_1^{lat} t^2 \,+\,
                        \gamma_2^{lat} t^3 \,+\, \gamma_3^{lat} t^4
                        \,+\, \ldots
   \slabel{fss:RGpertchi}
   \label{fss:RGpertxichi}
\end{subeqnarray}
we obtain
\begin{subeqnarray}
   w_0   & = & 2 A_{11}   \\
   w_1   & = & 2 (A_{21} - A_{10} A_{11})   \\
   \gamma_0   & = &  B_{11}
 \label{fss:RGpertvalues}
\end{subeqnarray}
and also recover the relations \reff{fss:AB_relations}.
%{\bf DA CONTROLLARE:  C'era prima un errore di segno su $\gamma_0$!!!}
% checked !!!
Conversely, if we make use of the well-known fact that the
coefficients $w_0$, $w_1$ and $\gamma_0$ are
{\em scheme-independent}\/ --- hence equal to their values
in the RG beta- and gamma-functions of the corresponding continuum
perturbation theory ---
we can recover the $1/x^2$ and $1/x^4$ terms in $F_\xi$,
and the $1/x^2$ term in $F_\chi$,
without the need for any lattice calculation other than
the trivial one leading to the prefactor $A$ in \reff{fss:xi1}.
We get
\begin{subeqnarray}
   F_\xi(x;s)   & = &
     s \left[ 1 \,-\,
                  (\smhalf w_0 \log s) \left( {A \over x} \right)^{\! 2}
                \,-\,  \left( \smfrac{1}{8} w_0^2 \log^2 s
                               + \smhalf w_1 \log s \right)
                       \left( {A \over x} \right)^{\! 4}
                \,+\,  O(x^{-6})
       \right]
   \nonumber \\ \slabel{fss:Fxi2} \\[3mm]
   F_\chi(x;s)   & = &
     s^2 \Biggl[ 1 \,-\,
     (\gamma_0 \log s) \left( {A \over x} \right)^{\! 2}
                \,+\,  O(x^{-4})
       \Biggr]
   \slabel{fss:Fchi2}
   \label{fss:Fxichi2}
\end{subeqnarray}
The subsequent terms can be determined
from the coefficients $w_2^{lat}, w_3^{lat}, \ldots$
and $\gamma_1^{lat}, \gamma_2^{lat}, \ldots$
{\em together with}\/ the coefficients $A_{n0}$ and $B_{n0}$.
%{\bf ANDREA: Is this correct????}
%YES.

\bigskip

\noindent
{\bf Remark.}
Our assumption that the expansions \reff{fss:xichi1}
are valid also in the double limit
$\beta,L \to \infty$ at fixed $x$
followed by expansion in powers of $1/x^2$
implies, in particular, the asymptotic scaling
\reff{xi_predicted2}--\reff{lambda_parameter}
of the infinite-volume correlation length and susceptibilities:
this can be deduced by applying our finite-size-scaling extrapolation
procedure (Section \ref{sec5.1}) {\em analytically}\/,
using the starting point \reff{fss:xichi1} and the
extrapolation functions \reff{fss:Fxichi}.
The validity of this assumption is thus as unproven as the
validity of asymptotic freedom itself;
and it has been explicitly questioned by
Patrascioiu and Seiler \cite{Pat-Seil_comment}.
All we can say is that our numerical data show good agreement
with the predictions \reff{fss:Fxichi}:
see Figures~\ref{fig_su3_fss_xif}, \ref{fig_su3_fss_chif},
\ref{fig_su3_fss_xia} and \ref{fig_su3_fss_chia}
in Section~\ref{sec5.2.1}.

Whatever the validity of this assumption at {\em leading}\/ order
in the double limit,
it is worth noting that this assumption is presumably {\em not}\/
valid at {\em next-to-leading}\/ order, that is,
as concerns the dominant {\em corrections}\/ to finite-size scaling.
This can be seen clearly in the exact solution of the
$N$-component mixed isovector/isotensor model
(with $r \equiv \beta_T/(\beta_V + \beta_T) \neq 0$)
at $N=\infty$ \cite{CP_LAT96,CP_1overN}:
\be
   {\xi_V^{(2nd)}(L)  \over  \xi_V^{(2nd)}(\infty)}
   \;=\;
   F_{\xi_V^{(2nd)}}(x)
   \left[ 1 \,+\, g_1(x) {\log L \over L^2}  \,+\, {g_2(x) \over L^2}
            \,+\, {g_3(x) \over L^2 [x^{-2} \log L + h(x)]}
            \,+\, \cdots
   \right]
 \label{large_N_FSS_exact}
\ee
where $F_{\xi_V^{(2nd)}}, g_1, g_2, g_3$ and $h$
are all explicitly computable functions;
moreover, $g_1, g_2, g_3$ and $h$ all have good large-$x$ asymptotic
expansions of the form
$C_0 + C_1 x^{-2} + C_2 x^{-4} + \ldots$
with leading behaviors
$g_1(x), g_3(x) \sim x^{-2}$  and  $g_2(x), h(x) \sim 1$.
The ``bad'' term in \reff{large_N_FSS_exact}
is the one involving $g_3$:
for $x,L \gg 1$ one gets {\em different}\/ expansions
depending on whether $x^2 \gg \log L$ or $x^2 \ll \log L$,
so the two limits $x\to\infty$ and $L\to\infty$ do not commute.
Indeed, in the finite-size-scaling limit $L \to \infty$ at fixed $x < \infty$,
this term behaves like $1/(L^2 \log L)$,
with a coefficient that tends to a constant at large $x$
and has a good asymptotic expansion in powers of $1/x^2$;
while in the finite-volume perturbative limit
$x \to \infty$ at fixed $L < \infty$, this term has an asymptotic expansion
in powers of $1/x^2$ with coefficients that are increasingly
{\em positive}\/ powers of $\log L$:
\be
   {1 \over L^2}
   \left[  {P_0(\log L) + o(1) \over x^2}  \,+\,
           {P_1(\log L) + o(1) \over x^4}  \,+\,
           {P_2(\log L) + o(1) \over x^6}  \,+\, \cdots \right]
\ee
where $P_k$ is a polynomial of degree $k$.
What happens, of course, is that the latter expansion {\em sums}\/
to the former;  but this resummation cannot be seen in any finite order
of perturbation theory.

\subsection{Perturbative Computations}   \label{appenB.2}

In an asymptotically free model, as noted in the preceding subsection,
the functions $F_{\scro}(x;s)$ at large $x$
can be computed in perturbation theory.
The starting point is the perturbative expansion for
the correlation function in a (fixed) periodic $L^d$ box.
In this computation we must take proper care of the zero mode. 
We will follow here the method used for the $N$-vector model
in \cite{Hasenfratz_84}. 

Let us first consider $U=\exp(i A)\in SU(N)$ and $V\in SU(N)$. We define
$A^V$ as 
\be
    \exp(i A^V) \equiv V \exp( iA)
\ee
Then let us use the standard Faddeev-Popov trick, rewriting the partition
function\footnote{It will be immediate to see that the same procedure
applies to any $SU(N)$-invariant correlation.} as
\begin{subeqnarray}
Z & \equiv & \int \prod_x dU_x \, e^{-\beta H} \\
  & = & \int \prod_x dU_x \, e^{-\beta H} 
     {\int dV \prod_a \delta( L^{-d} \sum_x (A^V_x)^a) \over
      \int dW \prod_a \delta( L^{-d} \sum_x (A^W_x)^a) }
\end{subeqnarray}
Then redefining $U' = VU$, $W'=WV^{-1}$ and using the two-sided invariance of 
the Haar measure and of the Hamiltonian, we get (after dropping primes)
\be
Z \;=\; \int \prod_x dU_x \, e^{-\beta H} 
           {\prod_a \delta( L^{-d} \sum_x A^a_x) \over
           \int dW \prod_a \delta( L^{-d} \sum_x (A^W_x)^a) }
  \;\,.
\ee
Let us now perform the $W$ integration.
When $\sum_x A_x^a=0$ (as is imposed by the delta function in the numerator),
the solution of $\sum_x(A_x^W)^a = 0$ is clearly $W=1$.
For $W = 1 + (\delta w^a) T^a$ with $\delta w^a$ infinitesimal, we have
\cite{Kawai_81,Rothe_92}
\be
 (A_x^W)^a \;=\;  A_x^a +\, (E(A)^{-1})_{ab} \delta w^b
\ee
with
\be
E(A) \;=\;  {\exp (i A_x^a T_A^a) - 1 \over i A_x^a T^a_A}
\ee
and $(T_A^a)_{bc} = - i f^{abc}$.
We therefore get
\be
\int\! dW \, \prod_a \delta\!\left( L^{-d} \sum_x (A_x^W)^a\right)
\;=\;  
\left| \hbox{det} \left[ L^{-d} \sum_x E(A_x)^{-1} \right] \right|^{-1}   \;.
\ee
This new term gives rise to a new set of vertices, formally vanishing
as $L\to\infty$. At one-loop order we will be only interested in the 
leading contribution, and we will thus write
\be
Z \;=\; \int \prod_x dU_x \, e^{-\beta H} \,
    \prod_a \delta\!\left( L^{-d} \sum_x A_x^a \right)
    \, \exp\!\left[{N\over 12 L^d} \sum_x A_x^a A_x^a + O(A^4) \right]   \;.
 \label{eqB.19}
\ee
The perturbative expansion is obtained as before. We get
\begin{subeqnarray}
   G_F(x;\beta,L)   &=& 
   1 \,-\, {N^2-1 \over{N \beta}} \, D_L^{(1)}(x)          \nonumber \\
& & \quad-\,  {N^2-1 \over {2 \beta^2}} \left[
      {N^2-2 \over 2dN^2} (1- L^{-d}) D_L^{(1)}(x) \,-\,
      {N^2-2 \over 2N^2} D_L^{(1)}(x)^2 \,-\,
       D_L^{(2)}(x) \right]                                  \nonumber \\
& &  \quad+\, O(1/\beta^3)                       
 \slabel{hasenfratz_GF_finite_L}   \\[4mm]
   G_A(x;\beta,L)   &=&
   1 \,-\, {2 N\over \beta} \, D_L^{(1)}(x)  \,-\,
   {N^2-2 \over 2 d \beta^2} 
       (1- L^{-d}) D_L^{(1)}(x) 
   \nonumber \\
   & & \quad+\,  {3 N^2\over2 \beta^2} {D_L^{(1)}(x)}^2 \,+\,
       {N^2\over\beta^2} D_L^{(2)}(x) 
    \;+\; O(1/\beta^3) 
 \slabel{hasenfratz_GA_finite_L} 
 \label{hasenfratz_GFA_finite_L}
\end{subeqnarray}
where
\be
   D_L^{(n)}(x)   \;\equiv\;
   {1 \over L^{dn}}
     \sum\limits_{p \neq 0}  {1-\cos p\cdot x  \over (\widehat{p}^2)^n}
   \;;
\ee
here the sum ranges over the momenta
$p_\mu = (2\pi/L) n_\mu$ with integers $0 \le n_\mu \le L-1$
(not all zero),
and $\widehat{p}^2 \equiv 4 \sum_\mu \sin^2 (p_\mu/2)$. 
An easy check of these expressions is provided by the fact 
that for the $SU(N)$ model with $N=2$ is equivalent to a 4-vector model.
We have verified that the expressions for
$G_F(x;\beta,L)$ and $G_A(x;\beta,L)$ at $N=2$ agree with the corresponding
expressions for the isovector \cite{Hasenfratz_84}
and isotensor \cite{CEPS_RPN} correlation functions of the
4-vector model.
It follows that (reverting now to the normalizations of $\chi_F$ and $\chi_A$
used in the main text) we have
\begin{subeqnarray}
   \chi_F(\beta,L)    & = &
    N L^d \Biggl\{ 1 \,-\, {N^2-1 \over {N\beta}} I_{1,L} 
       \nonumber \\
   & & \qquad \quad
       \,-\,  {N^2-1 \over {2 \beta^2}} 
       \left[ {N^2-2 \over {2d N^2}} (1 - L^{-d})  I_{1,L}
                 - {N^2-2 \over 2N^2} ( I_{1,L}^2 + I_{2,L} )
                 -  I_{2,L} \right]  \nonumber \\
   & & \qquad \quad
       \;+\; O(1/\beta^3) \Biggr\}
   \slabel{chiV_finiteV_PT}  
\\[4mm]
   \xi_F^{(2nd)}(\beta,L)^2    & = &
   {N \beta L^d \over N^2-1}
   \Biggl\{ 1 \,-\, {1 \over 2\beta}
      \Bigl[ {N^2-2 \over 2dN} (1 - L^{-d}) +
	     N I_{1,L}
   \nonumber \\
   & & \qquad \qquad  \quad
               + {N^2-2 \over 2N} \widehat{p}_0^2 L^d I_{3,L}
               + {N^2-2 \over N} {1 \over L^d \widehat{p}_0^2}   \Bigr]
      \;+\; O(1/\beta^2)
      \Biggr\}
  \slabel{xiV_finiteV_PT}   
\\[4mm]
   \chi_A(\beta,L)    & = &  (N^2 -1) L^d
   \Biggl\{ 1 \,-\, {2 N \over \beta} I_{1,L}        \nonumber \\
   & & \qquad \quad
   \,-\,   {1 \over 2\beta^2} 
      \left[ {N^2-2 \over d} (1 - L^{-d}) \, I_{1,L}
                 - 3 N^2 I_{1,L}^2
                 - 5 N^2 I_{2,L} \right]   \nonumber \\
   & & \qquad \quad   \,+\, O(1/\beta^3)
      \Biggr\}
   \slabel{chiT_finiteV_PT}  
\\[4mm]
   \xi_A^{(2nd)}(\beta,L)^2    & = &
   {\beta L^d \over 2N}
   \Biggl\{ 1 \,-\, {N \over 2 \beta} I_{1,L} \,-\,
           {N^2-2 \over 4d N \beta} \, (1 - L^{-d}) 
  \nonumber \\
  & & \qquad  \quad \,-\,    
       {3 N \over 4 \beta} \Bigl[ \widehat{p}_0^2 L^d I_{3,L} \,+\,
            {2 \over L^d \widehat{p}_0^2} \Bigr]
      \;+\; O(1/\beta^2)
      \Biggr\}
  \slabel{xiT_finiteV_PT}
  \label{all_finiteV_PT}
\end{subeqnarray}
where
\begin{subeqnarray}
   I_{1,L}   & = &
      {1 \over L^d} \sum\limits_{p \neq 0}  {1 \over \widehat{p}^2}  \\[3mm]
   I_{2,L}   & = &
      {1 \over L^{2d}} \sum\limits_{p \neq 0}  {1 \over (\widehat{p}^2)^2}
                                                                     \\[3mm]
   I_{3,L}   & = &
      {1 \over L^{2d}} \sum\limits_{p \neq 0,p_0}
          {1 \over \widehat{p}^2 (\widehat{p \!-\! p_0})^2}
\end{subeqnarray}
and $p_0 = (2\pi/L,0,\ldots,0)$.
In dimension $d=2$ the asymptotic behavior for large $L$ is as follows
\cite{CP_1overN}:
\begin{subeqnarray}
   I_{1,L}   & = &  {1 \over 2\pi} \log L  \,+\,  I_{1,fin}  \,+\,  
                    {1 \over L^2} I_1^{(1)}  \,+\,
                    O\!\left( {1 \over L^4} \right)  \\[2mm]
   I_{2,L}   & = &  I_{2,\infty}  \,+\, {1 \over 16\pi} {\log L \over L^2}
                     \,+\, {1 \over L^2} I_2^{(1)}  \,+\,
                    O\!\left( {1 \over L^4} \right)  \\[2mm]
   I_{3,L}   & = &  I_{3,\infty}  \,+\, {1 \over 16\pi} {\log L \over L^2} 
                     \,+\, {1 \over L^2} I_3^{(1)}  \,+\, 
                    O\!\left( {\log L \over L^4} \right)
\end{subeqnarray}
where
\begin{subeqnarray}
   I_{1,fin}  & = &  {1 \over 2\pi} \left[ \gamma_E - \log\pi
                            + \smhalf \log 2 - 2 \log \eta(i) \right]  \\[4mm]
   I_1^{(1)}  & = &  {\pi \over 72} - {1 \over 12} - {\pi \over 3} N_{-1,1}
                          + {2\pi^2 \over 3} (N_{-2,1} + N_{-2,2})     \\[4mm]
   I_{2,\infty}   & = &  {\zeta(3) \over 16\pi^3} + {1 \over 720}
        + {1 \over 8\pi^3} N_{3,1}  +  {1 \over 4\pi^2} (N_{2,1} + N_{2,2})
                                                            \nonumber \\
        & = &  {1 \over (2\pi)^4} \left( {11\pi^4 \over 180} \,+\,
     \pi^2 \sum\limits_{m=1}^\infty {1 \over m^2 \sinh^2 \pi m} \right) \\[4mm]
   I_2^{(1)}  & = &  {1 \over 8} I_{1,fin}  + {1 \over 4\pi} I_1^{(1)}  \\[4mm]
   I_{3,\infty}   & = &
      {1 \over (2\pi)^4} \Biggl[ {2\pi^2 \over 3} + 4\pi(1-2\log 2) + 2
           + 8\pi \scrn_{1,1} \Biggr]                                   \\[4mm]
   I_3^{(1)}  & = &  {\gamma_E - \log\pi  \over 16\pi}
                     - {2 + \log 2  \over 96\pi}  + {1 \over 72}
                     + {1 \over 24\pi^2}                      \nonumber \\
   & & \qquad    + {1 \over 12\pi} (N_{1,1} + \scrn_{1,1} - 2 \scrn_{-1,1})
                     + {1 \over 3} (\scrn_{-2,1} + \scrn_{-2,2})
 \label{I123_defs}
\end{subeqnarray}
Here
\be
   \eta(\tau)   \;=\;
   (e^{2\pi i\tau})^{1/24}
   \prod\limits_{n=1}^\infty \left( 1 - (e^{2\pi i\tau})^n \right)
\ee
is Dedekind's eta function \cite[Chapter 18]{Lang_87},
and
\begin{eqnarray}
   N_{p,q}  & = &  \sum\limits_{n=1}^\infty
                   {1 \over n^p \, (e^{2\pi n} - 1)^q}      \\[4mm]
   \scrn_{p,q}  & = &  \sum\limits_{n=1}^\infty
                   {1 \over (1-4n^2) \, n^p \, (e^{2\pi n} - 1)^q}
\end{eqnarray}
Numerically,
\begin{subeqnarray}
   I_{1,fin}    & \approx & \hphantom{-} 0.04876 56331 70141 30174 \\
   I_1^{(1)}    & \approx & -            0.02924 11947 95190 21443 \\
   I_{2,\infty} & \approx & \hphantom{-} 0.00386 69465 90737 21003 \\
   I_2^{(1)}    & \approx & \hphantom{-} 0.00376 87637 99483 90038 \\
   I_{3,\infty} & \approx & \hphantom{-} 0.00238 02586 56448 51979 \\
   I_3^{(1)}    & \approx & -            0.00226 83728 99086 75469
 \label{I123_numer}
\end{subeqnarray}
In Tables~\ref{table_I1}--\ref{table_I3}
we report the exact $I_{1,L}, I_{2,L}$ and $I_{3,L}$
for selected values of $L$
and compare with the asymptotic expansions.
The agreement is excellent, and we can even estimate numerically the
next terms in the expansions:
they are $\approx 0.121015/L^4$, $\approx 0.0052263/L^4$ and
$\approx 0.0245\log L/L^4 - 0.0132/L^4$, respectively.

In these expressions we can now take the finite-size-scaling limit
$\beta,L \to\infty$ with $x \equiv \xi_F^{(2nd)}(\beta,L)/L \;$ held fixed
and then expand in powers of $1/x^2$;
under the {\em assumption}\/ that \reff{hasenfratz_GFA_finite_L}
remain valid in this limit, we obtain for $d=2$
\begin{subeqnarray}
   {\chi_F(\beta,sL) \over \chi_F(\beta,L)}   & = &
   s^2 \Biggl[ 1 \,-\, {\log s \over 2\pi} \, x^{-2}         \nonumber \\
   & &  \qquad \,+\, {N^2-2 \over N^2-1} \left(
               {\log^2 s \over 16\pi^2} +
               \Bigl( {\pi \over 2} I_{3,\infty} +
                      {1 \over 16\pi^3} \Bigr) \log s
               \right)  x^{-4}  \,+\, O(x^{-6}) \Biggr]
   \slabel{chiF_FSS_PT_apen}                               \\[4mm]
   {\chi_A(\beta,sL) \over \chi_A(\beta,L)}   & = &
   s^2 \Biggl[ 1 \,-\, {\log s \over \pi} \, {N^2 \over N^2-1} \, x^{-2} \nonumber \\
   & & \quad \,+\, {N^2 \over (N^2-1)^2} \left(
               {3 \over 8\pi^2} \log^2 s  + (N^2-2)
               \Bigl( \pi I_{3,\infty} + {1 \over 8\pi^3} \Bigr) \log s
               \right)  x^{-4}  \,+\, O(x^{-6}) \Biggr]
        \nonumber \\
   \slabel{chiA_FSS_PT0}
   \label{all_FSS_PT0}
\end{subeqnarray}
and also
\be
   {\xi_F^{(2nd)}(\beta,L) \over \xi_A^{(2nd)}(\beta,L)}   \;=\;
   \left( {2N^2 \over N^2-1} \right) ^{\! 1/2} \,
   \left[ 1 \,+\,
     {N^2+1 \over N^2-1} \Bigl( \pi^2 I_{3,\infty} + {1 \over 8\pi^2} \Bigr)
         x^{-2}
              \,+\, O(x^{-4}) \right] \;.
   \label{xiVoverxiT_FSS_PT0}
\ee
	
Exploiting the renormalization group as discussed in the previous section,
we can obtain the finite-size-scaling function
for $\xi_F^{(2nd)}$ in terms of $x$ to order $1/x^4$, and for $\xi_A^{(2nd)}$ in terms of 
$x'\equiv \xi_A^{(2nd)}(\beta,L)/L \;$ to order $1/x'^4$:
\begin{subeqnarray}
   {\xi_F^{(2nd)}(\beta,sL) \over \xi_F^{(2nd)}(\beta,L)}
   & = &
   s \left[ 1 \,-\,  {w_0 \log s \over 2} \left( {A \over x} \right) ^{\! 2}
       \,-\,  \Biggl( {w_1 \log s \over 2} + {w_0^2 \log^2 s \over 8}
              \Biggr) \left( {A \over x} \right) ^{\! 4}
       \,+\,  O(x^{-6}) \right]
   \nonumber \\ \slabel{xiV_FSS_PT_RGimproved} \\[3mm]
   {\xi_A^{(2nd)}(\beta,sL) \over \xi_A^{(2nd)}(\beta,L)}
   & = &
   s \left[ 1 \,-\,  {w_0 \log s \over 2} \left( {A' \over x'} \right) ^{\! 2}
       \,-\,  \Biggl( {w_1 \log s \over 2} + {w_0^2 \log^2 s \over 8}
              \Biggr) \left( {A' \over x'} \right) ^{\! 4}
       \,+\,  O((x')^{-6}) \right]
   \nonumber \\ \slabel{xiT_FSS_PT_RGimproved}
   \label{xiVT_FSS_PT_RGimproved}
\end{subeqnarray}
where $A = [N/(N^2 - 1)]^{1/2}$,
$A' = (2N)^{-1/2}$, $w_0 = N/(4 \pi)$, $w_1 = N^2/(32 \pi^2)$. 
Of course, starting at order $1/x^6$ we expect the finite-size-scaling
functions for $\xi_F^{(2nd)}$ and $\xi_A^{(2nd)}$ to differ.
Finally, we can use \reff{xiVoverxiT_FSS_PT0} to express 
\reff{xiT_FSS_PT_RGimproved} in terms of $x$:
\begin{eqnarray}
   {\xi_A^{(2nd)}(\beta,sL) \over \xi_A^{(2nd)}(\beta,L)}
   & = &
   s \Biggl\{ 1 \,-\, {\log s \over 8\pi} \, {N^2 \over N^2-1} \, x^{-2}
   \nonumber \\
   & & \qquad
   - \; {N^2 \over (N^2-1)^2}
        \left[
        \left( (N^2+1) \left( {\pi \over 4} I_{3,\infty}
                              + {1 \over 32\pi^3} \right)
               + {N^2 \over 64\pi^2}
        \right) {\log s} \right. \nonumber \\
   & & \qquad \qquad \qquad \qquad
      \left.  \,+\,  {N^2 \log^2 s \over 128\pi^2}
        \right]  x^{-4}
    + \; O(x^{-6}) \Biggr\}
   \slabel{xiT_FSS_PT0}                        
\end{eqnarray}

%%%% \singlespace

%
% Beginning of Table 0 for SU3 article
% a_1 versus a'_1 \alpha_{-1}
%
%
\protect\normalsize
\begin{table}
\addtolength{\tabcolsep}{-1.0mm}
\begin{center}
\begin{tabular}{|r|r|r|} \hline
\multicolumn{1}{|c}{$N$}  &  \multicolumn{1}{|c}{$a_1/N$}  &
   \multicolumn{1}{|c|}{$a'_1 \alpha_{-1}/N$}    \\
\hline \hline
2  &  $-0.013188$  &  0.038174   \\
3  &  $-0.054914$  &  0.033343   \\
4  &  $-0.080255$  &  0.027781   \\
5  &  $-0.093870$  &  0.024527   \\
6  &  $-0.101766$  &  0.022580   \\
7  &  $-0.106696$  &  0.021344   \\
8  &  $-0.109965$  &  0.020517   \\
9  &  $-0.112237$  &  0.019939   \\
10  &  $-0.113878$  &  0.019520   \\
11  &  $-0.115101$  &  0.019207   \\
12  &  $-0.116035$  &  0.018967   \\
13  &  $-0.116765$  &  0.018780   \\
14  &  $-0.117346$  &  0.018630   \\
15  &  $-0.117816$  &  0.018509   \\
16  &  $-0.118202$  &  0.018410   \\
17  &  $-0.118522$  &  0.018327   \\
18  &  $-0.118790$  &  0.018258   \\
19  &  $-0.119017$  &  0.018199   \\
20  &  $-0.119212$  &  0.018149   \\
\multicolumn{1}{|c|}{$\vdots$}   &        &             \\
$\infty$   &  $-0.121019$   &  $0.017681$  \\
\hline
\end{tabular}
\end{center}
\caption{
   Comparison of 3-loop perturbative coefficients in the
   standard scheme ($a_1/N$)
   and in the ``energy-improved'' scheme ($a'_1 \alpha_{-1}/N$).
}
\label{table_improved}
\end{table}
\clearpage

%%%%%%%%%%%%%%%%%%%%%%%%%%%%%%%%%%%%%%%%%%%%%%%%%%%%%%%%%%

%%%%%%%%%%%%%%%%%%%%%%%%%%%%%%%%%%%%%%%%%%%%%%%%%%%%%%%%%%

%%%%%%%%%%%%%%%%%%%%%%%%%%%%%%%%%%%%%%%%%%%%%%%%%%%%%%%%%%%%

%%% THE NORMALIZATION OF \chi_A is RIGHT (twice the MC data).
% Beginning of Table 1 for SU3 article
% Static quantities: {chi_F, chi_A, xi_F, xi_A, E_F, E_A}
%
\protect\tiny
\begin{table}
\addtolength{\tabcolsep}{-1.3mm}
\begin{center}
\begin{tabular}{|r r|r r|r r|r r|r r|r r|r r|} \hline
$\beta$  & $L$ &
\multicolumn{2}{|c|}{$\chi_F$} &
\multicolumn{2}{|c|}{$\chi_A$} &
\multicolumn{2}{|c|}{$\xi_F$} &
\multicolumn{2}{|c|}{$\xi_A$} & 
\multicolumn{2}{|c|}{$E_F$} &
\multicolumn{2}{|c|}{$E_A$} \\
%\multicolumn{20}{|c|}{$d=2$ $SU(3)$ model Table LAST}
\hline
1.6000   &   32   &   106.180   & (0.460) & 36.186 & (0.064) &   4.4534   & (0.0233) &   1.6407   & (0.0095) &   0.4912228   & (0.0000772) &   0.2150757   & (0.0000662) \\
\hline
1.6500   &   8   &   65.220   & (0.022) & 54.064 & (0.030) &   3.5983   & (0.0014) &   1.8994   & (0.0009) &   0.5389755   & (0.0000656) &   0.2583347   & (0.0000643) \\
    &   16   &   130.603   & (0.185) & 60.006 & (0.104) &   5.2606   & (0.0071) &   2.5127   & (0.0040) &   0.5208573   & (0.0000780) &   0.2414590   & (0.0000715) \\
    &   32   &   149.314   & (1.082) & 46.200 & (0.178) &   5.5139   & (0.0420) &   2.1082   & (0.0184) &   0.5132461   & (0.0001078) &   0.2346354   & (0.0000987) \\
\hline
1.7000   &   32   &   214.259   & (1.248) & 62.884 & (0.268) &   6.9353   & (0.0386) &   2.8619   & (0.0190) &   0.5341336   & (0.0000884) &   0.2542263   & (0.0000835) \\
\hline
1.7500   &   8   &   73.865   & (0.019) & 67.554 & (0.032) &   3.9816   & (0.0013) &   2.1737   & (0.0009) &   0.5726885   & (0.0000561) &   0.2928524   & (0.0000602) \\
    &   16   &   176.848   & (0.145) & 95.368 & (0.122) &   6.4631   & (0.0059) &   3.3158   & (0.0036) &   0.5595000   & (0.0000589) &   0.2792882   & (0.0000602) \\
    &   32   &   298.994   & (1.014) & 89.280 & (0.292) &   8.5947   & (0.0271) &   3.8271   & (0.0142) &   0.5534794   & (0.0000536) &   0.2733057   & (0.0000531) \\
    &   64   &   299.882   & (0.968) & 73.222 & (0.090) &   8.3492   & (0.0349) &   3.0469   & (0.0147) &   0.5518565   & (0.0000217) &   0.2717081   & (0.0000216) \\
    &   128   &   298.185   & (0.792) & 72.672 & (0.054) &   8.1632   & (0.0767) &   2.9445   & (0.0478) &   0.5518589   & (0.0000118) &   0.2717117   & (0.0000117) \\
\hline
1.7750   &   32   &   342.269   & (0.712) & 105.482 & (0.232) &   9.3514   & (0.0187) &   4.2906   & (0.0098) &   0.5623785   & (0.0000353) &   0.2824141   & (0.0000361) \\
    &   64   &   361.421   & (1.005) & 84.134 & (0.096) &   9.3550   & (0.0308) &   3.4600   & (0.0122) &   0.5606244   & (0.0000165) &   0.2806474   & (0.0000168) \\
\hline
1.8000   &   32   &   385.818   & (0.694) & 123.924 & (0.262) &   10.0809   & (0.0178) &   4.7519   & (0.0097) &   0.5708396   & (0.0000326) &   0.2912637   & (0.0000341) \\
    &   64   &   439.150   & (0.826) & 97.766 & (0.088) &   10.5812   & (0.0216) &   3.9967   & (0.0090) &   0.5690047   & (0.0000101) &   0.2893803   & (0.0000104) \\
    &   128   &   433.613   & (0.760) & 94.744 & (0.048) &   10.4680   & (0.0436) &   3.6883   & (0.0243) &   0.5689154   & (0.0000063) &   0.2892876   & (0.0000065) \\
    &   256   &   434.198   & (1.702) & 94.682 & (0.106) &   10.8017   & (0.3262) &   3.5660   & (0.2547) &   0.5688939   & (0.0000092) &   0.2892665   & (0.0000095) \\
\hline
1.8250   &   32   &   430.306   & (0.374) & 145.158 & (0.162) &   10.8179   & (0.0096) &   5.2293   & (0.0053) &   0.5789228   & (0.0000174) &   0.2999071   & (0.0000186) \\
    &   64   &   528.604   & (1.091) & 114.556 & (0.132) &   11.8326   & (0.0249) &   4.6327   & (0.0107) &   0.5770031   & (0.0000098) &   0.2978886   & (0.0000104) \\
    &   128   &   519.784   & (0.991) & 108.454 & (0.062) &   11.5729   & (0.0451) &   4.1017   & (0.0232) &   0.5768699   & (0.0000062) &   0.2977490   & (0.0000065) \\
\hline
1.8500   &   8   &   81.233   & (0.018) & 80.898 & (0.034) &   4.3179   & (0.0013) &   2.4181   & (0.0009) &   0.6011965   & (0.0000504) &   0.3244052   & (0.0000578) \\
    &   16   &   214.142   & (0.122) & 134.086 & (0.136) &   7.3677   & (0.0053) &   3.9618   & (0.0033) &   0.5906575   & (0.0000491) &   0.3126732   & (0.0000546) \\
    &   32   &   470.662   & (0.354) & 166.786 & (0.172) &   11.4472   & (0.0091) &   5.6484   & (0.0052) &   0.5864995   & (0.0000164) &   0.3081675   & (0.0000179) \\
    &   64   &   637.906   & (1.299) & 136.050 & (0.184) &   13.2950   & (0.0263) &   5.4018   & (0.0123) &   0.5846955   & (0.0000096) &   0.3062333   & (0.0000104) \\
    &   128   &   626.666   & (1.278) & 124.436 & (0.080) &   12.9708   & (0.0460) &   4.5574   & (0.0224) &   0.5844757   & (0.0000060) &   0.3059981   & (0.0000065) \\
    &   256   &   622.604   & (2.810) & 124.372 & (0.152) &   12.9133   & (0.3147) &   4.7465   & (0.2098) &   0.5844637   & (0.0000084) &   0.3059870   & (0.0000091) \\
\hline
1.8750   &   32   &   510.097   & (0.331) & 190.144 & (0.180) &   12.0432   & (0.0086) &   6.0532   & (0.0050) &   0.5937732   & (0.0000153) &   0.3162469   & (0.0000171) \\
    &   64   &   763.471   & (1.760) & 163.174 & (0.296) &   14.8524   & (0.0328) &   6.2746   & (0.0162) &   0.5920621   & (0.0000108) &   0.3143766   & (0.0000118) \\
    &   128   &   753.186   & (1.927) & 143.374 & (0.122) &   14.4231   & (0.0554) &   5.1586   & (0.0250) &   0.5917596   & (0.0000066) &   0.3140472   & (0.0000073) \\
\hline
1.9000   &   16   &   230.282   & (0.067) & 153.840 & (0.082) &   7.7548   & (0.0030) &   4.2425   & (0.0019) &   0.6041531   & (0.0000265) &   0.3279628   & (0.0000305) \\
    &   32   &   547.737   & (0.311) & 214.660 & (0.188) &   12.6018   & (0.0082) &   6.4412   & (0.0049) &   0.6006624   & (0.0000147) &   0.3240371   & (0.0000167) \\
    &   64   &   900.729   & (2.122) & 196.618 & (0.422) &   16.4525   & (0.0370) &   7.2164   & (0.0189) &   0.5990848   & (0.0000115) &   0.3222796   & (0.0000129) \\
    &   128   &   909.462   & (2.779) & 165.368 & (0.180) &   16.1781   & (0.0665) &   5.7819   & (0.0272) &   0.5987348   & (0.0000068) &   0.3218904   & (0.0000076) \\
\hline
1.9250   &   16   &   237.991   & (0.065) & 163.958 & (0.084) &   7.9395   & (0.0030) &   4.3779   & (0.0019) &   0.6105105   & (0.0000259) &   0.3353310   & (0.0000302) \\
    &   32   &   583.050   & (0.299) & 239.702 & (0.198) &   13.1042   & (0.0080) &   6.7971   & (0.0048) &   0.6072621   & (0.0000140) &   0.3316204   & (0.0000162) \\
    &   64   &   1049.244   & (2.136) & 237.338 & (0.498) &   18.1022   & (0.0356) &   8.2014   & (0.0189) &   0.6058715   & (0.0000108) &   0.3300417   & (0.0000124) \\
    &   128   &   1102.266   & (3.550) & 192.392 & (0.250) &   18.1499   & (0.0705) &   6.6014   & (0.0286) &   0.6054381   & (0.0000065) &   0.3295529   & (0.0000075) \\
    &   256   &   1092.592   & (2.893) & 189.854 & (0.142) &   17.9768   & (0.1429) &   6.3931   & (0.0844) &   0.6054239   & (0.0000036) &   0.3295375   & (0.0000041) \\
\hline
1.9500   &   16   &   245.460   & (0.064) & 174.150 & (0.086) &   8.1184   & (0.0030) &   4.5087   & (0.0019) &   0.6166564   & (0.0000251) &   0.3425660   & (0.0000298) \\
    &   32   &   617.240   & (0.284) & 265.742 & (0.202) &   13.5848   & (0.0077) &   7.1375   & (0.0047) &   0.6135875   & (0.0000133) &   0.3390002   & (0.0000156) \\
    &   64   &   1194.664   & (2.208) & 283.412 & (0.590) &   19.6072   & (0.0360) &   9.1575   & (0.0193) &   0.6123274   & (0.0000104) &   0.3375481   & (0.0000122) \\
    &   128   &   1322.915   & (4.795) & 224.424 & (0.370) &   20.1813   & (0.0821) &   7.5869   & (0.0339) &   0.6118567   & (0.0000064) &   0.3370064   & (0.0000074) \\
\hline
1.9750   &   8   &   89.028   & (0.016) & 96.936 & (0.034) &   4.6891   & (0.0013) &   2.6896   & (0.0009) &   0.6311200   & (0.0000450) &   0.3598809   & (0.0000554) \\
    &   16   &   252.592   & (0.062) & 184.276 & (0.086) &   8.2897   & (0.0030) &   4.6340   & (0.0019) &   0.6225472   & (0.0000242) &   0.3495944   & (0.0000291) \\
    &   32   &   649.369   & (0.271) & 292.126 & (0.208) &   14.0154   & (0.0074) &   7.4520   & (0.0046) &   0.6196525   & (0.0000130) &   0.3461823   & (0.0000155) \\
    &   64   &   1337.681   & (2.070) & 334.646 & (0.634) &   21.0151   & (0.0335) &   10.0702   & (0.0183) &   0.6185086   & (0.0000099) &   0.3448418   & (0.0000117) \\
    &   128   &   1601.041   & (6.174) & 264.420 & (0.538) &   22.6762   & (0.0920) &   8.7306   & (0.0386) &   0.6180464   & (0.0000061) &   0.3443016   & (0.0000073) \\
\hline
1.9850   &   64   &   1390.642   & (2.082) & 355.578 & (0.668) &   21.5021   & (0.0333) &   10.3999   & (0.0182) &   0.6209157   & (0.0000098) &   0.3477107   & (0.0000117) \\
    &   128   &   1737.451   & (6.649) & 284.052 & (0.614) &   23.8641   & (0.0928) &   9.3045   & (0.0403) &   0.6204510   & (0.0000061) &   0.3471650   & (0.0000073) \\
\hline
2.0000   &   16   &   259.597   & (0.061) & 194.594 & (0.088) &   8.4570   & (0.0030) &   4.7572   & (0.0019) &   0.6282482   & (0.0000237) &   0.3564868   & (0.0000288) \\
    &   32   &   681.088   & (0.267) & 319.682 & (0.220) &   14.4468   & (0.0074) &   7.7643   & (0.0046) &   0.6255059   & (0.0000127) &   0.3532037   & (0.0000153) \\
    &   64   &   1477.094   & (1.945) & 390.562 & (0.682) &   22.3200   & (0.0313) &   10.9257   & (0.0176) &   0.6244452   & (0.0000095) &   0.3519448   & (0.0000114) \\
    &   128   &   1944.874   & (6.073) & 316.190 & (0.628) &   25.5228   & (0.0794) &   10.1983   & (0.0365) &   0.6239974   & (0.0000048) &   0.3514168   & (0.0000057) \\
    &   256   &   1908.004   & (6.260) & 293.300 & (0.298) &   24.9892   & (0.1477) &   8.8131   & (0.0745) &   0.6239548   & (0.0000033) &   0.3513668   & (0.0000039) \\
\hline
2.0120   &   32   &   696.055   & (0.259) & 333.252 & (0.220) &   14.6492   & (0.0073) &   7.9105   & (0.0046) &   0.6282516   & (0.0000123) &   0.3565320   & (0.0000150) \\
    &   64   &   1542.206   & (1.879) & 418.954 & (0.700) &   22.9143   & (0.0303) &   11.3251   & (0.0172) &   0.6272180   & (0.0000091) &   0.3552976   & (0.0000110) \\
    &   128   &   2109.703   & (8.410) & 344.006 & (0.922) &   26.7446   & (0.1038) &   10.9118   & (0.0490) &   0.6267917   & (0.0000058) &   0.3547902   & (0.0000070) \\
\hline
2.0250   &   16   &   266.221   & (0.060) & 204.694 & (0.090) &   8.6155   & (0.0030) &   4.8742   & (0.0019) &   0.6337170   & (0.0000233) &   0.3631777   & (0.0000287) \\
    &   32   &   711.390   & (0.257) & 347.596 & (0.226) &   14.8563   & (0.0073) &   8.0614   & (0.0046) &   0.6311148   & (0.0000123) &   0.3600304   & (0.0000150) \\
    &   64   &   1610.502   & (1.819) & 450.680 & (0.726) &   23.5007   & (0.0293) &   11.7383   & (0.0170) &   0.6301684   & (0.0000090) &   0.3588882   & (0.0000110) \\
    &   128   &   2323.057   & (8.888) & 379.398 & (1.056) &   28.3704   & (0.1053) &   11.7679   & (0.0502) &   0.6297447   & (0.0000058) &   0.3583793   & (0.0000070) \\
\hline
2.0370   &   32   &   725.143   & (0.254) & 360.854 & (0.228) &   15.0262   & (0.0073) &   8.1890   & (0.0046) &   0.6337699   & (0.0000120) &   0.3632885   & (0.0000148) \\
    &   64   &   1669.171   & (1.789) & 479.310 & (0.748) &   23.9831   & (0.0288) &   12.0785   & (0.0169) &   0.6328372   & (0.0000089) &   0.3621576   & (0.0000110) \\
    &   128   &   2502.593   & (9.227) & 411.874 & (1.192) &   29.6020   & (0.1043) &   12.4721   & (0.0511) &   0.6324092   & (0.0000056) &   0.3616399   & (0.0000068) \\
\hline
2.0500   &   16   &   272.727   & (0.059) & 214.894 & (0.090) &   8.7741   & (0.0030) &   4.9897   & (0.0019) &   0.6390405   & (0.0000226) &   0.3697781   & (0.0000282) \\
    &   32   &   740.767   & (0.249) & 376.128 & (0.232) &   15.2423   & (0.0072) &   8.3438   & (0.0045) &   0.6365752   & (0.0000119) &   0.3667528   & (0.0000148) \\
    &   64   &   1738.344   & (1.745) & 514.266 & (0.768) &   24.6064   & (0.0284) &   12.5067   & (0.0166) &   0.6356760   & (0.0000088) &   0.3656533   & (0.0000108) \\
    &   128   &   2759.170   & (9.709) & 459.300 & (1.350) &   31.5086   & (0.1060) &   13.5741   & (0.0522) &   0.6352783   & (0.0000056) &   0.3651698   & (0.0000068) \\
\hline
2.0620   &   32   &   754.856   & (0.246) & 390.290 & (0.234) &   15.4272   & (0.0071) &   8.4788   & (0.0045) &   0.6391212   & (0.0000118) &   0.3699156   & (0.0000147) \\
    &   64   &   1795.334   & (1.685) & 544.602 & (0.784) &   25.0612   & (0.0272) &   12.8264   & (0.0162) &   0.6382364   & (0.0000086) &   0.3688288   & (0.0000107) \\
    &   128   &   2985.634   & (10.167) & 504.588 & (1.556) &   33.0536   & (0.1077) &   14.4971   & (0.0544) &   0.6378740   & (0.0000055) &   0.3683857   & (0.0000068) \\
\hline
2.0750   &   16   &   279.162   & (0.058) & 225.344 & (0.092) &   8.9292   & (0.0030) &   5.1046   & (0.0019) &   0.6442409   & (0.0000222) &   0.3762899   & (0.0000280) \\
    &   32   &   768.768   & (0.242) & 404.696 & (0.236) &   15.5977   & (0.0070) &   8.6054   & (0.0045) &   0.6418070   & (0.0000116) &   0.3732738   & (0.0000145) \\
    &   64   &   1863.073   & (1.654) & 581.884 & (0.814) &   25.6495   & (0.0272) &   13.2415   & (0.0163) &   0.6409971   & (0.0000085) &   0.3722694   & (0.0000107) \\
    &   128   &   3226.608   & (10.533) & 556.628 & (1.750) &   34.6237   & (0.1093) &   15.4520   & (0.0565) &   0.6406263   & (0.0000054) &   0.3718133   & (0.0000067) \\
\hline
2.1000   &   8   &   95.710   & (0.012) & 112.276 & (0.028) &   5.0252   & (0.0011) &   2.9351   & (0.0007) &   0.6565474   & (0.0000316) &   0.3919404   & (0.0000410) \\
    &   16   &   285.320   & (0.050) & 235.602 & (0.082) &   9.0812   & (0.0026) &   5.2152   & (0.0017) &   0.6492320   & (0.0000189) &   0.3826249   & (0.0000241) \\
    &   32   &   796.666   & (0.240) & 434.538 & (0.246) &   15.9633   & (0.0071) &   8.8753   & (0.0045) &   0.6468963   & (0.0000113) &   0.3796861   & (0.0000144) \\
    &   64   &   1980.557   & (1.578) & 650.780 & (0.850) &   26.5699   & (0.0257) &   13.8946   & (0.0157) &   0.6461337   & (0.0000082) &   0.3787316   & (0.0000104) \\
    &   128   &   3690.326   & (10.761) & 668.022 & (2.076) &   37.4499   & (0.1068) &   17.2585   & (0.0562) &   0.6457853   & (0.0000052) &   0.3782972   & (0.0000066) \\
\hline
2.1120   &   32   &   809.972   & (0.237) & 449.240 & (0.248) &   16.1357   & (0.0071) &   9.0028   & (0.0045) &   0.6492912   & (0.0000112) &   0.3827288   & (0.0000143) \\
    &   64   &   2037.311   & (1.513) & 685.974 & (0.852) &   27.0201   & (0.0250) &   14.2143   & (0.0154) &   0.6485356   & (0.0000081) &   0.3817763   & (0.0000103) \\
    &   128   &   3940.610   & (10.102) & 731.578 & (2.110) &   39.0183   & (0.1000) &   18.2036   & (0.0537) &   0.6482214   & (0.0000051) &   0.3813825   & (0.0000065) \\
\hline
2.1250   &   32   &   823.270   & (0.236) & 464.228 & (0.254) &   16.3014   & (0.0071) &   9.1257   & (0.0045) &   0.6518132   & (0.0000110) &   0.3859503   & (0.0000142) \\
    &   64   &   2097.488   & (1.160) & 724.772 & (0.678) &   27.4922   & (0.0194) &   14.5589   & (0.0119) &   0.6510858   & (0.0000062) &   0.3850295   & (0.0000079) \\
    &   128   &   4187.871   & (8.579) & 800.986 & (1.892) &   40.4378   & (0.0841) &   19.1438   & (0.0447) &   0.6507901   & (0.0000041) &   0.3846575   & (0.0000053) \\
\hline
\end{tabular}
\caption{
   Our Monte Carlo data for the $SU(3)$ chiral model
   as a function of $\beta,L$.
   Errors are one standard deviation.
}
\label{su3_static}
\end{center}
\end{table} \clearpage \addtocounter{table}{-1}
\begin{table}
\addtolength{\tabcolsep}{-1.3mm}
%\protect\tiny
\begin{center}
\begin{tabular}{|r r|r r|r r|r r|r r|r r|r r|} \hline
$\beta$  & $L$ &
\multicolumn{2}{|c|}{$\chi_F$} &
\multicolumn{2}{|c|}{$\chi_A$} &
\multicolumn{2}{|c|}{$\xi_F$} &
\multicolumn{2}{|c|}{$\xi_A$} &
\multicolumn{2}{|c|}{$E_F$} &
\multicolumn{2}{|c|}{$E_A$} \\   \hline
2.1330   &   32   &   832.083   & (0.235) & 474.324 & (0.256) &   16.4152   & (0.0071) &   9.2094   & (0.0045) &   0.6533840   & (0.0000110) &   0.3879656   & (0.0000141) \\
    &   64   &   2132.065   & (1.477) & 747.888 & (0.884) &   27.7403   & (0.0247) &   14.7452   & (0.0152) &   0.6526391   & (0.0000080) &   0.3870197   & (0.0000102) \\
    &   128   &   4332.950   & (10.156) & 843.362 & (2.404) &   41.2531   & (0.0978) &   19.6641   & (0.0542) &   0.6523571   & (0.0000050) &   0.3866613   & (0.0000064) \\
\hline
2.1500   &   16   &   297.096   & (0.055) & 256.062 & (0.096) &   9.3701   & (0.0031) &   5.4275   & (0.0020) &   0.6587483   & (0.0000208) &   0.3948696   & (0.0000271) \\
    &   32   &   849.988   & (0.230) & 495.226 & (0.258) &   16.6545   & (0.0071) &   9.3843   & (0.0045) &   0.6566005   & (0.0000108) &   0.3921138   & (0.0000140) \\
    &   64   &   2208.862   & (1.131) & 800.390 & (0.714) &   28.3343   & (0.0191) &   15.1744   & (0.0119) &   0.6558975   & (0.0000061) &   0.3912145   & (0.0000079) \\
    &   128   &   4661.467   & (7.908) & 945.020 & (2.052) &   43.0964   & (0.0765) &   20.8693   & (0.0426) &   0.6556358   & (0.0000041) &   0.3908801   & (0.0000052) \\
    &   256   &   5800.368   & (28.591) & 753.756 & (2.170) &   47.4442   & (0.2408) &   18.6231   & (0.1082) &   0.6555141   & (0.0000028) &   0.3907252   & (0.0000036) \\
\hline
2.1750   &   16   &   302.786   & (0.054) & 266.330 & (0.096) &   9.5102   & (0.0030) &   5.5305   & (0.0019) &   0.6633039   & (0.0000205) &   0.4008229   & (0.0000269) \\
    &   32   &   875.396   & (0.228) & 525.918 & (0.266) &   16.9773   & (0.0071) &   9.6222   & (0.0045) &   0.6612342   & (0.0000107) &   0.3981387   & (0.0000139) \\
    &   64   &   2320.028   & (1.399) & 880.154 & (0.944) &   29.1901   & (0.0239) &   15.7892   & (0.0149) &   0.6605600   & (0.0000076) &   0.3972683   & (0.0000099) \\
    &   128   &   5136.621   & (9.109) & 1107.240 & (2.692) &   45.6605   & (0.0888) &   22.5935   & (0.0505) &   0.6603113   & (0.0000048) &   0.3969467   & (0.0000062) \\
\hline
2.2000   &   16   &   308.309   & (0.054) & 276.548 & (0.098) &   9.6479   & (0.0031) &   5.6317   & (0.0020) &   0.6677262   & (0.0000202) &   0.4066560   & (0.0000268) \\
    &   32   &   900.222   & (0.223) & 556.938 & (0.270) &   17.2957   & (0.0071) &   9.8549   & (0.0045) &   0.6657247   & (0.0000105) &   0.4040343   & (0.0000138) \\
    &   64   &   2424.608   & (1.389) & 960.220 & (0.996) &   29.9233   & (0.0237) &   16.3341   & (0.0150) &   0.6650824   & (0.0000075) &   0.4031963   & (0.0000099) \\
    &   128   &   5574.637   & (8.495) & 1274.160 & (2.810) &   47.8241   & (0.0827) &   24.0843   & (0.0481) &   0.6648547   & (0.0000047) &   0.4028996   & (0.0000062) \\
\hline
2.2163   &   32   &   915.878   & (0.220) & 577.170 & (0.272) &   17.4875   & (0.0070) &   9.9995   & (0.0045) &   0.6685961   & (0.0000102) &   0.4078351   & (0.0000136) \\
    &   64   &   2493.061   & (1.172) & 1014.850 & (0.874) &   30.4400   & (0.0204) &   16.7104   & (0.0129) &   0.6679553   & (0.0000064) &   0.4069924   & (0.0000085) \\
    &   128   &   5863.156   & (6.906) & 1392.120 & (2.428) &   49.2549   & (0.0674) &   25.0724   & (0.0389) &   0.6677350   & (0.0000038) &   0.4067032   & (0.0000050) \\
    &   256   &   9346.066   & (43.891) & 1248.620 & (4.972) &   63.1544   & (0.2830) &   27.2723   & (0.1422) &   0.6676388   & (0.0000026) &   0.4065776   & (0.0000034) \\
\hline
2.2500   &   16   &   318.851   & (0.053) & 296.690 & (0.100) &   9.9130   & (0.0031) &   5.8257   & (0.0020) &   0.6762227   & (0.0000194) &   0.4180181   & (0.0000262) \\
    &   32   &   948.196   & (0.216) & 620.144 & (0.280) &   17.9095   & (0.0071) &   10.3085   & (0.0045) &   0.6743138   & (0.0000100) &   0.4154666   & (0.0000135) \\
    &   64   &   2629.995   & (1.332) & 1130.070 & (1.068) &   31.4238   & (0.0236) &   17.4353   & (0.0150) &   0.6737359   & (0.0000072) &   0.4146986   & (0.0000096) \\
    &   128   &   6433.430   & (7.484) & 1648.840 & (3.088) &   51.8923   & (0.0724) &   26.9376   & (0.0441) &   0.6735334   & (0.0000045) &   0.4144290   & (0.0000061) \\
\hline
2.3000   &   8   &   104.676   & (0.010) & 135.234 & (0.028) &   5.5115   & (0.0011) &   3.2889   & (0.0007) &   0.6902831   & (0.0000278) &   0.4371946   & (0.0000387) \\
    &   16   &   328.898   & (0.052) & 316.712 & (0.102) &   10.1711   & (0.0031) &   6.0141   & (0.0020) &   0.6842134   & (0.0000188) &   0.4288765   & (0.0000259) \\
    &   32   &   994.279   & (0.212) & 684.642 & (0.294) &   18.5160   & (0.0072) &   10.7498   & (0.0046) &   0.6824760   & (0.0000097) &   0.4265184   & (0.0000132) \\
    &   64   &   2826.550   & (1.204) & 1308.600 & (1.062) &   32.8101   & (0.0220) &   18.4556   & (0.0140) &   0.6819260   & (0.0000064) &   0.4257741   & (0.0000087) \\
    &   128   &   7252.869   & (5.873) & 2068.860 & (2.842) &   55.5145   & (0.0580) &   29.5483   & (0.0356) &   0.6817342   & (0.0000035) &   0.4255148   & (0.0000048) \\
    &   256   &   14784.492   & (43.665) & 2339.380 & (8.236) &   82.8294   & (0.2470) &   39.4925   & (0.1339) &   0.6816631   & (0.0000024) &   0.4254191   & (0.0000033) \\
\hline
2.3500   &   8   &   106.658   & (0.010) & 140.678 & (0.028) &   5.6274   & (0.0011) &   3.3725   & (0.0007) &   0.6976482   & (0.0000270) &   0.4474960   & (0.0000381) \\
    &   16   &   338.631   & (0.050) & 336.828 & (0.104) &   10.4261   & (0.0032) &   6.1993   & (0.0020) &   0.6918567   & (0.0000182) &   0.4394306   & (0.0000254) \\
    &   32   &   1037.765   & (0.204) & 749.190 & (0.298) &   19.0713   & (0.0071) &   11.1583   & (0.0046) &   0.6901764   & (0.0000093) &   0.4371141   & (0.0000129) \\
    &   64   &   3016.538   & (1.529) & 1496.380 & (1.478) &   34.1722   & (0.0287) &   19.4541   & (0.0183) &   0.6896768   & (0.0000082) &   0.4364275   & (0.0000113) \\
    &   128   &   8033.751   & (7.911) & 2528.670 & (4.436) &   58.7903   & (0.0794) &   31.9539   & (0.0496) &   0.6894977   & (0.0000047) &   0.4361809   & (0.0000065) \\
\hline
2.4000   &   8   &   108.516   & (0.010) & 145.902 & (0.028) &   5.7361   & (0.0011) &   3.4512   & (0.0007) &   0.7046061   & (0.0000262) &   0.4573493   & (0.0000375) \\
    &   16   &   347.875   & (0.060) & 356.664 & (0.128) &   10.6711   & (0.0039) &   6.3774   & (0.0025) &   0.6991327   & (0.0000216) &   0.4496255   & (0.0000306) \\
    &   32   &   1079.687   & (0.156) & 814.664 & (0.240) &   19.6229   & (0.0056) &   11.5609   & (0.0036) &   0.6975099   & (0.0000070) &   0.4473557   & (0.0000099) \\
    &   64   &   3194.036   & (1.245) & 1686.080 & (1.298) &   35.3814   & (0.0239) &   20.3474   & (0.0152) &   0.6970186   & (0.0000066) &   0.4466706   & (0.0000093) \\
    &   128   &   8771.449   & (7.505) & 3016.330 & (4.794) &   61.7156   & (0.0765) &   34.0990   & (0.0487) &   0.6968607   & (0.0000045) &   0.4464494   & (0.0000063) \\
\hline
2.4500   &   8   &   110.324   & (0.010) & 151.112 & (0.028) &   5.8460   & (0.0011) &   3.5304   & (0.0007) &   0.7113191   & (0.0000255) &   0.4670023   & (0.0000370) \\
    &   16   &   356.622   & (0.048) & 376.076 & (0.106) &   10.9048   & (0.0032) &   6.5472   & (0.0021) &   0.7060263   & (0.0000172) &   0.4594249   & (0.0000247) \\
    &   32   &   1119.703   & (0.194) & 880.166 & (0.314) &   20.1449   & (0.0072) &   11.9409   & (0.0046) &   0.7044680   & (0.0000087) &   0.4572107   & (0.0000125) \\
    &   64   &   3367.182   & (1.470) & 1883.870 & (1.642) &   36.5663   & (0.0288) &   21.2168   & (0.0185) &   0.7040141   & (0.0000077) &   0.4565672   & (0.0000109) \\
    &   128   &   9474.514   & (9.176) & 3530.980 & (6.472) &   64.4080   & (0.0953) &   36.0784   & (0.0602) &   0.7038578   & (0.0000055) &   0.4563475   & (0.0000078) \\
\hline
2.5000   &   16   &   365.084   & (0.047) & 395.420 & (0.106) &   11.1422   & (0.0033) &   6.7183   & (0.0021) &   0.7125590   & (0.0000168) &   0.4688290   & (0.0000244) \\
    &   32   &   1157.698   & (0.150) & 945.280 & (0.252) &   20.6415   & (0.0057) &   12.3037   & (0.0036) &   0.7110889   & (0.0000067) &   0.4667133   & (0.0000097) \\
    &   64   &   3532.748   & (1.174) & 2085.280 & (1.396) &   37.6639   & (0.0239) &   22.0257   & (0.0153) &   0.7106662   & (0.0000062) &   0.4661069   & (0.0000089) \\
    &   128   &   10179.023   & (8.982) & 4094.380 & (6.946) &   67.1516   & (0.0962) &   38.0974   & (0.0612) &   0.7105248   & (0.0000053) &   0.4659036   & (0.0000077) \\
\hline
2.5500   &   8   &   113.678   & (0.009) & 161.068 & (0.028) &   6.0549   & (0.0012) &   3.6809   & (0.0008) &   0.7237782   & (0.0000243) &   0.4852234   & (0.0000361) \\
    &   16   &   373.150   & (0.046) & 414.406 & (0.108) &   11.3663   & (0.0033) &   6.8802   & (0.0021) &   0.7188319   & (0.0000163) &   0.4779650   & (0.0000240) \\
    &   32   &   1194.544   & (0.207) & 1010.970 & (0.364) &   21.1419   & (0.0081) &   12.6655   & (0.0052) &   0.7174061   & (0.0000090) &   0.4758909   & (0.0000132) \\
    &   64   &   3694.324   & (1.214) & 2293.600 & (1.532) &   38.8212   & (0.0251) &   22.8634   & (0.0161) &   0.7170100   & (0.0000063) &   0.4753159   & (0.0000093) \\
    &   128   &   10872.593   & (8.782) & 4698.330 & (7.456) &   69.8807   & (0.0949) &   40.0886   & (0.0609) &   0.7168815   & (0.0000053) &   0.4751282   & (0.0000077) \\
\hline
2.6000   &   8   &   115.258   & (0.009) & 165.892 & (0.028) &   6.1559   & (0.0012) &   3.7535   & (0.0008) &   0.7296174   & (0.0000237) &   0.4939109   & (0.0000355) \\
    &   16   &   380.853   & (0.045) & 433.038 & (0.108) &   11.5833   & (0.0033) &   7.0368   & (0.0021) &   0.7248022   & (0.0000158) &   0.4867659   & (0.0000234) \\
    &   32   &   1229.812   & (0.203) & 1076.310 & (0.372) &   21.6051   & (0.0082) &   13.0029   & (0.0053) &   0.7234513   & (0.0000089) &   0.4847783   & (0.0000131) \\
    &   64   &   3848.071   & (1.373) & 2502.680 & (1.832) &   39.8569   & (0.0293) &   23.6212   & (0.0188) &   0.7230687   & (0.0000071) &   0.4842156   & (0.0000105) \\
    &   128   &   11513.661   & (8.223) & 5303.290 & (7.516) &   72.2440   & (0.0920) &   41.8273   & (0.0586) &   0.7229631   & (0.0000051) &   0.4840607   & (0.0000075) \\
\hline
2.6500   &   8   &   116.781   & (0.009) & 170.628 & (0.028) &   6.2593   & (0.0012) &   3.8274   & (0.0008) &   0.7351712   & (0.0000231) &   0.5022599   & (0.0000350) \\
    &   16   &   388.390   & (0.044) & 451.748 & (0.108) &   11.8046   & (0.0034) &   7.1957   & (0.0022) &   0.7305643   & (0.0000155) &   0.4953540   & (0.0000233) \\
    &   32   &   1264.063   & (0.200) & 1142.080 & (0.380) &   22.0826   & (0.0083) &   13.3471   & (0.0053) &   0.7292410   & (0.0000086) &   0.4933832   & (0.0000129) \\
    &   64   &   3996.354   & (1.107) & 2714.490 & (1.552) &   40.8652   & (0.0242) &   24.3570   & (0.0156) &   0.7288651   & (0.0000057) &   0.4928232   & (0.0000085) \\
    &   128   &   12147.754   & (8.967) & 5940.830 & (8.824) &   74.5530   & (0.1020) &   43.5100   & (0.0657) &   0.7287631   & (0.0000055) &   0.4926733   & (0.0000082) \\
\hline
2.7000   &   8   &   118.222   & (0.009) & 175.182 & (0.028) &   6.3547   & (0.0012) &   3.8958   & (0.0008) &   0.7405211   & (0.0000226) &   0.5103857   & (0.0000346) \\
    &   16   &   395.501   & (0.043) & 469.824 & (0.110) &   12.0144   & (0.0034) &   7.3461   & (0.0022) &   0.7360461   & (0.0000151) &   0.5036077   & (0.0000230) \\
    &   32   &   1297.026   & (0.196) & 1207.480 & (0.384) &   22.5499   & (0.0084) &   13.6834   & (0.0054) &   0.7347644   & (0.0000084) &   0.5016786   & (0.0000127) \\
    &   64   &   4139.328   & (1.299) & 2927.980 & (1.906) &   41.8674   & (0.0293) &   25.0800   & (0.0188) &   0.7344162   & (0.0000066) &   0.5011559   & (0.0000100) \\
    &   128   &   12736.705   & (9.071) & 6574.620 & (9.440) &   76.6812   & (0.1066) &   45.0674   & (0.0677) &   0.7343188   & (0.0000055) &   0.5010097   & (0.0000083) \\
\hline
2.7750   &   8   &   120.323   & (0.008) & 181.962 & (0.028) &   6.5036   & (0.0012) &   4.0019   & (0.0008) &   0.7481788   & (0.0000218) &   0.5221505   & (0.0000338) \\
    &   16   &   405.798   & (0.042) & 496.756 & (0.110) &   12.3319   & (0.0035) &   7.5729   & (0.0022) &   0.7438488   & (0.0000146) &   0.5155007   & (0.0000224) \\
    &   32   &   1344.166   & (0.190) & 1304.740 & (0.390) &   23.2109   & (0.0084) &   14.1583   & (0.0054) &   0.7426646   & (0.0000081) &   0.5136930   & (0.0000124) \\
    &   64   &   4347.763   & (1.296) & 3256.200 & (2.026) &   43.3227   & (0.0301) &   26.1317   & (0.0194) &   0.7423202   & (0.0000064) &   0.5131666   & (0.0000098) \\
    &   128   &   13612.434   & (7.816) & 7582.630 & (8.912) &   79.8713   & (0.0955) &   47.3873   & (0.0616) &   0.7422121   & (0.0000047) &   0.5130026   & (0.0000071) \\
\hline
2.8500   &   8   &   122.246   & (0.008) & 188.302 & (0.028) &   6.6407   & (0.0012) &   4.0997   & (0.0008) &   0.7552800   & (0.0000211) &   0.5332061   & (0.0000332) \\
    &   16   &   415.352   & (0.041) & 522.530 & (0.110) &   12.6196   & (0.0035) &   7.7791   & (0.0023) &   0.7512008   & (0.0000140) &   0.5268598   & (0.0000219) \\
    &   32   &   1388.397   & (0.184) & 1400.020 & (0.394) &   23.8451   & (0.0085) &   14.6129   & (0.0055) &   0.7500500   & (0.0000078) &   0.5250796   & (0.0000121) \\
    &   64   &   4542.800   & (1.225) & 3581.880 & (2.030) &   44.6948   & (0.0293) &   27.1189   & (0.0190) &   0.7497428   & (0.0000062) &   0.5246044   & (0.0000096) \\
    &   128   &   14452.813   & (8.264) & 8630.020 & (10.208) &   82.9390   & (0.1065) &   49.6153   & (0.0682) &   0.7496480   & (0.0000051) &   0.5244567   & (0.0000079) \\
\hline
2.9250   &   8   &   124.097   & (0.008) & 194.534 & (0.026) &   6.7772   & (0.0013) &   4.1970   & (0.0008) &   0.7620488   & (0.0000205) &   0.5438826   & (0.0000325) \\
    &   16   &   424.495   & (0.040) & 547.922 & (0.110) &   12.9140   & (0.0036) &   7.9886   & (0.0023) &   0.7581258   & (0.0000137) &   0.5376944   & (0.0000216) \\
    &   32   &   1431.034   & (0.178) & 1495.590 & (0.398) &   24.4896   & (0.0086) &   15.0724   & (0.0055) &   0.7570512   & (0.0000075) &   0.5360119   & (0.0000118) \\
    &   64   &   4732.111   & (1.232) & 3915.350 & (2.152) &   46.1043   & (0.0310) &   28.1300   & (0.0200) &   0.7567380   & (0.0000061) &   0.5355216   & (0.0000096) \\
    &   128   &   15248.705   & (8.041) & 9694.170 & (10.658) &   85.8159   & (0.1041) &   51.7004   & (0.0669) &   0.7566473   & (0.0000048) &   0.5353791   & (0.0000076) \\
\hline
3.0000   &   8   &   125.860   & (0.008) & 200.580 & (0.026) &   6.9135   & (0.0013) &   4.2936   & (0.0008) &   0.7684608   & (0.0000198) &   0.5541019   & (0.0000318) \\
    &   16   &   433.226   & (0.039) & 572.832 & (0.110) &   13.2037   & (0.0036) &   8.1944   & (0.0023) &   0.7646672   & (0.0000131) &   0.5480557   & (0.0000209) \\
    &   32   &   1471.174   & (0.196) & 1588.900 & (0.454) &   25.0849   & (0.0098) &   15.4982   & (0.0063) &   0.7636364   & (0.0000082) &   0.5464188   & (0.0000130) \\
    &   64   &   4907.505   & (1.184) & 4239.350 & (2.170) &   47.3188   & (0.0307) &   29.0000   & (0.0198) &   0.7633401   & (0.0000058) &   0.5459503   & (0.0000093) \\
    &   128   &   16012.500   & (8.034) & 10782.600 & (11.342) &   88.6134   & (0.1081) &   53.7049   & (0.0699) &   0.7632664   & (0.0000048) &   0.5458335   & (0.0000076) \\
\hline
\end{tabular}
\caption{
   {\bf [Continued]}
}
\end{center}
\end{table} \clearpage \addtocounter{table}{-1}
\begin{table}
\addtolength{\tabcolsep}{-1.3mm}
%\protect\tiny
\begin{center}
\begin{tabular}{|r r|r r|r r|r r|r r|r r|r r|} \hline
$\beta$  & $L$ &
\multicolumn{2}{|c|}{$\chi_F$} &
\multicolumn{2}{|c|}{$\chi_A$} &
\multicolumn{2}{|c|}{$\xi_F$} &
\multicolumn{2}{|c|}{$\xi_A$} &
\multicolumn{2}{|c|}{$E_F$} &
\multicolumn{2}{|c|}{$E_A$} \\ 
\hline
3.0750   &   8   &   127.528   & (0.008) & 206.406 & (0.026) &   7.0477   & (0.0013) &   4.3885   & (0.0008) &   0.7744805   & (0.0000193) &   0.5638025   & (0.0000314) \\
    &   16   &   441.480   & (0.038) & 596.990 & (0.110) &   13.4818   & (0.0037) &   8.3918   & (0.0024) &   0.7708835   & (0.0000128) &   0.5580081   & (0.0000207) \\
    &   32   &   1508.936   & (0.193) & 1679.650 & (0.460) &   25.6650   & (0.0100) &   15.9091   & (0.0064) &   0.7698502   & (0.0000079) &   0.5563497   & (0.0000127) \\
    &   64   &   5079.512   & (1.163) & 4571.720 & (2.224) &   48.6312   & (0.0313) &   29.9381   & (0.0202) &   0.7695843   & (0.0000057) &   0.5559244   & (0.0000092) \\
    &   128   &   16751.266   & (7.928) & 11901.300 & (11.900) &   91.4303   & (0.1111) &   55.7398   & (0.0716) &   0.7695122   & (0.0000046) &   0.5558087   & (0.0000074) \\
\hline
3.1500   &   8   &   129.112   & (0.007) & 212.028 & (0.026) &   7.1759   & (0.0013) &   4.4792   & (0.0009) &   0.7802485   & (0.0000188) &   0.5731906   & (0.0000308) \\
    &   16   &   449.278   & (0.037) & 620.342 & (0.110) &   13.7492   & (0.0037) &   8.5813   & (0.0024) &   0.7767427   & (0.0000124) &   0.5674870   & (0.0000201) \\
    &   32   &   1545.692   & (0.189) & 1770.770 & (0.464) &   26.2414   & (0.0102) &   16.3184   & (0.0066) &   0.7757752   & (0.0000078) &   0.5659192   & (0.0000126) \\
    &   64   &   5243.404   & (1.119) & 4901.800 & (2.234) &   49.8794   & (0.0314) &   30.8266   & (0.0203) &   0.7755122   & (0.0000055) &   0.5654942   & (0.0000090) \\
    &   128   &   17464.784   & (7.801) & 13038.900 & (12.358) &   94.1656   & (0.1104) &   57.6953   & (0.0717) &   0.7754297   & (0.0000045) &   0.5653608   & (0.0000072) \\
\hline
3.2250   &   8   &   130.617   & (0.007) & 217.460 & (0.026) &   7.3028   & (0.0013) &   4.5689   & (0.0009) &   0.7856912   & (0.0000182) &   0.5821372   & (0.0000301) \\
    &   16   &   456.695   & (0.036) & 643.034 & (0.110) &   14.0111   & (0.0038) &   8.7665   & (0.0025) &   0.7822887   & (0.0000121) &   0.5765456   & (0.0000198) \\
    &   32   &   1580.734   & (0.150) & 1860.240 & (0.380) &   26.8115   & (0.0084) &   16.7222   & (0.0054) &   0.7813866   & (0.0000061) &   0.5750723   & (0.0000101) \\
    &   64   &   5397.960   & (1.117) & 5225.310 & (2.314) &   51.0566   & (0.0322) &   31.6666   & (0.0208) &   0.7811263   & (0.0000054) &   0.5746463   & (0.0000089) \\
    &   128   &   18127.984   & (7.514) & 14150.800 & (12.464) &   96.5991   & (0.1124) &   59.4401   & (0.0724) &   0.7810578   & (0.0000043) &   0.5745342   & (0.0000071) \\
\hline
3.3000   &   8   &   132.047   & (0.007) & 222.696 & (0.026) &   7.4273   & (0.0014) &   4.6567   & (0.0009) &   0.7908599   & (0.0000178) &   0.5907088   & (0.0000297) \\
    &   16   &   463.891   & (0.035) & 665.502 & (0.110) &   14.2822   & (0.0039) &   8.9571   & (0.0025) &   0.7875840   & (0.0000118) &   0.5852786   & (0.0000196) \\
    &   32   &   1614.179   & (0.179) & 1948.000 & (0.466) &   27.3826   & (0.0104) &   17.1238   & (0.0067) &   0.7867041   & (0.0000073) &   0.5838263   & (0.0000121) \\
    &   64   &   5544.951   & (1.095) & 5543.600 & (2.350) &   52.1957   & (0.0324) &   32.4698   & (0.0210) &   0.7864617   & (0.0000052) &   0.5834262   & (0.0000087) \\
    &   128   &   18768.903   & (7.349) & 15275.400 & (12.762) &   99.0107   & (0.1121) &   61.1518   & (0.0725) &   0.7864011   & (0.0000041) &   0.5833254   & (0.0000069) \\
\hline
3.3750   &   8   &   133.433   & (0.007) & 227.840 & (0.026) &   7.5529   & (0.0014) &   4.7449   & (0.0009) &   0.7958360   & (0.0000172) &   0.5990262   & (0.0000289) \\
    &   16   &   470.642   & (0.034) & 687.006 & (0.110) &   14.5351   & (0.0039) &   9.1354   & (0.0025) &   0.7925947   & (0.0000115) &   0.5936124   & (0.0000192) \\
    &   32   &   1645.385   & (0.177) & 2031.990 & (0.472) &   27.8936   & (0.0104) &   17.4849   & (0.0068) &   0.7917624   & (0.0000071) &   0.5922255   & (0.0000118) \\
    &   64   &   5688.593   & (1.057) & 5865.390 & (2.346) &   53.3511   & (0.0328) &   33.2879   & (0.0212) &   0.7915459   & (0.0000051) &   0.5918644   & (0.0000085) \\
    &   128   &   19387.428   & (7.047) & 16405.100 & (12.776) &   101.3455   & (0.1130) &   62.8175   & (0.0730) &   0.7914698   & (0.0000040) &   0.5917380   & (0.0000067) \\
\hline
3.4500   &   8   &   134.720   & (0.007) & 232.684 & (0.026) &   7.6715   & (0.0014) &   4.8282   & (0.0009) &   0.8004836   & (0.0000169) &   0.6068624   & (0.0000286) \\
    &   16   &   477.220   & (0.034) & 708.326 & (0.110) &   14.7918   & (0.0039) &   9.3157   & (0.0026) &   0.7974278   & (0.0000112) &   0.6017145   & (0.0000188) \\
    &   32   &   1675.883   & (0.173) & 2116.080 & (0.474) &   28.4156   & (0.0107) &   17.8532   & (0.0069) &   0.7966041   & (0.0000069) &   0.6003324   & (0.0000116) \\
    &   64   &   5824.061   & (1.024) & 6178.210 & (2.346) &   54.4221   & (0.0326) &   34.0458   & (0.0211) &   0.7963841   & (0.0000049) &   0.5999626   & (0.0000083) \\
    &   128   &   19986.030   & (6.855) & 17544.900 & (12.922) &   103.7616   & (0.1128) &   64.5286   & (0.0730) &   0.7963155   & (0.0000039) &   0.5998478   & (0.0000065) \\
\hline
3.5250   &   8   &   135.981   & (0.007) & 237.488 & (0.026) &   7.7909   & (0.0014) &   4.9120   & (0.0009) &   0.8050157   & (0.0000166) &   0.6145615   & (0.0000283) \\
    &   16   &   483.382   & (0.033) & 728.644 & (0.110) &   15.0330   & (0.0040) &   9.4846   & (0.0026) &   0.8019973   & (0.0000110) &   0.6094380   & (0.0000186) \\
    &   32   &   1704.968   & (0.139) & 2198.080 & (0.388) &   28.9257   & (0.0088) &   18.2127   & (0.0057) &   0.8012108   & (0.0000055) &   0.6081065   & (0.0000093) \\
    &   64   &   5954.719   & (1.014) & 6488.900 & (2.392) &   55.4890   & (0.0332) &   34.7985   & (0.0216) &   0.8009956   & (0.0000048) &   0.6077418   & (0.0000081) \\
    &   128   &   20546.909   & (7.109) & 18651.900 & (13.852) &   105.8729   & (0.1190) &   66.0306   & (0.0769) &   0.8009269   & (0.0000039) &   0.6076260   & (0.0000066) \\
\hline
3.6000   &   8   &   137.167   & (0.006) & 242.058 & (0.024) &   7.9048   & (0.0014) &   4.9920   & (0.0009) &   0.8092834   & (0.0000161) &   0.6218624   & (0.0000276) \\
    &   16   &   489.420   & (0.032) & 748.882 & (0.108) &   15.2812   & (0.0040) &   9.6585   & (0.0026) &   0.8063909   & (0.0000107) &   0.6169154   & (0.0000183) \\
    &   32   &   1733.083   & (0.165) & 2279.010 & (0.472) &   29.4442   & (0.0109) &   18.5754   & (0.0070) &   0.8056173   & (0.0000065) &   0.6155963   & (0.0000112) \\
    &   64   &   6082.096   & (1.005) & 6799.560 & (2.436) &   56.5707   & (0.0342) &   35.5541   & (0.0222) &   0.8053973   & (0.0000048) &   0.6152220   & (0.0000082) \\
    &   128   &   21105.728   & (6.794) & 19791.000 & (13.762) &   108.2497   & (0.1188) &   67.6945   & (0.0771) &   0.8053424   & (0.0000038) &   0.6151288   & (0.0000064) \\
\hline
3.6750   &   8   &   138.315   & (0.006) & 246.530 & (0.024) &   8.0208   & (0.0014) &   5.0731   & (0.0009) &   0.8133941   & (0.0000157) &   0.6289373   & (0.0000272) \\
    &   16   &   495.124   & (0.032) & 768.300 & (0.108) &   15.5208   & (0.0041) &   9.8262   & (0.0027) &   0.8105699   & (0.0000104) &   0.6240798   & (0.0000179) \\
    &   32   &   1759.858   & (0.134) & 2357.660 & (0.390) &   29.9407   & (0.0091) &   18.9232   & (0.0059) &   0.8098122   & (0.0000052) &   0.6227781   & (0.0000090) \\
    &   64   &   6203.522   & (0.979) & 7104.020 & (2.432) &   57.6513   & (0.0345) &   36.3129   & (0.0224) &   0.8096173   & (0.0000045) &   0.6224443   & (0.0000078) \\
    &   128   &   21634.998   & (6.641) & 20905.300 & (13.854) &   110.4495   & (0.1184) &   69.2402   & (0.0765) &   0.8095549   & (0.0000037) &   0.6223376   & (0.0000064) \\
\hline
3.7500   &   8   &   139.419   & (0.006) & 250.880 & (0.024) &   8.1353   & (0.0015) &   5.1531   & (0.0010) &   0.8173324   & (0.0000153) &   0.6357653   & (0.0000266) \\
    &   16   &   500.564   & (0.031) & 787.072 & (0.108) &   15.7489   & (0.0041) &   9.9858   & (0.0027) &   0.8145814   & (0.0000102) &   0.6310006   & (0.0000176) \\
    &   32   &   1785.608   & (0.161) & 2434.770 & (0.478) &   30.4259   & (0.0112) &   19.2631   & (0.0073) &   0.8138521   & (0.0000063) &   0.6297409   & (0.0000110) \\
    &   64   &   6318.153   & (0.962) & 7398.290 & (2.448) &   58.6223   & (0.0345) &   36.9932   & (0.0224) &   0.8136436   & (0.0000046) &   0.6293811   & (0.0000079) \\
    &   128   &   22144.794   & (6.649) & 22013.100 & (14.336) &   112.5532   & (0.1223) &   70.7189   & (0.0794) &   0.8135932   & (0.0000036) &   0.6292941   & (0.0000063) \\
\hline
3.8250   &   8   &   140.465   & (0.006) & 255.042 & (0.024) &   8.2449   & (0.0015) &   5.2297   & (0.0010) &   0.8210908   & (0.0000150) &   0.6423226   & (0.0000263) \\
    &   16   &   505.786   & (0.031) & 805.354 & (0.106) &   15.9740   & (0.0042) &   10.1432   & (0.0027) &   0.8184126   & (0.0000099) &   0.6376536   & (0.0000173) \\
    &   32   &   1810.521   & (0.128) & 2510.670 & (0.388) &   30.9094   & (0.0093) &   19.6001   & (0.0060) &   0.8176975   & (0.0000050) &   0.6364104   & (0.0000088) \\
    &   64   &   6431.746   & (0.940) & 7696.570 & (2.446) &   59.6574   & (0.0347) &   37.7196   & (0.0225) &   0.8175026   & (0.0000043) &   0.6360721   & (0.0000075) \\
    &   128   &   22636.136   & (6.543) & 23111.100 & (14.472) &   114.6291   & (0.1226) &   72.1756   & (0.0796) &   0.8174550   & (0.0000035) &   0.6359895   & (0.0000061) \\
\hline
3.9000   &   8   &   141.472   & (0.006) & 259.088 & (0.024) &   8.3532   & (0.0015) &   5.3054   & (0.0010) &   0.8246889   & (0.0000147) &   0.6486348   & (0.0000259) \\
    &   16   &   510.839   & (0.030) & 823.268 & (0.106) &   16.2062   & (0.0043) &   10.3047   & (0.0028) &   0.8220770   & (0.0000097) &   0.6440550   & (0.0000170) \\
    &   32   &   1834.277   & (0.154) & 2584.300 & (0.476) &   31.3848   & (0.0115) &   19.9308   & (0.0075) &   0.8214097   & (0.0000060) &   0.6428880   & (0.0000106) \\
    &   64   &   6538.744   & (0.931) & 7983.460 & (2.476) &   60.5608   & (0.0357) &   38.3547   & (0.0232) &   0.8212079   & (0.0000042) &   0.6425356   & (0.0000074) \\
    &   128   &   23111.072   & (6.218) & 24201.400 & (14.160) &   116.6178   & (0.1198) &   73.5761   & (0.0779) &   0.8211644   & (0.0000034) &   0.6424592   & (0.0000060) \\
\hline
3.9750   &   8   &   142.447   & (0.006) & 263.040 & (0.024) &   8.4630   & (0.0015) &   5.3819   & (0.0010) &   0.8281582   & (0.0000144) &   0.6547558   & (0.0000255) \\
    &   16   &   515.732   & (0.029) & 840.834 & (0.104) &   16.4320   & (0.0043) &   10.4618   & (0.0028) &   0.8256308   & (0.0000095) &   0.6502990   & (0.0000168) \\
    &   32   &   1856.951   & (0.151) & 2655.750 & (0.472) &   31.8260   & (0.0114) &   20.2397   & (0.0074) &   0.8249516   & (0.0000059) &   0.6491033   & (0.0000103) \\
    &   64   &   6643.349   & (0.905) & 8269.980 & (2.456) &   61.5834   & (0.0355) &   39.0685   & (0.0231) &   0.8247704   & (0.0000042) &   0.6487856   & (0.0000074) \\
    &   128   &   23571.870   & (6.199) & 25286.400 & (14.432) &   118.6558   & (0.1267) &   75.0049   & (0.0823) &   0.8247192   & (0.0000033) &   0.6486954   & (0.0000059) \\
\hline
4.0500   &   8   &   143.378   & (0.006) & 266.848 & (0.024) &   8.5686   & (0.0015) &   5.4554   & (0.0010) &   0.8314841   & (0.0000141) &   0.6606531   & (0.0000251) \\
    &   16   &   520.454   & (0.029) & 857.990 & (0.104) &   16.6579   & (0.0043) &   10.6189   & (0.0028) &   0.8290238   & (0.0000093) &   0.6562928   & (0.0000166) \\
    &   32   &   1879.347   & (0.150) & 2727.400 & (0.476) &   32.2969   & (0.0117) &   20.5684   & (0.0076) &   0.8283733   & (0.0000058) &   0.6551424   & (0.0000103) \\
    &   64   &   6743.618   & (0.885) & 8549.860 & (2.450) &   62.5236   & (0.0356) &   39.7214   & (0.0232) &   0.8281906   & (0.0000041) &   0.6548195   & (0.0000072) \\
    &   128   &   24021.313   & (6.086) & 26370.000 & (14.558) &   120.8127   & (0.1265) &   76.5033   & (0.0823) &   0.8281346   & (0.0000033) &   0.6547204   & (0.0000059) \\
\hline
4.1250   &   8   &   144.285   & (0.006) & 270.588 & (0.024) &   8.6761   & (0.0015) &   5.5301   & (0.0010) &   0.8347004   & (0.0000138) &   0.6663871   & (0.0000247) \\
    &   16   &   524.967   & (0.029) & 874.576 & (0.104) &   16.8763   & (0.0044) &   10.7708   & (0.0029) &   0.8322784   & (0.0000091) &   0.6620720   & (0.0000162) \\
    &   32   &   1900.451   & (0.145) & 2795.870 & (0.468) &   32.7399   & (0.0117) &   20.8754   & (0.0076) &   0.8316345   & (0.0000056) &   0.6609275   & (0.0000101) \\
    &   64   &   6841.328   & (0.878) & 8827.540 & (2.474) &   63.4853   & (0.0363) &   40.3874   & (0.0236) &   0.8314778   & (0.0000040) &   0.6606495   & (0.0000072) \\
    &   128   &   24446.725   & (5.868) & 27420.100 & (14.362) &   122.5206   & (0.1262) &   77.7064   & (0.0821) &   0.8314264   & (0.0000032) &   0.6605584   & (0.0000058) \\
\hline
4.2000   &   8   &   145.155   & (0.005) & 274.206 & (0.022) &   8.7820   & (0.0016) &   5.6036   & (0.0010) &   0.8377766   & (0.0000135) &   0.6718995   & (0.0000243) \\
    &   16   &   529.345   & (0.028) & 890.830 & (0.104) &   17.0906   & (0.0044) &   10.9195   & (0.0029) &   0.8354397   & (0.0000089) &   0.6677146   & (0.0000160) \\
    &   32   &   1921.307   & (0.144) & 2864.530 & (0.470) &   33.1993   & (0.0119) &   21.1947   & (0.0077) &   0.8347976   & (0.0000055) &   0.6665674   & (0.0000099) \\
    &   64   &   6934.739   & (0.864) & 9098.010 & (2.480) &   64.4375   & (0.0371) &   41.0506   & (0.0242) &   0.8346302   & (0.0000039) &   0.6662690   & (0.0000070) \\
    &   128   &   24863.001   & (5.990) & 28469.100 & (14.976) &   124.5811   & (0.1295) &   79.1329   & (0.0843) &   0.8345835   & (0.0000032) &   0.6661849   & (0.0000057) \\
\hline
4.2750   &   8   &   145.977   & (0.005) & 277.652 & (0.022) &   8.8806   & (0.0016) &   5.6722   & (0.0010) &   0.8407447   & (0.0000133) &   0.6772412   & (0.0000240) \\
    &   16   &   533.488   & (0.028) & 906.382 & (0.102) &   17.2944   & (0.0044) &   11.0613   & (0.0029) &   0.8384427   & (0.0000087) &   0.6730987   & (0.0000157) \\
    &   32   &   1941.074   & (0.142) & 2930.470 & (0.468) &   33.6517   & (0.0121) &   21.5069   & (0.0079) &   0.8378251   & (0.0000054) &   0.6719916   & (0.0000097) \\
    &   64   &   7024.695   & (0.851) & 9362.730 & (2.484) &   65.2749   & (0.0376) &   41.6358   & (0.0245) &   0.8376732   & (0.0000039) &   0.6717194   & (0.0000070) \\
    &   128   &   25255.723   & (5.878) & 29480.400 & (14.970) &   126.3164   & (0.1276) &   80.3430   & (0.0831) &   0.8376366   & (0.0000031) &   0.6716535   & (0.0000055) \\
\hline
4.3500   &   8   &   146.790   & (0.005) & 281.080 & (0.022) &   8.9832   & (0.0016) &   5.7433   & (0.0010) &   0.8436128   & (0.0000131) &   0.6824278   & (0.0000237) \\
    &   16   &   537.544   & (0.027) & 921.750 & (0.102) &   17.5079   & (0.0045) &   11.2088   & (0.0029) &   0.8413512   & (0.0000086) &   0.6783406   & (0.0000155) \\
    &   32   &   1960.189   & (0.139) & 2995.030 & (0.466) &   34.0655   & (0.0122) &   21.7946   & (0.0079) &   0.8407636   & (0.0000053) &   0.6772804   & (0.0000095) \\
    &   64   &   7111.600   & (0.846) & 9622.720 & (2.508) &   66.1570   & (0.0379) &   42.2494   & (0.0247) &   0.8406023   & (0.0000038) &   0.6769897   & (0.0000068) \\
    &   128   &   25661.107   & (5.622) & 30544.400 & (14.640) &   128.4454   & (0.1281) &   81.8288   & (0.0833) &   0.8405637   & (0.0000031) &   0.6769203   & (0.0000055) \\
\hline
\end{tabular}
\caption{
{\bf [Continued]}
}
\end{center}
\end{table}
\clearpage

%%%%%%%%%%%%%%%%%%%%%%%%%%%%%%%%%%%%%%%%%%%%%%%%%%%%%%%%%%%%%%%%%%%%%

%%%%%%%%%%%%%%%%%%%%%%%%%%%%%%%%%%%%%%%%%%%%%%%%%%%%%%%%%%%%%%%%%%%%%

%%%%%%%%%%%%%%%%%%%%%%%%%%%%%%%%%%%%%%%%%%%%%%%%%%%%%%%%%%%%%%%%%%%%%
%
% Beginning of Table 2 for SU3 article
% Dynamic quantities: {beta,L,sweeps,Tau_chi_F,Tau_chi_A,Tau_E_F,Tau_E_A}
%
\protect\footnotesize
%%%%\protect\tabcolsep 3.5pt
\begin{table}
\addtolength{\tabcolsep}{-1.3mm}
\begin{center}
\begin{tabular}{|r r|r|r r|r r|r r|r r|} \hline
\multicolumn{1}{|c}{$\beta$}  &
\multicolumn{1}{c|}{$L$} &
\multicolumn{1}{|c|}{Run Length} &
\multicolumn{2}{|c|}{$\tau_{int,{\cal M}_F^2}$} &
\multicolumn{2}{|c|}{$\tau_{int,{\cal M}_A^2}$} &
\multicolumn{2}{|c|}{$\tau_{int,{\cal E}_F}$} &
\multicolumn{2}{|c|}{$\tau_{int,{\cal E}_A}$} \\
%\multicolumn{20}{|c|}{$d=2$ $SU(3)$ model Table LAST}
\hline
 1.6000 &   32 &   260000 &  17.305 & (0.831) &  7.794 & (0.251) &  9.581 & (0.383) &   8.815 & (0.338)  \\  
 \hline
 1.6500 &    8 &  3000000 &   7.138 & (0.062) &  6.672 & (0.056) &  6.591 & (0.062) &   6.254 & (0.057)  \\  
  &   16 &  1000000 &  16.665 & (0.389) & 13.358 & (0.279) & 10.412 & (0.215) &   9.587 & (0.190)  \\ 
  &   32 &   200000 &  38.968 & (4.323) & 21.603 & (1.784) & 15.735 & (1.109) &  14.800 & (1.012)  \\ 
 \hline
 1.7000 &   32 &   220000 &  29.332 & (2.010) & 20.184 & (1.147) & 11.862 & (0.578) &  10.850 & (0.505)  \\  
 \hline
 1.7500 &    8 &  3000000 &   6.343 & (0.052) &  6.084 & (0.049) &  5.849 & (0.052) &   5.650 & (0.049)  \\  
  &   16 &  1000000 &  11.427 & (0.221) &  9.996 & (0.181) &  7.663 & (0.136) &   7.214 & (0.124)  \\ 
  &   32 &   500000 &  31.825 & (1.466) & 23.335 & (0.920) & 11.439 & (0.353) &  10.481 & (0.310)  \\ 
  &   64 &   600000 &  24.391 & (0.895) &  9.852 & (0.230) &  9.403 & (0.239) &   8.694 & (0.213)  \\ 
  &  128 &   500000 &  14.986 & (0.474) &  4.043 & (0.066) &  9.245 & (0.257) &   8.569 & (0.229)  \\ 
 \hline
 1.7750 &   32 &   960000 &  28.730 & (0.898) & 20.915 & (0.558) & 10.265 & (0.215) &   9.562 & (0.193)  \\  
  &   64 &  1000000 &  30.475 & (0.961) & 13.050 & (0.269) &  9.613 & (0.190) &   8.942 & (0.171)  \\ 
 \hline
 1.8000 &   32 &  1000000 &  27.259 & (0.813) & 20.773 & (0.541) &  9.672 & (0.192) &   9.061 & (0.174)  \\  
  &   64 &  2500000 &  35.528 & (0.761) & 17.646 & (0.266) &  9.443 & (0.117) &   8.774 & (0.104)  \\ 
  &  128 &  1500000 &  19.646 & (0.405) &  5.482 & (0.060) &  8.877 & (0.137) &   8.262 & (0.123)  \\ 
  &  256 &   200000 &  12.227 & (0.570) &  3.421 & (0.084) &  9.265 & (0.420) &   8.619 & (0.377)  \\ 
 \hline
 1.8250 &   32 &  2980000 &  23.598 & (0.377) & 18.498 & (0.262) &  8.790 & (0.096) &   8.277 & (0.088)  \\  
  &   64 &  2500000 &  43.423 & (1.028) & 23.780 & (0.417) &  9.356 & (0.115) &   8.756 & (0.104)  \\ 
  &  128 &  1500000 &  22.985 & (0.512) &  6.604 & (0.079) &  8.879 & (0.138) &   8.310 & (0.125)  \\ 
 \hline
 1.8500 &    8 &  3000000 &   6.021 & (0.048) &  5.844 & (0.046) &  5.539 & (0.048) &   5.384 & (0.046)  \\  
  &   16 &  1000000 &   8.984 & (0.154) &  8.294 & (0.136) &  6.423 & (0.104) &   6.173 & (0.098)  \\ 
  &   32 &  2980000 &  21.260 & (0.322) & 17.100 & (0.233) &  8.211 & (0.086) &   7.788 & (0.080)  \\ 
  &   64 &  2500000 &  45.814 & (1.114) & 28.145 & (0.536) &  9.274 & (0.113) &   8.722 & (0.103)  \\ 
  &  128 &  1500000 &  25.953 & (0.615) &  8.087 & (0.107) &  8.756 & (0.135) &   8.273 & (0.124)  \\ 
  &  256 &   200000 &  15.932 & (0.848) &  4.159 & (0.113) &  8.514 & (0.370) &   7.960 & (0.335)  \\ 
 \hline
 1.8750 &   32 &  3000000 &  18.984 & (0.271) & 15.672 & (0.203) &  7.563 & (0.076) &   7.219 & (0.071)  \\  
  &   64 &  1860000 &  48.280 & (1.399) & 32.170 & (0.761) &  9.092 & (0.128) &   8.552 & (0.117)  \\ 
  &  128 &  1140000 &  30.982 & (0.922) & 10.083 & (0.171) &  8.429 & (0.146) &   7.935 & (0.134)  \\ 
 \hline
 1.9000 &   16 &  2980000 &   8.485 & (0.081) &  8.009 & (0.075) &  6.168 & (0.056) &   5.966 & (0.054)  \\  
  &   32 &  2980000 &  16.959 & (0.230) & 14.487 & (0.181) &  7.298 & (0.072) &   6.998 & (0.068)  \\ 
  &   64 &  1500000 &  47.030 & (1.500) & 33.181 & (0.889) &  8.659 & (0.132) &   8.192 & (0.122)  \\ 
  &  128 &  1000000 &  36.960 & (1.284) & 13.366 & (0.279) &  8.042 & (0.146) &   7.655 & (0.135)  \\ 
 \hline
 1.9250 &   16 &  2980000 &   8.239 & (0.078) &  7.807 & (0.072) &  6.086 & (0.055) &   5.904 & (0.053)  \\  
  &   32 &  2980000 &  15.774 & (0.206) & 13.773 & (0.168) &  6.908 & (0.067) &   6.640 & (0.063)  \\ 
  &   64 &  1500000 &  42.913 & (1.307) & 31.477 & (0.821) &  8.126 & (0.120) &   7.745 & (0.112)  \\ 
  &  128 &  1000000 &  41.753 & (1.542) & 17.448 & (0.416) &  7.820 & (0.140) &   7.469 & (0.130)  \\ 
  &  256 &   800000 &  23.873 & (0.747) &  6.320 & (0.102) &  7.659 & (0.152) &   7.298 & (0.141)  \\ 
 \hline
 1.9500 &   16 &  2980000 &   8.075 & (0.075) &  7.679 & (0.070) &  5.960 & (0.053) &   5.804 & (0.051)  \\  
  &   32 &  2980000 &  14.523 & (0.182) & 12.792 & (0.150) &  6.549 & (0.062) &   6.323 & (0.058)  \\ 
  &   64 &  1500000 &  42.373 & (1.283) & 31.159 & (0.809) &  7.792 & (0.113) &   7.475 & (0.106)  \\ 
  &  128 &  1000000 &  51.348 & (2.103) & 24.058 & (0.674) &  7.694 & (0.136) &   7.373 & (0.128)  \\ 
 \hline
\end{tabular}
\end{center}
\caption{
   Dynamic data from our runs for the two-dimensional $SU(3)$ chiral model.
   A measurement is performed once every {\em two}\/ MGMC iterations;
   all times (both run lengths and autocorrelation times)
   are reported in units of measurements.
   The number of measurements discarded prior to
    beginning the analysis is always 20000;
   ``run length'' is the total number of measurements performed
   {\em after}\/ the discard interval.
   Error bar (one standard deviation) is shown in parentheses.
}
\label{su3_dyndata}
\end{table} \clearpage \addtocounter{table}{-1}

\begin{table}[p]
\addtolength{\tabcolsep}{-1.3mm}
\begin{center}
\begin{tabular}{|r r|r|r r|r r|r r|r r|} \hline
\multicolumn{1}{|c}{$\beta$}  &
\multicolumn{1}{c|}{$L$} &
\multicolumn{1}{|c|}{Run Length} &
\multicolumn{2}{|c|}{$\tau_{int,{\cal M}_F^2}$} &
\multicolumn{2}{|c|}{$\tau_{int,{\cal M}_A^2}$} &
\multicolumn{2}{|c|}{$\tau_{int,{\cal E}_F}$} &
\multicolumn{2}{|c|}{$\tau_{int,{\cal E}_A}$} \\
%\multicolumn{20}{|c|}{$d=2$ $SU(3)$ model Table LAST}
 \hline
 1.9750 &    8 &  3000000 &   5.696 & (0.045) &  5.589 & (0.043) &  5.275 & (0.044) &   5.174 & (0.043)  \\  
  &   16 &  3000000 &   7.833 & (0.072) &  7.480 & (0.067) &  5.753 & (0.051) &   5.608 & (0.049)  \\ 
  &   32 &  3000000 &  13.534 & (0.163) & 12.171 & (0.139) &  6.489 & (0.061) &   6.284 & (0.058)  \\ 
  &   64 &  1500000 &  36.539 & (1.027) & 27.368 & (0.666) &  7.397 & (0.105) &   7.112 & (0.099)  \\ 
  &  128 &  1000000 &  60.137 & (2.665) & 31.819 & (1.026) &  7.463 & (0.130) &   7.178 & (0.123)  \\ 
 \hline
 1.9850 &   64 &  1500000 &  36.103 & (1.009) & 27.220 & (0.660) &  7.313 & (0.103) &   7.037 & (0.097)  \\  
  &  128 &  1000000 &  60.704 & (2.703) & 33.529 & (1.109) &  7.539 & (0.132) &   7.226 & (0.124)  \\
\hline
\end{tabular}
\end{center}
\caption{
{\bf [Continued]}
}
\end{table} \clearpage \addtocounter{table}{-1}

\begin{table}[p]
\addtolength{\tabcolsep}{-1.3mm}
\begin{center}
\begin{tabular}{|r r|r|r r|r r|r r|r r|} \hline
\multicolumn{1}{|c}{$\beta$}  &
\multicolumn{1}{c|}{$L$} &
\multicolumn{1}{|c|}{Run Length} &
\multicolumn{2}{|c|}{$\tau_{int,{\cal M}_F^2}$} &
\multicolumn{2}{|c|}{$\tau_{int,{\cal M}_A^2}$} &
\multicolumn{2}{|c|}{$\tau_{int,{\cal E}_F}$} &
\multicolumn{2}{|c|}{$\tau_{int,{\cal E}_A}$} \\
%\multicolumn{20}{|c|}{$d=2$ $SU(3)$ model Table LAST}
\hline
 2.1500 &   16 &  2980000 &   7.051 & (0.062) &  6.869 & (0.059) &  5.338 & (0.045) &   5.243 & (0.044)  \\  
  &   32 &  2980000 &  10.500 & (0.112) &  9.993 & (0.104) &  5.613 & (0.049) &   5.505 & (0.047)  \\ 
  &   64 &  2500000 &  18.291 & (0.281) & 16.569 & (0.242) &  5.963 & (0.058) &   5.842 & (0.057)  \\ 
  &  128 &  1500000 &  47.581 & (1.526) & 37.061 & (1.049) &  6.279 & (0.082) &   6.129 & (0.079)  \\ 
  &  256 &   800000 &  80.627 & (4.637) & 47.039 & (2.066) &  6.139 & (0.109) &   5.999 & (0.105)  \\ 
\hline
 2.1750 &   16 &  3000000 &   6.979 & (0.060) &  6.805 & (0.058) &  5.363 & (0.046) &   5.272 & (0.044)  \\  
  &   32 &  3000000 &  10.461 & (0.111) &  9.958 & (0.103) &  5.692 & (0.050) &   5.575 & (0.048)  \\
\hline
\end{tabular}
\end{center}
\caption{
{\bf [Continued]}
}
\end{table} \clearpage \addtocounter{table}{-1}

\begin{table}[p]
\addtolength{\tabcolsep}{-1.3mm}
\begin{center}
\begin{tabular}{|r r|r|r r|r r|r r|r r|} \hline
\multicolumn{1}{|c}{$\beta$}  &
\multicolumn{1}{c|}{$L$} &
\multicolumn{1}{|c|}{Run Length} &
\multicolumn{2}{|c|}{$\tau_{int,{\cal M}_F^2}$} &
\multicolumn{2}{|c|}{$\tau_{int,{\cal M}_A^2}$} &
\multicolumn{2}{|c|}{$\tau_{int,{\cal E}_F}$} &
\multicolumn{2}{|c|}{$\tau_{int,{\cal E}_A}$} \\
%\multicolumn{20}{|c|}{$d=2$ $SU(3)$ model Table LAST}
 \hline
 2.5000 &   16 &  3000000 &   6.508 & (0.054) &  6.426 & (0.053) &  5.022 & (0.041) &   4.978 & (0.041)  \\  
  &   32 &  4940000 &   8.739 & (0.066) &  8.544 & (0.064) &  5.183 & (0.034) &   5.131 & (0.033)  \\ 
  &   64 &  1440000 &  12.243 & (0.203) & 11.831 & (0.193) &  5.117 & (0.061) &   5.065 & (0.061)  \\ 
  &  128 &   500000 &  19.447 & (0.700) & 18.306 & (0.640) &  5.220 & (0.109) &   5.160 & (0.107)  \\ 
 \hline
 2.5500 &    8 &  5000000 &   5.080 & (0.029) &  5.049 & (0.029) &  4.749 & (0.029) &   4.715 & (0.029)  \\  
  &   16 &  2980000 &   6.363 & (0.053) &  6.290 & (0.052) &  4.941 & (0.040) &   4.902 & (0.040)  \\ 
  &   32 &  2500000 &   8.607 & (0.091) &  8.442 & (0.088) &  4.987 & (0.045) &   4.948 & (0.044)  \\
\hline
\end{tabular}
\end{center}
\caption{
{\bf [Continued]}
}
\end{table} \clearpage \addtocounter{table}{-1}

\begin{table}[p]
\addtolength{\tabcolsep}{-1.3mm}
\begin{center}
\begin{tabular}{|r r|r|r r|r r|r r|r r|} \hline
\multicolumn{1}{|c}{$\beta$}  &
\multicolumn{1}{c|}{$L$} &
\multicolumn{1}{|c|}{Run Length} &
\multicolumn{2}{|c|}{$\tau_{int,{\cal M}_F^2}$} &
\multicolumn{2}{|c|}{$\tau_{int,{\cal M}_A^2}$} &
\multicolumn{2}{|c|}{$\tau_{int,{\cal E}_F}$} &
\multicolumn{2}{|c|}{$\tau_{int,{\cal E}_A}$} \\
%\multicolumn{20}{|c|}{$d=2$ $SU(3)$ model Table LAST}
\hline
 3.0750 &    8 &  5000000 &   4.925 & (0.028) &  4.909 & (0.028) &  4.552 & (0.028) &   4.535 & (0.027)  \\  
  &   16 &  3000000 &   5.955 & (0.048) &  5.915 & (0.047) &  4.677 & (0.037) &   4.659 & (0.037)  \\ 
  &   32 &  2000000 &   7.710 & (0.086) &  7.639 & (0.085) &  4.695 & (0.046) &   4.675 & (0.045)  \\ 
  &   64 &  1000000 &  10.245 & (0.187) & 10.067 & (0.183) &  4.822 & (0.068) &   4.797 & (0.067)  \\ 
  &  128 &   400000 &  13.333 & (0.447) & 13.173 & (0.439) &  4.819 & (0.109) &   4.807 & (0.108)  \\ 
 \hline
 3.1500 &    8 &  5000000 &   4.871 & (0.027) &  4.857 & (0.027) &  4.523 & (0.027) &   4.511 & (0.027)  \\  
  &   16 &  3000000 &   5.915 & (0.047) &  5.877 & (0.047) &  4.598 & (0.036) &   4.583 & (0.036)  \\
\hline
\end{tabular}
\end{center}
\caption{
{\bf [Continued]}
}
\end{table} \clearpage \addtocounter{table}{-1}

\begin{table}[p]
\addtolength{\tabcolsep}{-1.3mm}
\begin{center}
\begin{tabular}{|r r|r|r r|r r|r r|r r|} \hline
\multicolumn{1}{|c}{$\beta$}  &
\multicolumn{1}{c|}{$L$} &
\multicolumn{1}{|c|}{Run Length} &
\multicolumn{2}{|c|}{$\tau_{int,{\cal M}_F^2}$} &
\multicolumn{2}{|c|}{$\tau_{int,{\cal M}_A^2}$} &
\multicolumn{2}{|c|}{$\tau_{int,{\cal E}_F}$} &
\multicolumn{2}{|c|}{$\tau_{int,{\cal E}_A}$} \\
%\multicolumn{20}{|c|}{$d=2$ $SU(3)$ model Table LAST}
\hline
 3.7500 &    8 &  5000000 &   4.781 & (0.027) &  4.772 & (0.026) &  4.373 & (0.026) &   4.364 & (0.026)  \\
  &   16 &  3000000 &   5.717 & (0.045) &  5.703 & (0.045) &  4.534 & (0.035) &   4.526 & (0.035)  \\
  &   32 &  1980000 &   7.229 & (0.079) &  7.202 & (0.078) &  4.586 & (0.044) &   4.575 & (0.044)  \\
  &   64 &   980000 &   8.941 & (0.154) &  8.867 & (0.152) &  4.660 & (0.065) &   4.649 & (0.065)  \\
  &  128 &   400000 &  11.785 & (0.371) & 11.734 & (0.369) &  4.682 & (0.104) &   4.672 & (0.104)  \\
 \hline
 3.8250 &    8 &  5000000 &   4.777 & (0.026) &  4.768 & (0.026) &  4.398 & (0.026) &   4.391 & (0.026)  \\  
  &   16 &  3000000 &   5.677 & (0.044) &  5.656 & (0.044) &  4.504 & (0.035) &   4.492 & (0.035)  \\
\hline
\end{tabular}
\caption{
{\bf [Continued]}}
\end{center}
\end{table}
\clearpage

%%%%%%%%%%%%%%%%%%%%%%%%%%%%%%%%%%%%%%%%%%%%%%%%%%%%%%%%%%

%%%%%%%%%%%%%%%%%%%%%%%%%%%%%%%%%%%%%%%%%%%%%%%%%%%%%%%%%%

%%%%%%%%%%%%%%%%%%%%%%%%%%%%%%%%%%%%%%%%%%%%%%%%%%%%%%%%%%
%
% Beginning of Table 3 for SU3 article
% CPU timings
%
%
\protect\normalsize
\begin{table}
\addtolength{\tabcolsep}{-1.0mm}
\begin{center}
\begin{tabular}{|r|r|r|r|r|} \hline
\multicolumn{3}{|c|}{CPU timings (ms/measurement)} \\
\hline
     \multicolumn{1}{|c|}{$L$} &
     \multicolumn{1}{|c|}{Cray C-90} &
     \multicolumn{1}{c|}{IBM SP2}  \\
\hline \hline
  8 & 6 & 7  \\
 16 & 15 & 26   \\
 32 & 34 & 105 \\
 64 & 94 & 490 \\
128 & 270  & 2629 \\
256 & 911  & 15950 \\
\hline
\end{tabular}
\end{center}
\caption{
   CPU times in milliseconds per measurement for the $XY$-embedding
   MGMC algorithm for the two-dimensional $SU(3)$ chiral model.
   Each timing includes {\em two}\/ MGMC iterations
   (with $\gamma = 2$, $m_1 = 1$, $m_2 = 0$)
   followed by one measurement of all observables.
}
\label{cpu_timings}
\end{table}
\clearpage

%%%%%%%%%%%%%%%%%%%%%%%%%%%%%%%%%%%%%%%%%%%%%%%%%%%%%%%%%%

%%%%%%%%%%%%%%%%%%%%%%%%%%%%%%%%%%%%%%%%%%%%%%%%%%%%%%%%%%

%%%%%%%%%%%%%%%%%%%%%%%%%%%%%%%%%%%%%%%%%%%%%%%%%%%%%%%%%%
%
% BEGINNING OF STATIC FSS TABLES %%%%%%%%%%
%
% Beginning of Table 4 for SU3 article:
% Beginning of chi squared table for fitting curve of \xi_F

\protect\small
\begin{table}
%\addtolength{\tabcolsep}{-1.0mm}
\begin{center}
\begin{tabular}{|c||c|c|c|c|c|}
\hline
\multicolumn{6}{|c|}{ $\chi^2$ for the FSS fit of $\xi_{F}$}    \\
\hline
$x_{min}$  &  $n=11$  & $n=12$  & $n=13$  & $n=14$ & $n=15$  \\ \hline \hline
(0.50,0.40,0)                & \rm 180  718.80  & \rm 179  626.60  & \rm 178  560.20  & \rm 177  558.60  & \rm 176  558.30  \\
                             & \rm 3.99   0.0\% & \rm 3.50   0.0\% & \rm 3.15   0.0\% & \rm 3.16   0.0\% & \rm 3.17   0.0\% \\ \hline
($\infty$,0.40,0)            & \rm 154  673.80  & \rm 153  566.30  & \rm 152  533.00  & \rm 151  532.10  & \rm 150  531.80  \\
                             & \rm 4.38   0.0\% & \rm 3.70   0.0\% & \rm 3.51   0.0\% & \rm 3.52   0.0\% & \rm 3.55   0.0\% \\ \hline
($\infty$,$\infty$,0)        & \rm 108  236.00  & \rm 107  172.40  & \rm 106  154.80  & \rm 105  154.70  & \rm 104  153.40  \\
                             & \rm 2.19   0.0\% & \rm 1.61   0.0\% & \rm 1.46   0.1\% & \rm 1.47   0.1\% & \rm 1.48   0.1\% \\ \hline
(0.70,0.55,0.45)             & \rm 162  288.30  & \rm 161  219.20  & \rm 160  183.00  & \rm 159  182.50  & \rm 158  182.30  \\
                             & \rm 1.78   0.0\% & \rm 1.36   0.2\% & \rm 1.14  10.3\% & \rm 1.15   9.8\% & \rm 1.15   9.0\% \\ \hline
(0.75,0.60,0.50)             & \rm 150  222.40  & \rm 149  172.20  & \rm 148  129.90  & \rm 147  129.80  & \rm 146  129.80  \\
                             & \rm 1.48   0.0\% & \rm 1.16   9.4\% & \rm 0.88  85.6\% & \rm 0.88  84.3\% & \rm 0.89  82.9\% \\ \hline
(0.80,0.70,0.60)             & \rm 129  173.90  & \rm 128  135.00  & \sf 127   96.30  & \sf 126   96.28  & \sf 125   94.31  \\
                             & \rm 1.35   0.5\% & \rm 1.05  32.0\% & \sf 0.76  98.1\% & \sf 0.76  97.7\% & \sf 0.75  98.1\% \\ \hline
(0.95,0.85,0.60)             & \rm 111  150.30  & \rm 110  107.20  & \sf 109   77.62  & \sf 108   77.62  & \sf 107   75.67  \\
                             & \rm 1.35   0.8\% & \rm 0.97  55.8\% & \sf 0.71  99.0\% & \sf 0.72  98.8\% & \sf 0.71  99.1\% \\ \hline
(1.00,0.90,0.60)             & \rm 105  139.20  & \rm 104  100.90  & \sf 103   70.74  & \sf 102   70.73  & \sf 101   67.50  \\
                             & \rm 1.33   1.4\% & \rm 0.97  56.7\% & \sf 0.69  99.4\% & \sf 0.69  99.2\% & \sf 0.67  99.6\% \\ \hline
($\infty$,0.90,0.65)         & \rm  92  130.00  & \rm  91   77.01  & \it  90   60.85  & \sf  89   58.66  & \sf  88   58.31  \\
                             & \rm 1.41   0.6\% & \rm 0.85  85.2\% & \it 0.68  99.2\% & \sf 0.66  99.5\% & \sf 0.66  99.4\% \\ \hline
($\infty$,$\infty$,0.65)     & \rm  78   96.09  & \rm  77   56.51  & \sf  76   49.55  & \sf  75   46.63  & \sf  74   45.94  \\
                             & \rm 1.23   8.1\% & \rm 0.73  96.2\% & \sf 0.65  99.2\% & \sf 0.62  99.6\% & \sf 0.62  99.6\% \\ \hline
($\infty$,$\infty$,0.80)     & \rm 70     85.79 & \rm 69    51.89  & \sf 68    46.33  & \sf 67    43.42  & \sf 66    42.76  \\
                             & \rm 1.22   9.7\% & \rm 0.75  93.8\% & \sf 0.68  98.0\% & \sf 0.64  98.9\% & \sf 0.64  98.8\%  \\ \hline
($\infty$,$\infty$,$\infty$) & \rm  52   55.85  & \sf  51   25.23  & \sf  50   25.17  & \sf  49   24.11  & \sf  48   24.10  \\
                             & \rm 1.07  33.2\% & \sf 0.49  99.9\% & \sf 0.50  99.9\% & \sf 0.49  99.9\% & \sf 0.50  99.8\% \\ \hline
\end{tabular}
%\vspace*{1.5in}
\vspace*{0.5in}
%\doublespace
\caption{
  Degrees of freedom (DF), $\chi^2$, $\chi^2$/DF and confidence level
  for the $n^{th}$-order fit (\protect\ref{fss:gen}) of
  $\xi_F(\beta,2L)/\xi_F(\beta,L)$ versus $\xi_F(\beta,L)/L$.
  The indicated $x_{min}$ values apply to $L=8,16,32$, respectively;
  we always take $x_{min} = 0.14, 0$ for $L=64,128$.
  Our preferred fit is shown in {\em italics}\/;
   other good fits are shown in {\sf sans-serif};
   bad fits are shown in {\rm roman}.
}
\label{xiF_chisq_tab}
\end{center}
\end{table}

\clearpage

%%%%%%%%%%%%%%%%%%%%%%%%%%%%%%%%%%%%%%%%%%%%%%%%%%%%%%%%%%

%%%%%%%%%%%%%%%%%%%%%%%%%%%%%%%%%%%%%%%%%%%%%%%%%%%%%%%%%%

%%%%%%%%%%%%%%%%%%%%%%%%%%%%%%%%%%%%%%%%%%%%%%%%%%%%%%%%%%
% Beginning of Table 8 for SU3 article:
% Beginning of extrapolated \xi_F

\begin{table}
\addtolength{\tabcolsep}{-1.0mm}
\protect\footnotesize
\tabcolsep 2.5pt
\doublerulesep 1.5pt
\begin{center}  %%\hspace*{-4.0cm}
\begin{tabular}{|c||r@{\hspace{0.8mm}}r@{\hspace{0.3mm}}r|r@{\hspace{0.8mm}}r@{\hspace{0.3mm}}r|r@{\hspace{0.8mm}}r@{\hspace{0.3mm}}r|r@{\hspace{0.8mm}}r@{\hspace{0.3mm}}r|r@{\hspace{0.8mm}}r@{\hspace{0.3mm}}r|}
\hline
\multicolumn{1}{|c||}{$x_{min}$} & \multicolumn{3}{c|}{$\beta = 1.75$} & \multicolumn{3}{c|}{$\beta = 1.775$} & \multicolumn{3}{c|}{$\beta = 1.80$} & \multicolumn{3}{c|}{$\beta = 1.825$} & \multicolumn{3}{c|}{$\beta = 1.85$} \\ \hline
( 0.75,0.60,0.50) &    8.163 &    (0.076) & $    \times 10^{0}$  &    9.296 &    (0.031) & $    \times 10^{0}$  &    1.045 &    (0.002) & $    \times 10^{1}$  &    1.159 &    (0.003) & $    \times 10^{1}$  &    1.295 &    (0.003) & $    \times 10^{1}$  \\ \hline
(0.80,0.70,0.60) &    8.163 &    (0.077) & $    \times 10^{0}$  &    9.294 &    (0.030) & $    \times 10^{0}$  &    1.045 &    (0.002) & $    \times 10^{1}$  &    1.159 &    (0.003) & $    \times 10^{1}$  &    1.295 &    (0.003) & $    \times 10^{1}$  \\ \hline
(0.95,0.85,0.60) &    8.163 &    (0.077) & $    \times 10^{0}$  &    9.294 &    (0.030) & $    \times 10^{0}$  &    1.045 &    (0.002) & $    \times 10^{1}$  &    1.159 &    (0.003) & $    \times 10^{1}$  &    1.295 &    (0.003) & $    \times 10^{1}$  \\ \hline
(1.00,0.90,0.60) &    8.163 &    (0.076) & $    \times 10^{0}$  &    9.294 &    (0.030) & $    \times 10^{0}$  &    1.045 &    (0.002) & $    \times 10^{1}$  &    1.159 &    (0.003) & $    \times 10^{1}$  &    1.295 &    (0.003) & $    \times 10^{1}$  \\ \hline
\it ($\infty$,0.90,0.65) & \it 8.163 & \it (0.078) & $ \it \times 10^\hboxscript{0}$  & \it 9.291 & \it (0.030) & $ \it \times 10^\hboxscript{0}$  & \it 1.045 & \it (0.002) & $ \it \times 10^\hboxscript{1}$  & \it 1.158 & \it (0.003) & $ \it \times 10^\hboxscript{1}$  & \it 1.294 & \it (0.003) & $ \it \times 10^\hboxscript{1}$  \\ \hline
\sf ($\infty$,$\infty$,0.65) & \sf 8.163 & \sf (0.077) & $ \sf \times 10^\hboxsans{0}$  & \sf 9.292 & \sf (0.030) & $ \sf \times 10^\hboxsans{0}$  & \sf 1.045 & \sf (0.002) & $ \sf \times 10^\hboxsans{1}$  & \sf 1.158 & \sf (0.003) & $ \sf \times 10^\hboxsans{1}$  & \sf 1.294 & \sf (0.003) & $ \sf \times 10^\hboxsans{1}$  \\ \hline
\sf ($\infty$,$\infty$,0.80) & \sf 8.163 & \sf (0.078) & $ \sf \times 10^\hboxsans{0}$  & \sf 9.292 & \sf (0.030) & $ \sf \times 10^\hboxsans{0}$  & \sf 1.045 & \sf (0.002) & $ \sf \times 10^\hboxsans{1}$  & \sf 1.158 & \sf (0.003) & $ \sf \times 10^\hboxsans{1}$  & \sf 1.294 & \sf (0.003) & $ \sf \times 10^\hboxsans{1}$  \\ \hline
\sf ($\infty$,$\infty$,$\infty$) & \sf 8.163 & \sf (0.074) & $ \sf \times 10^\hboxsans{0}$  & \sf 9.296 & \sf (0.031) & $ \sf \times 10^\hboxsans{0}$  & \sf 1.045 & \sf (0.002) & $ \sf \times 10^\hboxsans{1}$  & \sf 1.159 & \sf (0.003) & $ \sf \times 10^\hboxsans{1}$  & \sf 1.295 & \sf (0.003) & $ \sf \times 10^\hboxsans{1}$  \\ \hline
\end{tabular} \end{center}
\par\vspace*{3mm}\par
\begin{center}  %%\hspace*{-4.0cm}
\begin{tabular}{|c||r@{\hspace{0.8mm}}r@{\hspace{0.3mm}}r|r@{\hspace{0.8mm}}r@{\hspace{0.3mm}}r|r@{\hspace{0.8mm}}r@{\hspace{0.3mm}}r|r@{\hspace{0.8mm}}r@{\hspace{0.3mm}}r|r@{\hspace{0.8mm}}r@{\hspace{0.3mm}}r|}
\hline
\multicolumn{1}{|c||}{$x_{min}$} & \multicolumn{3}{c|}{$\beta = 1.875$} & \multicolumn{3}{c|}{$\beta = 1.90$} &
\multicolumn{3}{c|}{$\beta = 1.925$} & \multicolumn{3}{c|}{$\beta = 1.95$} & \multicolumn{3}{c|}{$\beta = 1.975$} \\ \hline
( 0.75,0.60,0.50) &    1.444 &    (0.004) & $    \times 10^{1}$  &    1.613 &    (0.004) & $    \times 10^{1}$  &    1.804 &    (0.005) & $    \times 10^{1}$  &    2.003 &    (0.005) & $    \times 10^{1}$  &    2.230 &    (0.006) & $    \times 10^{1}$  \\ \hline
(0.80,0.70,0.60) &    1.444 &    (0.004) & $    \times 10^{1}$  &    1.614 &    (0.005) & $    \times 10^{1}$  &    1.804 &    (0.005) & $    \times 10^{1}$  &    2.003 &    (0.005) & $    \times 10^{1}$  &    2.229 &    (0.006) & $    \times 10^{1}$  \\ \hline
(0.95,0.85,0.60) &    1.444 &    (0.004) & $    \times 10^{1}$  &    1.614 &    (0.005) & $    \times 10^{1}$  &    1.804 &    (0.005) & $    \times 10^{1}$  &    2.003 &    (0.005) & $    \times 10^{1}$  &    2.229 &    (0.006) & $    \times 10^{1}$  \\ \hline
(1.00,0.90,0.60) &    1.444 &    (0.004) & $    \times 10^{1}$  &    1.614 &    (0.005) & $    \times 10^{1}$  &    1.804 &    (0.004) & $    \times 10^{1}$  &    2.003 &    (0.005) & $    \times 10^{1}$  &    2.229 &    (0.006) & $    \times 10^{1}$  \\ \hline
\it ($\infty$,0.90,0.65) & \it 1.444 & \it (0.004) & $ \it \times 10^\hboxscript{1}$  & \it 1.614 & \it (0.004) & $ \it \times 10^\hboxscript{1}$  & \it 1.804 & \it (0.005) & $ \it \times 10^\hboxscript{1}$  & \it 2.002 & \it (0.005) & $ \it \times 10^\hboxscript{1}$  & \it 2.227 & \it (0.007) & $ \it \times 10^\hboxscript{1}$  \\ \hline
\sf ($\infty$,$\infty$,0.65) & \sf 1.444 & \sf (0.004) & $ \sf \times 10^\hboxsans{1}$  & \sf 1.614 & \sf (0.004) & $ \sf \times 10^\hboxsans{1}$  & \sf 1.804 & \sf (0.005) & $ \sf \times 10^\hboxsans{1}$  & \sf 2.002 & \sf (0.005) & $ \sf \times 10^\hboxsans{1}$  & \sf 2.227 & \sf (0.007) & $ \sf \times 10^\hboxsans{1}$  \\ \hline
\sf ($\infty$,$\infty$,0.80) & \sf 1.444 & \sf (0.004) & $ \sf \times 10^\hboxsans{1}$  & \sf 1.614 & \sf (0.005) & $ \sf \times 10^\hboxsans{1}$  & \sf 1.804 & \sf (0.005) & $ \sf \times 10^\hboxsans{1}$  & \sf 2.002 & \sf (0.005) & $ \sf \times 10^\hboxsans{1}$  & \sf 2.227 & \sf (0.007) & $ \sf \times 10^\hboxsans{1}$  \\ \hline
\sf ($\infty$,$\infty$,$\infty$) & \sf 1.444 & \sf (0.004) & $ \sf \times 10^\hboxsans{1}$  & \sf 1.613 & \sf (0.004) & $ \sf \times 10^\hboxsans{1}$  & \sf 1.804 & \sf (0.005) & $ \sf \times 10^\hboxsans{1}$  & \sf 2.004 & \sf (0.005) & $ \sf \times 10^\hboxsans{1}$  & \sf 2.230 & \sf (0.007) & $ \sf \times 10^\hboxsans{1}$  \\ \hline
\end{tabular} \end{center}
\par\vspace*{3mm}\par
\begin{center}  %%\hspace*{-4.0cm}
\begin{tabular}{|c||r@{\hspace{0.8mm}}r@{\hspace{0.3mm}}r|r@{\hspace{0.8mm}}r@{\hspace{0.3mm}}r|r@{\hspace{0.8mm}}r@{\hspace{0.3mm}}r|r@{\hspace{0.8mm}}r@{\hspace{0.3mm}}r|r@{\hspace{0.8mm}}r@{\hspace{0.3mm}}r|}
\hline
\multicolumn{1}{|c||}{$x_{min}$} & \multicolumn{3}{c|}{$\beta = 1.985$} & \multicolumn{3}{c|}{$\beta = 2.00$} & \multicolumn{3}{c|}{$\beta = 2.012$} & \multicolumn{3}{c|}{$\beta = 2.025$} & \multicolumn{3}{c|}{$\beta = 2.037$} \\ \hline
( 0.75,0.60,0.50) &    2.326 &    (0.007) & $    \times 10^{1}$  &    2.489 &    (0.007) & $    \times 10^{1}$  &    2.613 &    (0.009) & $    \times 10^{1}$  &    2.762 &    (0.009) & $    \times 10^{1}$  &    2.889 &    (0.009) & $    \times 10^{1}$  \\ \hline
(0.80,0.70,0.60) &    2.325 &    (0.007) & $    \times 10^{1}$  &    2.487 &    (0.007) & $    \times 10^{1}$  &    2.612 &    (0.009) & $    \times 10^{1}$  &    2.761 &    (0.009) & $    \times 10^{1}$  &    2.888 &    (0.010) & $    \times 10^{1}$  \\ \hline
(0.95,0.85,0.60) &    2.324 &    (0.007) & $    \times 10^{1}$  &    2.487 &    (0.007) & $    \times 10^{1}$  &    2.612 &    (0.009) & $    \times 10^{1}$  &    2.761 &    (0.009) & $    \times 10^{1}$  &    2.888 &    (0.009) & $    \times 10^{1}$  \\ \hline
(1.00,0.90,0.60) &    2.324 &    (0.007) & $    \times 10^{1}$  &    2.487 &    (0.007) & $    \times 10^{1}$  &    2.612 &    (0.009) & $    \times 10^{1}$  &    2.761 &    (0.009) & $    \times 10^{1}$  &    2.888 &    (0.010) & $    \times 10^{1}$  \\ \hline
\it ($\infty$,0.90,0.65) & \it 2.322 & \it (0.007) & $ \it \times 10^\hboxscript{1}$  & \it 2.486 & \it (0.007) & $ \it \times 10^\hboxscript{1}$  & \it 2.610 & \it (0.009) & $ \it \times 10^\hboxscript{1}$  & \it 2.760 & \it (0.010) & $ \it \times 10^\hboxscript{1}$  & \it 2.889 & \it (0.010) & $ \it \times 10^\hboxscript{1}$  \\ \hline
\sf ($\infty$,$\infty$,0.65) & \sf 2.323 & \sf (0.007) & $ \sf \times 10^\hboxsans{1}$  & \sf 2.486 & \sf (0.007) & $ \sf \times 10^\hboxsans{1}$  & \sf 2.610 & \sf (0.009) & $ \sf \times 10^\hboxsans{1}$  & \sf 2.760 & \sf (0.009) & $ \sf \times 10^\hboxsans{1}$  & \sf 2.889 & \sf (0.010) & $ \sf \times 10^\hboxsans{1}$  \\ \hline
\sf ($\infty$,$\infty$,0.80) & \sf 2.323 & \sf (0.007) & $ \sf \times 10^\hboxsans{1}$  & \sf 2.486 & \sf (0.007) & $ \sf \times 10^\hboxsans{1}$  & \sf 2.611 & \sf (0.009) & $ \sf \times 10^\hboxsans{1}$  & \sf 2.760 & \sf (0.009) & $ \sf \times 10^\hboxsans{1}$  & \sf 2.889 & \sf (0.010) & $ \sf \times 10^\hboxsans{1}$  \\ \hline
\sf ($\infty$,$\infty$,$\infty$) & \sf 2.326 & \sf (0.007) & $ \sf \times 10^\hboxsans{1}$  & \sf 2.488 & \sf (0.007) & $ \sf \times 10^\hboxsans{1}$  & \sf 2.612 & \sf (0.009) & $ \sf \times 10^\hboxsans{1}$  & \sf 2.760 & \sf (0.009) & $ \sf \times 10^\hboxsans{1}$  & \sf 2.887 & \sf (0.009) & $ \sf \times 10^\hboxsans{1}$  \\ \hline
\end{tabular} \end{center}
\par\vspace*{3mm}\par
\begin{center}  %%\hspace*{-4.0cm}
\begin{tabular}{|c||r@{\hspace{0.8mm}}r@{\hspace{0.3mm}}r|r@{\hspace{0.8mm}}r@{\hspace{0.3mm}}r|r@{\hspace{0.8mm}}r@{\hspace{0.3mm}}r|r@{\hspace{0.8mm}}r@{\hspace{0.3mm}}r|r@{\hspace{0.8mm}}r@{\hspace{0.3mm}}r|}
\hline
\multicolumn{1}{|c||}{$x_{min}$} & \multicolumn{3}{c|}{$\beta = 2.05$} & \multicolumn{3}{c|}{$\beta = 2.062$} & \multicolumn{3}{c|}{$\beta = 2.075$} & \multicolumn{3}{c|}{$\beta = 2.10$} & \multicolumn{3}{c|}{$\beta = 2.112$} \\ \hline
( 0.75,0.60,0.50) &    3.077 &    (0.010) & $    \times 10^{1}$  &    3.229 &    (0.010) & $    \times 10^{1}$  &    3.423 &    (0.012) & $    \times 10^{1}$  &    3.775 &    (0.012) & $    \times 10^{1}$  &    3.986 &    (0.011) & $    \times 10^{1}$  \\ \hline
(0.80,0.70,0.60) &    3.077 &    (0.010) & $    \times 10^{1}$  &    3.231 &    (0.011) & $    \times 10^{1}$  &    3.425 &    (0.011) & $    \times 10^{1}$  &    3.777 &    (0.012) & $    \times 10^{1}$  &    3.979 &    (0.012) & $    \times 10^{1}$  \\ \hline
(0.95,0.85,0.60) &    3.078 &    (0.010) & $    \times 10^{1}$  &    3.231 &    (0.010) & $    \times 10^{1}$  &    3.425 &    (0.012) & $    \times 10^{1}$  &    3.778 &    (0.012) & $    \times 10^{1}$  &    3.979 &    (0.012) & $    \times 10^{1}$  \\ \hline
(1.00,0.90,0.60) &    3.078 &    (0.010) & $    \times 10^{1}$  &    3.232 &    (0.010) & $    \times 10^{1}$  &    3.425 &    (0.011) & $    \times 10^{1}$  &    3.778 &    (0.012) & $    \times 10^{1}$  &    3.979 &    (0.012) & $    \times 10^{1}$  \\ \hline
\it ($\infty$,0.90,0.65) & \it 3.079 & \it (0.010) & $ \it \times 10^\hboxscript{1}$  & \it 3.233 & \it (0.011) & $ \it \times 10^\hboxscript{1}$  & \it 3.428 & \it (0.012) & $ \it \times 10^\hboxscript{1}$  & \it 3.779 & \it (0.012) & $ \it \times 10^\hboxscript{1}$  & \it 3.979 & \it (0.012) & $ \it \times 10^\hboxscript{1}$  \\ \hline
\sf ($\infty$,$\infty$,0.65) & \sf 3.079 & \sf (0.010) & $ \sf \times 10^\hboxsans{1}$  & \sf 3.234 & \sf (0.011) & $ \sf \times 10^\hboxsans{1}$  & \sf 3.428 & \sf (0.012) & $ \sf \times 10^\hboxsans{1}$  & \sf 3.779 & \sf (0.012) & $ \sf \times 10^\hboxsans{1}$  & \sf 3.979 & \sf (0.012) & $ \sf \times 10^\hboxsans{1}$  \\ \hline
\sf ($\infty$,$\infty$,0.80) & \sf 3.079 & \sf (0.010) & $ \sf \times 10^\hboxsans{1}$  & \sf 3.233 & \sf (0.011) & $ \sf \times 10^\hboxsans{1}$  & \sf 3.427 & \sf (0.012) & $ \sf \times 10^\hboxsans{1}$  & \sf 3.778 & \sf (0.012) & $ \sf \times 10^\hboxsans{1}$  & \sf 3.979 & \sf (0.012) & $ \sf \times 10^\hboxsans{1}$  \\ \hline
\sf ($\infty$,$\infty$,$\infty$) & \sf 3.075 & \sf (0.010) & $ \sf \times 10^\hboxsans{1}$  & \sf 3.229 & \sf (0.011) & $ \sf \times 10^\hboxsans{1}$  & \sf 3.423 & \sf (0.012) & $ \sf \times 10^\hboxsans{1}$  & \sf 3.778 & \sf (0.012) & $ \sf \times 10^\hboxsans{1}$  & \sf 3.981 & \sf (0.012) & $ \sf \times 10^\hboxsans{1}$  \\ \hline
\end{tabular} \end{center}
\par\vspace*{3mm}\par
\caption{
  Estimated correlation lengths $\xi_{F,\infty}^{(2nd)}$ as a function of
  $\beta$, from various extrapolations.
  Error bar is one standard deviation (statistical errors only).
  All extrapolations use $s=2$ and $n=13$.
  The indicated $x_{min}$ values apply to $L=8,16,32$, respectively;
  we always take $x_{min} = 0.14, 0$ for $L=64,128$.
  Our preferred fit is shown in {\em italics}\/;
   other good fits are shown in {\sf sans-serif};
   bad fits are shown in {\rm roman}.
}
\label{estrap_xi_fund}

\end{table}
\clearpage \addtocounter{table}{-1}
\begin{table}
\addtolength{\tabcolsep}{-1.0mm}
\protect\footnotesize
\tabcolsep 2.5pt
\doublerulesep 1.5pt
\begin{center}  %%\hspace*{-4.0cm}

\begin{tabular}{|c||r@{\hspace{0.8mm}}r@{\hspace{0.3mm}}r|r@{\hspace{0.8mm}}r@{\hspace{0.3mm}}r|r@{\hspace{0.8mm}}r@{\hspace{0.3mm}}r|r@{\hspace{0.8mm}}r@{\hspace{0.3mm}}r|r@{\hspace{0.8mm}}r@{\hspace{0.3mm}}r|}
\hline
\multicolumn{1}{|c||}{$x_{min}$} & \multicolumn{3}{c|}{$\beta = 2.125$} & \multicolumn{3}{c|}{$\beta = 2.133$} & \multicolumn{3}{c|}{$\beta = 2.15$} & \multicolumn{3}{c|}{$\beta = 2.175$} & \multicolumn{3}{c|}{$\beta = 2.20$} \\ \hline
( 0.75,0.60,0.50) &    4.196 &    (0.010) & $    \times 10^{1}$  &    4.332 &    (0.011) & $    \times 10^{1}$  &    4.658 &    (0.012) & $    \times 10^{1}$  &    5.180 &    (0.017) & $    \times 10^{1}$  &    5.750 &    (0.018) & $    \times 10^{1}$  \\ \hline
(0.80,0.70,0.60) &    4.198 &    (0.011) & $    \times 10^{1}$  &    4.325 &    (0.013) & $    \times 10^{1}$  &    4.650 &    (0.013) & $    \times 10^{1}$  &    5.192 &    (0.019) & $    \times 10^{1}$  &    5.741 &    (0.022) & $    \times 10^{1}$  \\ \hline
(0.95,0.85,0.60) &    4.198 &    (0.011) & $    \times 10^{1}$  &    4.325 &    (0.013) & $    \times 10^{1}$  &    4.650 &    (0.013) & $    \times 10^{1}$  &    5.192 &    (0.019) & $    \times 10^{1}$  &    5.740 &    (0.021) & $    \times 10^{1}$  \\ \hline
(1.00,0.90,0.60) &    4.197 &    (0.011) & $    \times 10^{1}$  &    4.325 &    (0.013) & $    \times 10^{1}$  &    4.650 &    (0.013) & $    \times 10^{1}$  &    5.192 &    (0.019) & $    \times 10^{1}$  &    5.740 &    (0.021) & $    \times 10^{1}$  \\ \hline
\it ($\infty$,0.90,0.65) & \it 4.197 & \it (0.011) & $ \it \times 10^\hboxscript{1}$  & \it 4.323 & \it (0.014) & $ \it \times 10^\hboxscript{1}$  & \it 4.647 & \it (0.014) & $ \it \times 10^\hboxscript{1}$  & \it 5.189 & \it (0.020) & $ \it \times 10^\hboxscript{1}$  & \it 5.740 & \it (0.021) & $ \it \times 10^\hboxscript{1}$  \\
\hline
\end{tabular} \end{center}
\begin{center}  %%\hspace*{-4.0cm}
\par\vspace*{3mm}\par
\begin{tabular}{|c||r@{\hspace{0.8mm}}r@{\hspace{0.3mm}}r|r@{\hspace{0.8mm}}r@{\hspace{0.3mm}}r|r@{\hspace{0.8mm}}r@{\hspace{0.3mm}}r|r@{\hspace{0.8mm}}r@{\hspace{0.3mm}}r|r@{\hspace{0.8mm}}r@{\hspace{0.3mm}}r|}
\hline
\multicolumn{1}{|c||}{$x_{min}$} & \multicolumn{3}{c|}{$\beta = 2.2163$} & \multicolumn{3}{c|}{$\beta = 2.25$} & \multicolumn{3}{c|}{$\beta = 2.30$} & \multicolumn{3}{c|}{$\beta = 2.35$} & \multicolumn{3}{c|}{$\beta = 2.40$} \\ \hline
( 0.75,0.60,0.50) &    6.146 &    (0.018) & $    \times 10^{1}$  &    7.062 &    (0.022) & $    \times 10^{1}$  &    8.656 &    (0.022) & $    \times 10^{1}$  &    1.064 &    (0.004) & $    \times 10^{2}$  &    1.308 &    (0.004) & $    \times 10^{2}$  \\ \hline
(0.80,0.70,0.60) &    6.168 &    (0.020) & $    \times 10^{1}$  &    7.079 &    (0.027) & $    \times 10^{1}$  &    8.671 &    (0.028) & $    \times 10^{1}$  &    1.069 &    (0.005) & $    \times 10^{2}$  &    1.309 &    (0.006) & $    \times 10^{2}$  \\ \hline
(0.95,0.85,0.60) &    6.168 &    (0.020) & $    \times 10^{1}$  &    7.080 &    (0.027) & $    \times 10^{1}$  &    8.672 &    (0.028) & $    \times 10^{1}$  &    1.069 &    (0.005) & $    \times 10^{2}$  &    1.309 &    (0.006) & $    \times 10^{2}$  \\ \hline
(1.00,0.90,0.60) &    6.168 &    (0.021) & $    \times 10^{1}$  &    7.080 &    (0.027) & $    \times 10^{1}$  &    8.670 &    (0.028) & $    \times 10^{1}$  &    1.069 &    (0.005) & $    \times 10^{2}$  &    1.309 &    (0.006) & $    \times 10^{2}$  \\ \hline
\it ($\infty$,0.90,0.65) & \it 6.169 & \it (0.020) & $ \it \times 10^\hboxscript{1}$  & \it 7.082 & \it (0.027) & $ \it \times 10^\hboxscript{1}$  & \it 8.665 & \it (0.028) & $ \it \times 10^\hboxscript{1}$  & \it 1.068 & \it (0.006) & $ \it \times 10^\hboxscript{2}$  & \it 1.310 & \it (0.007) & $ \it \times 10^\hboxscript{2}$  \\
\hline
\end{tabular} \end{center}
\par\vspace*{3mm}\par
\begin{center}  %%\hspace*{-4.0cm}
\begin{tabular}{|c||r@{\hspace{0.8mm}}r@{\hspace{0.3mm}}r|r@{\hspace{0.8mm}}r@{\hspace{0.3mm}}r|r@{\hspace{0.8mm}}r@{\hspace{0.3mm}}r|r@{\hspace{0.8mm}}r@{\hspace{0.3mm}}r|r@{\hspace{0.8mm}}r@{\hspace{0.3mm}}r|}
\hline
\multicolumn{1}{|c||}{$x_{min}$} & \multicolumn{3}{c|}{$\beta = 2.45$} & \multicolumn{3}{c|}{$\beta = 2.50$} & \multicolumn{3}{c|}{$\beta = 2.55$} & \multicolumn{3}{c|}{$\beta = 2.60$} & \multicolumn{3}{c|}{$\beta = 2.65$} \\ \hline
( 0.75,0.60,0.50) &    1.594 &    (0.005) & $    \times 10^{2}$  &    1.950 &    (0.007) & $    \times 10^{2}$  &    2.406 &    (0.009) & $    \times 10^{2}$  &    2.930 &    (0.011) & $    \times 10^{2}$  &    3.588 &    (0.013) & $    \times 10^{2}$  \\ \hline
(0.80,0.70,0.60) &    1.596 &    (0.008) & $    \times 10^{2}$  &    1.947 &    (0.011) & $    \times 10^{2}$  &    2.409 &    (0.013) & $    \times 10^{2}$  &    2.936 &    (0.016) & $    \times 10^{2}$  &    3.589 &    (0.020) & $    \times 10^{2}$  \\ \hline
(0.95,0.85,0.60) &    1.596 &    (0.008) & $    \times 10^{2}$  &    1.946 &    (0.011) & $    \times 10^{2}$  &    2.411 &    (0.014) & $    \times 10^{2}$  &    2.937 &    (0.016) & $    \times 10^{2}$  &    3.587 &    (0.022) & $    \times 10^{2}$  \\ \hline
(1.00,0.90,0.60) &    1.596 &    (0.008) & $    \times 10^{2}$  &    1.946 &    (0.011) & $    \times 10^{2}$  &    2.412 &    (0.014) & $    \times 10^{2}$  &    2.936 &    (0.016) & $    \times 10^{2}$  &    3.587 &    (0.021) & $    \times 10^{2}$  \\ \hline
\it ($\infty$,0.90,0.65) & \it 1.594 & \it (0.009) & $ \it \times 10^\hboxscript{2}$  & \it 1.941 & \it (0.013) & $ \it \times 10^\hboxscript{2}$  & \it 2.422 & \it (0.018) & $ \it \times 10^\hboxscript{2}$  & \it 2.952 & \it (0.021) & $ \it \times 10^\hboxscript{2}$  & \it 3.601 & \it (0.026) & $ \it \times 10^\hboxscript{2}$  \\
\hline
\end{tabular} \end{center}
\par\vspace*{3mm}\par
\begin{center}  %%\hspace*{-4.0cm}
\begin{tabular}{|c||r@{\hspace{0.8mm}}r@{\hspace{0.3mm}}r|r@{\hspace{0.8mm}}r@{\hspace{0.3mm}}r|r@{\hspace{0.8mm}}r@{\hspace{0.3mm}}r|r@{\hspace{0.8mm}}r@{\hspace{0.3mm}}r|r@{\hspace{0.8mm}}r@{\hspace{0.3mm}}r|}
\hline
\multicolumn{1}{|c||}{$x_{min}$} & \multicolumn{3}{c|}{$\beta = 2.70$} & \multicolumn{3}{c|}{$\beta = 2.775$} & \multicolumn{3}{c|}{$\beta = 2.85$} & \multicolumn{3}{c|}{$\beta = 2.925$} & \multicolumn{3}{c|}{$\beta = 3.00$} \\ \hline
( 0.75,0.60,0.50) &    4.390 &    (0.018) & $    \times 10^{2}$  &    5.980 &    (0.023) & $    \times 10^{2}$  &    8.017 &    (0.034) & $    \times 10^{2}$  &    1.093 &    (0.005) & $    \times 10^{3}$  &    1.473 &    (0.006) & $    \times 10^{3}$  \\ \hline
(0.80,0.70,0.60) &    4.411 &    (0.026) & $    \times 10^{2}$  &    6.012 &    (0.034) & $    \times 10^{2}$  &    8.059 &    (0.050) & $    \times 10^{2}$  &    1.100 &    (0.007) & $    \times 10^{3}$  &    1.482 &    (0.009) & $    \times 10^{3}$  \\ \hline
(0.95,0.85,0.60) &    4.418 &    (0.029) & $    \times 10^{2}$  &    5.992 &    (0.038) & $    \times 10^{2}$  &    8.066 &    (0.061) & $    \times 10^{2}$  &    1.109 &    (0.008) & $    \times 10^{3}$  &    1.483 &    (0.012) & $    \times 10^{3}$  \\ \hline
(1.00,0.90,0.60) &    4.418 &    (0.029) & $    \times 10^{2}$  &    5.992 &    (0.038) & $    \times 10^{2}$  &    8.068 &    (0.059) & $    \times 10^{2}$  &    1.109 &    (0.008) & $    \times 10^{3}$  &    1.484 &    (0.012) & $    \times 10^{3}$  \\ \hline
\it ($\infty$,0.90,0.65) & \it 4.436 & \it (0.035) & $ \it \times 10^\hboxscript{2}$  & \it 6.021 & \it (0.045) & $ \it \times 10^\hboxscript{2}$  & \it 8.096 & \it (0.067) & $ \it \times 10^\hboxscript{2}$  & \it 1.115 & \it (0.009) & $ \it \times 10^\hboxscript{3}$  & \it 1.489 & \it (0.013) & $ \it \times 10^\hboxscript{3}$  \\
\hline
\end{tabular} \end{center}
\par\vspace*{3mm}\par
\caption{ {\bf [continued]}}
\end{table}
\clearpage \addtocounter{table}{-1}
\begin{table}
\addtolength{\tabcolsep}{-1.0mm}
\protect\footnotesize
\tabcolsep 2.5pt
\doublerulesep 1.5pt
\begin{center}  %%\hspace*{-4.0cm}
\begin{tabular}{|c||r@{\hspace{0.8mm}}r@{\hspace{0.3mm}}r|r@{\hspace{0.8mm}}r@{\hspace{0.3mm}}r|r@{\hspace{0.8mm}}r@{\hspace{0.3mm}}r|r@{\hspace{0.8mm}}r@{\hspace{0.3mm}}r|r@{\hspace{0.8mm}}r@{\hspace{0.3mm}}r|}
\hline
\multicolumn{1}{|c||}{$x_{min}$} & \multicolumn{3}{c|}{$\beta = 3.075$} & \multicolumn{3}{c|}{$\beta = 3.15$} & \multicolumn{3}{c|}{$\beta = 3.225$} & \multicolumn{3}{c|}{$\beta = 3.30$} & \multicolumn{3}{c|}{$\beta = 3.375$} \\ \hline
( 0.75,0.60,0.50) &    2.010 &    (0.009) & $    \times 10^{3}$  &    2.709 &    (0.012) & $    \times 10^{3}$  &    3.677 &    (0.018) & $    \times 10^{3}$  &    4.995 &    (0.024) & $    \times 10^{3}$  &    6.774 &    (0.036) & $    \times 10^{3}$  \\ \hline
(0.80,0.70,0.60) &    2.025 &    (0.012) & $    \times 10^{3}$  &    2.727 &    (0.017) & $    \times 10^{3}$  &    3.702 &    (0.024) & $    \times 10^{3}$  &    5.032 &    (0.032) & $    \times 10^{3}$  &    6.821 &    (0.048) & $    \times 10^{3}$  \\ \hline
(0.95,0.85,0.60) &    2.027 &    (0.018) & $    \times 10^{3}$  &    2.749 &    (0.025) & $    \times 10^{3}$  &    3.738 &    (0.037) & $    \times 10^{3}$  &    5.109 &    (0.049) & $    \times 10^{3}$  &    6.899 &    (0.072) & $    \times 10^{3}$  \\ \hline
(1.00,0.90,0.60) &    2.029 &    (0.018) & $    \times 10^{3}$  &    2.749 &    (0.025) & $    \times 10^{3}$  &    3.750 &    (0.037) & $    \times 10^{3}$  &    5.110 &    (0.051) & $    \times 10^{3}$  &    6.869 &    (0.073) & $    \times 10^{3}$  \\ \hline
\it ($\infty$,0.90,0.65) & \it 2.039 & \it (0.020) & $ \it \times 10^\hboxscript{3}$  & \it 2.761 & \it (0.027) & $ \it \times 10^\hboxscript{3}$  & \it 3.769 & \it (0.040) & $ \it \times 10^\hboxscript{3}$  & \it 5.132 & \it (0.055) & $ \it \times 10^\hboxscript{3}$  & \it 6.895 & \it (0.080) & $ \it \times 10^\hboxscript{3}$  \\
\hline
\end{tabular} \end{center}
\par\vspace*{3mm}\par
\begin{center}  %%\hspace*{-4.0cm}
\begin{tabular}{|c||r@{\hspace{0.8mm}}r@{\hspace{0.3mm}}r|r@{\hspace{0.8mm}}r@{\hspace{0.3mm}}r|r@{\hspace{0.8mm}}r@{\hspace{0.3mm}}r|r@{\hspace{0.8mm}}r@{\hspace{0.3mm}}r|r@{\hspace{0.8mm}}r@{\hspace{0.3mm}}r|}
\hline
\multicolumn{1}{|c||}{$x_{min}$} & \multicolumn{3}{c|}{$\beta = 3.45$} & \multicolumn{3}{c|}{$\beta = 3.525$} & \multicolumn{3}{c|}{$\beta = 3.60$} & \multicolumn{3}{c|}{$\beta = 3.675$} & \multicolumn{3}{c|}{$\beta = 3.75$} \\ \hline
( 0.75,0.60,0.50) &    9.199 &    (0.047) & $    \times 10^{3}$  &    1.240 &    (0.007) & $    \times 10^{4}$  &    1.694 &    (0.009) & $    \times 10^{4}$  &    2.295 &    (0.013) & $    \times 10^{4}$  &    3.126 &    (0.018) & $    \times 10^{4}$  \\ \hline
(0.80,0.70,0.60) &    9.268 &    (0.062) & $    \times 10^{3}$  &    1.248 &    (0.009) & $    \times 10^{4}$  &    1.707 &    (0.012) & $    \times 10^{4}$  &    2.311 &    (0.017) & $    \times 10^{4}$  &    3.149 &    (0.024) & $    \times 10^{4}$  \\ \hline
(0.95,0.85,0.60) &    9.411 &    (0.092) & $    \times 10^{3}$  &    1.268 &    (0.013) & $    \times 10^{4}$  &    1.736 &    (0.018) & $    \times 10^{4}$  &    2.352 &    (0.025) & $    \times 10^{4}$  &    3.206 &    (0.035) & $    \times 10^{4}$  \\ \hline
(1.00,0.90,0.60) &    9.365 &    (0.099) & $    \times 10^{3}$  &    1.257 &    (0.014) & $    \times 10^{4}$  &    1.730 &    (0.020) & $    \times 10^{4}$  &    2.337 &    (0.026) & $    \times 10^{4}$  &    3.182 &    (0.037) & $    \times 10^{4}$  \\ \hline
\it ($\infty$,0.90,0.65) & \it 9.407 & \it (0.105) & $ \it \times 10^\hboxscript{3}$  & \it 1.261 & \it (0.015) & $ \it \times 10^\hboxscript{4}$  & \it 1.738 & \it (0.021) & $ \it \times 10^\hboxscript{4}$  & \it 2.348 & \it (0.028) & $ \it \times 10^\hboxscript{4}$  & \it 3.191 & \it (0.040) & $ \it \times 10^\hboxscript{4}$  \\
\hline
\end{tabular} \end{center}
\begin{center}  %%\hspace*{-4.0cm}
\par\vspace*{3mm}\par
\begin{tabular}{|c||r@{\hspace{0.8mm}}r@{\hspace{0.3mm}}r|r@{\hspace{0.8mm}}r@{\hspace{0.3mm}}r|r@{\hspace{0.8mm}}r@{\hspace{0.3mm}}r|r@{\hspace{0.8mm}}r@{\hspace{0.3mm}}r|r@{\hspace{0.8mm}}r@{\hspace{0.3mm}}r|}
\hline
\multicolumn{1}{|c||}{$x_{min}$} & \multicolumn{3}{c|}{$\beta = 3.825$} & \multicolumn{3}{c|}{$\beta = 3.90$} & \multicolumn{3}{c|}{$\beta = 3.975$} & \multicolumn{3}{c|}{$\beta = 4.05$} & \multicolumn{3}{c|}{$\beta = 4.125$} \\ \hline
( 0.75,0.60,0.50) &    4.220 &    (0.024) & $    \times 10^{4}$  &    5.726 &    (0.037) & $    \times 10^{4}$  &    7.799 &    (0.049) & $    \times 10^{4}$  &    1.061 &    (0.007) & $    \times 10^{5}$  &    1.450 &    (0.010) & $    \times 10^{5}$  \\ \hline
(0.80,0.70,0.60) &    4.247 &    (0.031) & $    \times 10^{4}$  &    5.769 &    (0.046) & $    \times 10^{4}$  &    7.854 &    (0.060) & $    \times 10^{4}$  &    1.068 &    (0.009) & $    \times 10^{5}$  &    1.461 &    (0.012) & $    \times 10^{5}$  \\ \hline
(0.95,0.85,0.60) &    4.317 &    (0.045) & $    \times 10^{4}$  &    5.875 &    (0.065) & $    \times 10^{4}$  &    7.985 &    (0.084) & $    \times 10^{4}$  &    1.088 &    (0.012) & $    \times 10^{5}$  &    1.486 &    (0.016) & $    \times 10^{5}$  \\ \hline
(1.00,0.90,0.60) &    4.287 &    (0.049) & $    \times 10^{4}$  &    5.831 &    (0.070) & $    \times 10^{4}$  &    7.935 &    (0.092) & $    \times 10^{4}$  &    1.078 &    (0.013) & $    \times 10^{5}$  &    1.476 &    (0.017) & $    \times 10^{5}$  \\ \hline
\it ($\infty$,0.90,0.65) & \it 4.307 & \it (0.052) & $ \it \times 10^\hboxscript{4}$  & \it 5.859 & \it (0.077) & $ \it \times 10^\hboxscript{4}$  & \it 7.970 & \it (0.099) & $ \it \times 10^\hboxscript{4}$  & \it 1.090 & \it (0.015) & $ \it \times 10^\hboxscript{5}$  & \it 1.482 & \it (0.019) & $ \it \times 10^\hboxscript{5}$  \\
\hline
\end{tabular} \end{center}
\par\vspace*{3mm}\par
\begin{center}  %%\hspace*{-4.0cm}
\begin{tabular}{|c||r@{\hspace{0.8mm}}r@{\hspace{0.3mm}}r|r@{\hspace{0.8mm}}r@{\hspace{0.3mm}}r|r@{\hspace{0.8mm}}r@{\hspace{0.3mm}}r|r@{\hspace{0.8mm}}r@{\hspace{0.3mm}}r|r@{\hspace{0.8mm}}r@{\hspace{0.3mm}}r|}
\hline
\multicolumn{1}{|c||}{$x_{min}$} & \multicolumn{3}{c|}{$\beta = 4.20$} & \multicolumn{3}{c|}{$\beta = 4.275$} & \multicolumn{3}{c|}{$\beta = 4.35$} \\ \hline
( 0.75,0.60,0.50) &    1.979 &    (0.014) & $    \times 10^{5}$  &    2.683 &    (0.020) & $    \times 10^{5}$  &    3.639 &    (0.029) & $    \times 10^{5}$  \\ \hline
(0.80,0.70,0.60) &    1.991 &    (0.017) & $    \times 10^{5}$  &    2.704 &    (0.023) & $    \times 10^{5}$  &    3.662 &    (0.033) & $    \times 10^{5}$  \\ \hline
(0.95,0.85,0.60) &    2.029 &    (0.024) & $    \times 10^{5}$  &    2.753 &    (0.031) & $    \times 10^{5}$  &    3.726 &    (0.044) & $    \times 10^{5}$  \\ \hline
(1.00,0.90,0.60) &    2.010 &    (0.025) & $    \times 10^{5}$  &    2.734 &    (0.033) & $    \times 10^{5}$  &    3.696 &    (0.047) & $    \times 10^{5}$  \\ \hline
\it ($\infty$,0.90,0.65) & \it 2.018 & \it (0.028) & $ \it \times 10^\hboxscript{5}$  & \it 2.740 & \it (0.039) & $ \it \times 10^\hboxscript{5}$  & \it 3.717 & \it (0.055) & $ \it \times 10^\hboxscript{5}$  \\ \hline
\sf ($\infty$,$\infty$,0.65) & \sf 2.046 & \sf (0.038) & $ \sf \times 10^\hboxsans{5}$  & \sf 2.792 & \sf (0.052) & $ \sf \times 10^\hboxsans{5}$  & \sf 3.767 & \sf (0.077) & $ \sf \times 10^\hboxsans{5}$  \\ \hline
\sf ($\infty$,$\infty$,0.80) & \sf 2.057 & \sf (0.047) & $ \sf \times 10^\hboxsans{5}$  & \sf 2.800 & \sf (0.064) & $ \sf \times 10^\hboxsans{5}$  & \sf 3.785 & \sf (0.091) & $ \sf \times 10^\hboxsans{5}$  \\ \hline
\sf ($\infty$,$\infty$,$\infty$) & \sf 1.927 & \sf (0.074) & $ \sf \times 10^\hboxsans{5}$  & \sf 2.562 & \sf (0.102) & $ \sf \times 10^\hboxsans{5}$  & \sf 3.593 & \sf (0.148) & $ \sf \times 10^\hboxsans{5}$  \\ \hline
\end{tabular} \end{center}
\par\vspace*{3mm}\par
\caption{ {\bf [continued]}}
\end{table}
\clearpage

%%%%%%%%%%%%%%%%%%%%%%%%%%%%%%%%%%%%%%%%%%%%%%%%%%%%%%%%%%

%%%%%%%%%%%%%%%%%%%%%%%%%%%%%%%%%%%%%%%%%%%%%%%%%%%%%%%%%%

%%%%%%%%%%%%%%%%%%%%%%%%%%%%%%%%%%%%%%%%%%%%%%%%%%%%%%%%%%
%
% Beginning of Table 5 for SU3 article:
% Beginning of chi squared table for fitting curve of \chi_F

\protect\small
\begin{table}
\addtolength{\tabcolsep}{-1.0mm}
\begin{center}
\begin{tabular}{|c||c|c|c|c|c|} \hline
\multicolumn{6}{|c|}{ $\chi^2$ for the FSS fit of $\chi_{F}$}    \\
\hline
$L_{min}$ & $n=13$ & $n=14$ & $n=15$ & $n=16$ & $n=17$   \\ \hline \hline
($\infty$,$\infty$,0.4) & 100  175.80 &  99  175.60 &  98  173.50 &  97  173.40 &  96  171.80 \\ 
			& 1.76  0.0\% &  1.77   0.0\% & 1.77  0.0\% &  1.79   0.0\% &  1.79   0.0\% \\ \hline 
(1.0,0.95,0.65) &  97  503.80 &  96  495.40 &  95  474.20 &  94  474.00 &  93  473.70 \\ 
			& 5.19  0.0\% &  5.16   0.0\% & 4.99  0.0\% &  5.04   0.0\% &  5.09   0.0\% \\ \hline 
($\infty$,0.85,0.65) &  93  137.40 &  92  137.00 &  91  132.40 &  90  128.10 &  89  119.10 \\ 
			& 1.47  0.2\% &  1.49   0.2\% & 1.46  0.3\% &  1.42   0.5\% &  1.34   1.8\% \\ \hline 
($\infty$,0.95,0.65) &  87  124.00 &  86  123.60 &  85  119.00 &  84  114.70 &  83  104.50 \\ 
			& 1.42  0.6\% &  1.44   0.5\% & 1.40  0.9\% &  1.37   1.5\% &  1.26   5.5\% \\ \hline 
($\infty$,1.0,0.65) &  83  100.10 &  82  99.06 &  81  92.33 &  80  91.32 &  79  86.82 \\ 
			& 1.20  9.7\% &  1.21   9.7\% & 1.14  18.3\% &  1.14   18.2\% &  1.10   25.6\% \\ \hline 
($\infty$,$\infty$,0.65) &  76  82.99 &  75  80.29 &\sf  74  70.65 & \sf 73  68.96 & \sf 72  68.75 \\ 
			& 1.01  27.3\% &  1.07   31.7\% &\sf 0.95  58.9\% & \sf 0.94   61.2\% & \sf 0.95   58.7\% \\ \hline 
($\infty$,$\infty$,0.80) &   68  75.31 &   67  72.15 & \it  66  62.74 & \sf  65  59.63 & \sf  64  59.45 \\ 
		     &  1.10  25.4\% &   1.08   31.1\% & \it 0.95  59.1\% & \sf  0.92   66.5\% & \sf  0.93   63.8\% \\ \hline 
($\infty$,$\infty$,0.90) &  62  60.12 &  61  57.43 & \sf  60  50.74 & \sf  59  48.91 & \sf  58  48.66 \\ 
		    & 0.96  54.4\% &  0.94   60.6\% & \sf 0.85  79.7\% & \sf  0.83   82.3\% & \sf  0.84   80.4\% \\ \hline 
($\infty$,$\infty$,$\infty$) &  50  39.77 & \sf  49  35.96 & \sf  48  34.53 & \sf  47  32.95 & \sf  46  32.63 \\ 
		      & 0.79  85.0\% & \sf  0.73   91.7\% & \sf 0.72  92.8\% & \sf  0.70   94.0\% & \sf  0.71   93.1\% \\ \hline 
\end{tabular}
\end{center}
\caption{
  Degrees of freedom (DF), $\chi^2$, $\chi^2$/DF and confidence level
  for the $n^{th}$-order fit (\protect\ref{fss:gen}) of
  $ \chi_F (\beta,2L)/ \chi_F (\beta,L) $ versus $\xi_F(\beta,L)/L $.
  The indicated $x_{min}$ values apply to $L=8,16,32$, respectively;
  we always take $x_{min} = 0.14, 0$ for $L=64,128$.
  Our preferred fit is shown in {\em italics}\/;
   other good fits are shown in {\sf sans-serif};
   bad fits are shown in {\rm roman}.
}
\label{chiF_chisq_tab}
\end{table}
\clearpage

%%%%%%%%%%%%%%%%%%%%%%%%%%%%%%%%%%%%%%%%%%%%%%%%%%%%%%%%%%
%
%   TABLE OF EXTRAPOLATED DATA %%%%%%
%

%%%%%%%%%%%%%%%%%%%%%%%%%%%%%%%%%%%%%%%%%%%%%%%%%%%%%%%%%%

%%%%%%%%%%%%%%%%%%%%%%%%%%%%%%%%%%%%%%%%%%%%%%%%%%%%%%%%%%

%%%%%%%%%%%%%%%%%%%%%%%%%%%%%%%%%%%%%%%%%%%%%%%%%%%%%%%%%%
%
%   TABLE OF EXTRAPOLATED DATA %%%%%%
%
% Beginning of Table 9 for SU3 article:
% Beginning of extrapolated \chi_F

\begin{table}
\addtolength{\tabcolsep}{-1.0mm}
\protect\footnotesize
\tabcolsep 2.5pt
\doublerulesep 1.5pt
\begin{center}  %%\hspace*{-4.0cm}
\begin{tabular}{|c||r@{\hspace{0.8mm}}r@{\hspace{0.3mm}}r|r@{\hspace{0.8mm}}r@{\hspace{0.3mm}}r|r@{\hspace{0.8mm}}r@{\hspace{0.3mm}}r|r@{\hspace{0.8mm}}r@{\hspace{0.3mm}}r|r@{\hspace{0.8mm}}r@{\hspace{0.3mm}}r|}
\hline
\multicolumn{1}{|c||}{$x_{min}$} & \multicolumn{3}{c|}{ $\beta = 1.75$} & \multicolumn{3}{c|}{ $\beta = 1.775$} & \multicolumn{3}{c|}{ $\beta = 1.80$} & \multicolumn{3}{c|}{ $\beta = 1.825$} & \multicolumn{3}{c|}{ $\beta = 1.85$} \\ \hline
   ($\infty$,0.90,0.65) &    2.982 &    (0.008) & $    \times 10^{2}$  &    3.593 &    (0.010) & $    \times 10^{2}$  &    4.339 &    (0.006) & $    \times 10^{2}$  &    5.196 &    (0.008) & $    \times 10^{2}$  &    6.254 &    (0.010) & $    \times 10^{2}$  \\ \hline
($\infty$,1.0,0.65) &    2.982 &    (0.008) & $    \times 10^{2}$  &    3.594 &    (0.010) & $    \times 10^{2}$  &    4.340 &    (0.006) & $    \times 10^{2}$  &    5.197 &    (0.008) & $    \times 10^{2}$  &    6.255 &    (0.010) & $    \times 10^{2}$  \\ \hline
($\infty$,$\infty$,0.65) &    2.982 &    (0.008) & $    \times 10^{2}$  &    3.595 &    (0.011) & $    \times 10^{2}$  &    4.340 &    (0.006) & $    \times 10^{2}$  &    5.198 &    (0.008) & $    \times 10^{2}$  &    6.254 &    (0.010) & $    \times 10^{2}$  \\ \hline
\it ($\infty$,$\infty$,0.80) & \it 2.982 & \it (0.008) & $ \it \times 10^\hboxscript{2}$  & \it 3.595 & \it (0.010) & $ \it \times 10^\hboxscript{2}$  & \it 4.341 & \it (0.006) & $ \it \times 10^\hboxscript{2}$  & \it 5.199 & \it (0.008) & $ \it \times 10^\hboxscript{2}$  & \it 6.254 & \it (0.010) & $ \it \times 10^\hboxscript{2}$  \\ \hline
\sf ($\infty$,$\infty$,0.90) & \sf 2.982 & \sf (0.008) & $ \sf \times 10^\hboxsans{2}$  & \sf 3.595 & \sf (0.010) & $ \sf \times 10^\hboxsans{2}$  & \sf 4.340 & \sf (0.006) & $ \sf \times 10^\hboxsans{2}$  & \sf 5.198 & \sf (0.008) & $ \sf \times 10^\hboxsans{2}$  & \sf 6.254 & \sf (0.010) & $ \sf \times 10^\hboxsans{2}$  \\ \hline
\sf ($\infty$,$\infty$,$\infty$) & \sf 2.982 & \sf (0.008) & $ \sf \times 10^\hboxsans{2}$  & \sf 3.595 & \sf (0.010) & $ \sf \times 10^\hboxsans{2}$  & \sf 4.340 & \sf (0.006) & $ \sf \times 10^\hboxsans{2}$  & \sf 5.198 & \sf (0.008) & $ \sf \times 10^\hboxsans{2}$  & \sf 6.254 & \sf (0.010) & $ \sf \times 10^\hboxsans{2}$  \\ \hline
\end{tabular} \end{center}
\par\vspace*{3mm}\par
\begin{center}  %%\hspace*{-4.0cm}
\begin{tabular}{|c||r@{\hspace{0.8mm}}r@{\hspace{0.3mm}}r|r@{\hspace{0.8mm}}r@{\hspace{0.3mm}}r|r@{\hspace{0.8mm}}r@{\hspace{0.3mm}}r|r@{\hspace{0.8mm}}r@{\hspace{0.3mm}}r|r@{\hspace{0.8mm}}r@{\hspace{0.3mm}}r|}
\hline
\multicolumn{1}{|c||}{$x_{min}$} & \multicolumn{3}{c|}{ $\beta = 1.875$} & \multicolumn{3}{c|}{ $\beta = 1.90$} & \multicolumn{3}{c|}{ $\beta = 1.925$} & \multicolumn{3}{c|}{ $\beta = 1.95$} & \multicolumn{3}{c|}{ $\beta = 1.975$} \\ \hline
   ($\infty$,0.90,0.65) &    7.531 &    (0.015) & $    \times 10^{2}$  &    9.076 &    (0.021) & $    \times 10^{2}$  &    1.095 &    (0.002) & $    \times 10^{3}$  &    1.314 &    (0.004) & $    \times 10^{3}$  &    1.579 &    (0.005) & $    \times 10^{3}$  \\ \hline
($\infty$,1.0,0.65) &    7.530 &    (0.016) & $    \times 10^{2}$  &    9.075 &    (0.022) & $    \times 10^{2}$  &    1.095 &    (0.002) & $    \times 10^{3}$  &    1.315 &    (0.004) & $    \times 10^{3}$  &    1.579 &    (0.005) & $    \times 10^{3}$  \\ \hline
($\infty$,$\infty$,0.65) &    7.527 &    (0.016) & $    \times 10^{2}$  &    9.073 &    (0.022) & $    \times 10^{2}$  &    1.095 &    (0.002) & $    \times 10^{3}$  &    1.316 &    (0.004) & $    \times 10^{3}$  &    1.581 &    (0.005) & $    \times 10^{3}$  \\ \hline
\it ($\infty$,$\infty$,0.80) & \it 7.526 & \it (0.015) & $ \it \times 10^\hboxscript{2}$  & \it 9.073 & \it (0.022) & $ \it \times 10^\hboxscript{2}$  & \it 1.096 & \it (0.002) & $ \it \times 10^\hboxscript{3}$  & \it 1.317 & \it (0.004) & $ \it \times 10^\hboxscript{3}$  & \it 1.581 & \it (0.005) & $ \it \times 10^\hboxscript{3}$  \\ \hline
\sf ($\infty$,$\infty$,0.90) & \sf 7.527 & \sf (0.016) & $ \sf \times 10^\hboxsans{2}$  & \sf 9.073 & \sf (0.022) & $ \sf \times 10^\hboxsans{2}$  & \sf 1.095 & \sf (0.002) & $ \sf \times 10^\hboxsans{3}$  & \sf 1.316 & \sf (0.004) & $ \sf \times 10^\hboxsans{3}$  & \sf 1.581 & \sf (0.005) & $ \sf \times 10^\hboxsans{3}$  \\ \hline
\sf ($\infty$,$\infty$,$\infty$) & \sf 7.528 & \sf (0.016) & $ \sf \times 10^\hboxsans{2}$  & \sf 9.074 & \sf (0.022) & $ \sf \times 10^\hboxsans{2}$  & \sf 1.095 & \sf (0.002) & $ \sf \times 10^\hboxsans{3}$  & \sf 1.316 & \sf (0.004) & $ \sf \times 10^\hboxsans{3}$  & \sf 1.580 & \sf (0.005) & $ \sf \times 10^\hboxsans{3}$  \\ \hline
\end{tabular} \end{center}
\par\vspace*{3mm}\par
\begin{center}  %%\hspace*{-4.0cm}
\begin{tabular}{|c||r@{\hspace{0.8mm}}r@{\hspace{0.3mm}}r|r@{\hspace{0.8mm}}r@{\hspace{0.3mm}}r|r@{\hspace{0.8mm}}r@{\hspace{0.3mm}}r|r@{\hspace{0.8mm}}r@{\hspace{0.3mm}}r|r@{\hspace{0.8mm}}r@{\hspace{0.3mm}}r|}
\hline
\multicolumn{1}{|c||}{$x_{min}$} & \multicolumn{3}{c|}{ $\beta = 1.985$} & \multicolumn{3}{c|}{ $\beta = 2.00$} & \multicolumn{3}{c|}{ $\beta = 2.012$} & \multicolumn{3}{c|}{ $\beta = 2.025$} & \multicolumn{3}{c|}{ $\beta = 2.037$} \\ \hline
   ($\infty$,0.90,0.65) &    1.696 &    (0.005) & $    \times 10^{3}$  &    1.906 &    (0.004) & $    \times 10^{3}$  &    2.077 &    (0.006) & $    \times 10^{3}$  &    2.286 &    (0.007) & $    \times 10^{3}$  &    2.478 &    (0.008) & $    \times 10^{3}$  \\ \hline
($\infty$,1.0,0.65) &    1.696 &    (0.005) & $    \times 10^{3}$  &    1.905 &    (0.004) & $    \times 10^{3}$  &    2.077 &    (0.006) & $    \times 10^{3}$  &    2.285 &    (0.007) & $    \times 10^{3}$  &    2.476 &    (0.008) & $    \times 10^{3}$  \\ \hline
($\infty$,$\infty$,0.65) &    1.698 &    (0.005) & $    \times 10^{3}$  &    1.905 &    (0.004) & $    \times 10^{3}$  &    2.075 &    (0.006) & $    \times 10^{3}$  &    2.283 &    (0.007) & $    \times 10^{3}$  &    2.473 &    (0.008) & $    \times 10^{3}$  \\ \hline
\it ($\infty$,$\infty$,0.80) & \it 1.698 & \it (0.005) & $ \it \times 10^\hboxscript{3}$  & \it 1.906 & \it (0.004) & $ \it \times 10^\hboxscript{3}$  & \it 2.076 & \it (0.006) & $ \it \times 10^\hboxscript{3}$  & \it 2.283 & \it (0.007) & $ \it \times 10^\hboxscript{3}$  & \it 2.472 & \it (0.008) & $ \it \times 10^\hboxscript{3}$  \\ \hline
\sf ($\infty$,$\infty$,0.90) & \sf 1.698 & \sf (0.005) & $ \sf \times 10^\hboxsans{3}$  & \sf 1.906 & \sf (0.004) & $ \sf \times 10^\hboxsans{3}$  & \sf 2.076 & \sf (0.006) & $ \sf \times 10^\hboxsans{3}$  & \sf 2.283 & \sf (0.007) & $ \sf \times 10^\hboxsans{3}$  & \sf 2.474 & \sf (0.008) & $ \sf \times 10^\hboxsans{3}$  \\ \hline
\sf ($\infty$,$\infty$,$\infty$) & \sf 1.697 & \sf (0.005) & $ \sf \times 10^\hboxsans{3}$  & \sf 1.906 & \sf (0.004) & $ \sf \times 10^\hboxsans{3}$  & \sf 2.076 & \sf (0.006) & $ \sf \times 10^\hboxsans{3}$  & \sf 2.284 & \sf (0.008) & $ \sf \times 10^\hboxsans{3}$  & \sf 2.474 & \sf (0.008) & $ \sf \times 10^\hboxsans{3}$  \\ \hline
\end{tabular} \end{center}
\par\vspace*{3mm}\par
\begin{center}  %%\hspace*{-4.0cm}
\begin{tabular}{|c||r@{\hspace{0.8mm}}r@{\hspace{0.3mm}}r|r@{\hspace{0.8mm}}r@{\hspace{0.3mm}}r|r@{\hspace{0.8mm}}r@{\hspace{0.3mm}}r|r@{\hspace{0.8mm}}r@{\hspace{0.3mm}}r|r@{\hspace{0.8mm}}r@{\hspace{0.3mm}}r|}
\hline
\multicolumn{1}{|c||}{$x_{min}$} & \multicolumn{3}{c|}{ $\beta = 2.05$} & \multicolumn{3}{c|}{ $\beta = 2.062$} & \multicolumn{3}{c|}{ $\beta = 2.075$} & \multicolumn{3}{c|}{ $\beta = 2.10$} & \multicolumn{3}{c|}{ $\beta = 2.112$} \\ \hline
   ($\infty$,0.90,0.65) &    2.754 &    (0.010) & $    \times 10^{3}$  &    3.0 &    (0.010) & $    \times 10^{3}$  &    3.313 &    (0.011) & $    \times 10^{3}$  &    3.941 &    (0.013) & $    \times 10^{3}$  &    4.317 &    (0.015) & $    \times 10^{3}$  \\ \hline
($\infty$,1.0,0.65) &    2.752 &    (0.010) & $    \times 10^{3}$  &    2.999 &    (0.010) & $    \times 10^{3}$  &    3.313 &    (0.011) & $    \times 10^{3}$  &    3.944 &    (0.013) & $    \times 10^{3}$  &    4.321 &    (0.014) & $    \times 10^{3}$  \\ \hline
($\infty$,$\infty$,0.65) &    2.749 &    (0.010) & $    \times 10^{3}$  &    2.997 &    (0.011) & $    \times 10^{3}$  &    3.314 &    (0.011) & $    \times 10^{3}$  &    3.948 &    (0.013) & $    \times 10^{3}$  &    4.328 &    (0.015) & $    \times 10^{3}$  \\ \hline
\it ($\infty$,$\infty$,0.80) & \it 2.749 & \it (0.010) & $ \it \times 10^\hboxscript{3}$  & \it 2.997 & \it (0.011) & $ \it \times 10^\hboxscript{3}$  & \it 3.314 & \it (0.012) & $ \it \times 10^\hboxscript{3}$  & \it 3.949 & \it (0.014) & $ \it \times 10^\hboxscript{3}$  & \it 4.329 & \it (0.015) & $ \it \times 10^\hboxscript{3}$  \\ \hline
\sf ($\infty$,$\infty$,0.90) & \sf 2.749 & \sf (0.010) & $ \sf \times 10^\hboxsans{3}$  & \sf 2.997 & \sf (0.010) & $ \sf \times 10^\hboxsans{3}$  & \sf 3.314 & \sf (0.011) & $ \sf \times 10^\hboxsans{3}$  & \sf 3.948 & \sf (0.014) & $ \sf \times 10^\hboxsans{3}$  & \sf 4.327 & \sf (0.015) & $ \sf \times 10^\hboxsans{3}$  \\ \hline
\sf ($\infty$,$\infty$,$\infty$) & \sf 2.750 & \sf (0.010) & $ \sf \times 10^\hboxsans{3}$  & \sf 2.998 & \sf (0.010) & $ \sf \times 10^\hboxsans{3}$  & \sf 3.313 & \sf (0.011) & $ \sf \times 10^\hboxsans{3}$  & \sf 3.946 & \sf (0.014) & $ \sf \times 10^\hboxsans{3}$  & \sf 4.324 & \sf (0.016) & $ \sf \times 10^\hboxsans{3}$  \\ \hline
\end{tabular} \end{center}
\par\vspace*{3mm}\par
\caption{
  Estimated susceptibilities $\chi_{F,\infty}$ as a function of $\beta$,
  from various extrapolations.
  Error bar is one standard deviation (statistical errors only).
  All extrapolations use $s=2$ and $n=15$.
  The indicated $x_{min}$ values apply to $L=8,16,32$, respectively;
  we always take $x_{min} = 0.14, 0$ for $L=64,128$.
  Our preferred fit is shown in {\em italic}\/;
   other good fits are shown in {\sf sans-serif};
   bad fits are shown in {\rm roman}.
}
\label{estrap_chi_fund}

\end{table}
\clearpage \addtocounter{table}{-1}
\begin{table}
\addtolength{\tabcolsep}{-1.0mm}
\protect\footnotesize
\tabcolsep 2.5pt
\doublerulesep 1.5pt
\begin{center}  %%\hspace*{-4.0cm}
\begin{tabular}{|c||r@{\hspace{0.8mm}}r@{\hspace{0.3mm}}r|r@{\hspace{0.8mm}}r@{\hspace{0.3mm}}r|r@{\hspace{0.8mm}}r@{\hspace{0.3mm}}r|r@{\hspace{0.8mm}}r@{\hspace{0.3mm}}r|r@{\hspace{0.8mm}}r@{\hspace{0.3mm}}r|}
\hline
\multicolumn{1}{|c||}{$x_{min}$} & \multicolumn{3}{c|}{ $\beta = 2.125$} & \multicolumn{3}{c|}{ $\beta = 2.133$} & \multicolumn{3}{c|}{ $\beta = 2.15$} & \multicolumn{3}{c|}{ $\beta = 2.175$} & \multicolumn{3}{c|}{ $\beta = 2.20$} \\ \hline
   ($\infty$,0.90,0.65) &    4.742 &    (0.016) & $    \times 10^{3}$  &    5.004 &    (0.018) & $    \times 10^{3}$  &    5.686 &    (0.017) & $    \times 10^{3}$  &    6.886 &    (0.026) & $    \times 10^{3}$  &    8.225 &    (0.033) & $    \times 10^{3}$  \\ \hline
($\infty$,1.0,0.65) &    4.746 &    (0.016) & $    \times 10^{3}$  &    5.008 &    (0.019) & $    \times 10^{3}$  &    5.689 &    (0.017) & $    \times 10^{3}$  &    6.884 &    (0.026) & $    \times 10^{3}$  &    8.220 &    (0.034) & $    \times 10^{3}$  \\ \hline
($\infty$,$\infty$,0.65) &    4.754 &    (0.016) & $    \times 10^{3}$  &    5.015 &    (0.019) & $    \times 10^{3}$  &    5.694 &    (0.017) & $    \times 10^{3}$  &    6.881 &    (0.026) & $    \times 10^{3}$  &    8.209 &    (0.035) & $    \times 10^{3}$  \\ \hline
\it ($\infty$,$\infty$,0.80) & \it 4.755 & \it (0.017) & $ \it \times 10^\hboxscript{3}$  & \it 5.017 & \it (0.019) & $ \it \times 10^\hboxscript{3}$  & \it 5.694 & \it (0.018) & $ \it \times 10^\hboxscript{3}$  & \it 6.880 & \it (0.026) & $ \it \times 10^\hboxscript{3}$  & \it 8.207 & \it (0.034) & $ \it \times 10^\hboxscript{3}$  \\ \hline
\sf ($\infty$,$\infty$,0.90) & \sf 4.753 & \sf (0.016) & $ \sf \times 10^\hboxsans{3}$  & \sf 5.014 & \sf (0.019) & $ \sf \times 10^\hboxsans{3}$  & \sf 5.693 & \sf (0.017) & $ \sf \times 10^\hboxsans{3}$  & \sf 6.882 & \sf (0.026) & $ \sf \times 10^\hboxsans{3}$  & \sf 8.210 & \sf (0.034) & $ \sf \times 10^\hboxsans{3}$  \\ \hline
\sf ($\infty$,$\infty$,$\infty$) & \sf 4.750 & \sf (0.017) & $ \sf \times 10^\hboxsans{3}$  & \sf 5.012 & \sf (0.019) & $ \sf \times 10^\hboxsans{3}$  & \sf 5.692 & \sf (0.018) & $ \sf \times 10^\hboxsans{3}$  & \sf 6.884 & \sf (0.026) & $ \sf \times 10^\hboxsans{3}$  & \sf 8.216 & \sf (0.036) & $ \sf \times 10^\hboxsans{3}$  \\ \hline
\end{tabular} \end{center}
\par\vspace*{3mm}\par
\begin{center}  %%\hspace*{-4.0cm}
\begin{tabular}{|c||r@{\hspace{0.8mm}}r@{\hspace{0.3mm}}r|r@{\hspace{0.8mm}}r@{\hspace{0.3mm}}r|r@{\hspace{0.8mm}}r@{\hspace{0.3mm}}r|r@{\hspace{0.8mm}}r@{\hspace{0.3mm}}r|r@{\hspace{0.8mm}}r@{\hspace{0.3mm}}r|}
\hline
\multicolumn{1}{|c||}{$x_{min}$} & \multicolumn{3}{c|}{ $\beta = 2.2163$} & \multicolumn{3}{c|}{ $\beta = 2.25$} & \multicolumn{3}{c|}{ $\beta = 2.30$} & \multicolumn{3}{c|}{ $\beta = 2.35$} & \multicolumn{3}{c|}{ $\beta = 2.40$} \\ \hline
   ($\infty$,0.90,0.65) &    9.316 &    (0.033) & $    \times 10^{3}$  &    1.188 &    (0.005) & $    \times 10^{4}$  &    1.708 &    (0.006) & $    \times 10^{4}$  &    2.479 &    (0.016) & $    \times 10^{4}$  &    3.555 &    (0.024) & $    \times 10^{4}$  \\ \hline
($\infty$,1.0,0.65) &    9.310 &    (0.035) & $    \times 10^{3}$  &    1.188 &    (0.005) & $    \times 10^{4}$  &    1.709 &    (0.007) & $    \times 10^{4}$  &    2.478 &    (0.015) & $    \times 10^{4}$  &    3.554 &    (0.024) & $    \times 10^{4}$  \\ \hline
($\infty$,$\infty$,0.65) &    9.300 &    (0.034) & $    \times 10^{3}$  &    1.189 &    (0.005) & $    \times 10^{4}$  &    1.712 &    (0.007) & $    \times 10^{4}$  &    2.476 &    (0.015) & $    \times 10^{4}$  &    3.551 &    (0.024) & $    \times 10^{4}$  \\ \hline
\it ($\infty$,$\infty$,0.80) & \it 9.298 & \it (0.035) & $ \it \times 10^\hboxscript{3}$  & \it 1.189 & \it (0.005) & $ \it \times 10^\hboxscript{4}$  & \it 1.713 & \it (0.007) & $ \it \times 10^\hboxscript{4}$  & \it 2.476 & \it (0.015) & $ \it \times 10^\hboxscript{4}$  & \it 3.551 & \it (0.024) & $ \it \times 10^\hboxscript{4}$  \\ \hline
\sf ($\infty$,$\infty$,0.90) & \sf 9.302 & \sf (0.034) & $ \sf \times 10^\hboxsans{3}$  & \sf 1.189 & \sf (0.005) & $ \sf \times 10^\hboxsans{4}$  & \sf 1.712 & \sf (0.007) & $ \sf \times 10^\hboxsans{4}$  & \sf 2.476 & \sf (0.015) & $ \sf \times 10^\hboxsans{4}$  & \sf 3.551 & \sf (0.024) & $ \sf \times 10^\hboxsans{4}$  \\ \hline
\sf ($\infty$,$\infty$,$\infty$) & \sf 9.306 & \sf (0.035) & $ \sf \times 10^\hboxsans{3}$  & \sf 1.189 & \sf (0.005) & $ \sf \times 10^\hboxsans{4}$  & \sf 1.711 & \sf (0.007) & $ \sf \times 10^\hboxsans{4}$  & \sf 2.476 & \sf (0.015) & $ \sf \times 10^\hboxsans{4}$  & \sf 3.550 & \sf (0.024) & $ \sf \times 10^\hboxsans{4}$  \\ \hline
\end{tabular} \end{center}
\par\vspace*{3mm}\par
\doublerulesep 1.5pt
\begin{center}  %%\hspace*{-4.0cm}
\begin{tabular}{|c||r@{\hspace{0.8mm}}r@{\hspace{0.3mm}}r|r@{\hspace{0.8mm}}r@{\hspace{0.3mm}}r|r@{\hspace{0.8mm}}r@{\hspace{0.3mm}}r|r@{\hspace{0.8mm}}r@{\hspace{0.3mm}}r|r@{\hspace{0.8mm}}r@{\hspace{0.3mm}}r|}
\hline
\multicolumn{1}{|c||}{$x_{min}$} & \multicolumn{3}{c|}{ $\beta = 2.45$} & \multicolumn{3}{c|}{ $\beta = 2.50$} & \multicolumn{3}{c|}{ $\beta = 2.55$} & \multicolumn{3}{c|}{ $\beta = 2.60$} & \multicolumn{3}{c|}{ $\beta = 2.65$} \\ \hline
   ($\infty$,0.90,0.65) &    5.065 &    (0.043) & $    \times 10^{4}$  &    7.261 &    (0.067) & $    \times 10^{4}$  &    1.075 &    (0.011) & $    \times 10^{5}$  &    1.537 &    (0.016) & $    \times 10^{5}$  &    2.213 &    (0.024) & $    \times 10^{5}$  \\ \hline
($\infty$,1.0,0.65) &    5.068 &    (0.042) & $    \times 10^{4}$  &    7.264 &    (0.066) & $    \times 10^{4}$  &    1.075 &    (0.011) & $    \times 10^{5}$  &    1.537 &    (0.016) & $    \times 10^{5}$  &    2.215 &    (0.024) & $    \times 10^{5}$  \\ \hline
($\infty$,$\infty$,0.65) &    5.075 &    (0.043) & $    \times 10^{4}$  &    7.265 &    (0.067) & $    \times 10^{4}$  &    1.073 &    (0.011) & $    \times 10^{5}$  &    1.538 &    (0.016) & $    \times 10^{5}$  &    2.216 &    (0.024) & $    \times 10^{5}$  \\ \hline
\it ($\infty$,$\infty$,0.80) & \it 5.074 & \it (0.042) & $ \it \times 10^\hboxscript{4}$  & \it 7.262 & \it (0.067) & $ \it \times 10^\hboxscript{4}$  & \it 1.073 & \it (0.012) & $ \it \times 10^\hboxscript{5}$  & \it 1.545 & \it (0.018) & $ \it \times 10^\hboxscript{5}$  & \it 2.215 & \it (0.028) & $ \it \times 10^\hboxscript{5}$  \\ \hline
\sf ($\infty$,$\infty$,0.90) & \sf 5.074 & \sf (0.041) & $ \sf \times 10^\hboxsans{4}$  & \sf 7.259 & \sf (0.067) & $ \sf \times 10^\hboxsans{4}$  & \sf 1.073 & \sf (0.012) & $ \sf \times 10^\hboxsans{5}$  & \sf 1.544 & \sf (0.017) & $ \sf \times 10^\hboxsans{5}$  & \sf 2.214 & \sf (0.028) & $ \sf \times 10^\hboxsans{5}$  \\ \hline
\sf ($\infty$,$\infty$,$\infty$) & \sf 5.067 & \sf (0.041) & $ \sf \times 10^\hboxsans{4}$  & \sf 7.251 & \sf (0.065) & $ \sf \times 10^\hboxsans{4}$  & \sf 1.074 & \sf (0.012) & $ \sf \times 10^\hboxsans{5}$  & \sf 1.544 & \sf (0.018) & $ \sf \times 10^\hboxsans{5}$  & \sf 2.217 & \sf (0.028) & $ \sf \times 10^\hboxsans{5}$  \\ \hline
\end{tabular} \end{center}
\par\vspace*{3mm}\par
\begin{center}  %%\hspace*{-4.0cm}
\begin{tabular}{|c||r@{\hspace{0.8mm}}r@{\hspace{0.3mm}}r|r@{\hspace{0.8mm}}r@{\hspace{0.3mm}}r|r@{\hspace{0.8mm}}r@{\hspace{0.3mm}}r|r@{\hspace{0.8mm}}r@{\hspace{0.3mm}}r|r@{\hspace{0.8mm}}r@{\hspace{0.3mm}}r|}
\hline
\multicolumn{1}{|c||}{$x_{min}$} & \multicolumn{3}{c|}{ $\beta = 2.70$} & \multicolumn{3}{c|}{ $\beta = 2.775$} & \multicolumn{3}{c|}{ $\beta = 2.85$} & \multicolumn{3}{c|}{ $\beta = 2.925$} & \multicolumn{3}{c|}{ $\beta = 3.00$} \\ \hline
   ($\infty$,0.90,0.65) &    3.228 &    (0.037) & $    \times 10^{5}$  &    5.617 &    (0.063) & $    \times 10^{5}$  &    9.701 &    (0.118) & $    \times 10^{5}$  &    1.731 &    (0.022) & $    \times 10^{6}$  &    2.967 &    (0.041) & $    \times 10^{6}$  \\ \hline
($\infty$,1.0,0.65) &    3.228 &    (0.037) & $    \times 10^{5}$  &    5.619 &    (0.063) & $    \times 10^{5}$  &    9.706 &    (0.122) & $    \times 10^{5}$  &    1.732 &    (0.022) & $    \times 10^{6}$  &    2.969 &    (0.042) & $    \times 10^{6}$  \\ \hline
($\infty$,$\infty$,0.65) &    3.222 &    (0.038) & $    \times 10^{5}$  &    5.626 &    (0.066) & $    \times 10^{5}$  &    9.702 &    (0.122) & $    \times 10^{5}$  &    1.732 &    (0.023) & $    \times 10^{6}$  &    2.966 &    (0.042) & $    \times 10^{6}$  \\ \hline
\it ($\infty$,$\infty$,0.80) & \it 3.199 & \it (0.047) & $ \it \times 10^\hboxscript{5}$  & \it 5.622 & \it (0.091) & $ \it \times 10^\hboxscript{5}$  & \it 9.742 & \it (0.185) & $ \it \times 10^\hboxscript{5}$  & \it 1.739 & \it (0.038) & $ \it \times 10^\hboxscript{6}$  & \it 2.960 & \it (0.070) & $ \it \times 10^\hboxscript{6}$  \\ \hline
\sf ($\infty$,$\infty$,0.90) & \sf 3.202 & \sf (0.046) & $ \sf \times 10^\hboxsans{5}$  & \sf 5.625 & \sf (0.089) & $ \sf \times 10^\hboxsans{5}$  & \sf 9.759 & \sf (0.186) & $ \sf \times 10^\hboxsans{5}$  & \sf 1.742 & \sf (0.038) & $ \sf \times 10^\hboxsans{6}$  & \sf 2.956 & \sf (0.071) & $ \sf \times 10^\hboxsans{6}$  \\ \hline
\sf ($\infty$,$\infty$,$\infty$) & \sf 3.209 & \sf (0.046) & $ \sf \times 10^\hboxsans{5}$  & \sf 5.629 & \sf (0.091) & $ \sf \times 10^\hboxsans{5}$  & \sf 9.749 & \sf (0.180) & $ \sf \times 10^\hboxsans{5}$  & \sf 1.737 & \sf (0.037) & $ \sf \times 10^\hboxsans{6}$  & \sf 2.940 & \sf (0.069) & $ \sf \times 10^\hboxsans{6}$  \\ \hline
\end{tabular} \end{center}
\par\vspace*{3mm}\par
\begin{center}  %%\hspace*{-4.0cm}
\begin{tabular}{|c||r@{\hspace{0.8mm}}r@{\hspace{0.3mm}}r|r@{\hspace{0.8mm}}r@{\hspace{0.3mm}}r|r@{\hspace{0.8mm}}r@{\hspace{0.3mm}}r|r@{\hspace{0.8mm}}r@{\hspace{0.3mm}}r|r@{\hspace{0.8mm}}r@{\hspace{0.3mm}}r|}
\hline
\multicolumn{1}{|c||}{$x_{min}$} & \multicolumn{3}{c|}{ $\beta = 3.075$} & \multicolumn{3}{c|}{ $\beta = 3.15$} & \multicolumn{3}{c|}{ $\beta = 3.225$} & \multicolumn{3}{c|}{ $\beta = 3.30$} & \multicolumn{3}{c|}{ $\beta = 3.375$} \\ \hline
   ($\infty$,0.90,0.65) &    5.265 &    (0.079) & $    \times 10^{6}$  &    9.252 &    (0.146) & $    \times 10^{6}$  &    1.642 &    (0.028) & $    \times 10^{7}$  &    2.907 &    (0.052) & $    \times 10^{7}$  &    5.050 &    (0.095) & $    \times 10^{7}$  \\ \hline
($\infty$,1.0,0.65) &    5.263 &    (0.079) & $    \times 10^{6}$  &    9.274 &    (0.148) & $    \times 10^{6}$  &    1.640 &    (0.029) & $    \times 10^{7}$  &    2.911 &    (0.053) & $    \times 10^{7}$  &    5.066 &    (0.098) & $    \times 10^{7}$  \\ \hline
($\infty$,$\infty$,0.65) &    5.250 &    (0.082) & $    \times 10^{6}$  &    9.266 &    (0.150) & $    \times 10^{6}$  &    1.635 &    (0.029) & $    \times 10^{7}$  &    2.908 &    (0.052) & $    \times 10^{7}$  &    5.066 &    (0.097) & $    \times 10^{7}$  \\ \hline
\it ($\infty$,$\infty$,0.80) & \it 5.302 & \it (0.144) & $ \it \times 10^\hboxscript{6}$  & \it 9.330 & \it (0.256) & $ \it \times 10^\hboxscript{6}$  & \it 1.651 & \it (0.048) & $ \it \times 10^\hboxscript{7}$  & \it 2.940 & \it (0.084) & $ \it \times 10^\hboxscript{7}$  & \it 5.117 & \it (0.148) & $ \it \times 10^\hboxscript{7}$  \\ \hline
\sf ($\infty$,$\infty$,0.90) & \sf 5.305 & \sf (0.144) & $ \sf \times 10^\hboxsans{6}$  & \sf 9.376 & \sf (0.272) & $ \sf \times 10^\hboxsans{6}$  & \sf 1.629 & \sf (0.054) & $ \sf \times 10^\hboxsans{7}$  & \sf 2.833 & \sf (0.095) & $ \sf \times 10^\hboxsans{7}$  & \sf 4.981 & \sf (0.186) & $ \sf \times 10^\hboxsans{7}$  \\ \hline
\sf ($\infty$,$\infty$,$\infty$) & \sf 5.283 & \sf (0.142) & $ \sf \times 10^\hboxsans{6}$  & \sf 9.347 & \sf (0.275) & $ \sf \times 10^\hboxsans{6}$  & \sf 1.633 & \sf (0.053) & $ \sf \times 10^\hboxsans{7}$  & \sf 2.844 & \sf (0.096) & $ \sf \times 10^\hboxsans{7}$  & \sf 5.016 & \sf (0.191) & $ \sf \times 10^\hboxsans{7}$  \\ \hline
\end{tabular} \end{center}
\par\vspace*{3mm}\par
\caption{
{\bf [Continued]}
}

\end{table}
\clearpage \addtocounter{table}{-1}
\begin{table}
\addtolength{\tabcolsep}{-1.0mm}
\protect\footnotesize
\tabcolsep 2.5pt
\doublerulesep 1.5pt
\begin{center}  %%\hspace*{-4.0cm}
\begin{tabular}{|c||r@{\hspace{0.8mm}}r@{\hspace{0.3mm}}r|r@{\hspace{0.8mm}}r@{\hspace{0.3mm}}r|r@{\hspace{0.8mm}}r@{\hspace{0.3mm}}r|r@{\hspace{0.8mm}}r@{\hspace{0.3mm}}r|r@{\hspace{0.8mm}}r@{\hspace{0.3mm}}r|}
\hline
\multicolumn{1}{|c||}{$x_{min}$} & \multicolumn{3}{c|}{ $\beta = 3.45$} & \multicolumn{3}{c|}{ $\beta = 3.525$} & \multicolumn{3}{c|}{ $\beta = 3.60$} & \multicolumn{3}{c|}{ $\beta = 3.675$} & \multicolumn{3}{c|}{ $\beta = 3.75$} \\ \hline
   ($\infty$,0.90,0.65) &    8.966 &    (0.165) & $    \times 10^{7}$  &    1.559 &    (0.031) & $    \times 10^{8}$  &    2.817 &    (0.056) & $    \times 10^{8}$  &    4.977 &    (0.100) & $    \times 10^{8}$  &    8.799 &    (0.181) & $    \times 10^{8}$  \\ \hline
($\infty$,1.0,0.65) &    8.939 &    (0.177) & $    \times 10^{7}$  &    1.557 &    (0.032) & $    \times 10^{8}$  &    2.819 &    (0.061) & $    \times 10^{8}$  &    4.997 &    (0.114) & $    \times 10^{8}$  &    8.865 &    (0.215) & $    \times 10^{8}$  \\ \hline
($\infty$,$\infty$,0.65) &    8.953 &    (0.177) & $    \times 10^{7}$  &    1.558 &    (0.032) & $    \times 10^{8}$  &    2.820 &    (0.063) & $    \times 10^{8}$  &    4.989 &    (0.113) & $    \times 10^{8}$  &    8.837 &    (0.218) & $    \times 10^{8}$  \\ \hline
\it ($\infty$,$\infty$,0.80) & \it 9.041 & \it (0.262) & $ \it \times 10^\hboxscript{7}$  & \it 1.575 & \it (0.047) & $ \it \times 10^\hboxscript{8}$  & \it 2.847 & \it (0.089) & $ \it \times 10^\hboxscript{8}$  & \it 5.044 & \it (0.160) & $ \it \times 10^\hboxscript{8}$  & \it 8.923 & \it (0.301) & $ \it \times 10^\hboxscript{8}$  \\ \hline
\sf ($\infty$,$\infty$,0.90) & \sf 8.764 & \sf (0.342) & $ \sf \times 10^\hboxsans{7}$  & \sf 1.511 & \sf (0.064) & $ \sf \times 10^\hboxsans{8}$  & \sf 2.737 & \sf (0.118) & $ \sf \times 10^\hboxsans{8}$  & \sf 4.846 & \sf (0.209) & $ \sf \times 10^\hboxsans{8}$  & \sf 8.579 & \sf (0.382) & $ \sf \times 10^\hboxsans{8}$  \\ \hline
\sf ($\infty$,$\infty$,$\infty$) & \sf 8.803 & \sf (0.351) & $ \sf \times 10^\hboxsans{7}$  & \sf 1.511 & \sf (0.064) & $ \sf \times 10^\hboxsans{8}$  & \sf 2.713 & \sf (0.123) & $ \sf \times 10^\hboxsans{8}$  & \sf 4.819 & \sf (0.230) & $ \sf \times 10^\hboxsans{8}$  & \sf 8.328 & \sf (0.420) & $ \sf \times 10^\hboxsans{8}$  \\ \hline
\end{tabular} \end{center}
\par\vspace*{3mm}\par
\begin{center}  %%\hspace*{-4.0cm}
\begin{tabular}{|c||r@{\hspace{0.8mm}}r@{\hspace{0.3mm}}r|r@{\hspace{0.8mm}}r@{\hspace{0.3mm}}r|r@{\hspace{0.8mm}}r@{\hspace{0.3mm}}r|r@{\hspace{0.8mm}}r@{\hspace{0.3mm}}r|r@{\hspace{0.8mm}}r@{\hspace{0.3mm}}r|}
\hline
\multicolumn{1}{|c||}{$x_{min}$} & \multicolumn{3}{c|}{ $\beta = 3.825$} & \multicolumn{3}{c|}{ $\beta = 3.90$} & \multicolumn{3}{c|}{ $\beta = 3.975$} & \multicolumn{3}{c|}{ $\beta = 4.05$} & \multicolumn{3}{c|}{ $\beta = 4.125$} \\ \hline
   ($\infty$,0.90,0.65) &    1.550 &    (0.032) & $    \times 10^{9}$  &    2.762 &    (0.060) & $    \times 10^{9}$  &    4.919 &    (0.105) & $    \times 10^{9}$  &    8.900 &    (0.200) & $    \times 10^{9}$  &    1.581 &    (0.035) & $    \times 10^{10}$  \\ \hline
($\infty$,1.0,0.65) &    1.582 &    (0.039) & $    \times 10^{9}$  &    2.790 &    (0.076) & $    \times 10^{9}$  &    4.973 &    (0.135) & $    \times 10^{9}$  &    8.994 &    (0.253) & $    \times 10^{9}$  &    1.598 &    (0.044) & $    \times 10^{10}$  \\ \hline
($\infty$,$\infty$,0.65) &    1.578 &    (0.039) & $    \times 10^{9}$  &    2.774 &    (0.075) & $    \times 10^{9}$  &    4.942 &    (0.135) & $    \times 10^{9}$  &    8.965 &    (0.263) & $    \times 10^{9}$  &    1.601 &    (0.048) & $    \times 10^{10}$  \\ \hline
\it ($\infty$,$\infty$,0.80) & \it 1.594 & \it (0.054) & $ \it \times 10^\hboxscript{9}$  & \it 2.795 & \it (0.098) & $ \it \times 10^\hboxscript{9}$  & \it 4.993 & \it (0.173) & $ \it \times 10^\hboxscript{9}$  & \it 9.050 & \it (0.330) & $ \it \times 10^\hboxscript{9}$  & \it 1.619 & \it (0.060) & $ \it \times 10^\hboxscript{10}$  \\ \hline
\sf ($\infty$,$\infty$,0.90) & \sf 1.534 & \sf (0.067) & $ \sf \times 10^\hboxsans{9}$  & \sf 2.685 & \sf (0.124) & $ \sf \times 10^\hboxsans{9}$  & \sf 4.789 & \sf (0.220) & $ \sf \times 10^\hboxsans{9}$  & \sf 8.696 & \sf (0.417) & $ \sf \times 10^\hboxsans{9}$  & \sf 1.544 & \sf (0.074) & $ \sf \times 10^\hboxsans{10}$  \\ \hline
\sf ($\infty$,$\infty$,$\infty$) & \sf 1.488 & \sf (0.078) & $ \sf \times 10^\hboxsans{9}$  & \sf 2.547 & \sf (0.145) & $ \sf \times 10^\hboxsans{9}$  & \sf 4.649 & \sf (0.269) & $ \sf \times 10^\hboxsans{9}$  & \sf 8.275 & \sf (0.511) & $ \sf \times 10^\hboxsans{9}$  & \sf 1.462 & \sf (0.093) & $ \sf \times 10^\hboxsans{10}$  \\ \hline
\end{tabular} \end{center}
\par\vspace*{3mm}\par
\begin{center}  %%\hspace*{-4.0cm}
\begin{tabular}{|c||r@{\hspace{0.8mm}}r@{\hspace{0.3mm}}r|r@{\hspace{0.8mm}}r@{\hspace{0.3mm}}r|r@{\hspace{0.8mm}}r@{\hspace{0.3mm}}r|r@{\hspace{0.8mm}}r@{\hspace{0.3mm}}r|r@{\hspace{0.8mm}}r@{\hspace{0.3mm}}r|}
\hline
\multicolumn{1}{|c||}{$x_{min}$} & \multicolumn{3}{c|}{ $\beta = 4.20$} & \multicolumn{3}{c|}{ $\beta = 4.275$} & \multicolumn{3}{c|}{ $\beta = 4.35$} \\ \hline
   ($\infty$,0.90,0.65) &    2.851 &    (0.068) & $    \times 10^{10}$  &    5.045 &    (0.122) & $    \times 10^{10}$  &    9.045 &    (0.234) & $    \times 10^{10}$  \\ \hline
($\infty$,1.0,0.65) &    2.883 &    (0.083) & $    \times 10^{10}$  &    5.114 &    (0.149) & $    \times 10^{10}$  &    9.139 &    (0.275) & $    \times 10^{10}$  \\ \hline
($\infty$,$\infty$,0.65) &    2.925 &    (0.093) & $    \times 10^{10}$  &    5.220 &    (0.175) & $    \times 10^{10}$  &    9.298 &    (0.338) & $    \times 10^{10}$  \\ \hline
\it ($\infty$,$\infty$,0.80) & \it 2.953 & \it (0.114) & $ \it \times 10^\hboxscript{10}$  & \it 5.241 & \it (0.209) & $ \it \times 10^\hboxscript{10}$  & \it 9.376 & \it (0.387) & $ \it \times 10^\hboxscript{10}$  \\ \hline
\sf ($\infty$,$\infty$,0.90) & \sf 2.843 & \sf (0.140) & $ \sf \times 10^\hboxsans{10}$  & \sf 5.050 & \sf (0.254) & $ \sf \times 10^\hboxsans{10}$  & \sf 8.986 & \sf (0.472) & $ \sf \times 10^\hboxsans{10}$  \\ \hline
\sf ($\infty$,$\infty$,$\infty$) & \sf 2.620 & \sf (0.178) & $ \sf \times 10^\hboxsans{10}$  & \sf 4.485 & \sf (0.319) & $ \sf \times 10^\hboxsans{10}$  & \sf 8.523 & \sf (0.630) & $ \sf \times 10^\hboxsans{10}$  \\ \hline
\end{tabular} \end{center}
\par\vspace*{3mm}\par
\caption{
{\bf [Continued]}
}

\end{table}
\clearpage

%
% Beginning of Table 6 for SU3 article:
% Beginning of chi squared table for fitting curve of \xi_A

\protect\small
\begin{table}
\addtolength{\tabcolsep}{-1.0mm}
\begin{center}
\begin{tabular}{|c||c|c|c|c|c|} \hline
\multicolumn{6}{|c|}{ $\chi^2$ for the FSS fit of $\xi_{A}$}    \\
\hline
$L_{min}$ & $n=11$ & $n=12$ & $n=13$ & $n=14$ & $n=15$  \\ \hline \hline
(1.0,0.95,0.65) &  99  538.50 &  98  393.30 &  97  288.20 &  96  287.00 &  95  286.10 \\ 
		& 5.44  0.0\% & 4.01  0.0\% & 2.97  0.0\% & 2.99  0.0\% & 3.01  0.0 \% \\ \hline 
($\infty$,0.55,0.50) & 126  501.30 & 125  315.10 & 124  229.50 & 123  223.60 & 122  219.60 \\ 
		& 3.98  0.0\% & 2.52  0.0\% & 1.85  0.0\% & 1.82  0.0\% & 1.80  0.0 \% \\ \hline 
($\infty$,0.95,0.65) &  89  310.50 &  88  159.10 &  87  91.73 &  86  91.69 &  85  88.49 \\
		& 3.49  0.0\% & 1.81  0.0\% & 1.05  34.4\% & 1.07  31.7\% & 1.04  37.6 \% \\ \hline
($\infty$,$\infty$,0.50) &  91  276.20 &  90  138.70 &  89  89.89 &  88  86.40 &  87  85.56 \\ 
		& 3.04  0.0\% & 1.54  0.1\% & 1.01  45.4\% & 0.98  52.8\% & 0.98  52.4 \% \\ \hline 
($\infty$,$\infty$,0.65) &  78  231.60 &  77  101.80 &  76  68.66 &  75  68.32 &  74  67.24 \\ 
		& 2.97  0.0\% & 1.32  3.1\% & 0.90  71.3\% & 0.91  69.4\% & 0.91  69.8 \% \\ \hline 
($\infty$,$\infty$,$\infty$) &  52  139.70 &   51  39.97 & \it 50  32.82 & \sf  49  31.97 & \sf  48  29.94 \\ 
		& 2.69  0.0\% & 0.78  86.8\% & \it 0.66  97.1\% & \sf 0.65  97.1\% & \sf 0.62  98.1 \% \\ \hline 
\end{tabular}
\caption{
  Degrees of freedom (DF), $\chi^2$, $\chi^2$/DF and confidence level
  for the $n^{th}$-order fit (\protect\ref{fss:gen}) of
  $\xi_A(\beta,2L)/\xi_A(\beta,L)$ versus $\xi_F(\beta,L)/L$.
  The indicated $x_{min}$ values apply to $L=8,16,32$, respectively;
  we always take $x_{min} = 0.14, 0$ for $L=64,128$.
  Our preferred fit is shown in {\em italics}\/;
   other good fits are shown in {\sf sans-serif};
   bad fits are shown in {\rm roman}.
}
\label{xiA_chisq_tab}
\end{center}
\end{table}
\clearpage

%%%%%%%%%%%%%%%%%%%%%%%%%%%%%%%%%%%%%%%%%%%%%%%%%%%%%%%%%%

%%%%%%%%%%%%%%%%%%%%%%%%%%%%%%%%%%%%%%%%%%%%%%%%%%%%%%%%%%

%%%%%%%%%%%%%%%%%%%%%%%%%%%%%%%%%%%%%%%%%%%%%%%%%%%%%%%%%%

%%%%%%%%%%%%%%%%%%%%%%%%%%%%%%%%%%%%%%%%%%%%%%%%%%%%%%%%%%

%%%%%%%%%%%%%%%%%%%%%%%%%%%%%%%%%%%%%%%%%%%%%%%%%%%%%%%%%%

%%%%%%%%%%%%%%%%%%%%%%%%%%%%%%%%%%%%%%%%%%%%%%%%%%%%%%%%%%
%
%   TABLE OF EXTRAPOLATED DATA %%%%%%
%
% Beginning of Table 10 for SU3 article:
% Beginning of extrapolated \xi_A

\begin{table}
\addtolength{\tabcolsep}{-1.0mm}
\protect\footnotesize
\tabcolsep 2.5pt
\doublerulesep 1.5pt
\begin{center}  %%\hspace*{-4.0cm}
\begin{tabular}{|c||r@{\hspace{0.8mm}}r@{\hspace{0.3mm}}r|r@{\hspace{0.8mm}}r@{\hspace{0.3mm}}r|r@{\hspace{0.8mm}}r@{\hspace{0.3mm}}r|r@{\hspace{0.8mm}}r@{\hspace{0.3mm}}r|r@{\hspace{0.8mm}}r@{\hspace{0.3mm}}r|}
\hline
\multicolumn{1}{|c||}{$x_{min}$} & \multicolumn{3}{c|}{$\beta = 1.75$} & \multicolumn{3}{c|}{$\beta = 1.775$} & \multicolumn{3}{c|}{$\beta = 1.80$} & \multicolumn{3}{c|}{$\beta = 1.825$} & \multicolumn{3}{c|}{$\beta = 1.85$} \\ \hline
($\infty$,0.55,0.50) &    2.944 &    (0.048) & $    \times 10^{0}$  &    3.354 &    (0.012) & $    \times 10^{0}$  &    3.739 &    (0.010) & $    \times 10^{0}$  &    4.139 &    (0.013) & $    \times 10^{0}$  &    4.567 &    (0.015) & $    \times 10^{0}$  \\ \hline
($\infty$,$\infty$,0.50) &    2.944 &    (0.049) & $    \times 10^{0}$  &    3.351 &    (0.012) & $    \times 10^{0}$  &    3.734 &    (0.010) & $    \times 10^{0}$  &    4.133 &    (0.013) & $    \times 10^{0}$  &    4.562 &    (0.016) & $    \times 10^{0}$  \\ \hline
($\infty$,$\infty$,0.65) &    2.944 &    (0.049) & $    \times 10^{0}$  &    3.350 &    (0.013) & $    \times 10^{0}$  &    3.733 &    (0.010) & $    \times 10^{0}$  &    4.131 &    (0.013) & $    \times 10^{0}$  &    4.560 &    (0.016) & $    \times 10^{0}$  \\ \hline
\it ($\infty$,$\infty$,$\infty$) & \it 2.944 & \it (0.048) & $ \it \times 10^\hboxscript{0}$  & \it 3.350 & \it (0.013) & $ \it \times 10^\hboxscript{0}$  & \it 3.731 & \it (0.010) & $ \it \times 10^\hboxscript{0}$  & \it 4.130 & \it (0.013) & $ \it \times 10^\hboxscript{0}$  & \it 4.559 & \it (0.016) & $ \it \times 10^\hboxscript{0}$  \\ \hline
\end{tabular} \end{center}
\par\vspace*{3mm}\par
\begin{center}  %%\hspace*{-4.0cm}
\begin{tabular}{|c||r@{\hspace{0.8mm}}r@{\hspace{0.3mm}}r|r@{\hspace{0.8mm}}r@{\hspace{0.3mm}}r|r@{\hspace{0.8mm}}r@{\hspace{0.3mm}}r|r@{\hspace{0.8mm}}r@{\hspace{0.3mm}}r|r@{\hspace{0.8mm}}r@{\hspace{0.3mm}}r|}
\hline
\multicolumn{1}{|c||}{$x_{min}$} & \multicolumn{3}{c|}{$\beta = 1.875$} & \multicolumn{3}{c|}{$\beta = 1.90$} & \multicolumn{3}{c|}{$\beta = 1.925$} & \multicolumn{3}{c|}{$\beta = 1.95$} & \multicolumn{3}{c|}{$\beta = 1.975$} \\ \hline
($\infty$,0.55,0.50) &    5.088 &    (0.017) & $    \times 10^{0}$  &    5.707 &    (0.018) & $    \times 10^{0}$  &    6.443 &    (0.019) & $    \times 10^{0}$  &    7.223 &    (0.020) & $    \times 10^{0}$  &    7.998 &    (0.024) & $    \times 10^{0}$  \\ \hline
($\infty$,$\infty$,0.50) &    5.087 &    (0.017) & $    \times 10^{0}$  &    5.712 &    (0.019) & $    \times 10^{0}$  &    6.445 &    (0.019) & $    \times 10^{0}$  &    7.217 &    (0.020) & $    \times 10^{0}$  &    7.982 &    (0.025) & $    \times 10^{0}$  \\ \hline
($\infty$,$\infty$,0.65) &    5.085 &    (0.018) & $    \times 10^{0}$  &    5.714 &    (0.019) & $    \times 10^{0}$  &    6.448 &    (0.019) & $    \times 10^{0}$  &    7.217 &    (0.020) & $    \times 10^{0}$  &    7.975 &    (0.025) & $    \times 10^{0}$  \\ \hline
\it ($\infty$,$\infty$,$\infty$) & \it 5.086 & \it (0.017) & $ \it \times 10^\hboxscript{0}$  & \it 5.715 & \it (0.019) & $ \it \times 10^\hboxscript{0}$  & \it 6.450 & \it (0.019) & $ \it \times 10^\hboxscript{0}$  & \it 7.215 & \it (0.021) & $ \it \times 10^\hboxscript{0}$  & \it 7.971 & \it (0.026) & $ \it \times 10^\hboxscript{0}$  \\ \hline
\end{tabular} \end{center}
\par\vspace*{3mm}\par
\begin{center}  %%\hspace*{-4.0cm}
\begin{tabular}{|c||r@{\hspace{0.8mm}}r@{\hspace{0.3mm}}r|r@{\hspace{0.8mm}}r@{\hspace{0.3mm}}r|r@{\hspace{0.8mm}}r@{\hspace{0.3mm}}r|r@{\hspace{0.8mm}}r@{\hspace{0.3mm}}r|r@{\hspace{0.8mm}}r@{\hspace{0.3mm}}r|}
\hline
\multicolumn{1}{|c||}{$x_{min}$} & \multicolumn{3}{c|}{$\beta = 1.985$} & \multicolumn{3}{c|}{$\beta = 2.00$} & \multicolumn{3}{c|}{$\beta = 2.012$} & \multicolumn{3}{c|}{$\beta = 2.025$} & \multicolumn{3}{c|}{$\beta = 2.037$} \\ \hline
($\infty$,0.55,0.50) &    8.296 &    (0.027) & $    \times 10^{0}$  &    8.795 &    (0.031) & $    \times 10^{0}$  &    9.205 &    (0.036) & $    \times 10^{0}$  &    9.682 &    (0.037) & $    \times 10^{0}$  &    1.013 &    (0.004) & $    \times 10^{1}$  \\ \hline
($\infty$,$\infty$,0.50) &    8.277 &    (0.028) & $    \times 10^{0}$  &    8.778 &    (0.032) & $    \times 10^{0}$  &    9.189 &    (0.036) & $    \times 10^{0}$  &    9.673 &    (0.037) & $    \times 10^{0}$  &    1.013 &    (0.004) & $    \times 10^{1}$  \\ \hline
($\infty$,$\infty$,0.65) &    8.269 &    (0.028) & $    \times 10^{0}$  &    8.768 &    (0.032) & $    \times 10^{0}$  &    9.179 &    (0.037) & $    \times 10^{0}$  &    9.666 &    (0.038) & $    \times 10^{0}$  &    1.012 &    (0.004) & $    \times 10^{1}$  \\ \hline
\it ($\infty$,$\infty$,$\infty$) & \it 8.264 & \it (0.030) & $ \it \times 10^\hboxscript{0}$  & \it 8.767 & \it (0.032) & $ \it \times 10^\hboxscript{0}$  & \it 9.176 & \it (0.037) & $ \it \times 10^\hboxscript{0}$  & \it 9.665 & \it (0.037) & $ \it \times 10^\hboxscript{0}$  & \it 1.013 & \it (0.004) & $ \it \times 10^\hboxscript{1}$  \\ \hline
\end{tabular} \end{center}
\par\vspace*{3mm}\par
\begin{center}  %%\hspace*{-4.0cm}
\begin{tabular}{|c||r@{\hspace{0.8mm}}r@{\hspace{0.3mm}}r|r@{\hspace{0.8mm}}r@{\hspace{0.3mm}}r|r@{\hspace{0.8mm}}r@{\hspace{0.3mm}}r|r@{\hspace{0.8mm}}r@{\hspace{0.3mm}}r|r@{\hspace{0.8mm}}r@{\hspace{0.3mm}}r|}
\hline
\multicolumn{1}{|c||}{$x_{min}$} & \multicolumn{3}{c|}{$\beta = 2.05$} & \multicolumn{3}{c|}{$\beta = 2.062$} & \multicolumn{3}{c|}{$\beta = 2.075$} & \multicolumn{3}{c|}{$\beta = 2.10$} & \multicolumn{3}{c|}{$\beta = 2.112$} \\ \hline
($\infty$,0.55,0.50) &    1.080 &    (0.004) & $    \times 10^{1}$  &    1.139 &    (0.004) & $    \times 10^{1}$  &    1.221 &    (0.004) & $    \times 10^{1}$  &    1.365 &    (0.004) & $    \times 10^{1}$  &    1.436 &    (0.004) & $    \times 10^{1}$  \\ \hline
($\infty$,$\infty$,0.50) &    1.081 &    (0.004) & $    \times 10^{1}$  &    1.140 &    (0.004) & $    \times 10^{1}$  &    1.219 &    (0.004) & $    \times 10^{1}$  &    1.359 &    (0.004) & $    \times 10^{1}$  &    1.436 &    (0.004) & $    \times 10^{1}$  \\ \hline
($\infty$,$\infty$,0.65) &    1.082 &    (0.004) & $    \times 10^{1}$  &    1.141 &    (0.004) & $    \times 10^{1}$  &    1.220 &    (0.004) & $    \times 10^{1}$  &    1.360 &    (0.004) & $    \times 10^{1}$  &    1.434 &    (0.004) & $    \times 10^{1}$  \\ \hline
\it ($\infty$,$\infty$,$\infty$) & \it 1.082 & \it (0.004) & $ \it \times 10^\hboxscript{1}$  & \it 1.141 & \it (0.004) & $ \it \times 10^\hboxscript{1}$  & \it 1.221 & \it (0.004) & $ \it \times 10^\hboxscript{1}$  & \it 1.361 & \it (0.004) & $ \it \times 10^\hboxscript{1}$  & \it 1.434 & \it (0.004) & $ \it \times 10^\hboxscript{1}$  \\ \hline
\end{tabular} \end{center}
\par\vspace*{3mm}\par
\begin{center}  %%\hspace*{-4.0cm}
\begin{tabular}{|c||r@{\hspace{0.8mm}}r@{\hspace{0.3mm}}r|r@{\hspace{0.8mm}}r@{\hspace{0.3mm}}r|r@{\hspace{0.8mm}}r@{\hspace{0.3mm}}r|r@{\hspace{0.8mm}}r@{\hspace{0.3mm}}r|r@{\hspace{0.8mm}}r@{\hspace{0.3mm}}r|}
\hline
\multicolumn{1}{|c||}{$x_{min}$} & \multicolumn{3}{c|}{$\beta = 2.125$} & \multicolumn{3}{c|}{$\beta = 2.133$} & \multicolumn{3}{c|}{$\beta = 2.15$} & \multicolumn{3}{c|}{$\beta = 2.175$} & \multicolumn{3}{c|}{$\beta = 2.20$} \\ \hline
($\infty$,0.55,0.50) &    1.513 &    (0.004) & $    \times 10^{1}$  &    1.560 &    (0.004) & $    \times 10^{1}$  &    1.667 &    (0.005) & $    \times 10^{1}$  &    1.829 &    (0.007) & $    \times 10^{1}$  &    2.015 &    (0.007) & $    \times 10^{1}$  \\ \hline
($\infty$,$\infty$,0.50) &    1.511 &    (0.004) & $    \times 10^{1}$  &    1.558 &    (0.004) & $    \times 10^{1}$  &    1.663 &    (0.005) & $    \times 10^{1}$  &    1.826 &    (0.007) & $    \times 10^{1}$  &    2.013 &    (0.007) & $    \times 10^{1}$  \\ \hline
($\infty$,$\infty$,0.65) &    1.512 &    (0.004) & $    \times 10^{1}$  &    1.556 &    (0.004) & $    \times 10^{1}$  &    1.660 &    (0.005) & $    \times 10^{1}$  &    1.828 &    (0.007) & $    \times 10^{1}$  &    2.012 &    (0.008) & $    \times 10^{1}$  \\ \hline
\it ($\infty$,$\infty$,$\infty$) & \it 1.512 & \it (0.004) & $ \it \times 10^\hboxscript{1}$  & \it 1.555 & \it (0.005) & $ \it \times 10^\hboxscript{1}$  & \it 1.659 & \it (0.006) & $ \it \times 10^\hboxscript{1}$  & \it 1.828 & \it (0.007) & $ \it \times 10^\hboxscript{1}$  & \it 2.013 & \it (0.008) & $ \it \times 10^\hboxscript{1}$  \\ \hline
\end{tabular} \end{center}
\par\vspace*{3mm}\par
\begin{center}  %%\hspace*{-4.0cm}
\begin{tabular}{|c||r@{\hspace{0.8mm}}r@{\hspace{0.3mm}}r|r@{\hspace{0.8mm}}r@{\hspace{0.3mm}}r|r@{\hspace{0.8mm}}r@{\hspace{0.3mm}}r|r@{\hspace{0.8mm}}r@{\hspace{0.3mm}}r|r@{\hspace{0.8mm}}r@{\hspace{0.3mm}}r|}
\hline
\multicolumn{1}{|c||}{$x_{min}$} & \multicolumn{3}{c|}{$\beta = 2.2163$} & \multicolumn{3}{c|}{$\beta = 2.25$} & \multicolumn{3}{c|}{$\beta = 2.30$} & \multicolumn{3}{c|}{$\beta = 2.35$} & \multicolumn{3}{c|}{$\beta = 2.40$} \\ \hline
($\infty$,0.55,0.50) &    2.162 &    (0.007) & $    \times 10^{1}$  &    2.524 &    (0.008) & $    \times 10^{1}$  &    3.123 &    (0.008) & $    \times 10^{1}$  &    3.756 &    (0.015) & $    \times 10^{1}$  &    4.643 &    (0.016) & $    \times 10^{1}$  \\ \hline
($\infty$,$\infty$,0.50) &    2.165 &    (0.007) & $    \times 10^{1}$  &    2.529 &    (0.008) & $    \times 10^{1}$  &    3.123 &    (0.008) & $    \times 10^{1}$  &    3.757 &    (0.015) & $    \times 10^{1}$  &    4.665 &    (0.018) & $    \times 10^{1}$  \\ \hline
($\infty$,$\infty$,0.65) &    2.171 &    (0.008) & $    \times 10^{1}$  &    2.534 &    (0.010) & $    \times 10^{1}$  &    3.119 &    (0.009) & $    \times 10^{1}$  &    3.762 &    (0.017) & $    \times 10^{1}$  &    4.648 &    (0.023) & $    \times 10^{1}$  \\ \hline
\it ($\infty$,$\infty$,$\infty$) & \it 2.172 & \it (0.008) & $ \it \times 10^\hboxscript{1}$  & \it 2.535 & \it (0.010) & $ \it \times 10^\hboxscript{1}$  & \it 3.117 & \it (0.010) & $ \it \times 10^\hboxscript{1}$  & \it 3.760 & \it (0.018) & $ \it \times 10^\hboxscript{1}$  & \it 4.648 & \it (0.023) & $ \it \times 10^\hboxscript{1}$  \\ \hline
\end{tabular} \end{center}
\par\vspace*{3mm}\par
\caption{
  Estimated correlation lengths $\xi_{A,\infty}^{(2nd)}$ as a function
  of $\beta$, from different extrapolations. Error bar is one standard
  deviation (statistical errors only).
  All extrapolations use $s=2$, and $n=13$.
  The indicated $x_{min}$ values apply to $L=8,16,32$, respectively;
  we always take $x_{min} = 0.14, 0$ for $L=64,128$.
  Our preferred fit is shown in {\em italics}\/;
   bad fits are shown in {\rm roman}.
}

\label{estrap_xi_adj}

\end{table}
\clearpage \addtocounter{table}{-1}
\begin{table}
\addtolength{\tabcolsep}{-1.0mm}
\protect\footnotesize
\tabcolsep 2.5pt
\doublerulesep 1.5pt
\begin{center}  %%\hspace*{-4.0cm}
\begin{tabular}{|c||r@{\hspace{0.8mm}}r@{\hspace{0.3mm}}r|r@{\hspace{0.8mm}}r@{\hspace{0.3mm}}r|r@{\hspace{0.8mm}}r@{\hspace{0.3mm}}r|r@{\hspace{0.8mm}}r@{\hspace{0.3mm}}r|r@{\hspace{0.8mm}}r@{\hspace{0.3mm}}r|}
\hline
\multicolumn{1}{|c||}{$x_{min}$} & \multicolumn{3}{c|}{$\beta = 2.45$} & \multicolumn{3}{c|}{$\beta = 2.50$} & \multicolumn{3}{c|}{$\beta = 2.55$} & \multicolumn{3}{c|}{$\beta = 2.60$} & \multicolumn{3}{c|}{$\beta = 2.65$} \\ \hline
($\infty$,0.55,0.50) &    5.774 &    (0.018) & $    \times 10^{1}$  &    6.973 &    (0.025) & $    \times 10^{1}$  &    8.500 &    (0.034) & $    \times 10^{1}$  &    1.059 &    (0.004) & $    \times 10^{2}$  &    1.299 &    (0.004) & $    \times 10^{2}$  \\ \hline
($\infty$,$\infty$,0.50) &    5.793 &    (0.020) & $    \times 10^{1}$  &    6.969 &    (0.028) & $    \times 10^{1}$  &    8.546 &    (0.040) & $    \times 10^{1}$  &    1.064 &    (0.005) & $    \times 10^{2}$  &    1.299 &    (0.005) & $    \times 10^{2}$  \\ \hline
($\infty$,$\infty$,0.65) &    5.757 &    (0.028) & $    \times 10^{1}$  &    6.919 &    (0.037) & $    \times 10^{1}$  &    8.534 &    (0.057) & $    \times 10^{1}$  &    1.064 &    (0.007) & $    \times 10^{2}$  &    1.296 &    (0.007) & $    \times 10^{2}$  \\ \hline
\it ($\infty$,$\infty$,$\infty$) & \it 5.754 & \it (0.028) & $ \it \times 10^\hboxscript{1}$  & \it 6.911 & \it (0.036) & $ \it \times 10^\hboxscript{1}$  & \it 8.538 & \it (0.060) & $ \it \times 10^\hboxscript{1}$  & \it 1.067 & \it (0.007) & $ \it \times 10^\hboxscript{2}$  & \it 1.295 & \it (0.008) & $ \it \times 10^\hboxscript{2}$  \\ \hline
\end{tabular} \end{center}
\par\vspace*{3mm}\par
\begin{center}  %%\hspace*{-4.0cm}
\begin{tabular}{|c||r@{\hspace{0.8mm}}r@{\hspace{0.3mm}}r|r@{\hspace{0.8mm}}r@{\hspace{0.3mm}}r|r@{\hspace{0.8mm}}r@{\hspace{0.3mm}}r|r@{\hspace{0.8mm}}r@{\hspace{0.3mm}}r|r@{\hspace{0.8mm}}r@{\hspace{0.3mm}}r|}
\hline
\multicolumn{1}{|c||}{$x_{min}$} & \multicolumn{3}{c|}{$\beta = 2.70$} & \multicolumn{3}{c|}{$\beta = 2.775$} & \multicolumn{3}{c|}{$\beta = 2.85$} & \multicolumn{3}{c|}{$\beta = 2.925$} & \multicolumn{3}{c|}{$\beta = 3.00$} \\ \hline
($\infty$,0.55,0.50) &    1.560 &    (0.007) & $    \times 10^{2}$  &    2.174 &    (0.009) & $    \times 10^{2}$  &    2.883 &    (0.013) & $    \times 10^{2}$  &    3.969 &    (0.019) & $    \times 10^{2}$  &    5.383 &    (0.023) & $    \times 10^{2}$  \\ \hline
($\infty$,$\infty$,0.50) &    1.565 &    (0.008) & $    \times 10^{2}$  &    2.176 &    (0.012) & $    \times 10^{2}$  &    2.885 &    (0.016) & $    \times 10^{2}$  &    4.005 &    (0.028) & $    \times 10^{2}$  &    5.369 &    (0.033) & $    \times 10^{2}$  \\ \hline
($\infty$,$\infty$,0.65) &    1.562 &    (0.011) & $    \times 10^{2}$  &    2.176 &    (0.015) & $    \times 10^{2}$  &    2.879 &    (0.020) & $    \times 10^{2}$  &    4.004 &    (0.033) & $    \times 10^{2}$  &    5.352 &    (0.039) & $    \times 10^{2}$  \\ \hline
\it ($\infty$,$\infty$,$\infty$) & \it 1.557 & \it (0.013) & $ \it \times 10^\hboxscript{2}$  & \it 2.176 & \it (0.020) & $ \it \times 10^\hboxscript{2}$  & \it 2.879 & \it (0.027) & $ \it \times 10^\hboxscript{2}$  & \it 4.007 & \it (0.052) & $ \it \times 10^\hboxscript{2}$  & \it 5.314 & \it (0.059) & $ \it \times 10^\hboxscript{2}$  \\ \hline
\end{tabular} \end{center}
\par\vspace*{3mm}\par
\begin{center}  %%\hspace*{-4.0cm}
\begin{tabular}{|c||r@{\hspace{0.8mm}}r@{\hspace{0.3mm}}r|r@{\hspace{0.8mm}}r@{\hspace{0.3mm}}r|r@{\hspace{0.8mm}}r@{\hspace{0.3mm}}r|r@{\hspace{0.8mm}}r@{\hspace{0.3mm}}r|r@{\hspace{0.8mm}}r@{\hspace{0.3mm}}r|}
\hline
\multicolumn{1}{|c||}{$x_{min}$} & \multicolumn{3}{c|}{$\beta = 3.075$} & \multicolumn{3}{c|}{$\beta = 3.15$} & \multicolumn{3}{c|}{$\beta = 3.225$} & \multicolumn{3}{c|}{$\beta = 3.30$} & \multicolumn{3}{c|}{$\beta = 3.375$} \\ \hline
($\infty$,0.55,0.50) &    7.236 &    (0.040) & $    \times 10^{2}$  &    1.002 &    (0.005) & $    \times 10^{3}$  &    1.326 &    (0.008) & $    \times 10^{3}$  &    1.863 &    (0.011) & $    \times 10^{3}$  &    2.457 &    (0.015) & $    \times 10^{3}$  \\ \hline
($\infty$,$\infty$,0.50) &    7.258 &    (0.060) & $    \times 10^{2}$  &    1.004 &    (0.007) & $    \times 10^{3}$  &    1.334 &    (0.012) & $    \times 10^{3}$  &    1.868 &    (0.016) & $    \times 10^{3}$  &    2.454 &    (0.022) & $    \times 10^{3}$  \\ \hline
($\infty$,$\infty$,0.65) &    7.245 &    (0.070) & $    \times 10^{2}$  &    1.001 &    (0.008) & $    \times 10^{3}$  &    1.331 &    (0.014) & $    \times 10^{3}$  &    1.865 &    (0.018) & $    \times 10^{3}$  &    2.449 &    (0.025) & $    \times 10^{3}$  \\ \hline
\it ($\infty$,$\infty$,$\infty$) & \it 7.255 & \it (0.117) & $ \it \times 10^\hboxscript{2}$  & \it 1.003 & \it (0.014) & $ \it \times 10^\hboxscript{3}$  & \it 1.326 & \it (0.024) & $ \it \times 10^\hboxscript{3}$  & \it 1.836 & \it (0.033) & $ \it \times 10^\hboxscript{3}$  & \it 2.426 & \it (0.046) & $ \it \times 10^\hboxscript{3}$  \\ \hline
\end{tabular} \end{center}
\par\vspace*{3mm}\par
\begin{center}  %%\hspace*{-4.0cm}
\begin{tabular}{|c||r@{\hspace{0.8mm}}r@{\hspace{0.3mm}}r|r@{\hspace{0.8mm}}r@{\hspace{0.3mm}}r|r@{\hspace{0.8mm}}r@{\hspace{0.3mm}}r|r@{\hspace{0.8mm}}r@{\hspace{0.3mm}}r|r@{\hspace{0.8mm}}r@{\hspace{0.3mm}}r|}
\hline
\multicolumn{1}{|c||}{$x_{min}$} & \multicolumn{3}{c|}{$\beta = 3.45$} & \multicolumn{3}{c|}{$\beta = 3.525$} & \multicolumn{3}{c|}{$\beta = 3.60$} & \multicolumn{3}{c|}{$\beta = 3.675$} & \multicolumn{3}{c|}{$\beta = 3.75$} \\ \hline
($\infty$,0.55,0.50) &    3.412 &    (0.024) & $    \times 10^{3}$  &    4.559 &    (0.028) & $    \times 10^{3}$  &    6.257 &    (0.049) & $    \times 10^{3}$  &    8.576 &    (0.055) & $    \times 10^{3}$  &    1.141 &    (0.010) & $    \times 10^{4}$  \\ \hline
($\infty$,$\infty$,0.50) &    3.404 &    (0.036) & $    \times 10^{3}$  &    4.525 &    (0.042) & $    \times 10^{3}$  &    6.252 &    (0.079) & $    \times 10^{3}$  &    8.526 &    (0.090) & $    \times 10^{3}$  &    1.142 &    (0.016) & $    \times 10^{4}$  \\ \hline
($\infty$,$\infty$,0.65) &    3.404 &    (0.039) & $    \times 10^{3}$  &    4.517 &    (0.046) & $    \times 10^{3}$  &    6.251 &    (0.084) & $    \times 10^{3}$  &    8.498 &    (0.093) & $    \times 10^{3}$  &    1.140 &    (0.017) & $    \times 10^{4}$  \\ \hline
\it ($\infty$,$\infty$,$\infty$) & \it 3.358 & \it (0.076) & $ \it \times 10^\hboxscript{3}$  & \it 4.429 & \it (0.086) & $ \it \times 10^\hboxscript{3}$  & \it 6.077 & \it (0.164) & $ \it \times 10^\hboxscript{3}$  & \it 8.310 & \it (0.184) & $ \it \times 10^\hboxscript{3}$  & \it 1.094 & \it (0.031) & $ \it \times 10^\hboxscript{4}$  \\ \hline
\end{tabular} \end{center}
\par\vspace*{3mm}\par
\begin{center}  %%\hspace*{-4.0cm}
\begin{tabular}{|c||r@{\hspace{0.8mm}}r@{\hspace{0.3mm}}r|r@{\hspace{0.8mm}}r@{\hspace{0.3mm}}r|r@{\hspace{0.8mm}}r@{\hspace{0.3mm}}r|r@{\hspace{0.8mm}}r@{\hspace{0.3mm}}r|r@{\hspace{0.8mm}}r@{\hspace{0.3mm}}r|}
\hline
\multicolumn{1}{|c||}{$x_{min}$} & \multicolumn{3}{c|}{$\beta = 3.825$} & \multicolumn{3}{c|}{$\beta = 3.90$} & \multicolumn{3}{c|}{$\beta = 3.975$} & \multicolumn{3}{c|}{$\beta = 4.05$} & \multicolumn{3}{c|}{$\beta = 4.125$} \\ \hline
($\infty$,0.55,0.50) &    1.584 &    (0.012) & $    \times 10^{4}$  &    2.096 &    (0.018) & $    \times 10^{4}$  &    2.926 &    (0.026) & $    \times 10^{4}$  &    3.919 &    (0.033) & $    \times 10^{4}$  &    5.413 &    (0.055) & $    \times 10^{4}$  \\ \hline
($\infty$,$\infty$,0.50) &    1.586 &    (0.019) & $    \times 10^{4}$  &    2.090 &    (0.029) & $    \times 10^{4}$  &    2.914 &    (0.041) & $    \times 10^{4}$  &    3.901 &    (0.054) & $    \times 10^{4}$  &    5.412 &    (0.089) & $    \times 10^{4}$  \\ \hline
($\infty$,$\infty$,0.65) &    1.582 &    (0.020) & $    \times 10^{4}$  &    2.083 &    (0.031) & $    \times 10^{4}$  &    2.910 &    (0.043) & $    \times 10^{4}$  &    3.893 &    (0.058) & $    \times 10^{4}$  &    5.414 &    (0.092) & $    \times 10^{4}$  \\ \hline
\it ($\infty$,$\infty$,$\infty$) & \it 1.527 & \it (0.040) & $ \it \times 10^\hboxscript{4}$  & \it 1.978 & \it (0.057) & $ \it \times 10^\hboxscript{4}$  & \it 2.799 & \it (0.088) & $ \it \times 10^\hboxscript{4}$  & \it 3.715 & \it (0.108) & $ \it \times 10^\hboxscript{4}$  & \it 5.099 & \it (0.188) & $ \it \times 10^\hboxscript{4}$  \\ \hline
\end{tabular} \end{center}
\par\vspace*{3mm}\par
\begin{center}  %%\hspace*{-4.0cm}
\begin{tabular}{|c|| r r r | r r r | r r r |}
\hline
\multicolumn{1}{|c||}{$x_{min}$} & \multicolumn{3}{c|}{$\beta = 4.20$} & \multicolumn{3}{c|}{$\beta = 4.275$} & \multicolumn{3}{c|}{$\beta = 4.35$} \\ \hline
($\infty$,0.55,0.50) &    7.338 &    (0.064) & $    \times 10^{4}$  &    0.991 &    (0.011) & $    \times 10^{5}$  &    1.368 &    (0.013) & $    \times 10^{5}$  \\ \hline
($\infty$,$\infty$,0.50) &    7.362 &    (0.108) & $    \times 10^{4}$  &    1.009 &    (0.020) & $    \times 10^{5}$  &    1.367 &    (0.022) & $    \times 10^{5}$  \\ \hline
($\infty$,$\infty$,0.65) &    7.333 &    (0.110) & $    \times 10^{4}$  &    1.007 &    (0.020) & $    \times 10^{5}$  &    1.366 &    (0.023) & $    \times 10^{5}$  \\ \hline
\it ($\infty$,$\infty$,$\infty$) & \it 6.913 & \it (0.213) & $ \it \times 10^\hboxscript{4}$  & \it 0.911 & \it (0.037) & $ \it \times 10^\hboxscript{5}$  & \it 1.299 & \it (0.045) & $ \it \times 10^\hboxscript{5}$  \\ \hline
\end{tabular} \end{center}
\par\vspace*{3mm}\par
\caption{
{\bf [Continued]}}
\end{table}
\clearpage

%
% Beginning of Table 7 for SU3 article:
% Beginning of chi squared table for fitting curve of \chi_A

\begin{table}
\addtolength{\tabcolsep}{-1.0mm}
\begin{center}
\begin{tabular}{|c||c|c|c|c|c|} \hline
\multicolumn{6}{|c|}{ $\chi^2$ for the FSS fit of $\chi_{A}$}    \\
\hline
$L_{min}$ & $n=12$ & $n=13$ & $n=14$ & $n=15$ & $n=16$  \\ \hline \hline 
($\infty$,$\infty$,0.40) & 101  187.10 & 100  183.40 &  99  180.60 &  98  180.00 &  97  176.30 \\ 
		& 1.85  0.0\% &  1.83   0.0\% & 1.82  0.0\% &  1.84   0.0\% &  1.82   0.0 \% \\ \hline 
(1.0,0.95,0.65) &  98  723.90 &  97  723.30 &  96  720.30 &  95  706.00 &  94  692.30 \\ 
		& 7.39  0.0\% &  7.46   0.0\% & 7.50  0.0\% &  7.43   0.0\% &  7.36   0.0 \% \\ \hline 
($\infty$,0.95,0.65) &  88  150.30 &  87  149.50 &  86  137.70 &  85  137.70 &  84  135.70 \\ 
		& 1.71  0.0\% &  1.72   0.0\% & 1.60  0.0\% &  1.62   0.0\% &  1.62   0.0 \% \\ \hline 
($\infty$,1,0.65) &  84  122.00 &  83  121.80 &  82  103.30 &  81  103.30 &  80  97.79 \\ 
		& 1.45  0.4\% &  1.47   0.4\% & 1.26  5.6\% &  1.28   4.8\% &  1.22   8.6 \% \\ \hline 
($\infty$,1.0,0.9) &  70  93.73 &  69  93.40 &  68  79.16 &  67  78.48 &  66  74.02 \\ 
		& 1.34  3.1\% &  1.35   2.7\% & 1.16  16.7\% &  1.17   15.9\% &  1.12   23.3 \% \\ \hline 
($\infty$,$\infty$,0.65) &  77  90.56 &  76  87.86 &  75  78.06 &  74  77.98 &  73  63.79 \\ 
		& 1.18  13.8\% &  1.16   16.6\% & 1.04  38.2\% &  1.05   35.3\% &  0.87   77.1 \% \\ \hline 
($\infty$,$\infty$,0.80) &  70  93.02 &  69  79.17 &  68  76.42 &  67  68.85 &  66  68.67 \\ 
& 1.33  3.4\% &  1.15   18.9\% & 1.12  22.6\% &  1.03   41.5\% &  1.04   38.7 \% \\ \hline 
($\infty$,$\infty$,0.90) &  63  64.35 &  62  58.52 & \it 61  53.88 & \sf  60  53.34 & \sf  59  41.77 \\ 
		& 1.02  42.9\% &  0.94   60.2\% & \it 0.88  72.9\% & \sf  0.89   71.6\% & \sf  0.71   95.6 \% \\ \hline 
($\infty$,$\infty$,$\infty$) &  51  51.63 & \sf  50  35.76 & \sf  49  35.57 & \sf  48  32.40 & \sf  47  26.34 \\ 
		& 1.01  44.9\% & \sf  0.72   93.6\% & \sf 0.73  92.5\% & \sf  0.68   95.9\% & \sf  0.56   99.4 \% \\ \hline 
\end{tabular}
\end{center}
\caption{
  Degrees of freedom (DF), $\chi^2$, $\chi^2$/DF and confidence level
  for the $n^{th}$-order fit (\protect\ref{fss:gen}) of
  $ \chi_A (\beta,2L)/ \chi_A (\beta,L) $ versus $\xi_F(\beta,L)/L $.
  The indicated $x_{min}$ values apply to $L=8,16,32$, respectively;
  we always take $x_{min} = 0.14, 0$ for $L=64,128$.
  Our preferred fit is shown in {\em italics}\/;
   other good fits are shown in {\sf sans-serif};
   bad fits are shown in {\rm roman}.
}
\label{chiA_chisq_tab}
\end{table}
\clearpage

%%%%%%%%%%%%%%%%%%%%%%%%%%%%%%%%%%%%%%%%%%%%%%%%%%%%%%%%%%

%%%%%%%%%%%%%%%%%%%%%%%%%%%%%%%%%%%%%%%%%%%%%%%%%%%%%%%%%%

%%%%%%%%%%%%%%%%%%%%%%%%%%%%%%%%%%%%%%%%%%%%%%%%%%%%%%%%%%

%%%%%%%%%%%%%%%%%%%%%%%%%%%%%%%%%%%%%%%%%%%%%%%%%%%%%%%%%%

%%%%%%%%%%%%%%%%%%%%%%%%%%%%%%%%%%%%%%%%%%%%%%%%%%%%%%%%%%
%
%   TABLE OF EXTRAPOLATED DATA %%%%%%
%
% Beginning of Table 11 for SU3OB article:
% Beginning of extrapolated \chi_A

\begin{table}
\addtolength{\tabcolsep}{-1.0mm}
\protect\footnotesize
\tabcolsep 2.5pt
\doublerulesep 1.5pt
\begin{center}  %%\hspace*{-4.0cm} 
\begin{tabular}{|c||r@{\hspace{0.8mm}}r@{\hspace{0.3mm}}r|r@{\hspace{0.8mm}}r@{\hspace{0.3mm}}r|r@{\hspace{0.8mm}}r@{\hspace{0.3mm}}r|r@{\hspace{0.8mm}}r@{\hspace{0.3mm}}r|r@{\hspace{0.8mm}}r@{\hspace{0.3mm}}r|}
\hline
\multicolumn{1}{|c||}{$x_{min}$} & \multicolumn{3}{c|}{$\beta = 1.7500$} & \multicolumn{3}{c|}{$\beta = 1.7750$} & \multicolumn{3}{c|}{$\beta = 1.8000$} & \multicolumn{3}{c|}{$\beta = 1.8250$} & \multicolumn{3}{c|}{$\beta = 1.8500$} \\ \hline
($\infty$,1.0,0.65) &    7.267 &    (0.005) & $    \times 10^{1}$  &    8.295 &    (0.010) & $    \times 10^{1}$  &    9.475 &    (0.004) & $    \times 10^{1}$  &    1.084 &    (0.001) & $    \times 10^{2}$  &    1.243 &    (0.001) & $    \times 10^{2}$  \\ \hline
($\infty$,1.0,0.90) &    7.267 &    (0.006) & $    \times 10^{1}$  &    8.296 &    (0.011) & $    \times 10^{1}$  &    9.475 &    (0.004) & $    \times 10^{1}$  &    1.084 &    (0.001) & $    \times 10^{2}$  &    1.243 &    (0.001) & $    \times 10^{2}$  \\ \hline
($\infty$,$\infty$,0.65) &    7.267 &    (0.006) & $    \times 10^{1}$  &    8.296 &    (0.010) & $    \times 10^{1}$  &    9.475 &    (0.004) & $    \times 10^{1}$  &    1.084 &    (0.001) & $    \times 10^{2}$  &    1.243 &    (0.001) & $    \times 10^{2}$  \\ \hline
\it ($\infty$,$\infty$,0.90) & \it 7.267 & \it (0.006) & $ \it \times 10^\hboxscript{1}$  & \it 8.296 & \it (0.010) & $ \it \times 10^\hboxscript{1}$  & \it 9.475 & \it (0.004) & $ \it \times 10^\hboxscript{1}$  & \it 1.084 & \it (0.001) & $ \it \times 10^\hboxscript{2}$  & \it 1.243 & \it (0.001) & $ \it \times 10^\hboxscript{2}$  \\ \hline
\sf ($\infty$,$\infty$,$\infty$) & \sf 7.267 & \sf (0.006) & $ \sf \times 10^\hboxsans{1}$  & \sf 8.295 & \sf (0.010) & $ \sf \times 10^\hboxsans{1}$  & \sf 9.475 & \sf (0.004) & $ \sf \times 10^\hboxsans{1}$  & \sf 1.084 & \sf (0.001) & $ \sf \times 10^\hboxsans{2}$  & \sf 1.243 & \sf (0.001) & $ \sf \times 10^\hboxsans{2}$  \\ \hline
\end{tabular} \end{center}
%%\par\vspace*{3mm}\par
%%
\begin{center}  %%\hspace*{-4.0cm}
\begin{tabular}{|c||r@{\hspace{0.8mm}}r@{\hspace{0.3mm}}r|r@{\hspace{0.8mm}}r@{\hspace{0.3mm}}r|r@{\hspace{0.8mm}}r@{\hspace{0.3mm}}r|r@{\hspace{0.8mm}}r@{\hspace{0.3mm}}r|r@{\hspace{0.8mm}}r@{\hspace{0.3mm}}r|}
\hline
\multicolumn{1}{|c||}{$x_{min}$} & \multicolumn{3}{c|}{$\beta = 1.8750$} & \multicolumn{3}{c|}{$\beta = 1.9000$} & \multicolumn{3}{c|}{$\beta = 1.9250$} & \multicolumn{3}{c|}{$\beta = 1.9500$} & \multicolumn{3}{c|}{$\beta = 1.9750$} \\ \hline
($\infty$,1.0,0.65) &    1.431 &    (0.001) & $    \times 10^{2}$  &    1.646 &    (0.002) & $    \times 10^{2}$  &    1.900 &    (0.001) & $    \times 10^{2}$  &    2.196 &    (0.003) & $    \times 10^{2}$  &    2.532 &    (0.005) & $    \times 10^{2}$  \\ \hline
($\infty$,1.0,0.90) &    1.431 &    (0.001) & $    \times 10^{2}$  &    1.646 &    (0.002) & $    \times 10^{2}$  &    1.900 &    (0.001) & $    \times 10^{2}$  &    2.196 &    (0.003) & $    \times 10^{2}$  &    2.532 &    (0.005) & $    \times 10^{2}$  \\ \hline
($\infty$,$\infty$,0.65) &    1.430 &    (0.001) & $    \times 10^{2}$  &    1.646 &    (0.002) & $    \times 10^{2}$  &    1.900 &    (0.001) & $    \times 10^{2}$  &    2.196 &    (0.003) & $    \times 10^{2}$  &    2.532 &    (0.005) & $    \times 10^{2}$  \\ \hline
\it ($\infty$,$\infty$,0.90) & \it 1.431 & \it (0.001) & $ \it \times 10^\hboxscript{2}$  & \it 1.646 & \it (0.002) & $ \it \times 10^\hboxscript{2}$  & \it 1.900 & \it (0.001) & $ \it \times 10^\hboxscript{2}$  & \it 2.196 & \it (0.003) & $ \it \times 10^\hboxscript{2}$  & \it 2.532 & \it (0.005) & $ \it \times 10^\hboxscript{2}$  \\ \hline
\sf ($\infty$,$\infty$,$\infty$) & \sf 1.431 & \sf (0.001) & $ \sf \times 10^\hboxsans{2}$  & \sf 1.646 & \sf (0.002) & $ \sf \times 10^\hboxsans{2}$  & \sf 1.900 & \sf (0.001) & $ \sf \times 10^\hboxsans{2}$  & \sf 2.195 & \sf (0.003) & $ \sf \times 10^\hboxsans{2}$  & \sf 2.532 & \sf (0.004) & $ \sf \times 10^\hboxsans{2}$  \\ \hline
\end{tabular} \end{center}
%%\par\vspace*{3mm}\par
%%
\begin{center}  %%\hspace*{-4.0cm}
\begin{tabular}{|c||r@{\hspace{0.8mm}}r@{\hspace{0.3mm}}r|r@{\hspace{0.8mm}}r@{\hspace{0.3mm}}r|r@{\hspace{0.8mm}}r@{\hspace{0.3mm}}r|r@{\hspace{0.8mm}}r@{\hspace{0.3mm}}r|r@{\hspace{0.8mm}}r@{\hspace{0.3mm}}r|}
\hline
\multicolumn{1}{|c||}{$x_{min}$} & \multicolumn{3}{c|}{$\beta = 1.9850$} & \multicolumn{3}{c|}{$\beta = 2.0000$} & \multicolumn{3}{c|}{$\beta = 2.0120$} & \multicolumn{3}{c|}{$\beta = 2.0250$} & \multicolumn{3}{c|}{$\beta = 2.0370$} \\ \hline
($\infty$,1.0,0.65) &    2.679 &    (0.005) & $    \times 10^{2}$  &    2.930 &    (0.003) & $    \times 10^{2}$  &    3.135 &    (0.006) & $    \times 10^{2}$  &    3.383 &    (0.007) & $    \times 10^{2}$  &    3.616 &    (0.008) & $    \times 10^{2}$  \\ \hline
($\infty$,1.0,0.90) &    2.679 &    (0.005) & $    \times 10^{2}$  &    2.930 &    (0.003) & $    \times 10^{2}$  &    3.135 &    (0.006) & $    \times 10^{2}$  &    3.382 &    (0.007) & $    \times 10^{2}$  &    3.616 &    (0.008) & $    \times 10^{2}$  \\ \hline
($\infty$,$\infty$,0.65) &    2.679 &    (0.005) & $    \times 10^{2}$  &    2.930 &    (0.003) & $    \times 10^{2}$  &    3.135 &    (0.006) & $    \times 10^{2}$  &    3.382 &    (0.007) & $    \times 10^{2}$  &    3.616 &    (0.008) & $    \times 10^{2}$  \\ \hline
\it ($\infty$,$\infty$,0.90) & \it 2.679 & \it (0.005) & $ \it \times 10^\hboxscript{2}$  & \it 2.930 & \it (0.003) & $ \it \times 10^\hboxscript{2}$  & \it 3.135 & \it (0.006) & $ \it \times 10^\hboxscript{2}$  & \it 3.383 & \it (0.007) & $ \it \times 10^\hboxscript{2}$  & \it 3.616 & \it (0.008) & $ \it \times 10^\hboxscript{2}$  \\ \hline
\sf ($\infty$,$\infty$,$\infty$) & \sf 2.679 & \sf (0.005) & $ \sf \times 10^\hboxsans{2}$  & \sf 2.930 & \sf (0.003) & $ \sf \times 10^\hboxsans{2}$  & \sf 3.135 & \sf (0.006) & $ \sf \times 10^\hboxsans{2}$  & \sf 3.384 & \sf (0.007) & $ \sf \times 10^\hboxsans{2}$  & \sf 3.618 & \sf (0.007) & $ \sf \times 10^\hboxsans{2}$  \\ \hline
\end{tabular} \end{center}
%%\par\vspace*{3mm}\par
%%
\begin{center}  %%\hspace*{-4.0cm}
\begin{tabular}{|c||r@{\hspace{0.8mm}}r@{\hspace{0.3mm}}r|r@{\hspace{0.8mm}}r@{\hspace{0.3mm}}r|r@{\hspace{0.8mm}}r@{\hspace{0.3mm}}r|r@{\hspace{0.8mm}}r@{\hspace{0.3mm}}r|r@{\hspace{0.8mm}}r@{\hspace{0.3mm}}r|}
\hline
\multicolumn{1}{|c||}{$x_{min}$} & \multicolumn{3}{c|}{$\beta = 2.0500$} & \multicolumn{3}{c|}{$\beta = 2.0620$} & \multicolumn{3}{c|}{$\beta = 2.0750$} & \multicolumn{3}{c|}{$\beta = 2.1000$} & \multicolumn{3}{c|}{$\beta = 2.1120$} \\ \hline
($\infty$,1.0,0.65) &    3.925 &    (0.009) & $    \times 10^{2}$  &    4.208 &    (0.010) & $    \times 10^{2}$  &    4.571 &    (0.011) & $    \times 10^{2}$  &    5.289 &    (0.011) & $    \times 10^{2}$  &    5.686 &    (0.012) & $    \times 10^{2}$  \\ \hline
($\infty$,1.0,0.90) &    3.924 &    (0.009) & $    \times 10^{2}$  &    4.206 &    (0.010) & $    \times 10^{2}$  &    4.570 &    (0.011) & $    \times 10^{2}$  &    5.290 &    (0.011) & $    \times 10^{2}$  &    5.687 &    (0.012) & $    \times 10^{2}$  \\ \hline
($\infty$,$\infty$,0.65) &    3.925 &    (0.009) & $    \times 10^{2}$  &    4.207 &    (0.010) & $    \times 10^{2}$  &    4.570 &    (0.011) & $    \times 10^{2}$  &    5.290 &    (0.012) & $    \times 10^{2}$  &    5.687 &    (0.012) & $    \times 10^{2}$  \\ \hline
\it ($\infty$,$\infty$,0.90) & \it 3.925 & \it (0.009) & $ \it \times 10^\hboxscript{2}$  & \it 4.207 & \it (0.010) & $ \it \times 10^\hboxscript{2}$  & \it 4.570 & \it (0.011) & $ \it \times 10^\hboxscript{2}$  & \it 5.290 & \it (0.012) & $ \it \times 10^\hboxscript{2}$  & \it 5.686 & \it (0.012) & $ \it \times 10^\hboxscript{2}$  \\ \hline
\sf ($\infty$,$\infty$,$\infty$) & \sf 3.927 & \sf (0.009) & $ \sf \times 10^\hboxsans{2}$  & \sf 4.209 & \sf (0.010) & $ \sf \times 10^\hboxsans{2}$  & \sf 4.570 & \sf (0.011) & $ \sf \times 10^\hboxsans{2}$  & \sf 5.287 & \sf (0.011) & $ \sf \times 10^\hboxsans{2}$  & \sf 5.683 & \sf (0.012) & $ \sf \times 10^\hboxsans{2}$  \\ \hline
\end{tabular} \end{center}
%%\par\vspace*{3mm}\par
%%
\begin{center}  %%\hspace*{-4.0cm}
\begin{tabular}{|c||r@{\hspace{0.8mm}}r@{\hspace{0.3mm}}r|r@{\hspace{0.8mm}}r@{\hspace{0.3mm}}r|r@{\hspace{0.8mm}}r@{\hspace{0.3mm}}r|r@{\hspace{0.8mm}}r@{\hspace{0.3mm}}r|r@{\hspace{0.8mm}}r@{\hspace{0.3mm}}r|}
\hline
\multicolumn{1}{|c||}{$x_{min}$} & \multicolumn{3}{c|}{$\beta = 2.1250$} & \multicolumn{3}{c|}{$\beta = 2.1330$} & \multicolumn{3}{c|}{$\beta = 2.1500$} & \multicolumn{3}{c|}{$\beta = 2.1750$} & \multicolumn{3}{c|}{$\beta = 2.2000$} \\ \hline
($\infty$,1.0,0.65) &    6.142 &    (0.012) & $    \times 10^{2}$  &    6.420 &    (0.014) & $    \times 10^{2}$  &    7.112 &    (0.013) & $    \times 10^{2}$  &    8.282 &    (0.018) & $    \times 10^{2}$  &    9.600 &    (0.023) & $    \times 10^{2}$  \\ \hline
($\infty$,1.0,0.90) &    6.144 &    (0.012) & $    \times 10^{2}$  &    6.422 &    (0.014) & $    \times 10^{2}$  &    7.114 &    (0.013) & $    \times 10^{2}$  &    8.282 &    (0.017) & $    \times 10^{2}$  &    9.598 &    (0.023) & $    \times 10^{2}$  \\ \hline
($\infty$,$\infty$,0.65) &    6.143 &    (0.012) & $    \times 10^{2}$  &    6.421 &    (0.014) & $    \times 10^{2}$  &    7.113 &    (0.013) & $    \times 10^{2}$  &    8.281 &    (0.018) & $    \times 10^{2}$  &    9.598 &    (0.023) & $    \times 10^{2}$  \\ \hline
\it ($\infty$,$\infty$,0.90) & \it 6.143 & \it (0.012) & $ \it \times 10^\hboxscript{2}$  & \it 6.421 & \it (0.015) & $ \it \times 10^\hboxscript{2}$  & \it 7.113 & \it (0.013) & $ \it \times 10^\hboxscript{2}$  & \it 8.282 & \it (0.018) & $ \it \times 10^\hboxscript{2}$  & \it 9.599 & \it (0.023) & $ \it \times 10^\hboxscript{2}$  \\ \hline
\sf ($\infty$,$\infty$,$\infty$) & \sf 6.139 & \sf (0.012) & $ \sf \times 10^\hboxsans{2}$  & \sf 6.417 & \sf (0.014) & $ \sf \times 10^\hboxsans{2}$  & \sf 7.110 & \sf (0.013) & $ \sf \times 10^\hboxsans{2}$  & \sf 8.284 & \sf (0.018) & $ \sf \times 10^\hboxsans{2}$  & \sf 9.605 & \sf (0.022) & $ \sf \times 10^\hboxsans{2}$  \\ \hline
\end{tabular} \end{center}
%%\par\vspace*{3mm}\par
%%
\begin{center}  %%\hspace*{-4.0cm}
\begin{tabular}{|c||r@{\hspace{0.8mm}}r@{\hspace{0.3mm}}r|r@{\hspace{0.8mm}}r@{\hspace{0.3mm}}r|r@{\hspace{0.8mm}}r@{\hspace{0.3mm}}r|r@{\hspace{0.8mm}}r@{\hspace{0.3mm}}r|r@{\hspace{0.8mm}}r@{\hspace{0.3mm}}r|}
\hline
\multicolumn{1}{|c||}{$x_{min}$} & \multicolumn{3}{c|}{$\beta = 2.2163$} & \multicolumn{3}{c|}{$\beta = 2.2500$} & \multicolumn{3}{c|}{$\beta = 2.3000$} & \multicolumn{3}{c|}{$\beta = 2.3500$} & \multicolumn{3}{c|}{$\beta = 2.4000$} \\ \hline
($\infty$,1.0,0.65) &    1.063 &    (0.002) & $    \times 10^{3}$  &    1.309 &    (0.003) & $    \times 10^{3}$  &    1.777 &    (0.004) & $    \times 10^{3}$  &    2.420 &    (0.008) & $    \times 10^{3}$  &    3.297 &    (0.013) & $    \times 10^{3}$  \\ \hline
($\infty$,1.0,0.90) &    1.063 &    (0.002) & $    \times 10^{3}$  &    1.309 &    (0.003) & $    \times 10^{3}$  &    1.777 &    (0.004) & $    \times 10^{3}$  &    2.420 &    (0.008) & $    \times 10^{3}$  &    3.296 &    (0.013) & $    \times 10^{3}$  \\ \hline
($\infty$,$\infty$,0.65) &    1.063 &    (0.002) & $    \times 10^{3}$  &    1.309 &    (0.003) & $    \times 10^{3}$  &    1.777 &    (0.004) & $    \times 10^{3}$  &    2.420 &    (0.008) & $    \times 10^{3}$  &    3.297 &    (0.013) & $    \times 10^{3}$  \\ \hline
\it ($\infty$,$\infty$,0.90) & \it 1.063 & \it (0.002) & $ \it \times 10^\hboxscript{3}$  & \it 1.309 & \it (0.003) & $ \it \times 10^\hboxscript{3}$  & \it 1.777 & \it (0.004) & $ \it \times 10^\hboxscript{3}$  & \it 2.420 & \it (0.008) & $ \it \times 10^\hboxscript{3}$  & \it 3.296 & \it (0.014) & $ \it \times 10^\hboxscript{3}$  \\ \hline
\sf ($\infty$,$\infty$,$\infty$) & \sf 1.064 & \sf (0.002) & $ \sf \times 10^\hboxsans{3}$  & \sf 1.309 & \sf (0.003) & $ \sf \times 10^\hboxsans{3}$  & \sf 1.776 & \sf (0.004) & $ \sf \times 10^\hboxsans{3}$  & \sf 2.421 & \sf (0.008) & $ \sf \times 10^\hboxsans{3}$  & \sf 3.297 & \sf (0.013) & $ \sf \times 10^\hboxsans{3}$  \\ \hline
\end{tabular} \end{center}
%%\par\vspace*{3mm}\par
\caption{
  Estimated susceptibilities $\chi_{A,\infty}$ as a function of $\beta$,
  from various extrapolations.
  Error bar is one standard deviation (statistical errors only).
  All extrapolations use $s=2$ and $n=13$.
  The indicated $x_{min}$ values apply to $L=8,16,32$, respectively;
  we always take $x_{min} = 0.14, 0$ for $L=64,128$.
  Our preferred fit is shown in {\em italic}\/;
   other good fits are shown in {\sf sans-serif};
   bad fits are shown in {\rm roman}.
}
\label{estrap_chi_adj}

\end{table}
\clearpage \addtocounter{table}{-1}
\begin{table}
\addtolength{\tabcolsep}{-1.0mm}
\protect\footnotesize
\tabcolsep 2.5pt
\doublerulesep 1.5pt
\begin{center}  %%\hspace*{-4.0cm}
\begin{tabular}{|c||r@{\hspace{0.8mm}}r@{\hspace{0.3mm}}r|r@{\hspace{0.8mm}}r@{\hspace{0.3mm}}r|r@{\hspace{0.8mm}}r@{\hspace{0.3mm}}r|r@{\hspace{0.8mm}}r@{\hspace{0.3mm}}r|r@{\hspace{0.8mm}}r@{\hspace{0.3mm}}r|}
\hline
\multicolumn{1}{|c||}{$x_{min}$} & \multicolumn{3}{c|}{$\beta = 2.4500$} & \multicolumn{3}{c|}{$\beta = 2.5000$} & \multicolumn{3}{c|}{$\beta = 2.5500$} & \multicolumn{3}{c|}{$\beta = 2.6000$} & \multicolumn{3}{c|}{$\beta = 2.6500$} \\ \hline
($\infty$,1.0,0.65) &    4.490 &    (0.022) & $    \times 10^{3}$  &    6.106 &    (0.033) & $    \times 10^{3}$  &    8.515 &    (0.055) & $    \times 10^{3}$  &    1.168 &    (0.008) & $    \times 10^{4}$  &    1.605 &    (0.011) & $    \times 10^{4}$  \\ \hline
($\infty$,1.0,0.90) &    4.488 &    (0.022) & $    \times 10^{3}$  &    6.103 &    (0.034) & $    \times 10^{3}$  &    8.512 &    (0.059) & $    \times 10^{3}$  &    1.172 &    (0.009) & $    \times 10^{4}$  &    1.605 &    (0.013) & $    \times 10^{4}$  \\ \hline
($\infty$,$\infty$,0.65) &    4.490 &    (0.022) & $    \times 10^{3}$  &    6.105 &    (0.033) & $    \times 10^{3}$  &    8.512 &    (0.057) & $    \times 10^{3}$  &    1.168 &    (0.008) & $    \times 10^{4}$  &    1.604 &    (0.011) & $    \times 10^{4}$  \\ \hline
\it ($\infty$,$\infty$,0.90) & \it 4.489 & \it (0.022) & $ \it \times 10^\hboxscript{3}$  & \it 6.102 & \it (0.034) & $ \it \times 10^\hboxscript{3}$  & \it 8.508 & \it (0.059) & $ \it \times 10^\hboxscript{3}$  & \it 1.172 & \it (0.009) & $ \it \times 10^\hboxscript{4}$  & \it 1.604 & \it (0.013) & $ \it \times 10^\hboxscript{4}$  \\ \hline
\sf ($\infty$,$\infty$,$\infty$) & \sf 4.485 & \sf (0.022) & $ \sf \times 10^\hboxsans{3}$  & \sf 6.097 & \sf (0.033) & $ \sf \times 10^\hboxsans{3}$  & \sf 8.517 & \sf (0.059) & $ \sf \times 10^\hboxsans{3}$  & \sf 1.172 & \sf (0.009) & $ \sf \times 10^\hboxsans{4}$  & \sf 1.605 & \sf (0.013) & $ \sf \times 10^\hboxsans{4}$  \\ \hline
\end{tabular} \end{center}
\par\vspace*{3mm}\par
\begin{center}  %%\hspace*{-4.0cm}
\begin{tabular}{|c||r@{\hspace{0.8mm}}r@{\hspace{0.3mm}}r|r@{\hspace{0.8mm}}r@{\hspace{0.3mm}}r|r@{\hspace{0.8mm}}r@{\hspace{0.3mm}}r|r@{\hspace{0.8mm}}r@{\hspace{0.3mm}}r|r@{\hspace{0.8mm}}r@{\hspace{0.3mm}}r|}
\hline
\multicolumn{1}{|c||}{$x_{min}$} & \multicolumn{3}{c|}{$\beta = 2.7000$} & \multicolumn{3}{c|}{$\beta = 2.7750$} & \multicolumn{3}{c|}{$\beta = 2.8500$} & \multicolumn{3}{c|}{$\beta = 2.9250$} & \multicolumn{3}{c|}{$\beta = 3.0000$} \\ \hline
($\infty$,1.0,0.65) &    2.216 &    (0.017) & $    \times 10^{4}$  &    3.625 &    (0.028) & $    \times 10^{4}$  &    5.850 &    (0.049) & $    \times 10^{4}$  &    9.772 &    (0.090) & $    \times 10^{4}$  &    1.580 &    (0.016) & $    \times 10^{5}$  \\ \hline
($\infty$,1.0,0.90) &    2.205 &    (0.022) & $    \times 10^{4}$  &    3.628 &    (0.040) & $    \times 10^{4}$  &    5.878 &    (0.077) & $    \times 10^{4}$  &    9.816 &    (0.153) & $    \times 10^{4}$  &    1.575 &    (0.027) & $    \times 10^{5}$  \\ \hline
($\infty$,$\infty$,0.65) &    2.215 &    (0.017) & $    \times 10^{4}$  &    3.625 &    (0.029) & $    \times 10^{4}$  &    5.849 &    (0.049) & $    \times 10^{4}$  &    9.766 &    (0.092) & $    \times 10^{4}$  &    1.579 &    (0.016) & $    \times 10^{5}$  \\ \hline
\it ($\infty$,$\infty$,0.90) & \it 2.205 & \it (0.022) & $ \it \times 10^\hboxscript{4}$  & \it 3.627 & \it (0.040) & $ \it \times 10^\hboxscript{4}$  & \it 5.882 & \it (0.079) & $ \it \times 10^\hboxscript{4}$  & \it 9.820 & \it (0.159) & $ \it \times 10^\hboxscript{4}$  & \it 1.576 & \it (0.028) & $ \it \times 10^\hboxscript{5}$  \\ \hline
\sf ($\infty$,$\infty$,$\infty$) & \sf 2.209 & \sf (0.021) & $ \sf \times 10^\hboxsans{4}$  & \sf 3.629 & \sf (0.041) & $ \sf \times 10^\hboxsans{4}$  & \sf 5.877 & \sf (0.076) & $ \sf \times 10^\hboxsans{4}$  & \sf 9.801 & \sf (0.155) & $ \sf \times 10^\hboxsans{4}$  & \sf 1.570 & \sf (0.027) & $ \sf \times 10^\hboxsans{5}$  \\ \hline
\end{tabular} \end{center}
\par\vspace*{3mm}\par
\begin{center}  %%\hspace*{-4.0cm}
\begin{tabular}{|c||r@{\hspace{0.8mm}}r@{\hspace{0.3mm}}r|r@{\hspace{0.8mm}}r@{\hspace{0.3mm}}r|r@{\hspace{0.8mm}}r@{\hspace{0.3mm}}r|r@{\hspace{0.8mm}}r@{\hspace{0.3mm}}r|r@{\hspace{0.8mm}}r@{\hspace{0.3mm}}r|}
\hline
\multicolumn{1}{|c||}{$x_{min}$} & \multicolumn{3}{c|}{$\beta = 3.0750$} & \multicolumn{3}{c|}{$\beta = 3.1500$} & \multicolumn{3}{c|}{$\beta = 3.2250$} & \multicolumn{3}{c|}{$\beta = 3.3000$} & \multicolumn{3}{c|}{$\beta = 3.3750$} \\ \hline
($\infty$,1.0,0.65) &    2.624 &    (0.029) & $    \times 10^{5}$  &    4.383 &    (0.051) & $    \times 10^{5}$  &    7.267 &    (0.096) & $    \times 10^{5}$  &    1.227 &    (0.017) & $    \times 10^{6}$  &    2.015 &    (0.030) & $    \times 10^{6}$  \\ \hline
($\infty$,1.0,0.90) &    2.646 &    (0.053) & $    \times 10^{5}$  &    4.428 &    (0.095) & $    \times 10^{5}$  &    7.255 &    (0.179) & $    \times 10^{5}$  &    1.201 &    (0.031) & $    \times 10^{6}$  &    1.994 &    (0.058) & $    \times 10^{6}$  \\ \hline
($\infty$,$\infty$,0.65) &    2.621 &    (0.030) & $    \times 10^{5}$  &    4.376 &    (0.052) & $    \times 10^{5}$  &    7.259 &    (0.097) & $    \times 10^{5}$  &    1.225 &    (0.017) & $    \times 10^{6}$  &    2.016 &    (0.030) & $    \times 10^{6}$  \\ \hline
\it ($\infty$,$\infty$,0.90) & \it 2.647 & \it (0.055) & $ \it \times 10^\hboxscript{5}$  & \it 4.425 & \it (0.097) & $ \it \times 10^\hboxscript{5}$  & \it 7.245 & \it (0.184) & $ \it \times 10^\hboxscript{5}$  & \it 1.199 & \it (0.031) & $ \it \times 10^\hboxscript{6}$  & \it 1.991 & \it (0.058) & $ \it \times 10^\hboxscript{6}$  \\ \hline
\sf ($\infty$,$\infty$,$\infty$) & \sf 2.639 & \sf (0.054) & $ \sf \times 10^\hboxsans{5}$  & \sf 4.413 & \sf (0.098) & $ \sf \times 10^\hboxsans{5}$  & \sf 7.261 & \sf (0.182) & $ \sf \times 10^\hboxsans{5}$  & \sf 1.203 & \sf (0.032) & $ \sf \times 10^\hboxsans{6}$  & \sf 2.003 & \sf (0.060) & $ \sf \times 10^\hboxsans{6}$  \\ \hline
\end{tabular} \end{center}
\par\vspace*{3mm}\par
\begin{center}  %%\hspace*{-4.0cm}
\begin{tabular}{|c||r@{\hspace{0.8mm}}r@{\hspace{0.3mm}}r|r@{\hspace{0.8mm}}r@{\hspace{0.3mm}}r|r@{\hspace{0.8mm}}r@{\hspace{0.3mm}}r|r@{\hspace{0.8mm}}r@{\hspace{0.3mm}}r|r@{\hspace{0.8mm}}r@{\hspace{0.3mm}}r|}
\hline
\multicolumn{1}{|c||}{$x_{min}$} & \multicolumn{3}{c|}{$\beta = 3.4500$} & \multicolumn{3}{c|}{$\beta = 3.5250$} & \multicolumn{3}{c|}{$\beta = 3.6000$} & \multicolumn{3}{c|}{$\beta = 3.6750$} & \multicolumn{3}{c|}{$\beta = 3.7500$} \\ \hline
($\infty$,1.0,0.65) &    3.386 &    (0.054) & $    \times 10^{6}$  &    5.601 &    (0.090) & $    \times 10^{6}$  &    9.611 &    (0.170) & $    \times 10^{6}$  &    1.625 &    (0.029) & $    \times 10^{7}$  &    2.735 &    (0.055) & $    \times 10^{7}$  \\ \hline
($\infty$,1.0,0.90) &    3.337 &    (0.104) & $    \times 10^{6}$  &    5.463 &    (0.178) & $    \times 10^{6}$  &    9.370 &    (0.325) & $    \times 10^{6}$  &    1.584 &    (0.054) & $    \times 10^{7}$  &    2.666 &    (0.096) & $    \times 10^{7}$  \\ \hline
($\infty$,$\infty$,0.65) &    3.389 &    (0.053) & $    \times 10^{6}$  &    5.604 &    (0.090) & $    \times 10^{6}$  &    9.617 &    (0.175) & $    \times 10^{6}$  &    1.622 &    (0.029) & $    \times 10^{7}$  &    2.731 &    (0.055) & $    \times 10^{7}$  \\ \hline
\it ($\infty$,$\infty$,0.90) & \it 3.334 & \it (0.105) & $ \it \times 10^\hboxscript{6}$  & \it 5.471 & \it (0.182) & $ \it \times 10^\hboxscript{6}$  & \it 9.392 & \it (0.331) & $ \it \times 10^\hboxscript{6}$  & \it 1.587 & \it (0.054) & $ \it \times 10^\hboxscript{7}$  & \it 2.668 & \it (0.096) & $ \it \times 10^\hboxscript{7}$  \\ \hline
\sf ($\infty$,$\infty$,$\infty$) & \sf 3.348 & \sf (0.107) & $ \sf \times 10^\hboxsans{6}$  & \sf 5.473 & \sf (0.184) & $ \sf \times 10^\hboxsans{6}$  & \sf 9.323 & \sf (0.346) & $ \sf \times 10^\hboxsans{6}$  & \sf 1.579 & \sf (0.060) & $ \sf \times 10^\hboxsans{7}$  & \sf 2.600 & \sf (0.108) & $ \sf \times 10^\hboxsans{7}$  \\ \hline
\end{tabular} \end{center}
\par\vspace*{3mm}\par
\begin{center}  %%\hspace*{-4.0cm}
\begin{tabular}{|c||r@{\hspace{0.8mm}}r@{\hspace{0.3mm}}r|r@{\hspace{0.8mm}}r@{\hspace{0.3mm}}r|r@{\hspace{0.8mm}}r@{\hspace{0.3mm}}r|r@{\hspace{0.8mm}}r@{\hspace{0.3mm}}r|r@{\hspace{0.8mm}}r@{\hspace{0.3mm}}r|}
\hline
\multicolumn{1}{|c||}{$x_{min}$} & \multicolumn{3}{c|}{$\beta = 3.8250$} & \multicolumn{3}{c|}{$\beta = 3.9000$} & \multicolumn{3}{c|}{$\beta = 3.9750$} & \multicolumn{3}{c|}{$\beta = 4.0500$} & \multicolumn{3}{c|}{$\beta = 4.1250$} \\ \hline
($\infty$,1.0,0.65) &    4.682 &    (0.094) & $    \times 10^{7}$  &    7.846 &    (0.176) & $    \times 10^{7}$  &    1.345 &    (0.030) & $    \times 10^{8}$  &    2.310 &    (0.053) & $    \times 10^{8}$  &    3.949 &    (0.092) & $    \times 10^{8}$  \\ \hline
($\infty$,1.0,0.90) &    4.570 &    (0.159) & $    \times 10^{7}$  &    7.641 &    (0.281) & $    \times 10^{7}$  &    1.315 &    (0.048) & $    \times 10^{8}$  &    2.254 &    (0.082) & $    \times 10^{8}$  &    3.847 &    (0.144) & $    \times 10^{8}$  \\ \hline
($\infty$,$\infty$,0.65) &    4.668 &    (0.094) & $    \times 10^{7}$  &    7.812 &    (0.175) & $    \times 10^{7}$  &    1.336 &    (0.031) & $    \times 10^{8}$  &    2.305 &    (0.057) & $    \times 10^{8}$  &    3.952 &    (0.101) & $    \times 10^{8}$  \\ \hline
\it ($\infty$,$\infty$,0.90) & \it 4.568 & \it (0.160) & $ \it \times 10^\hboxscript{7}$  & \it 7.623 & \it (0.284) & $ \it \times 10^\hboxscript{7}$  & \it 1.304 & \it (0.049) & $ \it \times 10^\hboxscript{8}$  & \it 2.252 & \it (0.088) & $ \it \times 10^\hboxscript{8}$  & \it 3.843 & \it (0.154) & $ \it \times 10^\hboxscript{8}$  \\ \hline
\sf ($\infty$,$\infty$,$\infty$) & \sf 4.449 & \sf (0.191) & $ \sf \times 10^\hboxsans{7}$  & \sf 7.271 & \sf (0.341) & $ \sf \times 10^\hboxsans{7}$  & \sf 1.270 & \sf (0.062) & $ \sf \times 10^\hboxsans{8}$  & \sf 2.158 & \sf (0.111) & $ \sf \times 10^\hboxsans{8}$  & \sf 3.657 & \sf (0.200) & $ \sf \times 10^\hboxsans{8}$  \\ \hline
\end{tabular} \end{center}
\par\vspace*{3mm}\par
\begin{center}  %%\hspace*{-4.0cm}
\begin{tabular}{|c||r@{\hspace{0.8mm}}r@{\hspace{0.3mm}}r|r@{\hspace{0.8mm}}r@{\hspace{0.3mm}}r|r@{\hspace{0.8mm}}r@{\hspace{0.3mm}}r|}
\hline
\multicolumn{1}{|c||}{$x_{min}$} & \multicolumn{3}{c|}{$\beta = 4.2000$} & \multicolumn{3}{c|}{$\beta = 4.2750$} & \multicolumn{3}{c|}{$\beta = 4.3500$} \\ \hline
($\infty$,1.0,0.65) &    6.805 &    (0.160) & $    \times 10^{8}$  &    1.158 &    (0.029) & $    \times 10^{9}$  &    1.992 &    (0.050) & $    \times 10^{9}$  \\ \hline
($\infty$,1.0,0.90) &    6.615 &    (0.244) & $    \times 10^{8}$  &    1.128 &    (0.043) & $    \times 10^{9}$  &    1.941 &    (0.073) & $    \times 10^{9}$   \\ \hline
($\infty$,$\infty$,0.65) &    6.895 &    (0.184) & $    \times 10^{8}$  &    1.181 &    (0.034) & $    \times 10^{9}$  &    2.022 &    (0.063) & $    \times 10^{9}$  \\ \hline
\it ($\infty$,$\infty$,0.90) & \it 6.749 & \it (0.270) & $ \it \times 10^\hboxscript{8}$  & \it 1.151 & \it (0.048) & $ \it \times 10^\hboxscript{9}$  & \it 1.971 & \it (0.085) & $ \it \times 10^\hboxscript{9}$  \\ \hline
\sf ($\infty$,$\infty$,$\infty$) & \sf 6.280 & \sf (0.357) & $ \sf \times 10^\hboxsans{8}$  & \sf 1.033 & \sf (0.063) & $ \sf \times 10^\hboxsans{9}$  & \sf 1.880 & \sf (0.118) & $ \sf \times 10^\hboxsans{9}$  \\ \hline
\end{tabular} \end{center}
%%\par\vspace*{3mm}\par
%%
\caption{
{\bf [Continued]}}
\end{table}
\clearpage

%%%%%%%%%%%%%%%%%%%%%%%%%%%%%%%%%%%%%%%%%%%%%%%%%%%%%%%%%%%%%%%%%%%%%%%

%%%%%%%%%%%%%%%%%%%%%%%%%%%%%%%%%%%%%%%%%%%%%%%%%%%%%%%%%%%%%%%%%%%%%%%

%%%%%%%%%%%%%%%%%%%%%%%%%%%%%%%%%%%%%%%%%%%%%%%%%%%%%%%%%%%%%%%%%%%%%%%

% From pelisset@MAFALDA.PHYSICS.NYU.EDU Mon Sep 23 14:37:17 1996
% Received: from MAFALDA.PHYSICS.NYU.EDU by GUILLE.PHYSICS.NYU.EDU (5.61/1.34)
% 	id AA23704; Mon, 23 Sep 96 14:37:16 -0400
% Received: by MAFALDA.PHYSICS.NYU.EDU (4.1/1.34)
% 	id AA07864; Mon, 23 Sep 96 14:45:27 EDT
% Date: Mon, 23 Sep 96 14:45:27 EDT
% From: pelisset@MAFALDA.PHYSICS.NYU.EDU (Andrea Pelissetto)
% Message-Id: <9609231845.AA07864@MAFALDA.PHYSICS.NYU.EDU>
% To: sokal@guille.physics.nyu.edu
% Subject: table
% Status: R

\begin{table}
%\footnotesize
%\tabcolsep 4pt        % Less than the usual 6pt
%\doublerulesep 1.5pt  % Less than the usual 2pt
\begin{center}
\begin{tabular}{|rcrc|cccc|}
\hline
\multicolumn{1}{|c}{$L$} & $\beta$ &
   \multicolumn{1}{c}{$\xi_{F,\infty}(\beta)$} & $x$ & 
$-A_1(L)/\beta$ & $R_2(\beta,L)$ & est.\ $-A_{20}^{(F)}$ &
   $\vphantom{\Bigl(} \widetilde{R}_2(\beta,L)$
\\
\hline
8 & 2.10 & 38 & 0.63  & 
$-$0.227 & $-$0.0653 & $-$0.1719 & $-$0.06659
\\
8 & 2.30 & 87  & 0.69  & 
$-$0.207 & $-$0.0510 & $-$0.1540 & $-$0.06244
\\
8 & 3.00 & 1489  & 0.86  & 
$-$0.159 & $-$0.0264 & $-$0.1219 & $-$0.05503
\\
8 & 3.75 & 31910  & 1.02  & 
$-$0.127 & $-$0.0154 & $-$0.1009 & $-$0.05016
\\
8 & 4.35 & 371700  & 1.12  & 
$-$0.110 & $-$0.0113 & $-$0.0977 & $-$0.04943
\\ \hline
16 & 2.10 & 38 & 0.57  & 
$-$0.265 & $-$0.0951 & $-$0.2512 & $-$0.05458
\\
16 & 2.30 & 87  & 0.64  & 
$-$0.242 & $-$0.0733 & $-$0.2193 & $-$0.05044
\\
16 & 3.00 & 1489  & 0.83  & 
$-$0.186 & $-$0.0363 & $-$0.1580 & $-$0.04246
\\
16 & 3.75 & 31910  & 0.98  & 
$-$0.149 & $-$0.0214 & $-$0.1326 & $-$0.03915
\\
16 & 4.35 & 371700  & 1.09  & 
$-$0.128 & $-$0.0152 & $-$0.1189 & $-$0.03738
\\ \hline
32 & 2.10 & 38 & 0.50  & 
$-$0.304 & $-$0.1337 & $-$0.3618 & $-$0.04907
\\
32 & 2.30 & 87  & 0.58  & 
$-$0.278 & $-$0.0992 & $-$0.2972 & $-$0.04369
\\
32 & 3.00 & 1489  & 0.78  & 
$-$0.213 & $-$0.0480 & $-$0.2043 & $-$0.03595
\\
32 & 3.75 & 31910  & 0.95  & 
$-$0.170 & $-$0.0279 & $-$0.1641 & $-$0.03261
\\
32 & 4.35 & 371700  & 1.06  & 
$-$0.147 & $-$0.0196 & $-$0.1442 & $-$0.03095
\\ \hline
64 & 2.10 & 38 & 0.42  & 
$-$0.343 & $-$0.1887 & $-$0.5387 & $-$0.04812
\\
64 & 2.30 & 87  & 0.51  & 
$-$0.314 & $-$0.1344 & $-$0.4174 & $-$0.04111
\\
64 & 3.00 & 1489  & 0.74  & 
$-$0.240 & $-$0.0625 & $-$0.2691 & $-$0.03254
\\
64 & 3.75 & 31910  & 0.92  & 
$-$0.192 & $-$0.0353 & $-$0.2022 & $-$0.02867
\\
64 & 4.35 & 371700  & 1.03  & 
$-$0.166 & $-$0.0249 & $-$0.1767 & $-$0.02720
\\ \hline
128 & 2.10 & 38 & 0.29  & 
$-$0.383 & $-$0.2875 & $-$0.9015 & $-$0.05386
\\
128 & 2.30 & 87  & 0.43  & 
$-$0.349 & $-$0.1835 & $-$0.6042 & $-$0.04124
\\
128 & 3.00 & 1489  & 0.69  & 
$-$0.268 & $-$0.0794 & $-$0.3477 & $-$0.03034
\\
128 & 3.75 & 31910  & 0.88  & 
$-$0.214 & $-$0.0441 & $-$0.2542 & $-$0.02637
\\
128 & 4.35 & 371700  & 1.00  & 
$-$0.185 & $-$0.0295 & $-$0.1922 & $-$0.02374
\\ \hline
256 & 2.30 & 87  & 0.32  & 
$-$0.385 & $-$0.2662 & $-$0.9617 & $-$0.04579
\\
\hline
\end{tabular}
\end{center}
\caption{
   Comparison of Monte Carlo data with finite-volume perturbation theory
   for $\xi_F^{(2nd)}(\beta,L)$.
   Here $-A_1(L)/\beta$ is the first-order perturbative correction;
   $R_2(\beta,L)$ is the remainder to first-order perturbation theory;
   and ``est.\ $-A_{20}^{(F)}$'' denotes
   $\beta^2 R_2(\beta,L) + A_{22} \log^2 L + A_{21} \log L$,
   which as $\beta\to\infty$ at fixed $L$ should tend to
   $-A_{20}^{(F)} + O(\log^2 L/L^2)$.
}
\label{table_R2_xiF}
\end{table}

\clearpage

\begin{table}
%\footnotesize
%\tabcolsep 4pt        % Less than the usual 6pt
%\doublerulesep 1.5pt  % Less than the usual 2pt
\begin{center}
\begin{tabular}{|rcrc|cccc|}
\hline
\multicolumn{1}{|c}{$L$} & $\beta$ &
   \multicolumn{1}{c}{$\xi_{F,\infty}(\beta)$} & $x$ & 
$-A_1(L)/\beta$ & $R_2(\beta,L)$ & est.\ $-A_{20}^{(A)}$ &
   $\vphantom{\Bigl(} \widetilde{R}_2(\beta,L)$
\\
\hline
8 & 2.10 & 38 & 0.63  & 
$-$0.293 & $-$0.0872 & $-$0.2389 & $-$0.08890
\\
8 & 2.30 & 87  & 0.69  & 
$-$0.267 & $-$0.0688 & $-$0.2183 & $-$0.08413
\\
8 & 3.00 & 1489  & 0.86  & 
$-$0.205 & $-$0.0361 & $-$0.1795 & $-$0.07517
\\
8 & 3.75 & 31910  & 1.02  & 
$-$0.164 & $-$0.0213 & $-$0.1543 & $-$0.06934
\\
8 & 4.35 & 371700  & 1.12  & 
$-$0.141 & $-$0.0156 & $-$0.1489 & $-$0.06809
\\ \hline
16 & 2.10 & 38 & 0.57  & 
$-$0.326 & $-$0.1234 & $-$0.3364 & $-$0.07078
\\
16 & 2.30 & 87  & 0.64  & 
$-$0.297 & $-$0.0956 & $-$0.2978 & $-$0.06576
\\
16 & 3.00 & 1489  & 0.83  & 
$-$0.228 & $-$0.0477 & $-$0.2220 & $-$0.05589
\\
16 & 3.75 & 31910  & 0.98  & 
$-$0.182 & $-$0.0282 & $-$0.1886 & $-$0.05156
\\
16 & 4.35 & 371700  & 1.09  & 
$-$0.157 & $-$0.0200 & $-$0.1713 & $-$0.04931
\\ \hline
32 & 2.10 & 38 & 0.50  & 
$-$0.363 & $-$0.1686 & $-$0.4669 & $-$0.06191
\\
32 & 2.30 & 87  & 0.58  & 
$-$0.331 & $-$0.1264 & $-$0.3919 & $-$0.05566
\\
32 & 3.00 & 1489  & 0.78  & 
$-$0.254 & $-$0.0613 & $-$0.2748 & $-$0.04592
\\
32 & 3.75 & 31910  & 0.95  & 
$-$0.203 & $-$0.0355 & $-$0.2229 & $-$0.04159
\\
32 & 4.35 & 371700  & 1.06  & 
$-$0.175 & $-$0.0251 & $-$0.1979 & $-$0.03951
\\ \hline
64 & 2.10 & 38 & 0.42  & 
$-$0.401 & $-$0.2319 & $-$0.6701 & $-$0.05913
\\
64 & 2.30 & 87  & 0.51  & 
$-$0.366 & $-$0.1680 & $-$0.5362 & $-$0.05138
\\
64 & 3.00 & 1489  & 0.74  & 
$-$0.281 & $-$0.0784 & $-$0.3530 & $-$0.04080
\\
64 & 3.75 & 31910  & 0.92  & 
$-$0.225 & $-$0.0442 & $-$0.2694 & $-$0.03596
\\
64 & 4.35 & 371700  & 1.03  & 
$-$0.194 & $-$0.0311 & $-$0.2351 & $-$0.03398
\\ \hline
128 & 2.10 & 38 & 0.29  & 
$-$0.440 & $-$0.3318 & $-$1.0281 & $-$0.06216
\\
128 & 2.30 & 87  & 0.43  & 
$-$0.402 & $-$0.2252 & $-$0.7559 & $-$0.05060
\\
128 & 3.00 & 1489  & 0.69  & 
$-$0.308 & $-$0.0985 & $-$0.4509 & $-$0.03764
\\
128 & 3.75 & 31910  & 0.88  & 
$-$0.247 & $-$0.0546 & $-$0.3326 & $-$0.03262
\\
128 & 4.35 & 371700  & 1.00  & 
$-$0.213 & $-$0.0367 & $-$0.2584 & $-$0.02947
\\ \hline
256 & 2.30 & 87  & 0.32  & 
$-$0.438 & $-$0.3130 & $-$1.1307 & $-$0.05384
\\
\hline
\end{tabular}
\end{center}
\caption{
   Comparison of Monte Carlo data with finite-volume perturbation theory
   for $\xi_A^{(2nd)}(\beta,L)$.
   Here $-A_1(L)/\beta$ is the first-order perturbative correction;
   $R_2(\beta,L)$ is the remainder to first-order perturbation theory;
   and ``est.\ $-A_{20}^{(A)}$'' denotes
   $\beta^2 R_2(\beta,L) + A_{22} \log^2 L + A_{21} \log L$,
   which as $\beta\to\infty$ at fixed $L$ should tend to
   $-A_{20}^{(A)} + O(\log^2 L/L^2)$. 
}
\label{table_R2_xiA}
\end{table}

\clearpage

\begin{table}
%\footnotesize
%\tabcolsep 4pt        % Less than the usual 6pt
%\doublerulesep 1.5pt  % Less than the usual 2pt
\begin{center}
\begin{tabular}{|rcrc|crrrr|}
\hline
\multicolumn{1}{|c}{$L$} & $\beta$ &
   \multicolumn{1}{c}{$\xi_{F,\infty}(\beta)$} & $x$ &
$-B_1(L)/\beta$ & $-B_2(L)/\beta^2$ &  $S_3(\beta,L)$ &
    est.\ $-B_{30}^{(F)}$ &
   $\vphantom{\Bigl(} \widetilde{S}_3(\beta,L)$
\\
\hline
  8 &   2.10 &      38 &   0.63 &
 $-$0.482 &  $-$0.009 &  $-$0.01053 &  $-$0.03845 &  $-$0.01084
 \\
  8 &   2.30 &      87 &   0.69 &
 $-$0.440 &  $-$0.008 &  $-$0.00726 &  $-$0.02934 &  $-$0.00983
 \\
  8 &   3.00 &    1489 &   0.86 &
 $-$0.337 &  $-$0.005 &  $-$0.00275 &  $-$0.01522 &  $-$0.00826
 \\
  8 &   3.75 &   31908 &  1.02 &
 $-$0.270 &  $-$0.003 &  $-$0.00121 &  $-$0.00462 &  $-$0.00708
 \\
  8 &   4.35 &  371706 &  1.12 &
 $-$0.233 &  $-$0.002 &  $-$0.00077 &  $-$0.00475 &  $-$0.00709
 \\ \hline
 16 &   2.10 &      38 &   0.57 &
 $-$0.622 &   0.004 &  $-$0.01012 &  $-$0.05531 &  $-$0.00440
 \\
 16 &   2.30 &      87 &   0.64 &
 $-$0.568 &   0.003 &  $-$0.00685 &  $-$0.04494 &  $-$0.00391
 \\
 16 &   3.00 &    1489 &   0.83 &
 $-$0.435 &   0.002 &  $-$0.00226 &  $-$0.02250 &  $-$0.00286
 \\
 16 &   3.75 &   31908 &   0.98 &
 $-$0.348 &   0.001 &  $-$0.00101 &  $-$0.01493 &  $-$0.00250
 \\
 16 &   4.35 &  371706 &  1.09 &
 $-$0.300 &   0.001 &  $-$0.00061 &  $-$0.01190 &  $-$0.00236
 \\ \hline
 32 &   2.10 &      38 &   0.50 &
 $-$0.762 &   0.026 &  $-$0.00466 &  $-$0.04779 &  $-$0.00104
 \\
 32 &   2.30 &      87 &   0.58 &
 $-$0.696 &   0.022 &  $-$0.00224 &  $-$0.03194 &  $-$0.00066
 \\
 32 &   3.00 &    1489 &   0.78 &
 $-$0.534 &   0.013 &  $-$0.00037 &  $-$0.01464 &  $-$0.00024
 \\
 32 &   3.75 &   31908 &   0.95 &
 $-$0.427 &   0.008 &  $-$0.00010 &  $-$0.00987 &  $-$0.00013
 \\
 32 &   4.35 &  371706 &  1.06 &
 $-$0.368 &   0.006 &  $-$0.00003 &  $-$0.00736 &  $-$0.00007
 \\ \hline
 64 &   2.10 &      38 &   0.42 &
 $-$0.902 &   0.058 &   0.00587 &  $-$0.01717 &   0.00076
 \\
 64 &   2.30 &      87 &   0.51 &
 $-$0.824 &   0.048 &   0.00585 &  $-$0.00035 &   0.00099
 \\
 64 &   3.00 &    1489 &   0.74 &
 $-$0.632 &   0.028 &   0.00278 &   0.00363 &   0.00105
 \\
 64 &   3.75 &   31908 &   0.92 &
 $-$0.505 &   0.018 &   0.00143 &   0.00375 &   0.00105
 \\
 64 &   4.35 &  371706 &  1.03 &
 $-$0.436 &   0.013 &   0.00094 &   0.00617 &   0.00108
 \\ \hline
128 &   2.10 &      38 &   0.29 &
$-$1.043 &   0.098 &   0.01976 &   0.01929 &   0.00160
 \\
128 &   2.30 &      87 &   0.43 &
 $-$0.952 &   0.082 &   0.01786 &   0.05363 &   0.00190
 \\
128 &   3.00 &    1489 &   0.69 &
 $-$0.730 &   0.048 &   0.00760 &   0.04146 &   0.00180
 \\
128 &   3.75 &   31908 &   0.88 &
 $-$0.584 &   0.031 &   0.00367 &   0.02968 &   0.00169
 \\
128 &   4.35 &  371706 &  1.00 &
 $-$0.503 &   0.023 &   0.00256 &   0.04713 &   0.00185
 \\ \hline
256 &   2.30 &      87 &   0.32 &
$-$1.080 &   0.122 &   0.03279 &   0.11648 &   0.00234
 \\
\hline
\end{tabular}
\end{center}
\caption{
   Comparison of Monte Carlo data with finite-volume perturbation theory
   for $\chi_F(\beta,L)$.
   Here $-B_1(L)/\beta$ and $-B_2(L)/\beta^2$ are the
   first-order and second-order perturbative corrections;
   $S_3(\beta,L)$ is the remainder to second-order perturbation theory;
   and ``est.\ $-B_{30}^{(F)}$'' denotes
   $\beta^3 S_3(\beta,L) + B_{33} \log^3 L + B_{32} \log^2 L + B_{31} \log L$,
   which as $\beta\to\infty$ at fixed $L$ should tend to
   $-B_{30}^{(F)} + O(\log^3 L/L^2)$.
}
\label{table_S3_chiF}
\end{table}

\clearpage

\begin{table}
%\footnotesize
%\tabcolsep 4pt        % Less than the usual 6pt
%\doublerulesep 1.5pt  % Less than the usual 2pt
\begin{center}
\begin{tabular}{|rcrc|cccc|}
\hline
\multicolumn{1}{|c}{$L$} & $\beta$ &
   \multicolumn{1}{c}{$\xi_{F,\infty}(\beta)$} & $x$ &
$-B_1(L)/\beta$ & $-B_2(L)/\beta^2$ &  $S_3(\beta,L)$ &
   $\vphantom{\Bigl(} \widetilde{S}_3(\beta,L)$
\\
\hline
  8 &   2.10 &      38 &   0.63 &
$-$1.084 &   0.316 &      $-$0.013 &      $-$0.013
 \\
  8 &   2.30 &      87 &   0.69 &
 $-$0.989 &   0.263 &      $-$0.009 &      $-$0.013
 \\
  8 &   3.00 &    1489 &   0.86 &
 $-$0.759 &   0.155 &      $-$0.004 &      $-$0.013
 \\
  8 &   3.75 &   31908 &  1.02 &
 $-$0.607 &   0.099 &      $-$0.002 &      $-$0.012
 \\
  8 &   4.35 &  371706 &  1.12 &
 $-$0.523 &   0.074 &      $-$0.001 &      $-$0.013
 \\ \hline
 16 &   2.10 &      38 &   0.57 &
$-$1.400 &   0.562 &      $-$0.047 &      $-$0.021
 \\
 16 &   2.30 &      87 &   0.64 &
$-$1.278 &   0.469 &      $-$0.036 &      $-$0.020
 \\
 16 &   3.00 &    1489 &   0.83 &
 $-$0.980 &   0.275 &      $-$0.016 &      $-$0.020
 \\
 16 &   3.75 &   31908 &   0.98 &
 $-$0.784 &   0.176 &      $-$0.008 &      $-$0.020
 \\
 16 &   4.35 &  371706 &  1.09 &
 $-$0.676 &   0.131 &      $-$0.005 &      $-$0.020
 \\ \hline
 32 &   2.10 &      38 &   0.50 &
$-$1.715 &   0.885 &      $-$0.117 &      $-$0.026
 \\
 32 &   2.30 &      87 &   0.58 &
$-$1.566 &   0.738 &      $-$0.088 &      $-$0.026
 \\
 32 &   3.00 &    1489 &   0.78 &
$-$1.201 &   0.434 &      $-$0.039 &      $-$0.025
 \\
 32 &   3.75 &   31908 &   0.95 &
 $-$0.961 &   0.278 &      $-$0.020 &      $-$0.025
 \\
 32 &   4.35 &  371706 &  1.06 &
 $-$0.828 &   0.206 &      $-$0.013 &      $-$0.025
 \\ \hline
 64 &   2.10 &      38 &   0.42 &
$-$2.030 &  1.284 &      $-$0.234 &      $-$0.030
 \\
 64 &   2.30 &      87 &   0.51 &
$-$1.854 &  1.070 &      $-$0.177 &      $-$0.030
 \\
 64 &   3.00 &    1489 &   0.74 &
$-$1.421 &   0.629 &      $-$0.078 &      $-$0.029
 \\
 64 &   3.75 &   31908 &   0.92 &
$-$1.137 &   0.403 &      $-$0.040 &      $-$0.029
 \\
 64 &   4.35 &  371706 &  1.03 &
 $-$0.980 &   0.299 &      $-$0.025 &      $-$0.029
 \\ \hline
128 &   2.10 &      38 &   0.29 &
$-$2.346 &  1.757 &      $-$0.407 &      $-$0.033
 \\
128 &   2.30 &      87 &   0.43 &
$-$2.142 &  1.465 &      $-$0.307 &      $-$0.033
 \\
128 &   3.00 &    1489 &   0.69 &
$-$1.642 &   0.861 &      $-$0.137 &      $-$0.032
 \\
128 &   3.75 &   31908 &   0.88 &
$-$1.314 &   0.551 &      $-$0.070 &      $-$0.032
 \\
128 &   4.35 &  371706 &  1.00 &
$-$1.132 &   0.410 &      $-$0.044 &      $-$0.032
 \\ \hline
256 &   2.30 &      87 &   0.32 &
$-$2.429 &  1.922 &      $-$0.488 &      $-$0.035
 \\
\hline
\end{tabular}
\end{center}
\caption{
   Comparison of Monte Carlo data with finite-volume perturbation theory
   for $\chi_A(\beta,L)$.
   Here $-B_1(L)/\beta$ and $-B_2(L)/\beta^2$ are the
   first-order and second-order perturbative corrections;
   $S_3(\beta,L)$ is the remainder to second-order perturbation theory.
}
\label{table_S3_chiA}
\end{table}

\clearpage

%%%%%%%%%%%%%%%%%%%%%%%%%%%%%%%%%%%%%%%%%%%%%%%%%%%%%%%%%%%%%%%%%%%%%%%

%%%%%%%%%%%%%%%%%%%%%%%%%%%%%%%%%%%%%%%%%%%%%%%%%%%%%%%%%%%%%%%%%%%%%%%

%%%%%%%%%%%%%%%%%%%%%%%%%%%%%%%%%%%%%%%%%%%%%%%%%%%%%%%%%%%%%%%%%%%%%%%

\begin{table}
%\footnotesize
%\tabcolsep 4pt        % Less than the usual 6pt
%\doublerulesep 1.5pt  % Less than the usual 2pt
\begin{center}
\begin{tabular}{|r|ccc|c|}
\hline
    &              &  Asymptotic    &  Asymptotic          &  Deviation \\
$L$ &  Exact $I_{1,L}$   & through $O(1)$ &  through $O(L^{-2})$ &
                                                         $\times\, L^4$ \\
\hline
4   & 
0.268229166667 & 0.269401233323 & 0.267573658648 & 0.167810
 \\ 
8   &
0.379294686625 & 0.379719033399 & 0.379262139730 & 0.133312
 \\
16  &  
0.489924494596 & 0.490036833475 & 0.489922610058 & 0.123505
 \\ 
32  &
0.600326193679 & 0.600354633552 & 0.600326077697 & 0.121615
 \\ 
64  &
0.710665301887 & 0.710672433628 & 0.710665294665 & 0.121165
 \\ 
128 &
0.820988449414 & 0.820990233704 & 0.820988448964 & 0.121053
 \\ 
 256  &  
0.931307587624 & 0.931308033781 & 0.931307587596 & 0.121026
 \\
 512  &  
1.041625722310 & 1.041625833862 & 1.041625722313 & 0.121017
 \\ 
\hline
\end{tabular}
\end{center}
\caption{
   Exact $I_{1,L}$ compared with the asymptotic expansions
   through order 1 and through order $L^{-2}$.
   Last column is the deviation from the order-$L^{-2}$ expansion,
   multiplied by $L^4$.
}
\label{table_I1}
\end{table}

\begin{table}
%\footnotesize
%\tabcolsep 4pt        % Less than the usual 6pt
%\doublerulesep 1.5pt  % Less than the usual 2pt
\begin{center}
\begin{tabular}{|r|ccc|c|}
\hline
    &              &  Asymptotic    &  Asymptotic          &  Deviation \\
$L$ &  Exact $I_{2,L}$   & through $O(1)$ &  through $O(L^{-2})$ &
                                                         $\times\, L^4$ \\
\hline
 4  &  
0.00586615668403 & 0.00386694659074 & 0.00582620995440 & 0.01022642
 \\ 
 8  &  
0.00457375479608 & 0.00386694659074 & 0.00457222688493 & 0.00625832
 \\ 
 16  &  
0.00409721602477 & 0.00386694659074 & 0.00409713277760 & 0.00545569
 \\ 
 32  &  
0.00393796470343 & 0.00386694659074 & 0.00393795966578 & 0.00528235
 \\ 
 64  &  
0.00388806680392 & 0.00386694659074 & 0.00388806649158 & 0.00524022
 \\ 
 128  &  
0.00387306824345 & 0.00386694659074 & 0.00387306822397 & 0.00522975
 \\ 
 256  &  
0.00386868741477 & 0.00386694659074 & 0.00386868741355 & 0.00522714
 \\ 
 512  &  
0.00386743440014 & 0.00386694659074 & 0.00386743440007 & 0.00522655
 \\ 
\hline
\end{tabular}
\end{center}
\caption{
   Exact $I_{2,L}$ compared with the asymptotic expansions
   through order 1 and through order $L^{-2}$.
   Last column is the deviation from the order-$L^{-2}$ expansion,
   multiplied by $L^4$.
}
\label{table_I2}
\end{table}

\begin{table}
%\footnotesize
%\tabcolsep 4pt        % Less than the usual 6pt
%\doublerulesep 1.5pt  % Less than the usual 2pt
\begin{center}
\begin{tabular}{|r|ccc|c|}
\hline
    &              &  Asymptotic    &  Asymptotic          &  Deviation \\
$L$ &  Exact $I_{3,L}$   & through $O(1)$ &  through $O(L^{-2})$ &
                                                   $\times\, L^4/\log L$ \\
\hline
 4  &  
0.00406901041667 & 0.00238025865645 & 0.00396323566772 & 0.0195329
 \\ 
 8  &  
0.00300128408196 & 0.00238025865645 & 0.00299146736254 & 0.0193366
 \\ 
 16  &  
0.00258777659718 & 0.00238025865645 & 0.00258692694629 & 0.0200833
 \\ 
 32  &  
0.00244546104223 & 0.00238025865645 & 0.00244539225724 & 0.0208112
 \\ 
 64  &  
0.00239991399033 & 0.00238025865645 & 0.00239990868873 & 0.0213870
 \\ 
 128  &  
0.00238601321704 & 0.00238025865645 & 0.00238601282254 & 0.0218254
 \\ 
 256  &  
0.00238190764109 & 0.00238025865645 & 0.00238190761248 & 0.0221619
 \\ 
 512  &  
0.00238072350112 & 0.00238025865645 & 0.00238072349908 & 0.0224257
 \\
\hline
\end{tabular}
\end{center}
\caption{
   Exact $I_{3,L}$ compared with the asymptotic expansions
   through order 1 and through order $L^{-2}$.
   Last column is the deviation from the order-$L^{-2}$ expansion,
   multiplied by $L^4/\log L$.
   The deviation from the order-$L^{-2}$ expansion can be fitted
   approximately by $0.02454 \log L/L^4 - 0.01317/L^4$. 
 }
\label{table_I3}
\end{table}

\clearpage

%%%%%%%%%%%%%%%%%%%%%%%%%%%%%%%%%%%%%%%%%%%%%%%%%%%%%%%%%%%%%%%%%%%%%%%

%%%%%%%%%%%%%%%%%%%%%%%%%%%%%%%%%%%%%%%%%%%%%%%%%%%%%%%%%%%%%%%%%%%%%%%

%%%%%%%%%%%%%%%%%%%%%%%%%%%%%%%%%%%%%%%%%%%%%%%%%%%%%%%%%%%%%%%%%%%%%%%

% BEGINNING FIGURES

%%%%%%%% XI_F %%%%%%%%%%

% Figure 2: Deviazione curva  xi_F

\begin{figure}
\vspace*{0cm} \hspace*{-0cm}
\begin{center}
\epsfxsize = 0.9\textwidth
\leavevmode\epsffile{su3_dev_xif.ps}
%%\quad \vspace{5cm}  %% TEMPORARY UNTIL FILE IS THERE
\end{center}
\vspace*{-2cm}
\caption{
   Deviation of points from fit to $F_{{\xi}_F}$ with $s=2$, $n=13$,
   $x_{min} = (\infty,\infty,\infty,0.14,0)$.
   Symbols indicate $L=8$ ($\protect\fancyplus$), 16 ($\protect\fancycross$),
   32 ($+$), 64 ($\times$), 128 ($\Box$), 256 ($\Diamond$).
   Error bars are one standard deviation.
   Curves near zero indicate statistical error bars
   ($\pm$ one standard deviation) on the function $F_{\xi_F}(x)$.
}
\label{fig_su3_dev_xif}
\end{figure}

\clearpage

%
% Figure 3: FSS Plot \xi_F
%

\begin{figure}
\vspace*{0cm} \hspace*{-0cm}
\begin{center}
\epsfxsize = 0.9\textwidth
\leavevmode\epsffile{su3_fss_xif.ps}
%%\quad \vspace{5cm}  %% TEMPORARY UNTIL FILE IS THERE
\end{center}
\vspace*{-2cm}
\caption{
   $\xi_F(\beta,2L)/\xi_F(\beta,L)$ versus $\xi_F(\beta,L)/L$.
   Symbols indicate $L=8$ ($\protect\fancyplus$), 16 ($\protect\fancycross$),
   32 ($+$), 64 ($\times$), 128 ($\Box$).
   Error bars are one standard deviation.
   Solid curve is a thirteenth-order fit in \protect\reff{fss:gen},
   with $x_{min} = (\infty,0.90,0.65,0.14,0)$ for
   $L = (8,16,32,64,128)$.
   Dotted curves are the perturbative prediction
   (\protect\ref{xiF_FSS_PT})
   through orders $1/x^2$ (upper curve) and $1/x^4$ (lower curve).
}
\label{fig_su3_fss_xif}
\end{figure}
\clearpage

%%%%%%%% chi_F %%%%%%%%%%

% Figure 4: Deviazione curva  chi_F

\begin{figure}
\vspace*{0cm} \hspace*{-0cm}
\begin{center}
\epsfxsize = 0.9\textwidth
\leavevmode\epsffile{su3_dev_chif.ps}
%%\quad \vspace{5cm}  %% TEMPORARY UNTIL FILE IS THERE
\end{center}
\vspace*{-2cm}
\caption{
   Deviation of points from fit to $F_{{\chi}_F}$ with $s=2$, $n=15$,
   $x_{min} = (\infty,\infty,\infty,0.14,0)$.
   Symbols indicate $L=8$ ($\protect\fancyplus$), 16 ($\protect\fancycross$),
   32 ($+$).
   Error bars are one standard deviation.
   Curves near zero indicate statistical error bars
   ($\pm$ one standard deviation) on the function $F_{\chi_F}(x)$.
}
\label{fig_su3_dev_chif}
\end{figure}

\clearpage

%
% Figure 5: FSS Plot \chi_F
%

\begin{figure}
\vspace*{0cm} \hspace*{-0cm}
\begin{center}
\epsfxsize = 0.9\textwidth
\leavevmode\epsffile{su3_fss_chif.ps}
%%\quad \vspace{5cm}  %% TEMPORARY UNTIL FILE IS THERE
\end{center}
\vspace*{-2cm}
\caption{
   $\chi_F(\beta,2L)/\chi_F(\beta,L)$ versus $\xi_F(\beta,L)/L$.
   Symbols indicate $L=8$ ($\protect\fancyplus$), 16 ($\protect\fancycross$),
   32 ($+$), 64 ($\times$), 128 ($\Box$).
   Error bars are one standard deviation.
   Solid curve is a fifteenth-order fit in \protect\reff{fss:gen},
   with $x_{min} = (\infty,\infty,0.80,0.14,0)$ for
   $L = (8,16,32,64,128)$.
   Dotted curves are the perturbative prediction
   (\protect\ref{chiF_FSS_PT}) through orders
   $1/x^2$ (lower curve) and $1/x^4$ (upper curve).
}
\label{fig_su3_fss_chif}
\end{figure}
\clearpage

%%%%%%%% XI_A %%%%%%%%%%

% Figure 6: Deviazione curva  xi_A

\begin{figure}
\vspace*{0cm} \hspace*{-0cm}
\begin{center}
\epsfxsize = 0.9\textwidth
\leavevmode\epsffile{su3_dev_xia.ps}
%%\quad \vspace{5cm}  %% TEMPORARY UNTIL FILE IS THERE
\end{center}
\vspace*{-2cm}
\caption{
   Deviation of points from fit to $F_{{\xi}_A}$ with $s=2$, $n=13$,
   $x_{min} = (\infty,\infty,\infty,0.14,0)$.
   Symbols indicate $L=8$ ($\protect\fancyplus$), 16 ($\protect\fancycross$),
   32 ($+$).
   Error bars are one standard deviation.
   Curves near zero indicate statistical error bars
   ($\pm$ one standard deviation) on the function $F_{\xi_A}(x)$.
}
\label{fig_su3_dev_xia}
\end{figure}

\clearpage

%
% Figure 7: FSS Plot \xi_A
%

\begin{figure}
\vspace*{0cm} \hspace*{-0cm}
\begin{center}
\epsfxsize = 0.9\textwidth
\leavevmode\epsffile{su3_fss_xia.ps}
%%\quad \vspace{5cm}  %% TEMPORARY UNTIL FILE IS THERE
\end{center}
\vspace*{-2cm}
\caption{
   $\xi_A(\beta,2L)/\xi_A(\beta,L)$ versus $\xi_F(\beta,L)/L$.
   Symbols indicate $L=8$ ($\protect\fancyplus$), 16 ($\protect\fancycross$),
   32 ($+$), 64 ($\times$), 128 ($\Box$).
   Error bars are one standard deviation.
   Solid curve is a thirteenth-order fit in \protect\reff{fss:gen},
   with $x_{min} = (\infty,\infty,\infty,0.14,0)$ for
   $L = (8,16,32,64,128)$.
   Dotted curves are the perturbative prediction
   (\protect\ref{xiA_FSS_PT}) through orders
   $1/x^2$ (upper curve) and $1/x^4$ (lower curve).
}
\label{fig_su3_fss_xia}
\end{figure}
\clearpage

%%%%%%%% CHI_A %%%%%%%%%%

% Figure 8: Deviazione curva  chi_A

\begin{figure}
\vspace*{0cm} \hspace*{-0cm}
\begin{center}
\epsfxsize = 0.9\textwidth
\leavevmode\epsffile{su3_dev_chia.ps}
%%\quad \vspace{5cm}  %% TEMPORARY UNTIL FILE IS THERE
\end{center}
\vspace*{-2cm}
\caption{
   Deviation of points from fit to $F_{{\chi}_A}$ with $s=2$, $n=14$,
   $x_{min} = (\infty,\infty,\infty,0.14,0)$.
   Symbols indicate $L=8$ ($\protect\fancyplus$), 16 ($\protect\fancycross$),
   32 ($+$).
   Error bars are one standard deviation.
   Curves near zero indicate statistical error bars
   ($\pm$ one standard deviation) on the function $F_{\chi_A}(x)$.
}
\label{fig_su3_dev_chia}
\end{figure}

\clearpage

%
% Figure 9: FSS Plot \chi_A
%

\begin{figure}
\vspace*{0cm} \hspace*{-0cm}
\begin{center}
\epsfxsize = 0.9\textwidth
\leavevmode\epsffile{su3_fss_chia.ps}
%%\quad \vspace{5cm}  %% TEMPORARY UNTIL FILE IS THERE
\end{center}
\vspace*{-2cm}
\caption{
   $\chi_A(\beta,2L)/\chi_A(\beta,L)$ versus $\xi_F(\beta,L)/L$.
   Symbols indicate $L=8$ ($\protect\fancyplus$), 16 ($\protect\fancycross$),
   32 ($+$), 64 ($\times$), 128 ($\Box$).
   Error bars are one standard deviation.
   Solid curve is a fourteenth-order fit in \protect\reff{fss:gen},
   with $x_{min} = (\infty,\infty,0.90,0.14,0)$ for
   $L = (8,16,32,64,128)$.
   Dotted curves are the perturbative prediction
   (\protect\ref{chiA_FSS_PT}) through orders
   $1/x^2$ (lower curve) and $1/x^4$ (upper curve).
}
\label{fig_su3_fss_chia}
\end{figure}
\clearpage

%%%%%%%%%%%%%%%% XI_F / XI_A %%%%%%%%%%%%%%%%

% Figure 10: Deviazione curva  xi_F(L)/xi_A(L)

\clearpage

%
% Figure 11: FSS Plot \xi_F(L)/\xi_A(L) versus xi_F(L)/L
%

\clearpage

%
% Figure 12: (xi_F/xi_A)(2L)/(xi_F/xi_A)(L) vs xi_F(L)/L
%

\clearpage

%%%%%%%% RVTP %%%%%%%%%%

%
% Figures ?a and ?b:  bar{v}(z) and v(z)
%

\begin{figure}[p]
\vspace*{-0.5cm} \hspace*{-0cm}
\begin{center}
\epsfxsize = 4.3in
\leavevmode\epsffile{vbar.ps} \\
\vspace*{-5mm}
\epsfxsize = 4.3in
\leavevmode\epsffile{vfunction.ps}
\end{center}
\vspace*{-20mm}
\caption{
   The scaling functions $\bar{v}(z)$ and $v(z)$
   plotted versus $z \equiv \xi_{F,\infty}/L$,
   for the $SU(3)$ chiral model.
   Symbols indicate $L=8$ ($\protect\fancyplus$), 16 ($\protect\fancycross$),
   32 ($+$), 64 ($\times$), 128 ($\Box$), 256 ($\Diamond$).
   Note that $\bar{v}(z) \sim 1$ and $v(z) \sim \log^2 z$
   as $z \to\infty$.
}
\label{fig_bar_v_AND_v}
\end{figure}

\clearpage

% Figure ?:  \bar{g}_\Delta(z)

\clearpage

%

% Figure 1: RVTP

\clearpage

%
% Figure 13a and 13b: energy versus PT
%

\begin{figure}[p]
\vspace*{-0.5cm} \hspace*{-0cm}
\begin{center}
\epsfxsize = 4.3in
\leavevmode\epsffile{su3_enerF.ps} \\
\vspace*{-5mm}
\epsfxsize = 4.3in
\leavevmode\epsffile{su3_enerF-pert.ps}
\end{center}
\vspace*{-20mm}
\caption{
   (a) Fundamental energy $E_F$ versus $\beta$.
   Each point comes from the largest lattice 
   available at a given $\beta$: $L=64$ ($\times$),
   $L=128$ ($\Box$) or $L=256$ ($\Diamond$).
   Error bars (usually invisible) are statistical error 
   (one standard deviation)
   plus a conservative estimate of the systematic error due to
   finite-size corrections.
   Dashed curves are the perturbative prediction
   (\protect\ref{energy_F}) through orders
   $1/\beta$ (top curve), $1/\beta^2$ (middle curve),
   and $1/\beta^3$ (bottom curve). 
   (b)    Deviations of fundamental energy $E_F$ from
   three-loop perturbative prediction (\protect\ref{energy_F}), plotted
   versus $1/\beta^4$.
   Solid curve corresponds to the fit $E_F-E_F^{(3-loop)} = 
   k_4/\beta^4 + k_5/\beta^5$ for $\beta \ge 2.35$.
}
\label{fig_su3_enerF}
\end{figure}
\clearpage

%%%

%
% Figure 14a and 14b: Adjoint energy versus PT
%

\begin{figure}[p]
\vspace*{-0.5cm} \hspace*{-0cm}
\begin{center}
\epsfxsize = 4.3in
\leavevmode\epsffile{su3_enerA.ps} \\
\vspace*{-5mm}
\epsfxsize = 4.3in
\leavevmode\epsffile{su3_enerA-pert.ps}
\end{center}
\vspace*{-20mm}
\caption{
   (a) Adjoint energy $E_A$ versus $\beta$.
   Each point comes from the largest lattice
   available at a given $\beta$: $L=64$ ($\times$),
   $L=128$ ($\Box$) or $L=256$ ($\Diamond$).
   Error bars (usually invisible) are statistical error
   (one standard deviation)
   plus a conservative estimate of the systematic error due to
   finite-size corrections.
   Dashed curves are the perturbative prediction
   (\protect\ref{energy_A}) through orders
   $1/\beta$ (lower curve) and $1/\beta^2$ (upper curve).
   (b)    Deviations of fundamental energy $E_A$ from
   two-loop perturbative prediction (\protect\ref{energy_A}), plotted
   versus $1/\beta^3$.
   Solid curve corresponds to the fit $E_A-E_A^{(2-loop)} =
   k_3/\beta^3 + k_4/\beta^4$ for $\beta \ge 2.35$.
}
\label{fig_su3_enerA}
\end{figure}
\clearpage

%
% Figure 16: AS xi_F
%

\begin{figure}[p]
\vspace*{-0.5cm} \hspace*{-0cm}
\begin{center}
\epsfxsize = 4.3in
\leavevmode\epsffile{su3_scal_xif.ps} \\
\vspace*{-5mm}
\epsfxsize = 4.3in
\leavevmode\epsffile{su3_scal_xif_beta-2.ps}
\end{center}
\vspace*{-20mm}
\caption{
(a) $\xi^{(2nd)}_{F,\infty,estimate\,(\infty,0.90,0.65)} /
    \xi^{(exp)}_{F,\infty,theor}$
   versus $\beta$.
   Error bars are one standard deviation (statistical error only).
   There are four versions of $\xi^{(exp)}_{F,\infty,theor}$:
   standard perturbation theory in $1/\beta$ gives points
   $+$ (2-loop) and $\times$ (3-loop);
   ``improved'' perturbation theory in $1-E$ gives points
   $\Box$ (2-loop) and $\Diamond$ (3-loop). Dotted line is
   the Monte Carlo prediction
   $\widetilde{C}_{\xi_F^{(2nd)}} /
    \widetilde{C}_{\xi_F^{(exp)}} = 0.987 \pm 0.002$ \protect\cite{Rossi_94a}.
   (b) Same ratio plotted versus $1/\beta^2$.
   The lower solid line is the fit $\kappa_0 + \kappa_2/\beta^2$
   to the standard 3-loop estimates ($\times$) 
   for $\beta \ge 2.60$. The upper solid line
   is the constant fit $\kappa'_0$ to the ``improved'' 3-loop estimates ($\Diamond$)
   for $\beta \ge 2.60$. Dashed line is
   the Monte Carlo prediction
   $\widetilde{C}_{\xi_F^{(2nd)}} /
    \widetilde{C}_{\xi_F^{(exp)}} = 0.987 \pm 0.002$ \protect\cite{Rossi_94a}.
}
\label{fig_su3_scal_xif}
\end{figure}

\clearpage

%
% Figure 17: AS chi_F
%

\begin{figure}[p]
\vspace*{-0.5cm} \hspace*{-0cm}
\begin{center}
\epsfxsize = 4.3in
\leavevmode\epsffile{su3_scal_chif.ps} \\
\vspace*{-5mm}
\epsfxsize = 4.3in
\leavevmode\epsffile{su3_scal_chif_beta-2.ps}
\end{center}
\vspace*{-20mm}
\caption{
(a)
   $[\chi_{F,\infty,estimate\,(\infty,\infty,0.80)}] /
    [\chi_{F,\infty,theor} \protect\hbox{ without the prefactor } 
    \widetilde{C}_{\chi_F}]$
   versus $\beta$.
   Error bars are one standard deviation (statistical error only).
   There are four versions of $\chi_{F,\infty,theor}$:
   standard perturbation theory in $1/\beta$ gives points
   $+$ (2-loop) and $\times$ (3-loop);
   ``improved'' perturbation theory in $1-E$ gives points
   $\Box$ (2-loop) and $\Diamond$ (3-loop). 
   For clarity, error bars are shown only for the ``improved''
   three-loop estimates.
   (b) Same ratio plotted versus $1/\beta^2$.
   The lower solid line is the fit $\kappa_0 + \kappa_2/\beta^2$
   to the standard 3-loop estimates ($\times$)
   for $\beta \ge 2.65$. The upper solid line
   is the constant fit $\kappa'_0$ to the ``improved'' 3-loop 
   estimates ($\Diamond$) for $\beta \ge 2.55$.
}
\label{fig_su3_scal_chif}
\end{figure}

\clearpage

%
% Figure 19: AS chi_F/xi_F^2
%

\begin{figure}[p]
\vspace*{-0.5cm} \hspace*{-0cm}
\begin{center}
\epsfxsize = 3.5in
\leavevmode\epsffile{su3_scal_chif_ov_xif2.ps} \\
\vspace*{-8mm}
\epsfxsize = 3.5in
\leavevmode\epsffile{su3_scal_chif_ov_xif2_beta-3.ps}
\end{center}
\vspace*{-20mm}
\caption{
(a)
  $[(\chi_F/\xi^{(2nd)^2}_F)_{\infty,estimate(\infty,\infty,\infty)}]\,
   \times \,[\widetilde{C}^{\protect\B}_{\xi^{(exp)}_F}]^2 \, /$ \hfill\break
  $[(\chi_F/\xi^{(2nd)^2}_F)_{\infty,theor} \hbox{ without prefactors }
    \widetilde{C}_{\chi_F} \hbox{ and } \widetilde{C}_{\xi^{(2nd)}_F}]$
   versus $\beta$.
%   $\widetilde{C}_{\chi_F} \times $
%   $[ \protect\widetilde{C}^{(\protect\B)}_{\xi^{(exp)}_F} 
%    / \protect\widetilde{C}_{\xi^{(2nd)}_F} ]^2$
   Error bars are one standard deviation (statistical error only).
   There are six versions of $(\chi_F/\xi_F^2)_{\infty,theor}$:
   standard perturbation theory in $1/\beta$ gives points
   $+$ (2-loop), $\times$ (3-loop) and $\protect\fancyplus$  (4-loop);
   ``improved'' perturbation theory in $1-E$ gives points
   $\Box$ (2-loop), $\Diamond$ (3-loop) and $\bigcirc$ (4-loop).
   The standard two-loop perturbation theory ($+$) are off-scale
   above the graph.    For clarity, error bars are shown 
   only for the ``improved'' four-loop estimates.
   (b) Same quantity plotted versus $1/\beta^3$.
   The steeper solid line is the fit $\kappa_0 + \kappa_3/\beta^3$
   to the standard 4-loop estimates ($\protect\fancyplus$)
   for $\beta \ge 2.30$. The flatter solid line
   is the fit $\kappa'_0 + \kappa'_3 \beta^{-3}$ 
   to the ``improved'' 4-loop estimates ($\bigcirc$)
   for $\beta \ge 2.60$.
}
\label{fig_su3_scal_chif_ov_xif2}
  \end{figure}
\clearpage

%
% Figure 20: AS xi_A
%

\begin{figure}[p]
\vspace*{-0.5cm} \hspace*{-0cm}
\begin{center}
\epsfxsize = 4.3in
\leavevmode\epsffile{su3_scal_xia.ps} \\
\vspace*{-5mm}
\epsfxsize = 4.3in
\leavevmode\epsffile{su3_scal_xia_beta-2.ps}
\end{center}
\vspace*{-20mm}
\caption{
(a)
   $\xi^{(2nd)}_{A,\infty,estimate\,(\infty,\infty,\infty)}/
    \xi^{(exp)}_{A,\infty,theor} $
   versus $\beta$.
   Error bars are one standard deviation (statistical error only).
   There are four versions of $\xi^{(exp)}_{A,\infty,theor}$:
   standard perturbation theory in $1/\beta$ gives points
   $+$ (2-loop) and $\times$ (3-loop);
   ``improved'' perturbation theory in $1-E$ gives points
   $\Box$ (2-loop) and $\Diamond$ (3-loop). 
   For clarity, error bars are shown only for the ``improved''
   three-loop estimates.
   (b) Same ratio plotted versus $1/\beta^2$.
   The downward-tilting solid line is the fit $\kappa_0 + \kappa_2/\beta^2$
   to the standard 3-loop estimates ($\times$)
   for $3.15 \ge \beta \ge 2.40$. The upward-tilting solid line
   is the fit $\kappa'_0 + \kappa'_2/\beta^2$ 
   to the ``improved'' 3-loop estimates ($\Diamond$)
   for $\beta \ge 2.925$.}
\label{fig_su3_scal_xia}
\end{figure}

\clearpage

%
% Figure 21: AS chi_A
%

\begin{figure}
\vspace*{0cm} \hspace*{-0cm}
\begin{center}
\epsfxsize = 4.3in
\leavevmode\epsffile{su3_scal_chia.ps} \\
\vspace*{-5mm}
\epsfxsize = 4.3in
\leavevmode\epsffile{su3_scal_chia_beta-2.ps}
\end{center}
\vspace*{-20mm}
\caption{
(a)
   $[\chi_{A,\infty,estimate\,(\infty,\infty,0.90)}] /
    [\chi_{A,\infty,theor}\hbox{ without the prefactor }
    \widetilde{C}_{\chi_A}]$
   versus $\beta$.
   Error bars are one standard deviation (statistical error only).
   There are four versions of $\chi_{A,\infty,theor}$:
   standard perturbation theory in $1/\beta$ gives points
   $+$ (2-loop) and $\times$ (3-loop);
   ``improved'' perturbation theory in $1-E$ gives points
   $\Box$ (2-loop) and $\Diamond$ (3-loop).
   For clarity, error bars are shown only for the ``improved''
   three-loop estimates.
   (b) Same ratio plotted versus $1/\beta^2$.
   The flatter solid line is the fit $\kappa_0 + \kappa_2/\beta^2$
   to the standard 3-loop estimates ($\times$)
   for $\beta \ge 2.25$. The steeper solid line
   is the fit $\kappa'_0 + \kappa'_2/\beta^2$
   to the ``improved'' 3-loop estimates ($\Diamond$)
   for $\beta \ge 2.55$.
}
\label{fig_su3_scal_chia}
\end{figure}

\clearpage

%
% Figure 22: AS chi_A/xi_A^2
%

\begin{figure}[p]
\vspace*{-0.5cm} \hspace*{-0cm}
\begin{center}
\epsfxsize = 3.5in
\leavevmode\epsffile{su3_scal_chia_ov_xia2.ps} \\
\vspace*{-5mm}
\epsfxsize = 3.5in
\leavevmode\epsffile{su3_scal_chia_ov_xia2_beta-2.ps}
\end{center}
\vspace*{-20mm}
\caption{
(a)
   $[(\chi_A/\xi^{(2nd)^2}_A)_{\infty,estimate(\infty,\infty,\infty)}] \,
    \times \, [\widetilde{C}^{\protect\B}_{\xi^{(exp)}_A}]^2 \, /$ \hfill\break
   $[(\chi_A/\xi^{(2nd)^2}_A)_{\infty,theor}\hbox{ without prefactors }
    \widetilde{C}_{\chi_A} \hbox{ and } \widetilde{C}_{\xi^{(2nd)}_A}]$
   versus $\beta$.
   Error bars are one standard deviation (statistical error only).
   There are four versions of $(\chi_A/\xi_A^2)_{\infty,theor}$:
   standard perturbation theory in $1/\beta$ gives points
   $+$ (2-loop) and $\times$ (3-loop);
   ``improved'' perturbation theory in $1-E$ gives points
   $\Box$ (2-loop) and $\Diamond$ (3-loop).
   (b) Same quantity plotted 
   versus $1/\beta^2$.
   The steeper solid line is the fit $\kappa_0 + \kappa_2/\beta^2$
   to the standard 3-loop estimates ($\times$)
   for $\beta \ge 3.075$. The flatter solid line
   is the fit $\kappa'_0 + \kappa'_2/\beta^2$
   to the ``improved'' 3-loop estimates ($\Diamond$)
   for $\beta \ge 3.075$.
}

\label{fig_su3_scal_chia_ov_xia2}
  \end{figure}
\clearpage

%
% Figure 23: dynamic FSS
%

\begin{figure}
\vspace*{0cm} \hspace*{-0cm}
\begin{center}
\epsfxsize = 0.9\textwidth
\leavevmode\epsffile{su3_dynamic_fss.ps}
%%\quad \vspace{5cm}  %% TEMPORARY UNTIL FILE IS THERE
\end{center}
\vspace*{-2cm}
\caption{
   Dynamic finite-size-scaling plot of
   $\tau_{int,{\cal M}_F^2}/\xi_F^{(2nd)}(L)^{z_{int,{\cal M}_F^2}}$
   versus $\xi_F^{(2nd)}(L)/L$.
   Symbols indicate $L=16$ ($\protect\fancycross$),
   32 ($+$), 64 ($\times$), 128 ($\Box$) and 256 ($\Diamond$).
   Here $z_{int,{\cal M}_F^2} = 0.45$.
   We have included in the plot only those points satisfying 
   $\xi_F(L) \ge 8$.
}
\label{fig_su3_dynamic_fss}
\end{figure}

\clearpage
%
% Figure 24: dynamic FSS
%

\begin{figure}
\vspace*{0cm} \hspace*{-0cm}
\begin{center}
\epsfxsize = 0.9\textwidth
\leavevmode\epsffile{su3_dynamic_fss_adj.ps}
%%\quad \vspace{5cm}  %% TEMPORARY UNTIL FILE IS THERE
\end{center}
\vspace*{-2cm}
\caption{
   Dynamic finite-size-scaling plot of
   $\tau_{int,{\cal M}_A^2}/\xi_F^{(2nd)}(L)^{z_{int,{\cal M}_A^2}}$
   versus $\xi_F^{(2nd)}(L)/L$.
   Symbols indicate $L=16$ ($\protect\fancycross$),
   32 ($+$), 64 ($\times$), 128 ($\Box$) and 256 ($\Diamond$).
   Here $z_{int,{\cal M}_A^2} = 0.45$.
   We have included in the plot only those points satisfying
   $\xi_F(L) \ge 8$.
}
\label{fig_su3_dynamic_fss_adj}
\end{figure}

\clearpage

\end{document}